\documentclass[traditabstract]{aa}

\usepackage{graphicx}
\usepackage{natbib}
\usepackage{colortbl}
\usepackage{color}
\usepackage{array}
\usepackage{amsfonts,amssymb,amsmath}
\usepackage{multirow}
\usepackage{txfonts}
\usepackage{lscape}
\usepackage{aalongtable}

\newcommand{\dtm}{$\mathcal{{DT\!\!M}}$}

\newcommand{\hi}{\mbox{H\,{\sc i}}}

\newcommand{\mgii}{\mbox{Mg\,{\sc ii}}} 
\newcommand{\siii}{\mbox{Si\,{\sc ii}}}

\newcommand{\suii}{\mbox{S\,{\sc ii}}}

\newcommand{\znii}{\mbox{Zn\,{\sc ii}}}

\newcommand{\ci}{\mbox{C\,{\sc i}}}  
\newcommand{\cii}{\mbox{C\,{\sc ii}}}

\newcommand{\nhi}{$N(\mbox{\hi{}})$}
\newcommand{\lognhi}{$\log N(\mbox{\hi{}})$}

\begin{document}

\title{Dust-depletion sequences in damped Lyman-$\alpha$ absorbers:}
\subtitle{A unified picture from low-metallicity systems to the Galaxy\thanks{Based on observations carried out at the European Organisation for Astronomical Research in the Southern Hemisphere under ESO programmes 065.P-0038, 065.O-0063, 066.A-0624, 067.A-0078, and 068.A-0600.}}

 \author{A. De Cia \inst{1,2}, C. Ledoux \inst{3}, L. Mattsson \inst{4}, P. Petitjean \inst{5}, R. Srianand \inst{6}, I. Gavignaud \inst{7} and E.~B. Jenkins \inst{8}} 
 
\institute{
 European Southern Observatory, Karl-Schwarzschild Str. 2, 85748 Garching bei München, Germany\\
 \email{adecia@eso.org}
\and
Department of Particle Physics and Astrophysics, The Weizmann Institute of Science, Rehovot 76100, Israel 
\and
European Southern Observatory, Alonso de C\'ordova 3107, Casilla 19001, Vitacura, Santiago 19, Chile
\and 
Nordita, KTH Royal Institute of Technology \& Stockholm University, Roslagstullsbacken 23, SE-106 91, Stockholm, Sweden
\and
IAP, CNRS and Universit\'e Paris 6, 98bis Boulevard Arago, 75014 Paris, France
\and
IUCAA, Post Bag 4, Ganesh Khind, Pune 411 007, India
\and
Departamento de Ciencias Fisicas, Facultad de Ingenieria, Universidad Andres Bello, Santiago, Chile
\and
Princeton University Observatory, Princeton, NJ 08544-1001, USA
}
   
   \date{Received Month dd, 2015; accepted Month dd, 2015}

    \abstract{We study metal depletion due to dust in the interstellar medium (ISM) to infer the properties of dust grains and characterize the metal and dust content of galaxies down to low metallicity and intermediate redshift $z$. We provide metal column densities and abundances of a sample of 70 damped Lyman-$\alpha$ absorbers (DLAs) towards quasars, observed at high spectral resolution with the Very Large Telescope (VLT) Ultraviolet and Visual Echelle Spectrograph (UVES). This is the largest sample of phosphorus abundances measured in DLAs so far. We use literature measurements for Galactic clouds to cover the high-metallicity end. We discover tight (scatter $\lesssim0.2$~dex) correlations between [Zn/Fe] and the observed relative abundances from dust depletion. This implies that grain growth in the ISM is an important process of dust production. These sequences are continuous in [Zn/Fe] from dust-free to dusty DLAs, and to Galactic clouds, suggesting that the availability of refractory metals in the ISM is crucial for dust production, regardless of the star formation history. We observe [S/Zn] up to $\sim0.25$~dex in DLAs, which is broadly consistent with Galactic stellar abundances. Furthermore, we find a good agreement between the nucleosynthetic pattern of Galactic halo stars and our observations of the least dusty DLAs. This supports recent star formation in low-metallicity DLAs. The derived depletions of Zn, O, P, S, Si, Mg, Mn, Cr, and Fe correlate with [Zn/Fe], with steeper slopes for more refractory elements. P is mostly not affected by dust depletion. We present canonical depletion patterns to be used as reference in future studies of relative abundances and depletion. We derive the total (dust-corrected) metallicity, typically $-2\lesssim$[$M$/H]$_{\rm tot}\lesssim0$ for DLAs, and scattered around solar metallicity for the Galactic ISM. The dust-to-metal ratio (\dtm{}) increases with metallicity, again supporting the importance of grain growth for dust production. The dust extinction $A_V$ derived from the depletion is typically $<0.2$~mag in DLAs. Finally, we derive elemental abundances in dust, which is key to understanding the dust composition and its evolution. We observe similar abundances of Mg, Si, and Fe in dust; this suggests that grain species such as pyroxenes and iron oxides are more important than olivine, but this needs to be confirmed by more detailed analysis. Overall, we characterize dust depletion, nucleosynthesis, and dust-corrected metallicity in DLAs, providing a unified picture from low-metallicity systems to the Galactic ISM.}


\titlerunning{Dust-depletion sequences in DLAs and the Galaxy}

\authorrunning{De Cia et~al.}

   \maketitle

\section{Introduction}

The study of chemical abundances in Galactic and extragalactic environments, down to the lowest metallicities, is key to understanding the chemical evolution of galaxies, the environmental effect on nucleosynthesis, the properties of cosmic dust, its role in the formation of molecules, and its importance during planet and star formation. Through absorption-line spectroscopy, the column densities of different ions in the gas phase can be determined, thus providing the (relative) abundances in the interstellar medium (ISM) of a galaxy. These gas-phase relative abundances can vary depending on the star formation and nucleosynthesis history, or if part of the metals are locked into dust grains, i.e. dust depletion. Disentangling these two effects requires a better understanding of nucleosynthesis and dust depletion at low metallicities. The study of dust depletion can in turn provide new insights into the origin and composition of dust grains \citep[e.g.][]{Savage96,Jenkins13,Jenkins14}.

In the Galaxy, refractory elements (i.e. elements with the highest condensation temperatures) are typically more strongly depleted into dust grains, indicating that dust is indeed responsible for lowering the observed abundances, down to $-2.5$ dex in the most extreme cases \citep[see][for a review]{Savage96}. Having recognized that the amount of depletion depends on the environment, \citet{Savage96} identified similarities in the depletion \textit{patterns} of four types of clouds in the Galactic ISM, i.e. those in the cool disk, warm disk, warm halo, and in a combination of the disk and halo. Metals tend to be more heavily depleted in the denser and cooler disk environment than in the halo, where dust destruction could also be more effective. These are fixed patterns of abundances that require a fixed composition of the dust grains for a specific environment. However, these reference depletion patterns were derived from only 13 absorbing systems. 

More recently, \citet{Jenkins09} showed that there is a wide range of abundances of several elements in a sample of 243 Galactic lines of sights and that the observed abundances of 17 elements tightly correlate with each other. This is due to an increasing amount of dust depletion in the ISM, which can be described by a single parameter per system, the depletion strength $F_*$. The slopes of these linear depletion \textit{sequences} observed between the abundances and $F_*$ are different from element to element; the steeper slope the stronger is the tendency to be depleted into dust grains. This study offered new insight and knowledge of dust depletion. However, only Galactic lines of sight were considered, while low-metallicity and high-redshift systems are potentially fundamental pieces of the puzzle.

Damped Lyman $\alpha$ (DLA) systems towards bright sources such as quasars (QSOs) and long-duration ($t>2s$) gamma-ray bursts (GRBs) can be observed and studied in detail in \textit{absorption} out to high redshift \citep[e.g. for GRB/,130606A at $z\sim6$,][]{Hartoog15}. The largest reservoirs of neutral gas in the Universe \citep[e.g.][]{SanchezRamirez16}, DLAs have column densities of neutral hydrogen $\log N(\mbox{\hi{}}) > 20.3$ \citep[e.g.][]{Wolfe05}; DLAs typically span from less then 1\% of solar to solar metallicity and evolve with redshift \citep[e.g.][]{Ledoux02,Prochaska03,Rafelski12}. Studying the galaxy counterparts of QSO-DLAs in emission is difficult because of the bright QSO in the background. Nevertheless, there is increasing evidence that they are associated with typically faint and low-mass galaxies \citep[e.g.][and references therein]{Djorgovski96,Moller98,Moller04,Fynbo10,Fynbo11,Noterdaeme12,Schulze12,Noterdaeme12,Krogager13,Christensen14}, whereas DLAs at the higher end of their metallicity distribution have more massive counterparts \citep[e.g. $M\sim10^{10}M_\odot$][]{Fynbo13,Ma15}. Therefore, DLAs are  a unique laboratory for extending our understanding of dust depletion and nucleosynthesis in extragalactic sources, at low metallicity and high redshift.

While \citet{Jenkins09} showed the homogeneity of dust depletion in the Galaxy, the situation is still very unclear for extragalactic sources. There are several difficulties while studying dust depletion in DLAs. The first one is that while the gas in the Galactic ISM can be assumed to have overall solar abundance, the total metallicity of DLAs can be very low and it is not known a priori. In turn, differences in the observed abundances could be due to true differences in the total metallicities but also to differences in dust depletion and nucleosynthesis. Moreover, at the low metallicities that are typical for DLAs the effects of nucleosynthesis and dust depletion are less understood. A possible way to disentangle these effects is to investigate the abundances of elements with different nucleosynthetic and refractory properties. Overcoming these difficulties thus requires large samples to provide significant results. A number of studies on QSO-DLAs have shown evidence for some correlations in the relative abundances due to dust depletion \citep[e.g.][]{Ledoux02,Prochaska02,Vladilo02,Dessauges-Zavadsky06,Rodriguez06,Meiring06,Som13}. In particular, the relative abundance of zinc with respect to iron [Zn/Fe] correlates with [Si/Fe], [Si/Ti] (but with a small number of Ti measurements), metal column densities, and metallicity. Thus, [Zn/Fe] has been widely used as an indicator of dust depletion, and is often referred to as ``depletion factor''. However, due to the small number of available observables, these studies are often limited in their capability of disentangling the extent of dust depletion from nucleosynthetic effects. 

In \citet{DeCia13} we assumed that depletion sequences of Zn, Fe, and Si exist in DLAs as well as in the Galaxy \citep[as observed by][]{Jenkins09}. In this paper we expand our analysis of relative abundances to a large sample of DLAs with wide range of metal-line absorption (column density measurements of nine different metals, in a total of 70 DLAs) observed with the Very Large Telescope (VLT) Ultraviolet and Visual Echelle Spectrograph (UVES) and include the Galactic lines of sight studied by \citet{Jenkins09}. Our aim here is to answer the question of the existence of depletion sequences from DLAs to the Galactic clouds, and to characterize the absorbing systems in terms of their metal and dust content. This has strong implications on the origin of dust and its composition.

We describe our observations in Section \ref{observations} and present the properties of our sample, along with the observed relative abundances in Sect. \ref{sample}. In Sect. \ref{sequences} we characterize the correlations among the observed relative abundances, and then convert these abundances to depletions, assuming, first, a trend of the depletion of Zn and, second, over- or underabundances from nucleosynthesis processes. We discuss our results in Sect. \ref{discussion}, present new depletion patterns, and derive the dust-corrected abundances and metallicity. We further derive elemental abundances in dust (which are crucial to study the dust composition), the dust-to-metal ratio, and the expected dust extinction based on the depletion. We finally summarize and conclude in Section \ref{conclusions}.

Throughout the paper we use a linear unit for the column densities $N$ of ions cm$^{-2}$. We refer to relative abundances of elements $X$ and $Y$ as $\left[X/Y\right] \equiv \log{\frac{N(X)}{N(Y)}} - \log{\frac{N(X)_\odot}{N(Y)_\odot}}$, where reference solar abundances are reported in Table \ref{tab solar}. We report $1\,\sigma$ and $3\,\sigma$ significance levels for the quoted errors and limits, respectively, unless otherwise stated.

\begin{table}
\begin{minipage}[t]{\columnwidth}
\caption{Adopted solar abundances.}
\label{tab solar}
\centering
\renewcommand{\footnoterule}{}
\begin{tabular}{ccc}
\hline\hline
\rule[-0.2cm]{0mm}{0.8cm}
Element & $\log ({\rm X}/{\rm H})_\odot +12$\footnote{Abundances of \citet{Asplund09}, following the recommendations of \citet{Lodders09} on whether to rely on photospheric (s), meteoritic (m) abundances, or their average (a).}\\
\hline
H         & 12.0   & s \\
O         & 8.69   & s \\ 
Mg       & 7.565 & a \\
Si        & 7.51    & m \\
P         & 5.42    & a \\
S         & 7.135  & a \\
Cr       & 5.64    & a \\
Mn      & 5.48    & m \\
Fe       & 7.475  & a \\
Zn      & 4.63    & m \\

\hline \hline
\end{tabular}
\end{minipage}
\end{table}

\section{Observations and data reduction}
\label{observations}

   \begin{figure*}
   \centering
   \includegraphics[width=185mm,angle=0]{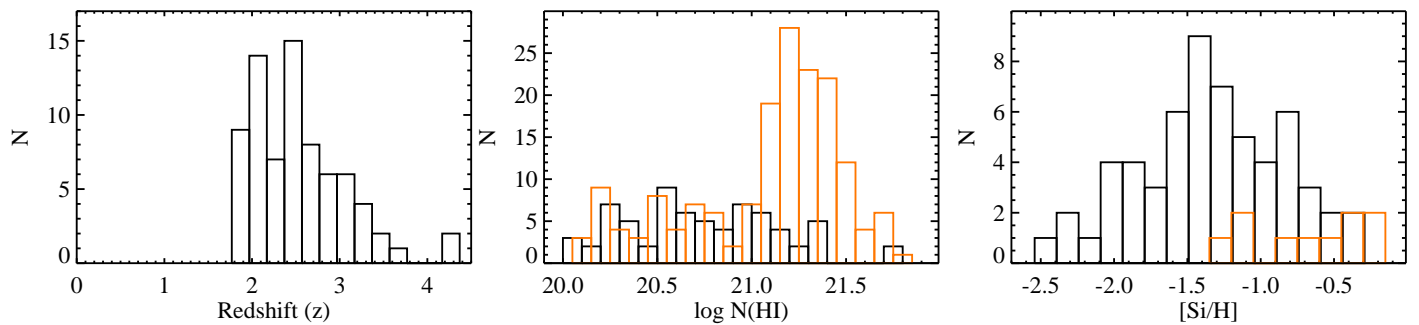}
   \caption{Distribution of redshift (left), \nhi{} (centre) and observed [Si/H] (right) for our DLA sample (black) and the Galactic absorbers (orange). All Galactic absorbers ($z=0$) are not shown in the redshift histogram for a clearer visualization.}
              \label{fig histo}
    \end{figure*}

We initially built up
a sample of 55 DLA (\lognhi{}$\ge 20.3$) and 13 sub-DLA
($19.7\le\log N($H\,{\sc i}$)<20.3$) absorbers at $z_{\rm abs}>1.7$. Most
of the absorbers were originally selected from the statistical DLA sample of
the Large Bright QSO Survey \citep[LBQS;][]{Wolfe95}. The
observations took place at the VLT between 2000 and 2003 in the course of
a large and systematic survey for molecular hydrogen in high-redshift DLA
systems \citep{Petitjean00,Ledoux03,Srianand05}, leading to a
large collection of high spectral resolution, high-quality UVES spectra of
quasars. A few additional high-quality UVES spectra were also retrieved
from the ESO public archive, either to complete our observations of the
same systems or to increase the size of this DLA abundance database. Our final sample is composed of 70 absorbers with  \lognhi{}$\ge 20.0$, 61 of which are DLAs. A system with $\log N($H\,{\sc i}$) = 19.95\pm0.07$ is included too. We exclude the systems with lower \lognhi{} to avoid ionization effects. In DLAs the singly ionized state of most elements is dominant \citep[e.g.][]{Wolfe05,Viegas95, Peroux07}, while for lower \hi{} column densities metals are less shielded from ionizing radiation \citep{Howk99,DeCia12,Vreeswijk13}. Our sample is the same as in \citet{Ledoux06b}. 
A few (4) absorbing systems are located with $\sim5000$ km~s$^{-1}$ from the QSO. We do not avoid these proximate DLAs (PDLA) because we do not observe any differences between DLAs and PDLAs and we intend to study the dust-depletion properties of the overall DLA population. For simplicity, we refer to our sample as a DLA sample. The metallicities and oxygen abundances of part of this sample were published in \citet{Ledoux06b} and \citet{Petitjean08}, respectively.

The data were reduced using a modified version of the UVES
pipeline \citep{Ballester00}, which is
available as a context of MIDAS, the ESO data reduction system. The
main characteristics of the UVES pipeline are to perform a precise inter-order
background subtraction for science frames and master flat-fields, as well as
an optimal 2-D extraction of the object signal rejecting cosmic ray impacts and subtracting the sky spectrum simultaneously.

We worked with this modified data reduction pipeline in semi-automatic mode
using a script handling the organization of the raw data files. The
pipeline products were checked step by step. The same script also calls
dedicated sub-routines to convert
the wavelength scale of the spectra reduced by the pipeline to vacuum-heliocentric values and to co-add
the individual 1-D spectra and their overlapping parts with
appropriate scaling, weighting, and kappa-sigma clipping. The associated
variance spectra were scaled and weighted in the same manner. During this
merging process, the spectra were rebinned to a common (thus constant)
wavelength step (typically 0.038 \AA\ pix$^{-1}$) using the smallest mean step
of the individual spectra originating from different instrument settings. This
yielded, after combination, signal-to-noise ratios in the ranges 14--33
and 54--10 in the blue and  red spectroscopic arm of UVES, respectively.

\subsection{Abundance measurements and sample characteristics}
\label{sample}

The absorption line analysis was homogeneously performed using standard Voigt-profile fitting techniques, using the \textsc{MIDAS/FITLYMAN} software \citep{Fontana95}. We adopted the oscillator strengths compiled by \citet{Morton03}, with the exception of \znii{} and \suii{}, for which we adopted the recently reassessed oscillator strengths of \citet{Kisielius15} and \citet{Kisielius14}, respectively. This has the effect of lowering the Zn column densities by 0.1~dex, and increasing the S column density by 0.04~dex with respect to the previous values.

We identifed the absorption lines by making sure that their velocity profiles are consistent among different transitions. We decomposed the absorption-line profiles into different velocity components having their own $z$, $N,$ and broadening $b$ values. We used different transitions from the same ion when available. All lines in the system were fitted simultaneously tying together $z$ and $b$ values (turbulent broadening) of all species. The decomposition of blended lines was possible when additional lines of the same species were available. The detection of lines located in the Ly-$\alpha$ forest (bluer than Ly-$\alpha$) is confirmed by their kinematical similarities with the uncontaminated absorption lines detected outside of the forest. This is possible thanks to the high spectral resolution and signal to noise achieved in our spectra. The multi-component structure of the line-profile fit is determined from lines outside the forest.

By definition, DLAs have a high column density of neutral hydrogen. This neutral gas acts as a shield against ionization and excitation of the metals \citep[e.g.][]{Vreeswijk13}, and several metals (C, Si, Fe, Zn, etc.) are mostly in their singly ionized state. Thus, no ionization corrections or thermal equilibrium assumptions are needed to derive metal abundances. A unique characteristic of this large dataset is that it samples equally well both the low and high ends of the DLA metallicity distribution, from [Si/H$]\approx -2.6$ up to about half of solar. Figure \ref{fig histo} shows the redshift, \nhi,{} and [Si/H] distribution of our DLA sample.

We provide ionic column densities for each velocity component of the line profiles (Table \ref{tab columns vel}) and the average abundances along the lines of sight (Table \ref{tab abundances}) for all the systems in our sample. For the sake of compactness, when indicating a QSO name in figures, we append letters to the QSO names when there are multiple absorption systems at different redshifts along the same line of sight. The Voigt-profile fits of the velocity profiles of selected low-ionization transition lines from the DLA systems are shown in Appendix \ref{sect velocity profile}. In case of blended or nearby lines, the combined Voigt-profile fit of both transitions is represented. The column densities are measured through this work, except for the systems at z$_{\rm abs} = 3.390$ towards Q\,0000$-$263 \citep{Molaro01}, z$_{\rm abs} =  3.025$ and $2.087$ towards Q\,0347$-$383 and Q\,1444$+$014, respectively \citep{Ledoux03}, z$_{\rm abs} = 1.962$ towards Q\,0551$-$366 \citep{Ledoux02}, and z$_{\rm abs} = 4.224$ towards Q\,1441$+$276 \citep{Ledoux06}. We use the total column densities (integrated along the line profiles) for the analysis on the dust depletion.

\subsection{Comparison of metal column densities with the literature}
\label{sect comp lit}

A number of DLAs in this paper also have published metal column densities in the literature. A full comparison of the dataset is beyond the scope of this paper. Nevertheless, we compare a subset of DLAs with the literature measurements, as a sanity check. From our DLA sample we select the first ten (alphabetically) DLAs and QSOs that have published column densities in the literature, and excluding those with identical N values, published in the past by members of our team. Among these ten DLAs\footnote{Namely, 
Q\,0013$-$004 ($z_{\rm abs}=1.973$), Q\,0112$-$306 ($z_{\rm abs}=2.418$), Q\,0112$+$030 ($z_{\rm abs}=2.423$), Q\,0528$+$250 ($z_{\rm abs}=2.811$), Q\,0913$+$072 ($z_{\rm abs}=2.618$), Q\,1111$-$152 ($z_{\rm abs}=3.266$), Q\,1157$+$014 ($z_{\rm abs}=1.944$), Q\,1209$+$093 ($z_{\rm abs}=2.584$), Q\,1232$+$082 ($z_{\rm abs}=3.266$), Q\,1337$+$113 ($z_{\rm abs}=2.796$).}, we find that the metal column densities are in a good agreement with what has previously been published (within $\leq0.1$~dex from our measurements, and $\leq0.06$~dex for the vast majority).

There are three exceptions to the above, which we further investigated and concluded that it is likely that they were inaccurately estimated in the literature. \textit{a)} The first is the Zn column density in the system at $z_{\rm abs}=2.811$ towards Q\,0528$-$250, for which our measurements of $N$(Zn) agrees with \citet{Lu96}, but differs by 0.16~dex from \citet{Centurion03}. In the latter work, however, the Voigt-profile fit of the lines in this system is not shown. \textit{b)} We find a discrepancy for Q\,0013$-$004 ($z_{\rm abs}=1.973$) in column densities of 0.1--0.2~dex with respect to \citet{Petitjean02}, but they did not consider the velocity components at $\sim -300$~km~s$^{-1}$. We find a 0.6~dex difference for Mn, but no figure is shown in the paper. \textit{c)} The metal column densities measured by \citet{Srianand00} for Q\,1232$+$082 ($z_{\rm abs}=2.338$) are higher than our measurements by 0.05, 0.12, and 0.17~dex for Si, Mg, and Fe, respectively, which is possibly caused by a different decomposition of the line profiles. The different measurements are however consistent within their uncertainties. 

Phosphorus measurements are rare, because they are difficult to obtain (lines in the Ly-$\alpha$ forest). We compare some of our measurements with the literature compilation of \citet{Dessauges-Zavadsky04} and \citet{Dessauges-Zavadsky06}. We can compare $N$(P) for seven systems.\footnote{Q\,0100$+$130 ($z_{\rm abs}=2.309$), Q\,1331$+$170 ($z_{\rm abs}=1.776$), Q\,0450$-$131 ($z_{\rm abs}=2.067$), Q\,0841$+$129 ($z_{\rm abs}=2.375$ and $z_{\rm abs}=2.476$), Q\,1157$+$140 ($z_{\rm abs}=1.944$), and Q\,2230$+$025 ($z_{\rm abs}=1.864$).} Among these, four show a very good agreement in the $N$(P) measurements, within $1\sigma$ (differences of $0$, $0.02$, $0.06$ and $0.12$~dex), one system shows a difference of $0.14$~dex (consistent within $1.15\sigma$), and for two systems the $N$(P) measurements of \citet{Dessauges-Zavadsky06} are $0.21$ and $0.32$~dex higher than our values, but within $2.44\sigma$ and $1.97\sigma$, respectively.

\section{Sequences of observed relative abundances}
\label{sequences}
   \begin{figure}[t!]
   \centering
   \includegraphics[width=90mm,angle=0]{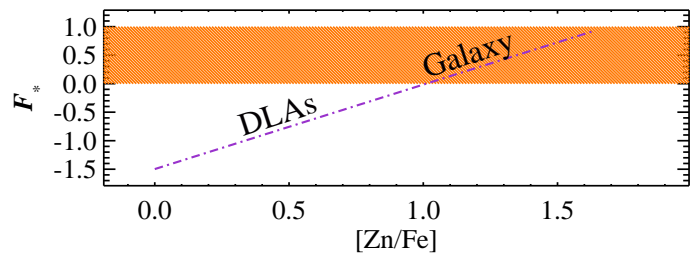}
   \caption{Relation between the depletion strength factor $F_*$ and [Zn/Fe], as derived from Eq. 10 of \citet{Jenkins09}: $F_* = 1.48\times {\rm[Zn/Fe]} - 1.50$. We highlight the region between $F_*=0$ and $F_*=1$, values, which were defined to represent the lowest and highest depletion states in the Galaxy, respectively.}
              \label{fig f* znfe}
    \end{figure}
    
   \begin{figure*}
   \centering
   \includegraphics[width=190mm,angle=0]{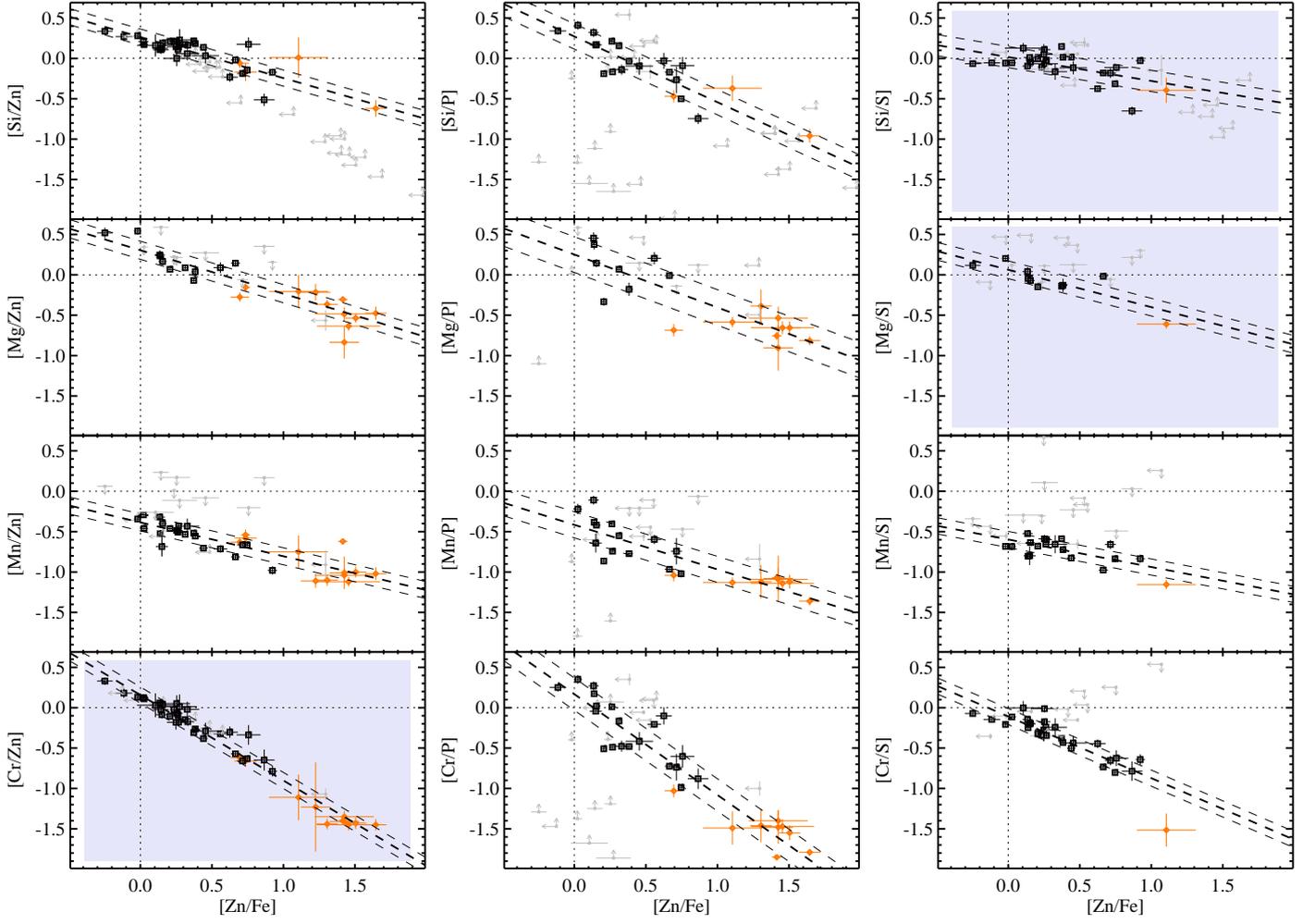}
   \caption{Relative abundances of an element X with respect to Zn (left column), P (central), and S (right), against the [Zn/Fe]. The black squares are for DLAs and the orange diamonds are for the Galactic lines of sight. The abundances are summed among individual velocity components; see Table \ref{tab abundances}. The dashed curves show the linear fit to the data and the intrinsic scatter $\sigma_{\rm int}$. The sequences become steeper for elements that are more strongly depleted in dust grains. The panels with shaded areas show relative abundances of elements that share similar nucleosynthetic history.}
              \label{fig znfe}
    \end{figure*}         
    \begin{figure*}
   \centering
   \includegraphics[width=190mm,angle=0]{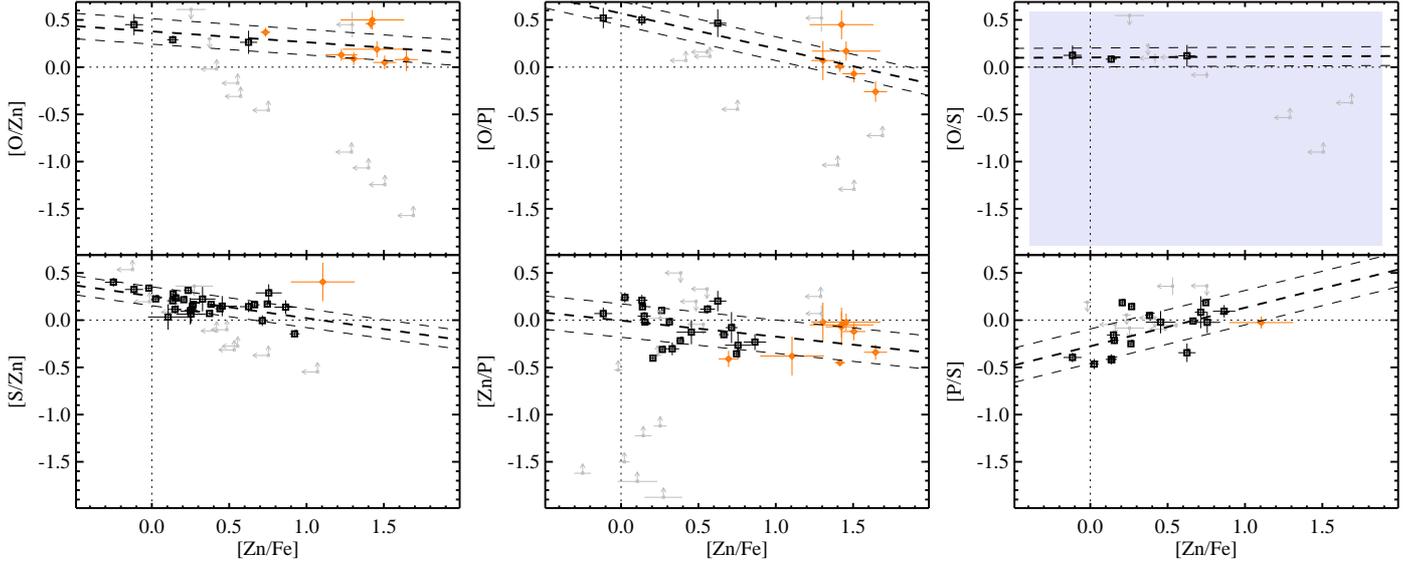}
   \caption{Same as Fig. \ref{fig znfe}, but for non-refractory elements.}
              \label{fig znfe non depleted}
    \end{figure*}
Since DLAs have a wide range of metallicities, typically subsolar, it is not straightforward to compare the abundances of different absorbers on a common scale. On the other hand, relative abundances of one element $X$ with respect to another $Y$, [$X/Y$] can be compared among each other,  and with the solar $[X/Y]=0$. The deviation from solar are mainly due to either nucleosynthesis effects or dust depletion.  

We aim to answer the question of the existence and properties of dust-depletion sequences in DLAs. As an indicator of the global amount of dust in the absorbing systems, we used the observed [Zn/Fe], which is often referred to as the depletion factor. This is a purely observational quantity, as opposed to the depletion strength factor $F_*$, which was defined to characterize the overall intensity of depletion for each individual system and derived by the simultaneous fitting of the depletion sequences (correlations among abundances) observed in the Galaxy by \citet{Jenkins09}. Figure \ref{fig f* znfe} shows the relation between $F_*$ and [Zn/Fe], as formally derived from Eq. 10 of \citet{Jenkins09}. The relation between $F_*$ and [Zn/Fe] can be derived by combining the relation between $F_*$ and [Zn/H] and the relation between $F_*$ and [Fe/H], which are reported in \cite{Jenkins09}. In this paper we rely on observed quantities such as [Zn/Fe], and we only show a comparison with $F_*$ for completeness. A negative $F_*$ is expected for most DLAs; see also \citet{Wiseman16}. 
 
[Zn/Fe] is a reliable tracer of the dust content in a galaxy \citep[see][]{DeCia13}, unless amorphous carbonaceous grains are the dominant dust species. The use of [Zn/Fe] as a tracer of dust is further supported by its relation to other dust tracers, such as [Si/Ti], [Mg/S], and [Si/S]. In Sect. \ref{sect znfe} we further discuss the use of [Zn/Fe] as a dust indicator and the possible small intrinsic scatter in [Zn/Fe] due to Zn and Fe nucleosynthesis. 

We include the Galactic absorbers from \citet{Jenkins09} in our analysis for reference, completeness, and to possibly extend the range of metallicity
and depletion strength of our sample. We consider the 169 systems with $\log N (\mbox{\hi{}})\geqslant20$. 

With the aim to increase the visibility of the effects of dust depletion, we consider the relative abundance of an element $X$ with respect to a non-refractory element, such as Zn, S, or P. Figure \ref{fig znfe} shows the relative abundances [$X$/Zn] (left column), [$X$/P] (central column), and [$X$/S] (right column) for elements that are strongly depleted into dust grains. Figure \ref{fig znfe non depleted} shows the same quantities, but for elements that are not strongly depleted. Black squares refer to our DLA sample and orange diamonds to the Galactic absorbers from \citet{Jenkins09}. The rows are arranged from top to bottom in a rough order of how much of an element is depleted into dust grains. Lower and upper limits on the relative abundances are shown in grey in Figs. \ref{fig znfe} and \ref{fig znfe non depleted}, but they are excluded from further analysis, as discussed in Appendix \ref{sect limits}.

Sequences of relative abundances, from DLAs to the Galaxy, appear evident directly from the comparison of the observed data (Fig. \ref{fig znfe}). We fit the sequences with a linear least-squares approximation in one dimension ($y = a + b x$) that considers errors on both $x$ and $y$ data ($\sigma_x$ and $\sigma_y$) and includes intrinsic scatter, using the \textsc{IDL} routine \textsc{MPFITEXY} \citep{Williams10}. \textsc{MPFITEXY} utilizes the \textsc{MPFIT} package \citep{Markwardt09}. Each data point is weighted as $1/\sqrt{\sigma_x^2 + b^2\sigma_y^2 + \sigma_{\rm int}^2}$, where $\sigma_{\rm int}$ is the intrinsic scatter of the data around the model \citep[“Nukers’ Estimate”;][]{Tremaine02}. The value $\sigma_{\rm int}$ is automatically scaled to produce a reduced $\chi^2_\nu\sim1$. We use an initial guess for $\sigma_{\rm int}$ of 0.1~dex. In this way the variables $x$ and $y$ are treated symmetrically. Table \ref{tab coeff seq} lists the coefficients resulting from the linear fit to the observed sequences of relative abundances,
\begin{equation}
[X/Y] = A_1 + B_1 \times {\rm [Zn/Fe]} \mbox{.}
\label{eq seq rel ab}
\end{equation}
The values $\sigma_{\rm [X/Y]}$ are the estimates of the intrinsic scatter in the correlations and are reported in Table \ref{tab coeff seq}. The best fit and  ranges of intrinsic scatter are shown in Figs. \ref{fig znfe} and \ref{fig znfe non depleted}. The intrinsic scatters are $\lesssim0.2$~dex,and $\lesssim0.1$~dex for the relations with [$X$/Zn]. The errors on the relative abundances are typically smaller, suggesting that there is some level of intrinsic variation in the relative abundances, such as nucleosynthesis effects or peculiar abundances, up to $\lesssim0.2$~dex. See Appendix \ref{sect znfe} for a discussion on [Zn/Fe].   

We report the Pearson correlation coefficients for the sequences of relative abundances in Table \ref{tab coeff seq}. For the refractory elements (Fig. \ref{fig znfe}), most of the $r$ values are $\leq-0.83$, and down to $-0.98$, indicating very strong linear relations. Two exceptions are for [Si/S] and [Mn/S], for which $r=-0.60$ and $r=-0.69$, respectively, indicating moderately strong linear relations, however on less well-sampled datasets. For the non-refractory elements (Fig. \ref{fig znfe non depleted}), the $r$ values are lower, but this is expected given that dust depletion is much lower for such elements (i.e. flatter sequences of relative abundances).

We observe linear sequences of relative abundances for DLAs and the Galaxy, fitting the two samples  together. As a sanity check, we fit [$X$/Zn] versus [Zn/Fe] for DLAs and Galactic absorbers independently, for [Mg/Zn], [Mn/Zn], and [Cr/Zn], which have the most complete datasets for both samples. The independent fits give results that are very similar to those obtained when considering DLAs and Galactic clouds altogether. This also supports our approach of treating DLAs and Galactic absorbers together, i.e. as generic ISM clouds.

\begin{table}
\centering
\caption{Coefficients of the sequences of observed relative abundances shown in Figs. \ref{fig znfe} and \ref{fig znfe non depleted}, resulting from a linear fit of $[X/Y] = A_1 + B_1\times {\rm [Zn/Fe]}$. Degrees of freedom of the fit $\nu$, intrinsic scatter $\sigma_{[X/Y]}$ in the relation, and the Pearson correlation coefficients $r$ are reported.}
\begin{tabular}{@{} l | r r r r r}
\hline \hline
\rule[-0.2cm]{0mm}{0.8cm}
$[X/Y]$    &    \multicolumn{1}{c}{$A_1$}    &  \multicolumn{1}{c}{$B_1$}    &   \multicolumn{1}{c}{$\nu$} & \multicolumn{1}{c}{$\sigma_{[X/Y]}$} & \multicolumn{1}{c}{$r$}\\ 

\hline

 [O/Zn] & $ 0.38\pm 0.10$ & $-0.11\pm 0.09$ & $   9$ & $0.13$ & $-0.42 $ \\

 [S/Zn] & $ 0.25\pm 0.03$ & $-0.23\pm 0.07$ & $  27$ & $0.10$ & $-0.31 $ \\

[Si/Zn] & $ 0.26\pm 0.03$ & $-0.51\pm 0.06$ & $  34$ & $0.10$ & $-0.83 $ \\

[Mg/Zn] & $ 0.30\pm 0.04$ & $-0.54\pm 0.05$ & $  20$ & $0.11$ & $-0.92 $ \\

[Mn/Zn] & $-0.39\pm 0.03$ & $-0.42\pm 0.04$ & $  31$ & $0.11$ & $-0.87 $ \\

[Cr/Zn] & $ 0.16\pm 0.03$ & $-1.06\pm 0.04$ & $  43$ & $0.10$ & $-0.98 $ \\

\hline

 [Zn/P] & $-0.00\pm 0.06$ & $-0.17\pm 0.07$ & $  27$ & $0.18$ & $-0.37 $ \\

  [O/P] & $ 0.57\pm 0.10$ & $-0.37\pm 0.09$ & $   7$ & $0.13$ & $-0.78 $ \\

 [Si/P] & $ 0.27\pm 0.06$ & $-0.81\pm 0.10$ & $  19$ & $0.16$ & $-0.87 $ \\

 [Mg/P] & $ 0.25\pm 0.10$ & $-0.66\pm 0.11$ & $  15$ & $0.22$ & $-0.85 $ \\

 [Mn/P] & $-0.42\pm 0.06$ & $-0.55\pm 0.07$ & $  21$ & $0.16$ & $-0.86 $ \\

 [Cr/P] & $ 0.17\pm 0.07$ & $-1.25\pm 0.09$ & $  27$ & $0.20$ & $-0.94 $ \\

 [Fe/P] & $ 0.01\pm 0.06$ & $-1.22\pm 0.08$ & $  27$ & $0.17$ & $-0.95 $ \\

\hline

  [O/S] & $ 0.10\pm 0.09$ & $ 0.01\pm 0.28$ & $   1$ & $0.10$ & $ 0.08 $ \\

  [P/S] & $-0.28\pm 0.07$ & $ 0.41\pm 0.15$ & $  16$ & $0.18$ & $ 0.57 $ \\

 [Si/S] & $ 0.02\pm 0.04$ & $-0.30\pm 0.09$ & $  26$ & $0.13$ & $-0.60 $ \\

 [Mg/S] & $ 0.06\pm 0.05$ & $-0.46\pm 0.11$ & $   8$ & $0.11$ & $-0.85 $ \\

 [Mn/S] & $-0.60\pm 0.04$ & $-0.34\pm 0.08$ & $  17$ & $0.10$ & $-0.69 $ \\

 [Cr/S] & $-0.11\pm 0.03$ & $-0.76\pm 0.08$ & $  27$ & $0.10$ & $-0.88 $ \\
 
 [Fe/S] & $-0.23\pm 0.03$ & $-0.83\pm 0.07$ & $  27$ & $0.10$ & $-0.93 $ \\

\hline\hline

 \end{tabular}

\label{tab coeff seq}
\end{table}

\subsection{From relative abundances to depletion}
\label{abs dep}
The depletion of one element $X$, $\delta_X$, is defined as the amount of $X$ that is missing from the observed gas phase $[X/{\rm H}]$ with respect to its total abundance $[X/{\rm H}]_{\rm tot}$. This difference is typically attributed to the presence of dust, the missing metals being locked into dust grains \citep{Savage96}. $\delta_X$ is thus defined as 
\begin{equation}
\delta_X=[X/{\rm H}]-[X/{\rm H}]_{\rm tot} - \alpha_X \mbox{,}
\label{eq delta}
\end{equation}
where $\alpha_X$ is the over- or underabundance of $X$ with respect to Fe, purely due to nucleosynthesis effects (i.e. for the Galaxy these are the stellar [$X$/Fe]). The value $\alpha_X$ is likely to be metallicity dependent, as observed in the Galaxy. This term is typically neglected in the standard definition for the Galactic ISM \citep[e.g.][]{Joseph88}. It is expressed on a logarithmic scale, and is discussed further in the next section. 

Eq. \ref{eq delta} can only be used directly if the reference total abundances $[X/{\rm H}]_{\rm tot}$ are known, as is the case for the Galaxy, where reference total abundances are solar. However, they are not known in advance for DLAs, which can have subsolar $[X/{\rm H}]_{\rm tot}$. In such cases, $\delta_X$ can be derived from the observed [$X$/Zn], by correcting it for the zinc abundance [Zn/H], as follows: 
\begin{equation}
\begin{array}{lcl}
\delta_X &=& [X/{\rm Zn}] + [{\rm Zn/H}] - [X/{\rm H}]_{\rm tot} - \alpha_X \\
         &=& [X/{\rm Zn}] + \delta_{\rm Zn} - \alpha_X \mbox{,}
\end{array}     
\label{eq delta [X/Zn]}
\end{equation}
where [X/Zn] are the observed values for DLA and Galactic clouds, and $\delta_{\rm Zn}$ is the depletion of Zn. Below we discuss how $\delta_{\rm Zn}$ and $\alpha_X$ can be constrained. 

\begin{figure}
   \centering
   \includegraphics[width=90mm,angle=0]{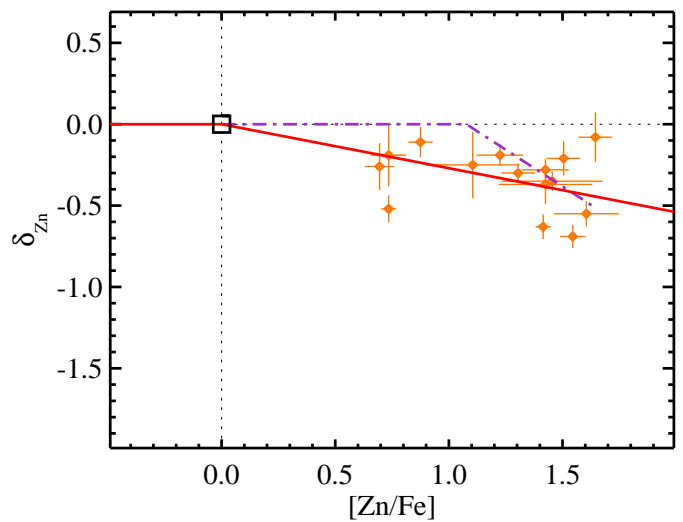}
   \caption{Observed depletion of Zn in Galactic individual absorbers, taken from \citet{Jenkins09}, which is the observed [Zn/H] for the Galaxy. The solid red curve shows a linear fit to the data with a slope of $B_{\delta_{\rm Zn}}=-0.27\pm0.03$, which we assume for the derivation of the depletion. The dashed curve shows the expected relation from Eq. 10 of \citet{Jenkins09}. The depletion of Zn is bound to be zero for DLAs with [Zn/Fe] $=0$.}
              \label{fig znh Gal}
    \end{figure}

In the Galaxy, the depletion of Zn, $\delta_{\rm Zn}$, is also the observed zinc abundance, [Zn/H] ($\delta_{\rm Zn} = [{\rm Zn/ H}]-[{\rm Zn/ H}]_{\rm tot}$, and $[{\rm Zn/H}]_{\rm tot}=0$ in the Galaxy. Figure \ref{fig znh Gal} shows the observed [Zn/H] with respect to [Zn/Fe] for individual Galactic absorbers (orange diamonds). We estimate the trend of $\delta_{\rm Zn}$ with [Zn/Fe] from a fit to the [Zn/H] observed in the Galaxy. We cannot include DLAs in this fit because their intrinsically low metallicities naturally produce a wide range of [Zn/H], regardless of dust depletion. Nevertheless, we can use the observational fact that [Zn/Fe] reaches zero at low metallicity for DLAs to assume no depletion of Zn at [Zn/Fe] $=0$ and, therefore, bind the relation to $\delta_{\rm Zn}=0$. This method is purely observational, but the scatter in the observed [Zn/H] is very high. We fit a linear model to the data, considering errors along both axes, and allowing for intrinsic scatter in this relation using the \textsc{MPFITEXY} procedure described above. The solid red curve in Fig. \ref{fig znh Gal} shows our fit to the observed [Zn/H] in the Galaxy, where the best-fit slope is $B_{\delta_{\rm Zn}} = -0.27\pm0.03$, with an intrinsic scatter of $0.16$~dex. The intercept is fixed to $\delta_{\rm Zn}=0$, but similar results are achieved without this assumption (a slightly negative intercept). While this relation should not be trusted for a one-to-one conversion of individual data points, we use it here merely to convert the [$X$/Zn] scale to a scale of [$X$/H].  

As a sanity check, we compare our results with the [Zn/H] trend with the depletion strength factor $F_*$ obtained by \citet{Jenkins09} for the Galaxy \citep[see fig. 7 of][]{Jenkins09}. We derive the expected relation between [Zn/H] and [Zn/Fe] by inverting Eq. 10 of \citet{Jenkins09} for [Zn/H] and [Fe/H], respectively. The dash-dotted curve in Fig. \ref{fig znh Gal} shows this relation, assuming that [Zn/H] remains zero (i.e. no depletion of Zn) for low [Zn/Fe]. The two curves differ by a variable amount, at most $\sim0.2$~dex, which is comparable to the scatter of the relation. In this paper we rely as much as possible on the observed data and therefore we choose to adopt the fit to the data for our assumption on the relation between $\delta_{\rm Zn}$ and [Zn/Fe].

The zero intercept of the $\delta_{\rm Zn}$ versus [Zn/Fe] relation is $\delta_{\rm Zn}=0$. However, if [Zn/Fe] would be slightly supersolar in these regime owing to small nucleosynthesis differences between Zn and Fe \citep[e.g. $\sim0.24$~dex,][]{Barbuy15}, the $\delta_{\rm Zn}$ curve could be shifted by this small amount along the x-axis (to the right). This would have no effect on our results on the depletion. In fact, even if we would assume the $\delta_{\rm Zn}$ versus [Zn/Fe] purely derived from \citet[][dash-dotted curve in Fig. \ref{fig znh Gal}]{Jenkins09}, whose knee is shifted by $\sim1$~dex with respect to our assumption, the effect on the depletion sequences would be limited, as shown in Sect. \ref{sec depl seq} and in the bottom right panel of Fig. \ref{fig abs dep}.

The derivation of $\delta_X$ is somewhat uncertain because it depends on the adopted $\delta_{\rm Zn}$, which is not known a priori for DLAs. Nevertheless, the observations in the Galaxy constrain the $\delta_{\rm Zn}$ to be small. Our assumption on the $\delta_{\rm Zn}$ versus [Zn/Fe] relation cannot heavily affect our results on the depletion (effects much smaller than $0.2$~dex for most systems). Indeed, depletion effects can be as strong as up to $\sim2$~dex. Thus, as a first approximation, these results are useful to determine the depletion. One advantage of this technique of deriving the depletions from the relative abundances is that it is independent from \hi{} measurements in DLAs. In addition, by construction this method can be applied to systems with non-solar metallicities. 

Taking together Eq. \ref{eq seq rel ab}, Eq. \ref{eq delta [X/Zn]}, and the fit shown in Fig. \ref{fig znh Gal}, the depletion can be empirically described as
\begin{equation}
 \alpha_X + \delta_X =   A_1 + B_1\times {\rm [Zn/Fe]}  + ( B_{\delta_{\rm Zn}} \times {\rm [Zn/Fe]}) \mbox{,}
   \label{eq alpha delta}
\end{equation}
where [Zn/Fe] are the observed relative abundances, the coefficients $A_1$ and $B_1$ are reported in Table \ref{tab coeff seq}, $B_{[{\rm Zn}/H]}$ is the slope of the fit in Fig. \ref{fig znh Gal}, and $\alpha_X$ is the nucleosynthetic over- or underabundance, which we discuss below.

\subsection{Correcting for nucleosynthesis effects}
\label{sec nucleo corr}

          \begin{figure}[!t]
   \centering
   \includegraphics[width=90mm,angle=0]{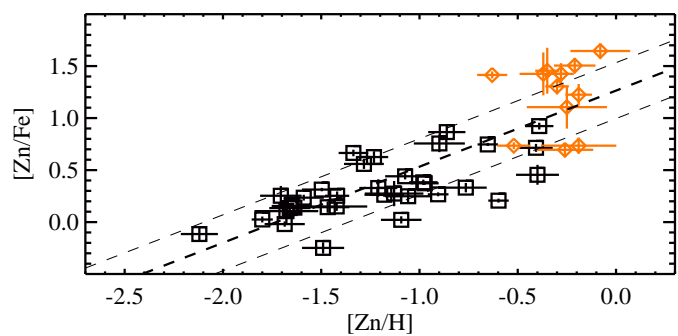}
   \caption{Observed abundances of Zn, with respect to H and Fe. The trend (dashed curve, [Zn/Fe] $=0.73 \times [{\rm Zn/H}]  + 1.26$ with an intrinsic scatter of 0.3~dex and a Pearson correlation coefficient of 0.81, from a linear fit to the DLA, black squares, and Galactic data, orange diamonds) is due to dust depletion, because Fe is more heavily depleted than Zn. This relation is used here only to allow a more reasonable assumption on the nucleosynthesis, i.e. to convert the literature information on nucleosynthesis - often expressed as a function of metallicity - to a scale of [Zn/Fe].}
              \label{fig met znfe}
    \end{figure}
The depletion $\delta_X$ should in principle  be a negative number because it represents the metals that are missing from the gas phase. However, the sequences of relative abundances for $\alpha$ elements reach positive values for [Zn/Fe] $=0$. This is not surprising because we expect some nucleosynthesis effects, such as $\alpha$-elements enhancement. Indeed, for example \citet{Dessauges-Zavadsky06} and \citet{Rafelski12} estimated an $\alpha$-element enhancement of $\sim0.25$--$0.35$~dex in silicon for the DLAs with low [Zn/Fe]. 

The nucleosynthesis effects are likely to be metallicity dependent and stronger at lower metallicities, as observed from the stellar [$X$/Fe] in the Galaxy \citep[e.g.][]{Jonsell05,Lambert87,Wheeler89,McWilliam97,Bergemann08,Becker12,Mishenina15,Battistini15}. These studies indicate a plateau in the nucleosynthetic over- or underabundances below a certain metallicity (around $\sim-1$ for $\alpha$ elements, and similar for Mn), and then this plateau vanishes around solar metallicity. However, these results are mostly based on Galactic stars, and they may not be entirely appropriate for the gas in DLAs. This is because any change in nucleosynthesis is instantaneously visible on stellar abundances, while recycling of the metals in the ISM happens on more extended timescales and physical processes. For Mn, \citet{Bergemann08} showed that at low metallicities, the stellar [Mn/Fe] estimates can be underestimated when assuming local thermal equilibrium, which is mostly the case in the literature.

         \begin{figure*}[!h!t]
   \centering
   \includegraphics[width=190mm,angle=0]{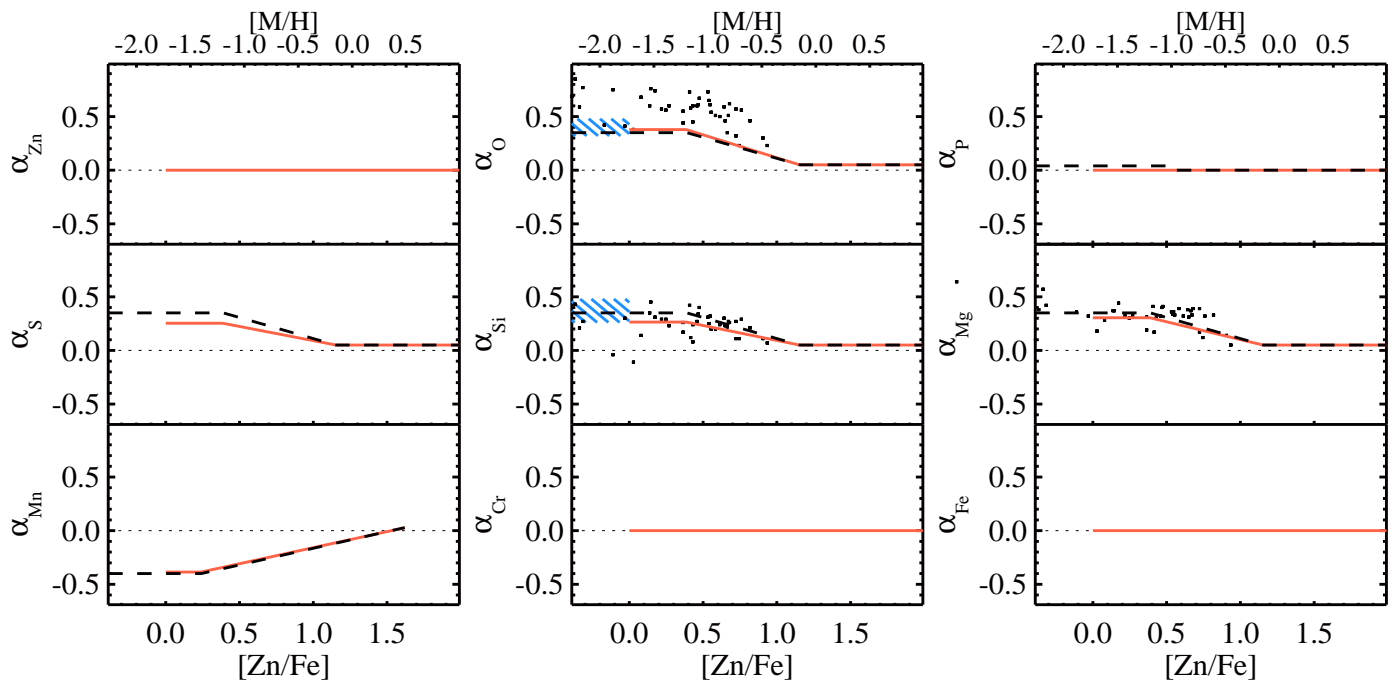}
   \caption{Assumed nucleosynthetic over- or underabundances (red solid curves), in units of [Zn/Fe]. The values of $\alpha_{X, 0}$ (at [Zn/Fe] $=0$) are adopted from the observed intercepts of the sequences of relative abundances ($\alpha_{\rm O,0} = 0.38\pm 0.10$, $\alpha_{\rm P,0} = 0.00\pm 0.06$, $\alpha_{\rm S,0} = 0.25\pm 0.03$, $\alpha_{\rm Si,0} = 0.26\pm 0.03$, $\alpha_{\rm Mg,0} = 0.30\pm 0.04$, and $\alpha_{\rm Mn,0} = -0.39\pm 0.03$). The shape of the curves are adopted from Galactic observations in the literature, from \citet{Lambert87} and \citet{McWilliam97} for $\alpha$ elements (dashed black curves), from \citet{Wheeler89}, \citet{Mishenina15}, and \citet{Battistini15} for Mn, and from \citet{Caffau16} and \citet{Roederer14} for P. These are presented in units of stellar metallicity [$M$/H], where the scaling to ISM [Zn/Fe] units assumes a linear relation of the data shown in Fig. \ref{fig met znfe}. This is only a scale conversion that allows us to use the nucleosynthetic over- or underabundances in the literature, and should not be used as a one-to-one conversion between [Zn/Fe] and [$M$/H]. The shaded region corresponds to the $\alpha$-element enhancement measured by \citet{Becker12} for metal-poor DLAs (within $1 \sigma$ of the mean), while the small dots are derived from \citet{Jonsell05} for metal-poor stars (see Appendix \ref{sect oxygen}).}
              \label{fig nucleosynthesis}
    \end{figure*}
Thus, we adopt the overall distribution of the nucleosynthetic over- or underabundances with metallicity from the literature (stellar [$X$/Fe] in the Galaxy), but scale the values for the low-metallicity systems according to the DLA observations. In practice, for systems with low [Zn/Fe] we derive nucleosynthetic over- or underabundances from the zero intercepts of the sequences of relative abundances ([$X$/Zn] at [Zn/Fe] $=0$, see Table \ref{tab coeff seq}), for $\alpha$ elements, P and Mn. These are $\alpha_{\rm O,0} = 0.38\pm 0.10$, $\alpha_{\rm P,0} = 0.00\pm 0.06$, $\alpha_{\rm S,0} = 0.25\pm 0.03$, $\alpha_{\rm Si,0} = 0.26\pm 0.03$, $\alpha_{\rm Mg,0} = 0.30\pm 0.04$, and $\alpha_{\rm Mn,0} = -0.39\pm 0.03$. These are values observed in DLAs.

As for the shape of the nucleosynthesis curves, for the $\alpha$ elements we refer to the nucleosynthetic overabundances observed by \citet{Lambert87} and \citet{McWilliam97}, and also confirmed by \citet{Wheeler89}, for Mg, Si, and O in Galactic stars. These studies found $[\alpha/{\rm Fe}]=0.35$ for metallicities $[M/{\rm H}]\leqslant -1.2$, $[\alpha/{\rm Fe}]=0.05$ for metallicities $[M/{\rm H}]\geqslant -0.15$, and in between the metallicities decreased linearly. For the nucleosynthesis of Mn, we refer to the observations of [Mn/Fe] for Galactic stars of \citet{Wheeler89}, \citet{Mishenina15} and \citet{Battistini15}. These indicate that [Mn/Fe] is subsolar until $[M/{\rm H}]\leqslant -1.4$ and increasing towards higher metallicity. The nucleosynthetic abundances of Mn are not constrained for [$M$/H] $> 0.5$ in the Galaxy. This has no effect on our results because only two measurements of Mn are constrained in this high [Zn/Fe] regime.

We convert these literature results into the units of [Zn/Fe] of the ISM using a linear fit between [Zn/H] and [Zn/Fe], as shown in Fig. \ref{fig met znfe}. While such conversion is not recommended for individual measurements, given the scatter in the [Zn/Fe] versus [Zn/H] relation, we use it here to convert the metallicity scale to a [Zn/Fe] scale to allow reasonable assumptions to be made on the nucleosynthesis pattern of DLAs. For this reason, we do not consider the [Zn/Fe] versus [Zn/H] relation to be part of our methodology. A distribution of the nucleosythesis over- or underabundances could be assumed regardless of such relation. Furthermore, we do not intend to use [Zn/Fe] as a metallicity indicator, given the scatter in Fig. \ref{fig met znfe}. 

The resulting curves of nucleosynthetic over- or underabundances with respect to iron are shown in Fig. \ref{fig nucleosynthesis}. As a sanity check, we compare these curves to yet other literature observations, from \citet{Jonsell05} and \citet{Becker12} for metal-poor stars (dots in Fig \ref{fig nucleosynthesis}) and metal-poor DLAs (shaded area), respectively. There is a general good agreement among the $\alpha_X$ in the Galaxy and DLAs, and from different observations. One debatable exception is for oxygen, which we discuss in Appendix \ref{sect oxygen}.

 \begin{figure*}[!h!t]
   \centering
   \includegraphics[width=190mm,angle=0]{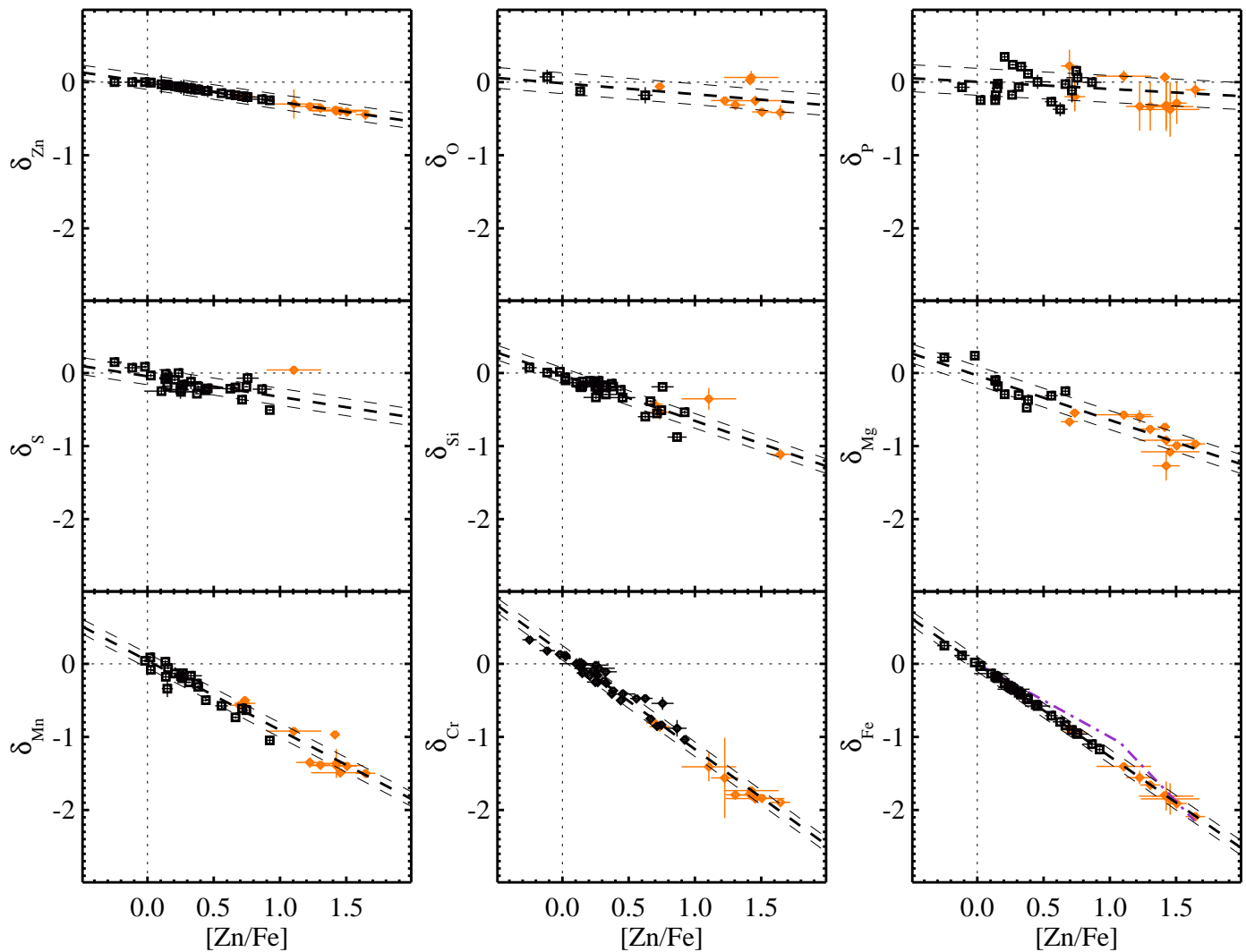}
   \caption{Distribution of the depletion $\delta_X$ with [Zn/Fe] (depletion sequences), derived from the observed [$X$/Zn] (see Sect. \ref{abs dep}). The black squares are for DLAs and the orange diamonds are for the Galactic lines of sight. The distributions of Zn and Fe depletion are shown here for clarity. They have no scatter by construction. The dash-dotted curve shows what the depletion of Fe would have been if we had assumed the expected [Zn/H] trend with [Zn/Fe] based on Eq. 10 of \citet{Jenkins09}; see Fig. \ref{fig znh Gal}. This is shown only for Fe, as an example.}
   \label{fig abs dep}
    \end{figure*}
The nucleosynthesis curves that we adopt are shown in Fig. \ref{fig nucleosynthesis} (red solid curves). We subtract these curves from the uncorrected depletions and obtain estimates of the depletions (Sect. \ref{sec depl seq}). Therefore, these derived depletions are subject to our assumptions on the nucleosynthesis effects. However, such effects must be limited because of the relatively small zero intercepts observed in the sequences of relative abundances (Fig. \ref{fig znfe}), and  because of the existence of such tight correlations, for DLAs and also among abundances in the Galaxy \citep{Jenkins09}, which are due to dust depletion. Nucleosynthetic over- or underabundances can be moderate ($\lesssim 0.4$~dex), but they are secondary when considering the strong effects of dust depletion, which can change the gas-phase metal abundances by up to 2~dex for the most refractory elements, and create a correlation between observed relative abundances. In particular, while our description of nucleosynthesis in DLAs is simplified, we note that, first, the absolute values of the nucleosynthetic over- or underabundances at zero depletion are constrained by the observations and, second, the effects of nucleosynthesis and dust depletion act in different regimes; that is, nucleosynthesis effects are not important (zero) for the dustiest systems, and depletion vanishes at [Zn/Fe] $=0$. Therefore our simplified description of nucleosynthesis in DLAs can be used to both safely determine dust depletion and to constrain the observed nucleosynthesis $\alpha_{X, 0}$ at zero depletion.

\subsection{The dust-depletion sequences}
\label{sec depl seq}

Figure \ref{fig abs dep} shows the depletion sequences of several elements ($\delta_X$ vs [Zn/Fe]) for DLAs and Galactic lines of sights, after subtracting the correction for nucleosynthesis described above. We empirically characterize the trends of dust depletions by fitting them linearly, 
 \begin{equation}
\delta_X =  A_2 + B_2\times {\rm [Zn/Fe]}\mbox{,}
\label{eq delta emp}
\end{equation}
again considering errors on both axis and including an estimate of the intrinsic scatter of the correlations (see \textsc{MPFITEXY} description above). The best-fit coefficients are reported in Table \ref{tab coeff abs depl} along with the intrinsic scatter $\sigma_{\delta_X}$ and a Pearsons coefficients $r$.
\begin{table}[!h]
\centering
\caption{Coefficients of the depletion sequences ($\delta_X = A_2 + B_2 \times$ [Zn/Fe]) shown in Fig. \ref{fig abs dep}. The last two columns list the internal scatter and the Pearson correlation coefficients. $^a$ $r=-1$ by construction.}
\begin{tabular}{ l | r r r r}
\hline \hline
\rule[-0.2cm]{0mm}{0.8cm}
 
$X$   &   \multicolumn{1}{c}{$A_2$}    &  \multicolumn{1}{c}{$B_2$} & \multicolumn{1}{c}{$\sigma_{\delta_{X}}$} & \multicolumn{1}{c}{$r$}\\ 

\hline
   $\delta_{\rm Zn} $ & $ 0.00    \pm 0.01$ & $-0.27\pm 0.03$ & $0.10$ & $-1.00^a $ \\
    $\delta_{\rm O} $ & $-0.02    \pm 0.10$ & $-0.15\pm 0.09$ & $0.14$ & $-0.50 $ \\
    $\delta_{\rm P} $ & $ 0.01    \pm 0.06$ & $-0.10\pm 0.07$ & $0.18$ & $-0.32 $ \\
    $\delta_{\rm S} $ & $-0.04    \pm 0.03$ & $-0.28\pm 0.08$ & $0.12$ & $-0.50 $ \\
   $\delta_{\rm Si} $ & $-0.03    \pm 0.03$ & $-0.63\pm 0.06$ & $0.10$ & $-0.88 $ \\
   $\delta_{\rm Mg} $ & $-0.03    \pm 0.05$ & $-0.61\pm 0.05$ & $0.13$ & $-0.93 $ \\
   $\delta_{\rm Mn} $ & $ 0.04    \pm 0.03$ & $-0.95\pm 0.04$ & $0.10$ & $-0.97 $ \\
   $\delta_{\rm Cr} $ & $ 0.15    \pm 0.03$ & $-1.32\pm 0.04$ & $0.10$ & $-0.99 $ \\
   $\delta_{\rm Fe} $ & $-0.01    \pm 0.03$ & $-1.26\pm 0.04$ & $0.10$ & $-1.00^a $ \\
\hline\hline

 \end{tabular}
\label{tab coeff abs depl}
\end{table}

       \begin{figure*}[!h!t]
   \centering
   \includegraphics[width=120mm,angle=0]{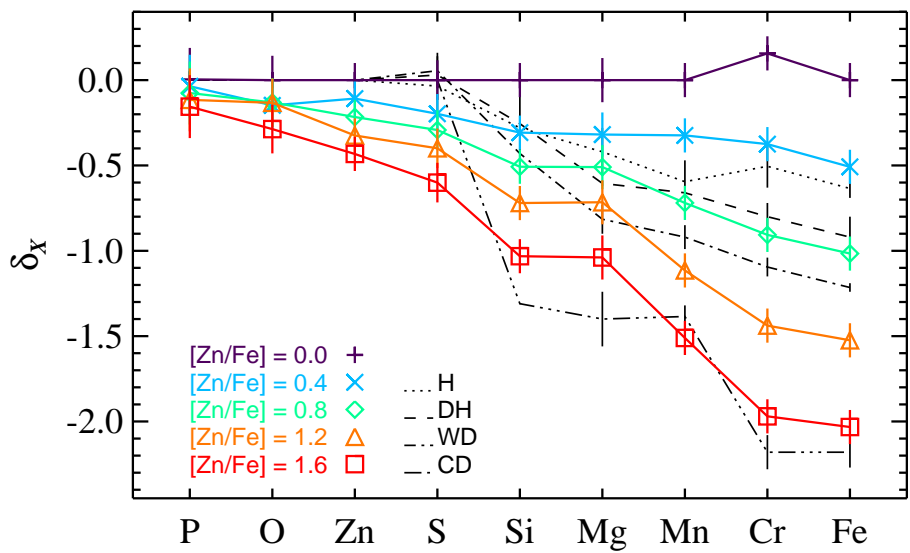}
   \caption{Depletion patterns for different levels of [Zn/Fe]; see Table \ref{tab depl patterns} for values and uncertainties. The four depletion patterns of \citet{Savage96} for Galactic halo (H), disk + halo (DH), warm disk (WD), and cool disk (CD) are shown for comparison.}
              \label{fig depl patterns}
    \end{figure*}
 \begin{table*}
\centering
\caption{Depletion patterns for different levels of [Zn/Fe]. $\sigma_{\delta_X}$ is the internal scatter of the depletion sequences of Fig. \ref{fig abs dep}, taken from Table \ref{tab coeff abs depl}.}
\begin{tabular}{ c | r r r r r r r r r }
\hline \hline
\rule[-0.2cm]{0mm}{0.8cm}
[Zn/Fe]  &   $\delta_{\rm Zn}$  &   $\delta_{\rm O}$ &  $\delta_{\rm P}$  &  $\delta_{\rm S}$  &  $\delta_{\rm Si}$  &  $\delta_{\rm Mg}$  &$\delta_{\rm Mn}$  &  $\delta_{\rm Cr}$ &  $\delta_{\rm Fe}$  \\
\hline

    0.00
 & $ 0.00$ 
 & $ 0.00$ 
 & $ 0.00$ 
 & $ 0.00$ 
 & $ 0.00$ 
 & $ 0.00$ 
 & $ 0.00$ 
 & $ 0.16$ 
 & $ 0.00$ 
\\
    0.40
 & $-0.11$ 
 & $-0.15$ 
 & $-0.04$ 
 & $-0.20$ 
 & $-0.31$ 
 & $-0.32$ 
 & $-0.32$ 
 & $-0.37$ 
 & $-0.51$ 
\\
    0.80
 & $-0.22$ 
 & $-0.13$ 
 & $-0.08$ 
 & $-0.29$ 
 & $-0.51$ 
 & $-0.51$ 
 & $-0.72$ 
 & $-0.91$ 
 & $-1.02$ 
\\
    1.20
 & $-0.32$ 
 & $-0.13$ 
 & $-0.12$ 
 & $-0.40$ 
 & $-0.72$ 
 & $-0.72$ 
 & $-1.12$ 
 & $-1.44$ 
 & $-1.52$ 
\\
    1.60
 & $-0.43$ 
 & $-0.29$ 
 & $-0.16$ 
 & $-0.60$ 
 & $-1.03$ 
 & $-1.04$ 
 & $-1.51$ 
 & $-1.97$ 
 & $-2.03$ 
\\
\hline
$\sigma_{\delta_X}$ &  0.10 & 0.14 & 0.18 &  0.12 & 0.10 & 0.13 & 0.10 &0.10 & 0.10 \\

\hline \hline
 \end{tabular}
\label{tab depl patterns}
\end{table*}
For Zn (Fe), we derive the depletion $\delta_{\rm Zn}$ ($\delta_{\rm Fe}$) using Eq. \ref{eq alpha delta}, given that $\alpha_{\rm Zn}=0$ ($\alpha_{\rm Fe}=0$) and that  [Zn/Zn] $= 0 \times$ [Zn/Fe], i.e. $A_1=0$ and $B_1=0$ ([Fe/Zn] $= -1 \times$ [Zn/Fe], i.e. $A_1=0$ and $B_1=-1$), which has no scatter by construction. Therefore no scatter is shown in Fig. \ref{fig abs dep} for Zn and Fe (first and last panel).

\subsection{Canonical depletion patterns}
\label{dep patterns}

In the previous sections, we derived the depletion of several metals into dust grains as a continuous function of [Zn/Fe]. Here we slice these depletion sequences at five (arbitrary) values of [Zn/Fe], thereby providing five canonical depletion patterns corresponding to [Zn/Fe] $=0.0,\,0.4,\,0.8,\,1.2,1.6$. The values of the depletion of several elements at these depletion factors are listed in Table \ref{tab depl patterns} and visualized in Fig. \ref{fig depl patterns}. These depletion patterns can be used as a new reference for studies of relative abundances, keeping in mind that this is a small subset, because the distributions of $\delta_X$ are continuous in [Zn/Fe]. The uncertainties $\sigma_{\delta_X}$ are the internal scatter of the depletion sequences shown in Fig. \ref{fig abs dep}, taken from Table \ref{tab coeff abs depl}. These values represent the average scatter around $\delta_X$, and are typically comprised between 0.1 and 0.2~dex, as also reported in Table \ref{tab depl patterns}.

\section{Discussion}
\label{discussion}

We observe correlations between [$X/Y$] and [Zn/Fe], where $X$ is a refractory and $Y$ is a non-refractory element, which we call sequences of relative abundances. Dust depletion must be the primary driver of such correlations because their slopes depend on the refractory properties of the metals; see Sect. \ref{sec disc depletion}. The scatter in these sequences is $\lesssim0.2$~dex, suggesting that nucleosynthesis effects should be much smaller than the effects of dust depletion. This is not straightforward, considering the variety of systems included in our sample, from low-metallicity, high-redshift DLAs to the Galaxy, with potentially very different star formation histories.

We separate and characterize the effects of dust depletion and nucleosynthesis on the observed relative abundances. This is possible thanks to the fact that the effects of dust depletion and nucleosynthesis dominate in different metallicity regimes, and by studying the relative abundances of several elements with different refractory and nucleosynthetic properties. Below we discuss our results on the non-refractory elements, nucleosynthesis, and dust depletion, and further characterize the absorbers in terms of total (dust-corrected) metallicity, elemental abundances in dust, dust-to-metal ratio, and dust extinction. 

\subsection{The non-refractory elements}

\subsubsection{Phosphorus, zinc, and oxygen}

Non-refractory elements are fundamental for accurate dust-independent estimates of the metallicity of DLAs or gas clouds. Among the non-refractory elements, P appears to be the least depleted into dust grains. Indeed, [Zn/P] decreases mildly with an increasing dust content, although with a large scatter. Furthermore, the slopes of the observed [$X$/P] sequences with [Zn/Fe] are all steeper than [$X$/Zn]. Finally, all the [Zn/P] values measured in Galactic clouds are negative (Fig. \ref{fig znfe non depleted}). These facts suggest that P is less depleted into dust grains than Zn. P measurements are rare or difficult to obtain in DLAs. This study is by far the largest sample of P abundance measurements in DLAs to date. Also, the depletion sequence of P has the largest scatter ($0.18$~dex) among the elements considered in this study. We find a lower depletion of P than could be expected from its relatively high condensation temperature. We discuss this further in Sect. \ref{sec disc depletion}.

In principle, P is the best metallicity indicator among the metals considered here, but caution should be exerted because of the difficulty in measuring P abundances. We thus recommend the use of P abundances, if available, as a first estimate of the total metallicity, in combination with Zn. For a more accurate metallicity estimate, we recommend correcting the abundances of different metals for dust depletion; see Sect. \ref{dust_corrected met}.

Oxygen is the second-least depleted metal in our sample. While O is slightly less depleted than Zn, the two elements contribute very differently to the absolute dust composition, where dust grains bear much more oxygen because of the intrinsically high O abundance ($10^4$ times more abundant than Zn). Silicate grains and iron oxides contain a significant amount of oxygen (up to 50\%). We discuss elemental abundances in dust in Sect. \ref{epsilon}.

Zinc is among the least depleted metals in our sample. In our analysis we adopt the recently improved estimates of oscillator strengths for \znii{} lines presented by \citet{Kisielius15}. These authors found that Zn column densities are lower by 0.1~dex with respect to previous estimates. The depletion of Zn increases with dust content, up to $\sim$0.4--0.5\,dex for the dustiest systems in the Galaxy \citep[see also][]{Jenkins09}. Zn measurements are very useful metallicity estimators, especially at low [Zn/Fe], but can suffer the effect of dust depletion at high [Zn/Fe].

\subsubsection{Sulfur}
\label{sulfur}

In this paper we adopt the oscillator strengths of \suii{} lines of \citet{Kisielius14}. The new values of [S/Zn] are therefore higher by 0.14~dex with respect to previous estimates. We observe [S/Zn] ratios up to 0.25~dex for the least dusty DLAs, contrary to previous work of \citet{Centurion00} and \citet{Dessauges-Zavadsky06}, who found nearly solar [S/Zn] in DLAs, but using earlier oscillator strength values. We therefore see that there is now less tension between the [S/Zn] observed in DLAs and in the Galaxy \citep[$\sim0.35$~dex; e.g.][]{Lambert87,McWilliam97}.

Sulfur is commonly believed to be a non-refractory element that never depletes into dust grains. However, S is often considered a troublesome element because it sporadically shows signs of depletion \citep[e.g.][]{Jenkins09}. In fact, \citet{Jenkins09} showed that S may be depleted up to 1 dex in the dustiest lines of sight in the Galaxy. Our results indicate that S is mildly depleted into dust grains for DLAs, similarly to Zn. The $\delta_{\rm S}$ trend that we find is determined almost completely by DLA abundances because of the paucity of simultaneous S, Zn, and Fe column density measurements in the Galaxy. The scatter in the depletion sequence of S is very large, and more observations of S depletion at high [Zn/Fe] will possibly further constrain the trend in the future. Dust species, such as FeS and FeS$_2$, could exist in space and contribute to the depletion of sulfur. Thus, we consider sulfur a mildly depleted element, and we recommend including it in studies of relative abundances in combination with other non-refractory elements.

\subsection{Nucleosynthesis}
\label{disc nucleo}

The comparison of the observed abundances relative to elements with different nucleosynthetic properties is crucial to disentangle the effects of nucleosynthesis from dust depletion. In particular, we consider the trends of [$X$/Zn], [$X$/S], and [$X$/P] with [Zn/Fe]; see Figs \ref{fig znfe} and \ref{fig znfe non depleted}. The panels showing [Si/S], [Mg/S], [O/S], and [Cr/Zn] are less affected by nucleosynthesis effects because of the similar nucleosynthetic history between Si, Mg, O, and S, which are all $\alpha$ elements, and those with affinity to the iron group, such as Cr and Zn. Zn is not strictly an iron-group element, but its production traces Fe within a small scatter; see Appendix \ref{sect znfe}.

The small difference between the [$X$/Zn], [$X$/S], and [$X$/P] trends with [Zn/Fe], as well as the limited scatter (see Table \ref{tab coeff seq}) of the sequences, already imply that the nucleosynthesis effects must be much smaller than the effects of dust depletion.

For the DLAs at [Zn/Fe] $=0$, the effects of dust depletion are negligible. In this regime, the nucleosynthesis effects are most evident. The observed zero intercept (at [Zn/Fe] $=0$) of the sequences of relative abundances ([$X$/Zn] vs [Zn/Fe]) thus provide dust-free estimates of the nucleosynthetic over- or underabundances of $X$, with respect to Zn and Fe, for the least dusty DLAs. We observe $\alpha_{\rm O,0} = 0.38\pm 0.10$, $\alpha_{\rm P,0} = 0.00\pm 0.06$, $\alpha_{\rm S,0} = 0.25\pm 0.03$, $\alpha_{\rm Si,0} = 0.26\pm 0.03$, $\alpha_{\rm Mg,0} = 0.30\pm 0.04$, and $\alpha_{\rm Mn,0} = -0.39\pm 0.03$. The $\alpha_{X, 0}$ are observed values, which do not depend on our methodology or assumptions. These values are remarkably similar to the nucleosynthesis pattern of the Galaxy, at low metallicity, as can be seen in Fig. \ref{fig nucleosynthesis}. This is rather surprising, given the potentially very different star formation histories of low-metallicity systems such as DLAs and the Galaxy. The $\alpha$-element enhancement supports the idea that core-collapse supernovae 
  are important contributors to the enrichment of the gas. These objects evolve on a timescale of 1--10 Myr \citep[while type Ia supernovae can have much longer lifetimes, around 1 Gyr, e.g.][]{Kobayashi09}. These results suggest the presence of a young stellar population, and thus ongoing or recent star formation in these DLAs.

When converting the relative abundances to depletions, we assumed the overall distribution of the nucleosynthetic over- or underabundances with metallicity from Galactic measurements in the literature, but scaled it for the $\alpha_{X, 0}$ values observed for DLAs at [Zn/Fe] $=0$, where no depletion is expected. At this stage, we cannot rule out that the overall distribution of $\alpha_{X}$ with [Zn/Fe] in DLAs is different than from the Galaxy. Indeed, the metallicity where nucleosynthesis over- or underabundances vary (the knees in Fig. \ref{fig nucleosynthesis}) may depend on the efficiency of star formation in producing metals. Nevertheless, the $\alpha_{X, 0}$ observed in DLAs limit the possible extent of nucleosynthesis effects. For the dustiest systems, i.e. for the Galactic absorbers in our sample, we can safely assume zero (or little) nucleosynthesis effects. Our assumption on the $\alpha_{X}$ distribution has minimal influence on our results on dust depletion.

\subsection{Dust depletion}
\label{sec disc depletion}
\citet{Jenkins09} showed that the observed abundances of different elements in Galactic lines of sight correlate with each other and parametrized them in depletion sequences. Here we include QSO-DLAs as well, and thus consider relative abundances, rather than abundances (with respect to H). We find that the abundances relative to elements that are not heavily depleted into dust grains [$X$/Zn,P,S] correlate with [Zn/Fe]. Moreover, the slope of these correlations are steeper for refractory elements. This implies that the main driver of these trends is indeed dust depletion, and that nucleosynthesis effects must be much weaker than that caused by depletion effects. This suggests that the depletion of dust is more affected by ISM processes than the production of metals in stars. Thus, a first possible interpretation of the mere existence of the sequences of relative abundances, from DLAs to the Galaxy, is that grain growth in the ISM must be an important process of dust formation. The dust fraction and the level of depletion varies significantly even within a galaxy, and this is likely because of dust condensation in the ISM \citep{Jenkins09,Mattsson12,Mattsson12b,Mattsson14b,Tchernyshyov15}. Furthermore, the sequences of observed relative abundances are continuous in [Zn/Fe] all the way from DLAs to Galactic absorbers. This suggests that the availability of refractory metals in the ISM is a crucial driver of dust production, regardless of the star formation history. In fact, DLA galaxies may have a wide range of star formation histories, which in principle are also different from those of the Galaxy.

We made two assumptions while deriving the depletion, on $\delta_{\rm Zn}$ and $\alpha_X$; see Eq. \ref{eq alpha delta}. The estimate of $\delta_{\rm Zn}$ is somewhat uncertain because the depletion of Zn is only known for the Galaxy, but not for DLAs. Nevertheless, it is constrained to be small by the observations and null for [Zn/Fe] $=0$. The value $\alpha_X$ is constrained at [Zn/Fe] $=0$ by the zero intercepts of the sequences of relative abundances, which are small and in agreement with other estimates in the literature. While $\delta_{\rm Zn}$ is negligible at low [Zn/Fe], $\alpha_X$ vanishes at high [Zn/Fe]. Moreover, their values are much smaller than the effects that dust depletion of refractory elements can have. Therefore, these assumptions cannot significantly affect the derivation of $\delta_X$. Thus, we provide an estimate for the dust depletion of different metals $\delta_X$ as a function of [Zn/Fe] (Table \ref{tab coeff abs depl},  Fig. \ref{fig abs dep}), from DLAs to the Galaxy.  

We compare our results with the abundances and depletion $\delta_X$ derived by \citet{Tchernyshyov15} for the Magellanic Clouds (MC; see Appendix \ref{sect comparison MC}). We find that the relative abundances in the MC are consistent with the sequences of relative abundances that we found for DLAs and the Galaxy. As a further sanity check, we compare our estimates of depletion with those of \citet{Vladilo02}. We use the [Zn/Fe] of the first ten DLAs listed in their Table 3, and compute the depletion of iron $\delta_{\rm Fe}$ (one of the most highly depleted elements), based on Eq. \ref{eq delta emp} and the coefficients in Table \ref{tab coeff abs depl}. In eight out of ten cases, our $\delta_{\rm Fe}$ agrees well (with a difference $\leq0.05$~dex) with the iron depletion estimated by \citet{Vladilo02} for their 'S00 \& S11' models, i.e. where the dust-corrected [Zn/Fe] $=0$. In the remaining two cases, our $\delta_{\rm Fe}$ lie between their estimates for the 'S00 \& S11' and the 'E00 \& E11' models, the latter referring to models with dust-corrected [Zn/Fe] $=0.1$. We note that our method to estimate the depletion does not assume a fixed grain composition (or fractions of an element in dust), and is based on the observed sequences of relative abundances. A second cross-comparison is with the yet independent method of estimating the depletion by \citet{Vladilo11}. While our results for solar metallicity are in good agreement with \citet{Jenkins09}, we find severely more negative depletions than the typical depletion derived by \citet{Vladilo11}\footnote{We find $\delta_{\rm Mg} \sim -1$, $\delta_{\rm Si} \sim -1$, and $\delta_{\rm Fe} \sim -2$ for the dustiest systems, e.g. Galactic clouds at solar metallicity (see Fig. \ref{fig abs dep}). The depletion of Mg, Si, and Fe derived by \citet{Vladilo11} are shown in their figure 5 as a function of dust-free [Fe/H], i.e. $-0.5\leq \delta_{\rm Mg} \leq0.0$, $-0.7\leq \delta_{\rm Si} \leq 0.0$, and $-1.2 \leq \delta_{\rm Fe} \leq -0.5$.}.

\begin{figure}
   \centering
   \includegraphics[width=89mm,angle=0]{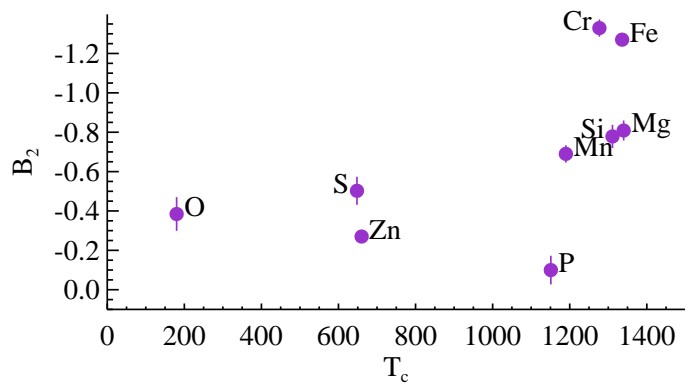}
   \caption{Slopes of the depletion sequences $B_2$ tend to be steeper for elements with higher condensation temperature $T_c$.}
              \label{fig condensationT}
    \end{figure}

Remarkably, the slopes of the depletion sequences tend to be steeper for the most refractory elements. This is shown in Fig. \ref{fig condensationT}, where the condensation temperatures ($T_c$, at which 50\% of the element are removed from the gas phase, at a pressure of $10^{-4}$~atm) are those compiled by \citet{Savage96}. We note that the $T_c$ values of \citet{Lodders03} provide similar results.
This confirms that indeed dust depletion is the process responsible for driving the depletion sequences and the sequences of relative abundances. Condensation could be an important process of dust formation, although the actual picture is likely more complex. 

Fig. \ref{fig condensationT} shows some correlation between the condensation temperature and the slope of the depletion sequences. However, even though phosphorus has a fairly high condensation temperature, we find that its depletion is small. The scatter in this trend is large and similar results for phosphorus were derived by \citet[][see their figure 2]{Welty94} for the ISM in the Galaxy, indicating that condensation is not the sole process regulating dust formation. Additional processes may be at play, such as the formation of stable molecules such as PN, which would inhibit the condensation of P in dust grains \citep{Joseph88}. In this case the formation of PN molecules could contribute to the depletion of P from the gas phase. 

\subsection{Canonical depletion patterns}

In Sect. \ref{dep patterns} we derived a new set of depletion patterns, for P, O, Zn, S, Si, Mg, Mn, Cr, and Fe at different depletion factors ([Zn/Fe] $=0.0,\,0.4,\,0.8,\,1.2,\,1.6$). These are the results of a linear fit of the sequences of observed [$X$/Zn] with respect to [Zn/Fe] (Figs. \ref{fig znfe} and \ref{fig znfe non depleted}) and their conversion to dust depletion $\delta_X$ (Sect. \ref{abs dep}). This depends on the assumption on $\delta_{\rm Zn}$ and the nucleosynthesis effects (Sect. \ref{sec nucleo corr}), which are small corrections and cannot heavily affect our results. This analysis includes high-resolution spectra of a total of 70 DLA and 169 Galactic absorbers, making it the largest study on depletion patterns to date. 

Figure \ref{fig depl patterns} shows a comparison between the new depletion patterns with the four previous standard depletion patterns of \citet{Savage96}. This reveals some similarities between the old and new sets of patterns at depletion factors above [Zn/Fe] $=0.6$. The DLAs in our sample have mostly [Zn/Fe] $\lesssim1$, possibly indicating that their depletion patterns are more similar to what has been observed for halo (H) and disk plus halo (DH) galactic lines of sight. This has been recognized before, for example by \citet{Ledoux02}, \citet{Prochaska02}, and \citet{Dessauges-Zavadsky06}.

We also note some level of disagreement. For instance,  \citet{Savage96} mostly observe a larger depletion of Mg than of Si, up to $\sim0.3$~dex larger for DH and warm disk (WD) Galactic lines of sight; but we obtain similar depletion of Mg and Si at all [Zn/Fe]. However, the four patterns of \citet{Savage96} were derived from only 13 Galactic lines of sight and deviations are expected given the scatter in the observed depletion sequences. The major discrepancy is for the non-refractory elements (such as for Zn and S), for which we, as well as \citet{Jenkins09}, find a stronger depletion than what has been measured or assumed by \citet{Savage96}. We note that S is considered to be a troublesome element because it is mostly assumed to be undepleted, but is often observed to be otherwise depleted. In fact, \citet{Jenkins09} measured up to 1~dex of sulfur depletion in the dustiest lines of sight in the Galaxy. Our fit to the depletion sequence of S is admittedly poor at high [Zn/Fe] (see Fig. \ref{fig abs dep}) and therefore we suggest exerting some caution when referring to the lack of S depletion in dusty systems.  

Moreover, the four discrete depletion patterns of \citet{Savage96} were selected by qualitatively identifying four types of environments, but both our study and \citet{Jenkins09} show that there are more variations in the environmental types. In particular, we find that [Zn/Fe] is a fundamental parameter to describe the depletion properties of a given environment. While we find depletion sequences from DLAs to the Galaxy that are continuous in [Zn/Fe], we slice these sequences in five depletion patterns to ease their application. Thus, we recommend the use of these new depletion patterns as a reference. They can be used to recover the full depletion-pattern curve, given a subset of observed relative abundances in a DLA. The best-fitting curve can then be used to infer the [Zn/Fe] ratio in the system, in a similar way that, for example 'halo' or 'disk plus halo' environments can be inferred with the previous depletion patterns of \citet{Savage96}. Furthermore, these results can be used to derive the total dust-corrected metallicity, dust-to-metal ratio, and dust extinction, as discussed in the following sections.

\subsection{Dust-corrected metallicity}
\label{dust_corrected met}

       \begin{figure}
   \centering
   \includegraphics[width=89mm,angle=0]{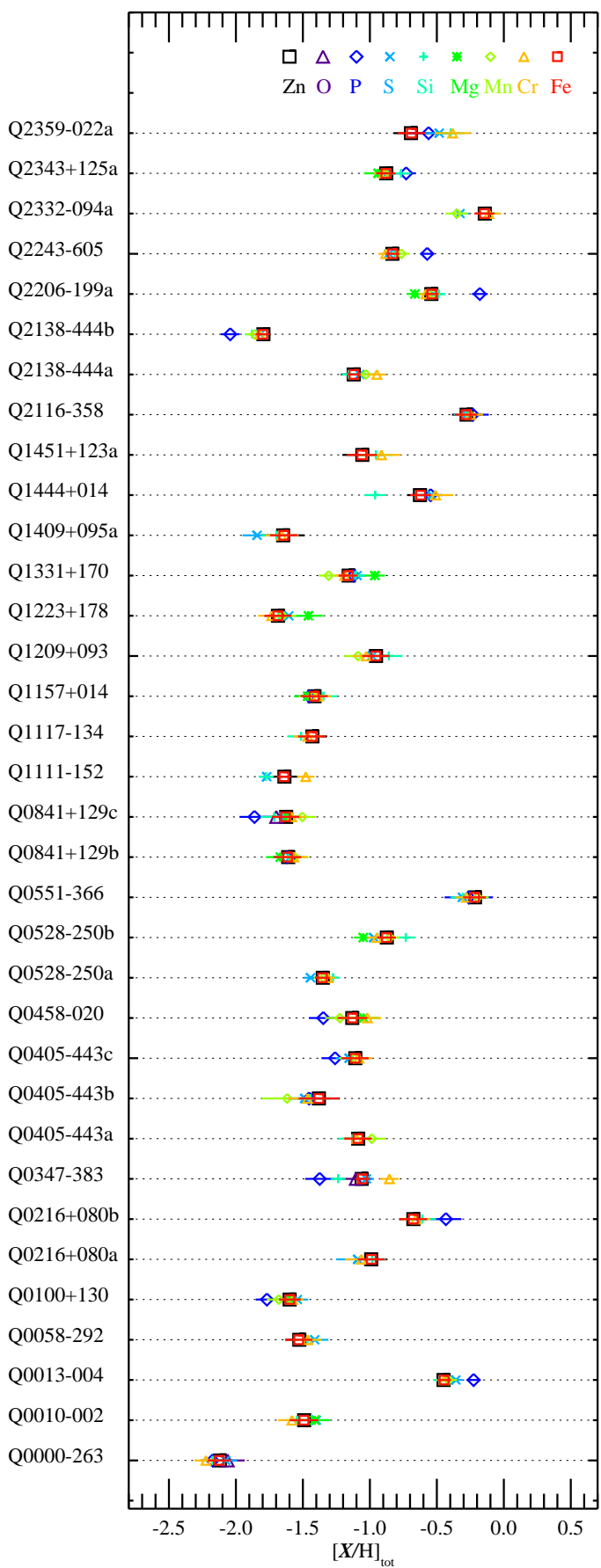}
   \caption{Total abundances, corrected for dust depletion and nucleosynthesis effects, of the DLA systems in our sample. Letters appended to a QSO name refer to different absorption systems along the same line of sight; see Table \ref{tab columns vel}.}
              \label{fig int met}
    \end{figure}
The total abundances [$X$/H]$_{\rm tot}$ of different elements can be derived by inverting Eq. \ref{eq delta}, where $[X/{\rm H}]$ are the observed gaseous abundances, $\alpha_X$ is the nucleosynthetic over- or underabundances (see Fig. \ref{fig nucleosynthesis}), and $\delta_X$ is the dust depletion of an element $X$ (Fig. \ref{fig abs dep}). If [Zn/Fe] is known, the sum $\alpha_X+\delta_X$ can be easily derived empirically from Eq. \ref{eq alpha delta} and does not depend on the assumptions made to isolate $\delta_X$. In Fig. \ref{fig int met} we show the total abundances of DLAs after correction for dust depletion and nucleosynthesis, only for systems with constrained column densities of Zn, Fe, and at least a third element. There is a good agreement in the total abundances of different metals.\footnote{We note that P has the largest scatter around the average metallicity. At low metallicities, 30\% of the estimated [P/H]$_{\rm tot}$ are slightly below the other [$X$/H]$_{\rm tot}$. In this regime the effects of depletion are minimal. At relatively high metallicities, 30\% of the [P/H]$_{\rm tot}$ estimated in DLAs are slightly above the [$X$/H]$_{\rm tot}$ for the other elements; 40\% of the estimated [P/H]$_{\rm tot}$ closely match the [$M$/H]$_{\rm tot}$ for the other metals, over a wide range of metallicities. We test, with negative results, whether a systematic overestimation of P column densities by $1\sigma$ in our measurements could influence these results. We interpret the under and overestimate of [P/H]$_{\rm tot}$ as originating from the observed scatter in the sequences of relative abundances.} This is an important cross-check because abundances of each metal in DLAs are now calculated independently, while our method was developed to use the observed sample as a whole. This confirms the robustness of our method. The dust- and nucleosynthesis-corrected relative abundances in the DLAs that we derive are typically solar. 
       \begin{figure}[!t]
   \centering
   \includegraphics[width=89mm,angle=0]{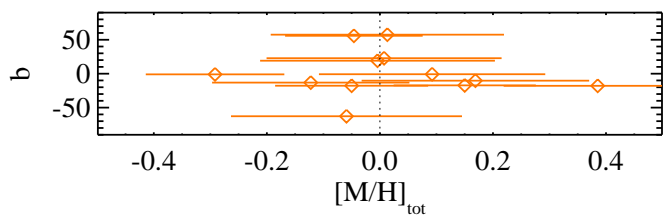}
   \caption{Total (dust-corrected) metallicity for Galactic clouds vs. their Galactic latitude.}
              \label{fig int met Gal}
    \end{figure}

We calculate the total (dust-corrected) metallicity [$M$/H]$_{\rm tot}$ as the average of the total abundances $[X/{\rm H}]_{\rm tot}$ over different elements $X$, for each system.  We consider the uncertainty on the metallicity to be the sum in quadrature of the mean of the errors on $[X/{\rm H}]_{\rm tot}$ and the standard deviation of the $[X/{\rm H}]_{\rm tot}$ distribution. The mean error on the average metallicity is 0.13~dex; the analogous analysis for the Galactic systems leads to an error on the average metallicity of 0.17~dex.

We calculate the total abundances in DLAs and in the Galaxy with the same procedure. The dust corrections are stronger in the case of the Galaxy because dust depletion is typically stronger at these metallicities than for most DLAs. The [$M$/H]$_{\rm tot}$ that we derive for the Galaxy are scattered around the solar metallicity, with large errors, as shown in Fig. \ref{fig int met Gal}.

\subsection{The elemental abundances in dust}
\label{epsilon}

 While the depletion $\delta_X$ is a relative measure of how much of an element $X$ is locked in dust grains, the overall abundance of an element in dust also depends on its cosmic abundance. For example, a small depletion of oxygen is sufficient to result in a large abundance of oxygen in dust because of the high cosmic abundance of this element. We define the elemental abundances in dust $\epsilon_X$ as the following:
 \begin{equation}
\begin{array}{lcl}
 
 \epsilon_X &=& \log \frac{N(X)_{\rm dust}}{N({\rm H})_{\rm tot}} = \\
            &=& \log \left( \frac{N(X)_{\rm dust}}{N({\rm X})_{\rm tot}} \times \frac{N(X)_{\rm tot}}{N({\rm H})_{\rm tot}} \right) = \\ 
            &=& \log  \left(  (1 - 10^{\delta_X}) \times 10^{\left(  [X/{\rm H}]_{\rm tot} + \log \left( \frac{N(X)}{N({\rm H})} \right)_\odot \right)}  \right) \mbox{,}
\end{array} 
 \label{eq epsilon}
 \end{equation}
         \begin{figure}[!t]
   \centering
      \includegraphics[width=90mm,angle=0]{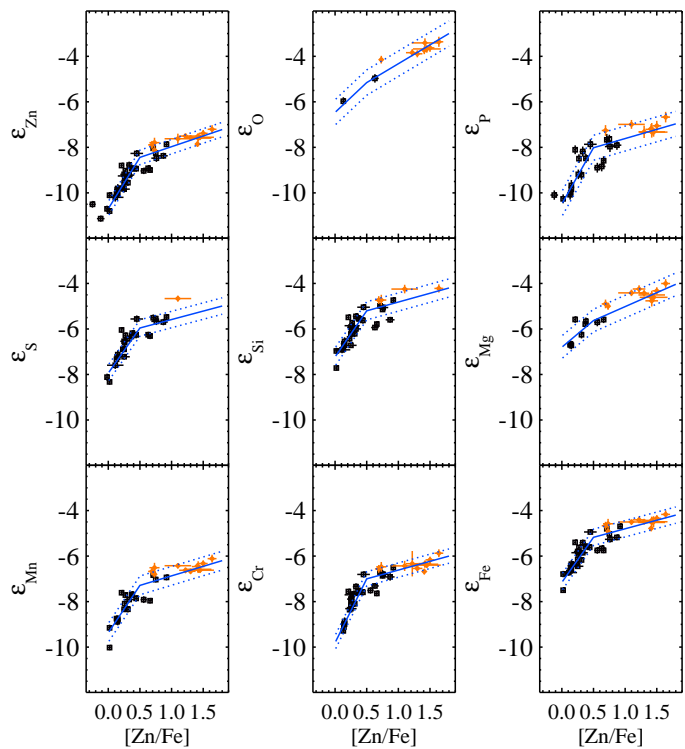}
   \caption{Elemental abundances in dust, $\epsilon_X$, for DLAs (black squares) and Galactic lines of sight (orange diamonds), as derived from Eq. \ref{eq epsilon}. The blue solid curves show a linear segment fit to the data, where the region within a standard deviation of the fit residuals is confined by the dotted curves. These uncertainties are also reported in Table \ref{tab epsilon patterns}.}
              \label{fig epsilon all}
    \end{figure}
 where $N(X)_{\rm dust}/N({\rm X})_{\rm tot} = dtm = (1 - 10^{\delta_X})$ (the dust-to-metal ratio defined in Sect. \ref{dtm}) and $[X/\mbox{H}]_{\rm tot} = \log \frac{N(X)_{\rm tot}}{N({\rm H})_{\rm tot}} - \log \left( \frac{N(X)}{N({\rm H})} \right)_\odot$ from fundamental relations. The value $[X/{\rm H}]_{\rm tot}$ is the total (dust-corrected) abundance of element $X$ in an individual system, and this can computed for each element and each DLA system as described in Sect. \ref{dust_corrected met}. Eq. \ref{eq epsilon} can in principle be simplified in the Galaxy, where $[X/{\rm H}]_{\rm tot}$ is zero. Nevertheless, we compute all $\epsilon_X$ in a consistent manner to avoid biasing our results\footnote{If we assume exact solar abundances $[X/{\rm H}]_{\rm tot}\equiv 0$ for all Galactic clouds, the distribution of the elemental abundances in dust for the Galaxy would be slightly flatter, but still mostly within the scatter of the trends observed otherwise. We do not make this assumption because we expect that Galactic abundances can be scattered around the solar values.}. 
  
\begin{figure*}[!ht]
   \centering
   \includegraphics[width=120mm,angle=0]{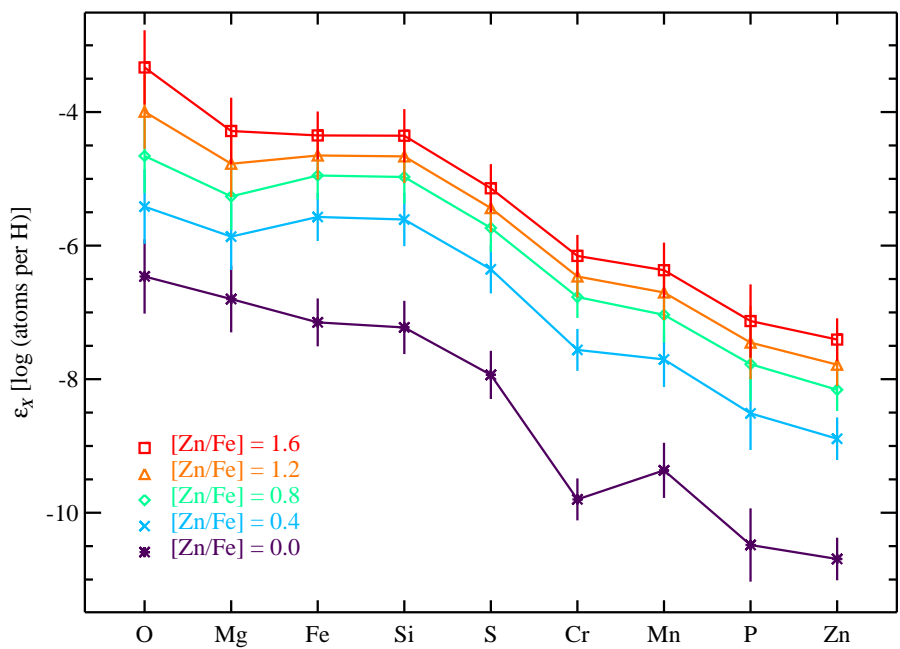}
 \caption{Elemental abundances in dust, $\epsilon_X$, for different depletion factors [Zn/Fe], listed in Table \ref{tab epsilon patterns}.}
              \label{fig epsilon patterns}
    \end{figure*}

 Figure \ref{fig epsilon all} shows the $\epsilon_X$ for the DLA and Galactic clouds, which are derived using the equation above. We fit two linear segment to the data to characterize the typical values of $\epsilon_X$ at different depletion factors. Given the number of free parameters and the paucity of data in some cases, such fits should only be considered as a way to estimate the typical values of $\epsilon_X$, avoiding the possible effects introduced by fitting a second-order polynomial or binning the data. The abundances in dust $\epsilon_X$ for five different values of [Zn/Fe] are shown in Fig. \ref{fig epsilon patterns} and listed in Table \ref{tab epsilon patterns}, where the errors are the standard deviation of the fit residuals. 

The elemental abundances in dust can be used to investigate the dust composition at different levels of depletion. The value$10^{\epsilon_X}$ is indeed in units of atoms of an element $X$ in dust per hydrogen atom. The main dust constituent among the elements considered here is oxygen, which is about ten times more abundant than the other metals for the most depleted systems. The other most important dust constituents are, Mg, Fe, and Si, with a similar abundance, and then S. On the other hand, Cr, Mn, P, and Zn play a minor role, although their relative abundances grow with [Zn/Fe]. For the main constituents, the relative abundances in dust (among different metals) evolve mildly with [Zn/Fe]. In general, we observe similar abundances of Mg, Si, and Fe in dust, i.e. ${\rm Mg: Si: Fe}\sim 1$. The similarity between $\epsilon_{\rm Si}$ and $\epsilon_{\rm Fe}$ shows that the elemental abundances in dust are more sensitive to the cosmic abundance of an element than to its depletion. The Mg: Si: Fe ratio may indicate that grain species such as pyroxenes (e.g. enstatite Mg SiO$_3$) and iron oxides (e.g. w\"ustite FeO) may be dominant, while olivine (fosterite Mg$_2$ SiO$_4$, fayalite Fe$_2$ SiO$_4$) may be less important. Sulfur seems about ten times less abundant in dust than Mg, Si, and Fe. We stress that the depletion of sulfur has been controversial \citep[see Sect. \ref{sulfur} and][]{Jenkins09}.  
  \begin{table*}
\centering
\caption{Elemental abundances in dust for different levels of [Zn/Fe] derived from the linear segments fit of Fig. \ref{fig epsilon all}.}
\begin{tabular}{ c | r r r r r r r r r }
\hline \hline
\rule[-0.2cm]{0mm}{0.8cm}
[Zn/Fe]  &   $\epsilon_{\rm Zn}$  &   $\epsilon_{\rm O}$ &  $\epsilon_{\rm P}$  &  $\epsilon_{\rm S}$  &  $\epsilon_{\rm Si}$  &  $\epsilon_{\rm Mg}$  &$\epsilon_{\rm Mn}$  &  $\epsilon_{\rm Cr}$ &  $\epsilon_{\rm Fe}$  \\
\hline

   0.00
 & $-10.69$ 
 & $ -6.46$ 
 & $-10.48$ 
 & $ -7.94$ 
 & $ -7.23$ 
 & $ -6.80$ 
 & $ -9.36$ 
 & $ -9.80$ 
 & $ -7.15$ 
\\
    0.40
 & $ -8.89$ 
 & $ -5.42$ 
 & $ -8.51$ 
 & $ -6.35$ 
 & $ -5.61$ 
 & $ -5.86$ 
 & $ -7.70$ 
 & $ -7.56$ 
 & $ -5.57$ 
\\
    0.80
 & $ -8.16$ 
 & $ -4.66$ 
 & $ -7.78$ 
 & $ -5.73$ 
 & $ -4.97$ 
 & $ -5.26$ 
 & $ -7.04$ 
 & $ -6.77$ 
 & $ -4.95$ 
\\
    1.20
 & $ -7.78$ 
 & $ -3.99$ 
 & $ -7.45$ 
 & $ -5.44$ 
 & $ -4.66$ 
 & $ -4.77$ 
 & $ -6.70$ 
 & $ -6.46$ 
 & $ -4.65$ 
\\
    1.60
 & $ -7.41$ 
 & $ -3.33$ 
 & $ -7.13$ 
 & $ -5.14$ 
 & $ -4.35$ 
 & $ -4.28$ 
 & $ -6.37$ 
 & $ -6.15$ 
 & $ -4.35$ 
\\

\hline
$\sigma_{\epsilon_X}$ & 0.32 &     0.56 &     0.55 &     0.36 &     0.40  &     0.50  &     0.41  &   0.31  &  0.36 \\

\hline \hline
 \end{tabular}
\label{tab epsilon patterns}
\end{table*}

For a more detailed study on the dust composition and evolution based on these elemental abundances, from low-metallicity DLAs to the Galaxy, see Mattsson et al. in preparation. Future observations could improve the accuracy of the derivation of the elemental abundances. One caveat is the contribution of carbon in the dust composition. Carbon is indeed one of the candidate carriers for the 2175 \AA{} bump observed in the extinction curve of the Galaxy and in a few DLAs \citep{Eliasdottir09,Ma15,Ledoux15}. C is not considered in our study because \cii{} typically has only few but strong lines in the rest-frame UV, so it is mostly saturated and therefore difficult to constrain. Carbonaceous grains could contribute significantly in the overall dust content of a galaxy.

Finally,  the elemental abundances in dust do not strongly vary with [Zn/Fe] relative to each other. On the other hand, we know that DLAs tend to have different extinction curves than the Galaxy \citep{Zafar11,Schady12}. This suggests that either the presence of carbon or the grain-size distribution is the main driver shaping different extinction curves. This supports the idea that carbonaceous grains are the likely carriers of the 2175 \AA{} bump observed in Galactic extinction curves.

\begin{table}
\small
\centering
\caption{The properties of the DLA sample derived from the depletion: total (dust-corrected) metallicity, dust-to-metal ratio (normalized by the Galactic value), and dust extinction.}
\begin{tabular}{@{} l @{\hspace{0.5mm}} | @{\hspace{1mm}} c @{\hspace{2mm}} c @{\hspace{2mm}} c @{\hspace{2mm}} c@{}}

\hline\hline
\rule[-0.2cm]{0mm}{0.8cm}
Quasar & [Zn/Fe] & [$M$/H]$_{\rm tot}$ & \dtm{}  & $A_V$\\

\hline
  Q0000-263 & $-0.12\pm0.06$
 & $-2.14\pm0.11$
 & $< 0.59$
            & ...
\\
  Q0010-002 & $-0.25\pm0.05$
 & $-1.49\pm0.13$
 & $< 0.42$
            & ...
\\
  Q0013-004 & $ 0.75\pm0.01$
 & $-0.40\pm0.10$
 & $ 0.92\pm0.03$
 & $ 0.11\pm0.03$
\\
  Q0058-292 & $ 0.23\pm0.03$
 & $-1.48\pm0.11$
 & $ 0.51\pm0.12$
 & $ 0.01\pm0.00$
\\
  Q0100+130 & $ 0.14\pm0.02$
 & $-1.62\pm0.11$
 & $ 0.34\pm0.15$
 & $ 0.01\pm0.00$
\\
 Q0102-190a       & $< 0.16$
            & ...
 & $ 0.39\pm0.14$
            & ...
\\
 Q0102-190b       & $< 1.69$
            & ...
 & $ 1.03\pm0.00$
            & ...
\\
 Q0135-273a       & $<-0.02$
            & ...
 & $< 0.68$
            & ...
\\
 Q0135-273b       & $<-0.13$
            & ...
 & $< 0.58$
            & ...
\\
 Q0216+080a & $ 0.25\pm0.06$
 & $-1.02\pm0.13$
 & $ 0.53\pm0.11$
 & $ 0.00\pm0.00$
\\
 Q0216+080b & $ 0.33\pm0.05$
 & $-0.61\pm0.15$
 & $ 0.64\pm0.09$
 & $ 0.02\pm0.01$
\\
  Q0336-017       & $< 0.38$
            & ...
 & $ 0.70\pm0.07$
            & ...
\\
  Q0347-383 & $ 0.63\pm0.06$
 & $-1.10\pm0.18$
 & $ 0.87\pm0.04$
 & $ 0.02\pm0.01$
\\
 Q0405-443a & $ 0.02\pm0.03$
 & $-1.08\pm0.12$
 & $ 0.06\pm0.22$
 & $ 0.00\pm0.01$
\\
 Q0405-443b & $ 0.15\pm0.04$
 & $-1.46\pm0.18$
 & $ 0.37\pm0.15$
 & $ 0.01\pm0.01$
\\
 Q0405-443c & $ 0.26\pm0.02$
 & $-1.14\pm0.12$
 & $ 0.56\pm0.11$
 & $ 0.02\pm0.01$
\\
 Q0405-443d       & $< 1.57$
            & ...
 & $ 1.02\pm0.00$
            & ...
\\
  Q0450-131       & $< 0.53$
            & ...
 & $ 0.82\pm0.05$
            & ...
\\
  Q0458-020 & $ 0.56\pm0.05$
 & $-1.15\pm0.16$
 & $ 0.83\pm0.04$
 & $ 0.13\pm0.06$
\\
 Q0528-250a & $ 0.25\pm0.04$
 & $-1.35\pm0.08$
 & $ 0.54\pm0.11$
 & $ 0.01\pm0.00$
\\
 Q0528-250b & $ 0.37\pm0.01$
 & $-0.90\pm0.12$
 & $ 0.69\pm0.08$
 & $ 0.09\pm0.03$
\\
  Q0551-366 & $ 0.71\pm0.04$
 & $-0.25\pm0.11$
 & $ 0.91\pm0.03$
 & $ 0.11\pm0.04$
\\
 Q0841+129b & $ 0.15\pm0.03$
 & $-1.61\pm0.11$
 & $ 0.38\pm0.15$
 & $ 0.00\pm0.00$
\\
 Q0841+129c & $ 0.13\pm0.04$
 & $-1.65\pm0.14$
 & $ 0.34\pm0.16$
 & $ 0.00\pm0.00$
\\
  Q0913+072       & $< 1.51$
            & ...
 & $ 1.02\pm0.00$
            & ...
\\
  Q1036-229       & $< 0.48$
            & ...
 & $ 0.78\pm0.06$
            & ...
\\
 Q1108-077a       & $< 1.40$
            & ...
 & $ 1.02\pm0.00$
            & ...
\\
  Q1111-152 & $ 0.25\pm0.09$
 & $-1.66\pm0.14$
 & $ 0.54\pm0.11$
 & $ 0.01\pm0.00$
\\
  Q1117-134 & $ 0.14\pm0.05$
 & $-1.46\pm0.11$
 & $ 0.35\pm0.15$
 & $ 0.00\pm0.00$
\\
  Q1157+014 & $ 0.31\pm0.02$
 & $-1.42\pm0.11$
 & $ 0.62\pm0.09$
 & $ 0.07\pm0.03$
\\
  Q1209+093 & $ 0.44\pm0.03$
 & $-0.98\pm0.13$
 & $ 0.75\pm0.06$
 & $ 0.09\pm0.03$
\\
  Q1223+178 & $-0.02\pm0.03$
 & $-1.64\pm0.14$
 & $< 0.68$
            & ...
\\
  Q1331+170 & $ 0.66\pm0.03$
 & $-1.14\pm0.12$
 & $ 0.89\pm0.03$
 & $ 0.04\pm0.01$
\\
 Q1337+113a       & $< 1.40$
            & ...
 & $ 1.02\pm0.00$
            & ...
\\
 Q1337+113b       & $< 0.57$
            & ...
 & $ 0.84\pm0.04$
            & ...
\\
  Q1340-136       & $< 0.55$
            & ...
 & $ 0.83\pm0.05$
            & ...
\\
 Q1409+095a & $ 0.11\pm0.13$
 & $-1.70\pm0.15$
 & $ 0.27\pm0.17$
 & $ 0.00\pm0.00$
\\
 Q1409+095b       & $< 1.29$
            & ...
 & $ 1.01\pm0.01$
            & ...
\\
  Q1444+014 & $ 0.87\pm0.07$
 & $-0.63\pm0.19$
 & $ 0.95\pm0.02$
 & $ 0.02\pm0.01$
\\
 Q1451+123a & $ 0.27\pm0.12$
 & $-1.00\pm0.15$
 & $ 0.57\pm0.10$
 & $ 0.01\pm0.00$
\\
 Q1451+123b       & $< 1.46$
            & ...
 & $ 1.02\pm0.00$
            & ...
\\
 Q1451+123c       & $< 1.98$
            & ...
 & $ 1.03\pm0.00$
            & ...
\\
 Q2059-360a       & $< 1.07$
            & ...
 & $ 0.99\pm0.01$
            & ...
\\
 Q2059-360b       & $< 0.42$
            & ...
 & $ 0.73\pm0.07$
            & ...
\\
  Q2116-358 & $ 0.45\pm0.10$
 & $-0.27\pm0.10$
 & $ 0.76\pm0.06$
 & $ 0.02\pm0.01$
\\
 Q2138-444a & $ 0.33\pm0.08$
 & $-1.07\pm0.11$
 & $ 0.64\pm0.09$
 & $ 0.01\pm0.00$
\\
 Q2138-444b & $ 0.02\pm0.03$
 & $-1.86\pm0.10$
 & $ 0.07\pm0.21$
 & $ 0.00\pm0.00$
\\
 Q2152+137b       & $< 0.70$
            & ...
 & $ 0.90\pm0.03$
            & ...
\\
 Q2206-199a & $ 0.21\pm0.02$
 & $-0.51\pm0.15$
 & $ 0.47\pm0.13$
 & $ 0.03\pm0.01$
\\
 Q2206-199b       & $< 0.46$
            & ...
 & $ 0.77\pm0.06$
            & ...
\\
  Q2243-605 & $ 0.27\pm0.02$
 & $-0.79\pm0.12$
 & $ 0.56\pm0.11$
 & $ 0.02\pm0.01$
\\
 Q2332-094a & $ 0.92\pm0.03$
 & $-0.20\pm0.14$
 & $ 0.97\pm0.02$
 & $ 0.03\pm0.01$
\\
 Q2332-094b       & $< 0.75$
            & ...
 & $ 0.92\pm0.03$
            & ...
\\
 Q2343+125a & $ 0.38\pm0.03$
 & $-0.86\pm0.10$
 & $ 0.70\pm0.08$
 & $ 0.01\pm0.00$
\\
 Q2359-022a & $ 0.76\pm0.08$
 & $-0.53\pm0.18$
 & $ 0.92\pm0.03$
 & $ 0.05\pm0.03$
\\
 Q2359-022b       & $< 1.43$
            & ...
 & $ 1.02\pm0.00$
            & ...
\\
\hline\hline

\end{tabular}
\label{tab znfe dtm met av}
\end{table}

\subsection{Dust-to-metal ratio}
\label{dtm}
\begin{figure}[b!]
   \centering
   \includegraphics[width=90mm,angle=0]{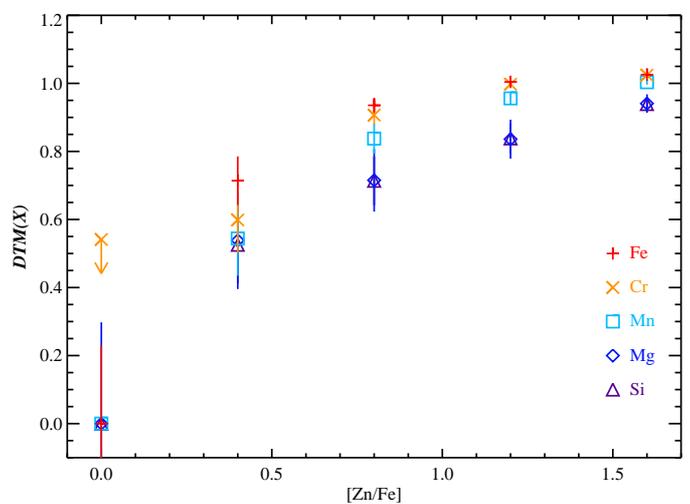}
   \caption{Dust-to-metal ratio derived using Eq. \ref{eq dtm} and the depletion of Si, Mg, Mn, Cr, and Fe reported in Table \ref{tab depl patterns}. \dtm{} based on non-refractory elements are typically $\sim$50\% of \dtm{}$_{\rm Fe, Cr}$ and are not recommended.}
              \label{fig dtm}
    \end{figure}

The dust-to-metal ratio $dtm$ can be defined as the fraction of a metal in the dust phase, written as  \begin{equation}
\begin{array}{lcl}
dtm =   \frac{N(X)_{\rm dust}}{N(X)_{tot}} &=& 1 - 10^{\delta_X}\mbox{,}\\
\mbox{\dtm{}} &=& dtm / dtm({\rm Gal})
\end{array}
\label{eq dtm}
\end{equation}
\citep[see][for further details]{DeCia13}, where \dtm{} is the dust-to-metal ratio normalized by the Galactic value $dtm({\rm Gal})=0.98$\footnote{Derived using Eq. C.3 of \citet{DeCia13}, which is based on the formalization of \citet{Jenkins09}, and assuming an average [Fe/Zn]$_{\rm Gal} = -1.22$, as we measure in the Galactic clouds studied by \cite{Jenkins09}} and the $\delta_X$ is the depletion of an element $X$ discussed above. This \dtm{} definition is not by mass but by column density ratios. For each element, we can calculate the expected value of the \dtm{} and its distribution with [Zn/Fe]. Only elements that are depleted into dust grains are suitable dust tracers for this analysis. Figure \ref{fig dtm} shows the \dtm{} derived using Eq. \ref{eq dtm} for the depletion patterns of  Si, Mg, Mn, Cr, and Fe (resulting from the fit of the depletion sequences, see Table \ref{tab depl patterns}). 

First, it is evident that the values of \dtm{} tend to increase with [Zn/Fe], and thus with the metallicity and metal column. This is strong evidence for the growth of dust grains in the ISM to be an important process of dust formation. This is because the amount of SN-produced dust does not have a strong dependence on metallicity, as discussed and modelled in \citet{Mattsson14}. The \dtm{} distribution for all elements converges to the Galactic value at high [Zn/Fe]. Also,  the values of \dtm{} derived from Fe and Cr are very similar and are significantly higher than the \dtm{} derived from Si and Mg, which trace each other almost perfectly. The measurements for Mn are somewhere in between. One possible explanation is that iron-group elements start to condensate in dust grains earlier than Si and Mg. The higher condensation temperature of Fe with respect to Si supports this idea. Measurements of \dtm{} based on Fe or Cr are recommended, when available. The \dtm{} can be calculated from the depletion of the non-refractory elements as well and its values are lower than the \dtm{} calculated for the refractory elements (roughly half of the \dtm{} based on Si). We do not recommend the use of the non-refractory elements alone to calculate the \dtm{}. 

       \begin{figure}[b!]
   \centering
   \includegraphics[width=90mm,angle=0]{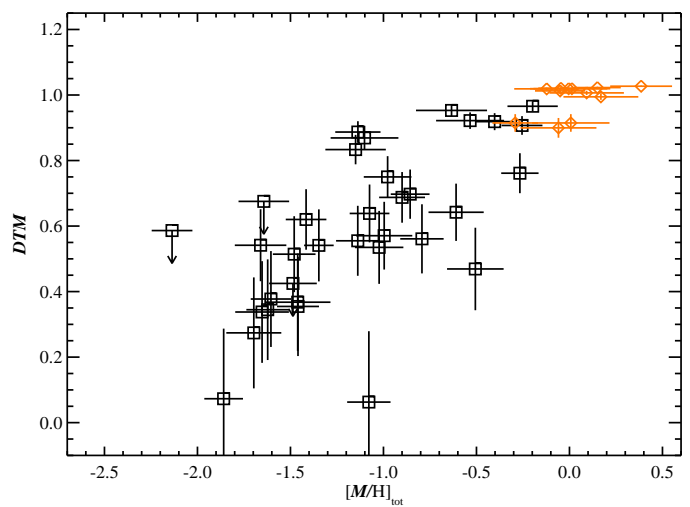}
   \caption{Distribution of \dtm{} with respect to the total metallicity. The \dtm{} values are derived from the observed depletion of Fe $\delta_{\rm Fe}$. Black squares are for DLAs, diamonds for Galactic absorbers.}
              \label{fig dtm obs}
    \end{figure}
We also derive the \dtm{} values directly from the observed depletion, rather than from the fit of the depletion sequences, choosing $\delta_{\rm Fe}$ as a reference to calculate \dtm{}. Figure \ref{fig dtm obs} shows the distribution of the observed \dtm{} with the total metallicity for DLAs and the Galaxy. We confirm that the \dtm{} and [$M$/H]$_{\rm tot}$ values for the Galactic absorbers cluster around the Galactic \dtm{} and solar metallicity, while the \dtm{} increases with metallicity for DLAs. The \dtm{} distribution is consistent with the values of \citet{DeCia13}, which were obtained with a different method that is based only on the observed [Zn/Fe] and assumes certain slopes of the depletion sequences of Zn and Fe. Similar \dtm{} trends have been found for GRB DLAs by \citet{Wiseman16}, who calculated the \dtm{} through a sum of the column densities of multiple metals.

\subsection{$A_V$ estimate}
\label{av}
The optical dust extinction $A_V$ can be derived from the observed abundances as follows:
\begin{equation}
\begin{array}{lcl}
\frac{A_V}{N_{\rm metals}} = \frac{dtm}{dtm{\rm (Gal)}}  \times \left(\frac{A_V}{N(H)}\right)_{\rm Gal} \mbox{,}\\
A_V = \mbox{\dtm{}} \times  \left(\frac{A_V}{N(H)}\right)_{\rm Gal} \times N(\mbox{\hi{}}) \times 10^{[M/{\rm H}]}\mbox{,}
\end{array}
\label{eq av}
\end{equation}
where $N_{\rm metals}$ is the overall column density of metals, $\left(\frac{A_V}{N(H)}\right)_{\rm Gal}=0.45 \times 10^{-21}$ mag cm$^2$ is the Galactic value from \citet{Watson11}, \dtm{} is the dust-to-metal ratio normalized by the Galaxy, and [$M$/H]$_{\rm tot}$ is the total (dust-corrected) metallicity discussed above. The derivation of the $A_V$ in Eq. \ref{eq av} is the inversion of the relation to calculate the \dtm{} from the observed $A_V$, as also described in \citet{Savaglio01} and \citet{Savaglio04}, with the difference that here we derive the \dtm{} from our depletion analysis. We do not assume $\left(\frac{A_V}{N(H)}\right)_{\rm Gal}$ for DLAs, but scale it using the \dtm{} derived from the observed depletion.

\begin{figure}
   \centering
   \includegraphics[width=90mm,angle=0]{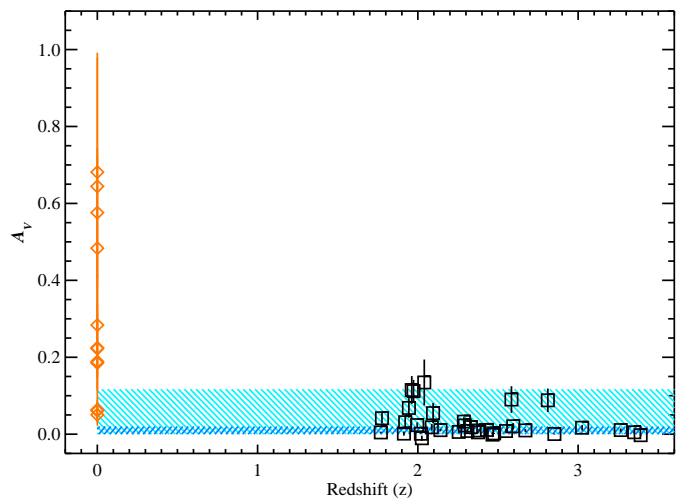}
   \caption{$A_V$ estimated from dust depletion using Eq. \ref{eq av}, with respect to the redshift. Black squares are for DLAs, diamonds for Galactic absorbers. The light blue shaded area shows the optical extinction estimates of \citet{Ellison05}, while the dark blue shaded area shows those from \citet{Vladilo08} and \citet{Khare12}.}
              \label{fig av}
    \end{figure}

The estimated $A_V$ are shown in Fig. \ref{fig av}. The values of extinction that we derive for QSO-DLAs are generally small ($A_V \lesssim 0.2$~mag) and mostly consistent with the $A_V$ estimated in the literature by studying the colours or extinction curves of DLAs, for example, $A_V \lesssim 0.12$~mag \citep{Ellison05} and $A_V \lesssim 0.02$~mag \citep{Vladilo08,Khare12}. The selection of QSOs is different among these samples, which have different spectral resolutions and thus require different QSO minimum brightnesses. Nevertheless, recent studies have tried to quantify this bias and find that DLAs towards reddened QSOs are intrinsically rare (less than 1\%; see Krogager et al. 2016, in review). Absorbing systems with neutral (\ci{}) or molecular gas, strong \mgii{} absorption, or prominent 2175 \AA{} features tend to have higher dust extinction than typical DLAs \citep{York06,Menard08,Srianand08,Noterdaeme09,Ledoux15}. 

We find that Galactic absorbers can have a higher extinction than DLAs, as expected from the higher levels of depletion observed in those systems. We compare the $A_V$ that we derive from the depletion observed in Galactic clouds with the $A_V$ measured from the reddening \citep[$A_V = E(B-V) \times R_V$, where we assume $R_V=3.08$ for the Galaxy;][]{Pei92}. In a few cases the $A_V$ derived from the $E(B-V)$ reported by \citet{Jenkins09} is higher than what we estimate from the depletion. One possibility to explain this discrepancy is that pure carbonaceous grains may in principle cause dust extinction, while not producing any signature in the relative abundances of the other metals. 

In the case of DLAs, a system-by-system comparison is not straightforward because the intrinsic colour of each QSO is not known a priori. \citet{Krogager16} measured a high depletion ([Zn/Fe] $=1.22$) from the absorption lines of a DLA towards a reddened quasar for which they estimated $A_V=0.28\pm0.07$ from the reddening in the continuum. For this case, using our Eqs. \ref{eq delta emp}, \ref{eq dtm}, and \ref{eq av}, we estimate $A_V=0.24\pm0.07$, which is in pretty good agreement with the extinction estimated from the reddening. \citet{Vladilo06} measured the $A_V$ from the spectral energy distribution (SED) of a few individual QSOs, three of which have [Zn/Fe] and metallicity estimates that we can use to compute the $A_V$ based on the measured depletion. They derived $A_V=0.16^{+0.04}_{-0.06}$ and $A_V=0.14^{+0.04}_{-0.06}$~mag from the SED for Q0013-004 and Q1157+014, respectively. These two QSOs are also part of our sample and the $A_V$ estimates are consistent; we measure $A_V=0.11\pm 0.03$ and $A_V=0.07\pm0.03$~mag from the depletion. J1323-0021 hosts a super-solar metallicity DLA at $z=0.716$, and while \citet{Vladilo06} found $A_V=0.44^{+0.07}_{-0.11}$~mag from the SED, we estimate $A_V\sim0.55$~mag from the depletion. In eight cases \citet{Vladilo06} estimated $A_V$ limits from the SED on individual DLAs. For seven out of these eight systems, we find consistent $A_V$ from the depletion; the exception is J1010+0003 for which we derive $A_V\sim0.21$~mag, but \citet{Vladilo06} found $A_V<0.13$. The comparison between the $A_V$ estimated from the depletion and the $A_V$ estimated from the reddening shows consistent results in the majority of suitable DLAs, and small deviations ($\sim0.1$~dex) for two cases. A more detailed comparison will be discussed in a future paper.

Caution should be exerted in using these results, at least until a direct comparison of $A_V$ that is derived with different methods is established. We stress that there may be dust species, such as carbon grains, which could produce significant extinction without altering the depletion properties considered in this work. The estimate of $A_V$ is carried out here only for a subset of our absorber sample, i.e. those with observed $N$(\hi{}), [Zn/Fe], [$M$/H]$_{\rm tot}$ and constrained \dtm{}. In principle it is possible to extend this work by making assumptions on [Zn/Fe] given other observed abundances, which is beyond the scope of this paper. 

\section{Summary and conclusions}
 \label{conclusions}
By studying the relative abundances of a large sample of QSO-DLAs and Galactic absorbers, which are all observed at high resolution, we show homogeneous properties of dust depletion in the ISM, all the way from low-metallicity DLAs to Galactic clouds. The observed abundances of refractory metals relative to non-refractory elements are strongly correlated with [Zn/Fe]. The slopes of these sequences of relative abundances depend on how strongly a metal tends to deplete into dust grains (its condensation temperature). Thus, dust depletion is the factor that shapes these relations.

The mere existence of such sequences implies that nucleosynthesis effects are much smaller than the effects that dust depletion can have on observed elemental abundances (up to 2 dex). Moreover, it implies that the depletion of dust is primarily controlled by ISM processes and not as much by stellar yields. This suggests that grain growth in the ISM is an important process of dust production. The sequences of relative abundances are continuous in [Zn/Fe] from DLAs to Galactic clouds. This suggests that the availability of refractory metals in the ISM is crucial to regulate dust production, regardless of the star formation history. 

We observe [S/Zn] up to $\sim0.24$~dex in DLAs, using the new oscillator strengths for \znii{} and \suii{} lines \citep[$0.14$~dex higher than in previous works;][]{Kisielius14,Kisielius15}. These results indicate a broad consistency between the nucleosynthesis of S in DLAs and in the Galaxy.

We observationally constrain the nucleosynthetic over- or underabundances of O, P, S, Si, Mg, O, and Mn in the least dusty DLAs ($\alpha_{\rm O,0} = 0.38\pm 0.10$, $\alpha_{\rm P,0} = 0.00\pm 0.06$, $\alpha_{\rm S,0} = 0.25\pm 0.03$, $\alpha_{\rm Si,0} = 0.26\pm 0.03$, $\alpha_{\rm Mg,0} = 0.30\pm 0.04$, and $\alpha_{\rm Mn,0} = -0.39\pm 0.03$). Remarkably, these values are consistent with the nucleosynthetic patterns (stellar [$X$/Fe]) observed in the Galaxy. These results suggest the presence of a young stellar population in these DLA galaxies, and thus a recent or on-going star formation.

We make two assumptions to convert the observed relative abundances into dust depletions: first, a mild trend of $\delta_{\rm Zn}$ from a fit to the observed data and, second, a distribution of nucleosynthetic over- or underabundances from literature studies on Galactic stars, but using the observations of DLAs at [Zn/Fe] $=0$. These are the two main assumptions used in this paper, and they can be improved if better observational constraints are available in the future. Nevertheless, these corrections are bound by the observations to be small and much weaker than the effects dust depletion can have. Therefore, these assumptions do not significantly affect the results of this paper.

We derive the dust depletion $\delta_X$ of Zn, O, P, S, Si, Mg, Mn, Cr, and Fe. The depletions $\delta_X$ are also strongly correlated with [Zn/Fe], and can be described with continuous functions of [Zn/Fe], from DLAs to the Galactic ISM. By slicing these depletion sequences, we find depletion patterns at five levels of [Zn/Fe]. We recommend using these canonical depletion patterns as a reference for the study of relative abundances and depletion. P is found to be the least depleted element into dust grains, but its measurements are the most uncertain. Therefore we recommend using this element as a first metallicity indicator, when available, in combination with Zn and O. 

Furthermore, we provide a prescription on how to derive the total (dust-corrected) abundances [$X$/H]$_{\rm tot}$ and metallicity [$M$/H]$_{\rm tot}$, the dust-to-metal ratio \dtm{}, and dust extinction $A_V$ based on the depletion, and thereby characterize the metal and dust properties of DLAs and Galactic absorbers. The QSO-DLAs have total metallicities spanning from less than 1\% of solar to nearly solar metallicity. Galactic absorbers, on the other hand, cluster around solar metallicity. We find that accurate corrections for dust depletion are important for DLAs and for the Galaxy. The \dtm{} increases with [Zn/Fe] and metallicity, and flattens around the Galactic value close to solar metallicity. This again indicates that grain growth must be an important process of dust production. Given the values of \dtm{} derived from the depletion of different elements, Fe and Cr are more sensitive tracers of dust than Si and Mg. The $A_V$ derived from the depletion properties is typically $<0.2$~mag in QSO-DLAs, while Galactic clouds often show higher amounts of dust.

Finally, we present a powerful technique to extract useful information on the dust composition from the observed relative abundances. We derive elemental abundances in dust $\epsilon_X$ from the depletion and the total (dust-corrected) metallicity, for given values of [Zn/Fe]. We find that that Mg, Fe, and Si are about equally abundant in dust grains, at all levels of [Zn/Fe]. They are the main metallic constituents of dust grains, after O (and possibly C, which is not covered in this paper). The small variation of the relative $\epsilon_X$ in different environments suggests that carbonaceous grains and/or the grain-size distribution is responsible for the observed variety in extinction curves between DLAs and the Galaxy, i.e. their slope and the extent of the 2175 \AA{} bump. Our results constrain the possible mixes of dust grains, suggesting that pyroxene and iron oxides are more important dust species than olivine. For a more detailed study of dust composition and its evolution, based on these elemental abundances in dust, see Mattsson et al., in preparation.

In this paper we use [Zn/Fe] as a proxy for dust, and our results rely on the observed relative abundances. While we make no formal use of the depletion strength factor $F_*$ of \citet{Jenkins09}, we note that $F_*$ is proportional to [Zn/Fe] and is expected to be negative for most DLAs. 

This study presents a unified picture of dust depletion from QSO-DLAs to the Galaxy. We separate the effects of dust depletion, nucleosynthesis, and metallicity, and provide important guidelines to characterize the metal and dust properties of galaxies down to low metallicity and at high redshift.

\begin{acknowledgements} 
We thank the anonymous referee for a detailed, constructive, and useful report. We thank Patricia Schady, Trystyn Berg, Sara Ellison, Palle M{\o}ller, and Wolfgang Kerzendorf for insightful discussions. We thank Dan Welty for alerting us on the discrepancies in MC literature studies. ADC acknowledges support by the Weizmann Institute of Science Dean of Physics Fellowship and the Koshland Center for Basic Research. This research has made use of NASA's Astrophysics Data System.
\end{acknowledgements}

\bibliographystyle{aa} 

\bibliography{biblio}

\newpage
\appendix

\section{On the use of [Zn/Fe] as dust tracer}
\label{sect znfe}
Our study relies on [Zn/Fe] as a tracer of dust depletion. This depends on, first, how representative the depletion of Fe is for the bulk of the dust and, second, whether Zn and Fe have a similar nucleosynthetic history, i.e. how well the abundances of Zn and Fe trace each other. Because Fe is a highly refractory and abundant element, [Zn/Fe] traces most observed mixes of dust grains with the potential exception of amorphous carbonaceous grains \citep[see][and Mattsson et al. in prep. for further discussions]{DeCia13}. 

Regarding the nucleosynthesis of Zn and Fe, \citet{Sneden91} observed that [Zn/Fe] $\sim0$ for metal-poor stars with $-3\lesssim [M/{\rm H}] \lesssim 0$. \citet{Barbuy15} and \citet{Saito09} showed that [Zn/Fe] is close to solar for $-2\lesssim [M/{\rm H}] \lesssim 0$. Outside this metallicity range they find that [Zn/Fe] deviates strongly from solar, to positive values for very metal-poor stars \citep[see also][]{Primas00}, and to negative [Zn/Fe] for metal-rich red giants. The results presented here are mostly for the metallicity range where stellar [Zn/Fe] are solar (Zn and Fe trace each other). We recommend using caution and parallel techniques to those presented here, when using [Zn/Fe] outside the metallicity range $-2\lesssim [M/{\rm H}] \lesssim 0$. 

The fact that Zn overall traces Fe is also supported by the observation that [Zn/Fe] correlates with the metallicity in DLAs; see Fig. \ref{fig met znfe}, which suggests that the higher the availability of metals the higher is the amount of metals locked into dust grains. This correlation has a large scatter, which is not unexpected, given that [$M$/H] also depends on the hydrogen column density, and that this is not directly related to the formation of dust. The observed trend of [Zn/Fe] with [$M$/H] converges to zero for low-metallicity DLAs, again suggesting no nucleosynthesis deviations between Zn and Fe. 

Moreover, we observe tight correlations between [$X/Y$] and [Zn/Fe], where $Y$ is a non-refractory element for several elements $X$ (see Figs. \ref{fig znfe} and \ref{fig znfe non depleted}). These correlations are due to dust depletion and they could not have such narrow scatter if Zn would not trace Fe overall. When fitting these correlations we allow for a measure of intrinsic scatter, which can account for $\lesssim0.2$ dex due to nucleosynthesis peculiarity of Zn relative to Fe.

If the intrinsic reference [Zn/Fe] were not solar, then this could in principle affect our estimates of the nucleosynthesis effects, but would not ultimately affect our results on the depletion. Indeed, we use the observed relative abundances of DLAs at [Zn/Fe] $=0$ (where we expect no depletion) to estimate the nucleosynthesis effects (Fig. \ref{fig nucleosynthesis}). However, \citet{Barbuy15} found [Zn/Fe] $=0.24 \pm 0.02$ in Galactic bulge field red giants with $-1.3 < [{\rm Fe/H}] < -0.5$. If this would be the true "zero-depletion" reference level, then our estimates of the nucleosynthesis effects $\alpha_X$ would be slightly lower, i.e. lower by $0.02$,  $0.04$, $0.14$, $0.13$, and $0.10$~dex for O, S, Si, Mg, and Mn, respectively (calculated by shifting the sequences of relative abundances by 0.24~dex). However, the observed sequences of relative abundances extend all the way down to [Zn/Fe] $=0$. Therefore we do not see any evidence in DLAs for super-solar [Zn/Fe] at zero-depletion. One possible explanation for the apparent discrepancy between the [Zn/Fe] measurements of \citet{Barbuy15} and the DLA observations is an observational bias. In the \citet{Barbuy15} study, only Galactic bulge field red giants are considered, where neither gas nor star formation plays a role, while the ISM in DLAs has been reprocessed with metals coming from a larger variety of stellar populations.

\begin{figure}
   \centering
   \includegraphics[width=90mm,angle=0]{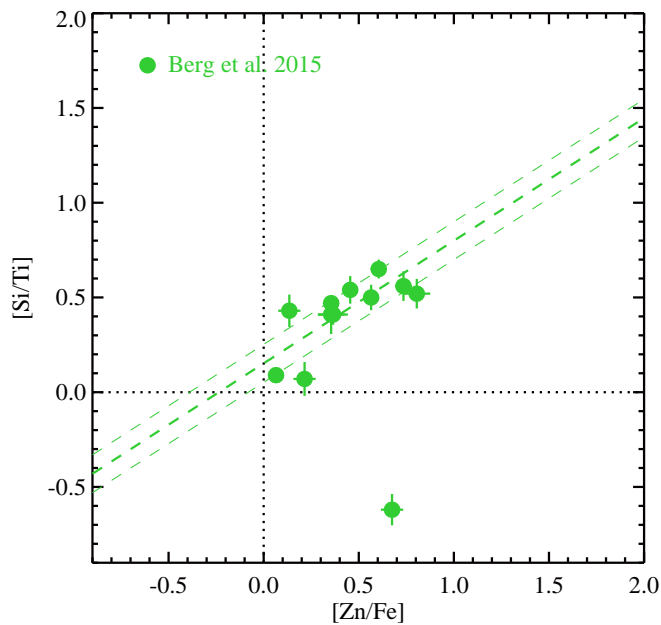}
   \caption{Correlation between the [Si/Ti] and [Zn/Fe], using the measurements of \citet{Berg15}, illustrate the effect of dust depletion. Si and Ti are both $\alpha$ elements, but Ti is more refractory than Si. The measurements are not expected to lie on the one-to-one (dotted) line. The slope depends on the relative depletion of Si with respect to Ti, and, on the other hand, on the relative depletion of Zn with respect to Fe, which are independent from each other. The outlying data point is Q 0013-004 (SDSS 0016-0012), for which \citet{Herbert06} remarked that the Ti measurement is not reliable and likely a misunderstanding (velocity profile spanning $\sim 1000$ km~s$^{-1}$).}
              \label{fig SiTi}
    \end{figure}

\citet{Berg15} suggested that, based on the argued non-correlation of [Si/Ti] and [Si/Ca] with [Zn/Fe],  Zn does not trace Fe in nucleosynthesis. However, when only constrained measurements (no upper/lower limits) from their analysis are included, then the relative abundances [Si/Ti] and [Si/Ca]  show a correlation with [Zn/Fe], the slope of which depends on the refractory properties of the metals involved. Figure \ref{fig SiTi} shows this correlation, using the measurements compiled by \citet{Berg15}. The linear fit (including $x$ and $y$ error) is [Si/Ti] $ = 0.15 +  0.65 \times$ [Zn/Fe]. We argue that if there were a large intrinsic scatter in the relations between the considered relative abundances, this should be already evident from the constrained measurements. The tension therefore seems to arise from inconsistent literature estimates of column density upper limits in the case of non-detected lines, in particular for weak lines such as for Ti. In Appendix \ref{sect limits} we discuss the (un)reliability of upper and lower limits. 

Finally, \citet{Berg15} argue that [Zn/Fe] in DLAs may in fact be subsolar, similar to dwarf spheroidal galaxies \citep[$-0.8\leq {\rm [Zn/Fe]} \leq0.2$][]{Shetrone03, Sbordone07}. However, if DLAs had such a wide range of nucleosynthetic subsolar [Zn/Fe], then we could not observe the tight sequences of relative abundances (Fig. \ref{fig znfe}). Two DLAs however show a slightly negative [Zn/Fe] $=-0.12\pm0.06$ and  $-0.25\pm0.05$. These relative abundances of these two DLAs lie on the sequences of relative abundances. Thus, we cannot exclude that (some) DLAs may have a slightly subsolar [Zn/Fe]. This effect however cannot be stronger than $-0.2$ dex. We cannot further constrain the reality of this effect, given that this is based on only two data points. Nevertheless, strong negative values of [Zn/Fe] are not observed in DLAs, including dust-free measurements. Therefore the abundances observed in DLAs are not overall similar to dwarf spheroidals.

While we use [Zn/Fe] as a tracer of dust, we remark that [Zn/Fe] correlates with [Si/Ti] (Fig. \ref{fig SiTi}), [Mg/S], and [Si/S] (Fig. \ref{fig znfe}). Si, Ti, Mg, and S are all $\alpha$ elements, and therefore their relative abundances are not influenced by nucleosynthesis effects. There is no evidence for [Zn/Fe] overabundance in the DLA data, i.e. there is no offset observed in the correlations between [Zn/Fe] and [Si/Ti] or [Mg/S], or [Si/S]. The use of [Zn/Fe] as a tracer of dust is therefore further supported by its relation with other dust tracers, such as [Si/Ti], [Mg/S], and [Si/S].

We therefore recommend the use of [Zn/Fe] as a tracer of dust depletion in the metallicity range $-2\lesssim [M/{\rm H}] \lesssim 0$, keeping in mind that small ($\lesssim0.2$ dex) deviations might be due to nucleosynthesis of Zn and Fe. The observed relative abundances of different metals at zero depletion show no evidence of such deviations in DLAs.

\section{On the exclusion of upper and lower limits on the abundances}
\label{sect limits}

In this paper we present the relative abundances of several metals, which are derived from the measurements of the column densities through Voigt-profile fitting of absorption lines in the spectra. Our analysis and results include only the constrained measurements, and exclude upper and lower limits. We choose to do so for several reasons. First, the limits that we estimate for our sample are mostly unconstraining. They are derived from the spectra of undetected weak lines, when strong ones are not available, or saturated lines. The vast majority of the limits are consistent with the observed correlations of relative abundances, and there is no real tension between limits and constrained values.They are shown in Figs. \ref{fig znfe} and \ref{fig znfe non depleted} for completeness.

Moreover, in general limits are most often calculated with techniques that are different and independent from the measurements of the detected lines, such as equivalent width (EW) measurement of the noise continuum. This is not a straightforward measurement because it depends on the oscillator strength of one individual selected available line, where the stronger line available provides the more constraining limit. Besides, the EW strongly depends on the aperture used for the integration. The aperture should in principle match the line profile of the detected lines, but often this is not the case in the literature. Limits derived from Voigt-profile fitting techniques are often too conservative because of error underestimations. In general, limits are less reliable than the constrained measurements, which are based on the simultaneous detection and modelling of several absorption lines of the same ion. While we can monitor the determination of lower and upper limits in our measurements, we do not consider such estimates given in the literature as reliable.

\section{On the nucleosynthetic overabundance of oxygen}
\label{sect oxygen}
In Sect. \ref{sec nucleo corr} we find a good agreement between the oxygen overabundance $\alpha_{\rm O}$ in DLAs and the stellar measurements of \citet{Lambert87}, \citet{McWilliam97}, and \citet{Nissen97}. However, these values of $\alpha_{\rm O}$ in DLAs are lower (by $\sim0.2$~dex) than the values that have been measured by \citet{Jonsell05} also in Galactic stars. \citet{Garcia06} has shown that measuring O abundances from the IR triplet produces higher (by $\sim0.19$~dex) abundances than when using the forbidden 6300 \AA{} line, or the OH UV molecular lines. The O estimates are further complicated by LTE/NLTE effects. Moreover, \citet{Nissen02} has shown that applying hydrodynamical models that include stellar granulation produces lower O abundances, extending to [O/Fe] $\simeq$ 0.5 at [Fe/H] $=-2.5$. Thus, the discrepancy that we see between the oxygen abundances in DLAs and some measurements in the Galaxy relates to an open debate on the oxygen abundances at low metallicities for stars in the Galaxy, which may involve technical issues that are not settled yet. For DLAs, we rely on the observed zero intercept of the [O/Zn] versus [Fe/Zn] relation, which yields $\alpha_{\rm O}=0.38\pm0.1$. This value is consistent with the oxygen measurements for metal-poor DLAs of \citet{Becker12}. If the real underlying $\alpha_{\rm O, 0}$ in DLAs were higher than what we measure by $\sim0.2$~dex, then we would derive a similar depletion of oxygen at high [Zn/Fe] (Fig. \ref{fig abs dep}), but with a $\sim0.2$~dex zero intercept and a slightly flatter slope. This would mostly affect the abundances of oxygen in dust $\epsilon_{\rm O}$ for systems with [Zn/Fe] $<0.5$ and will be discussed in Mattsson et al., in preparation. Until the debate on the oxygen abundances for low-metallicities stars in the Galaxy has been settled, we cannot further discuss the possibility of a different $\alpha_{\rm O, 0}$ .

\section{Comparison with the depletion in the Magellanic Clouds}
\label{sect comparison MC}

We compare some of our results with the recent work of \citet{Tchernyshyov15} on the abundances and dust depletion in the Small Magellanic Cloud (SMC) and Large Magellanic Cloud (LMC). We test whether the observed relative abundances in the SMC and LMC are consistent with our sequences of relative abundances. We find a good agreement for [Si/Zn] and [Cr/Zn] versus [Fe/Zn], i.e. the SMC and LMC data points line up along these sequences. The sequences of relative abundances observed in DLAs, the Galaxy, and the Magellanic Clouds are thus consistent. This supports the scenario in which these sequences are produced by depletion of dust grains, which are growing in the ISM of a galaxy and hardly depend on the star formation history, but rather on the availability of metals in the gas.

The [Si/P] and [Cr/P] estimates for the SMC and LMC tend to lie above the expected relation with [Zn/Fe], roughly by a few times 0.1~dex. We find weaker depletion of P than has been derived by \citet{Tchernyshyov15} for the SMC. However, the observed values of [Zn/P] in the SMC are below 0.2~dex. These values argue against a much stronger depletion of P with respect to Zn, unless significant nucleosynthesis effects are at play. We find that a third of the $N$(\hi{}) in their FUSE LMC sample should be upper limits (e.g. towards Sk-70 69, Sk-69 104, etc.), but are reported (and presumably treated) like constrained measurements in \citet{Tchernyshyov15}. Moreover,  the depletion of P derived by \citet{Tchernyshyov15} using their Eq. 6 is 0.4~dex too low for the SMC, and 0.4~dex too high for the LMC because the reference P abundances for the SMC (4.7) and LMC (5.1) have been swapped, as also confirmed by Tchernyshyov (private communication). Taking this into account likely solves the apparent discrepancy on the P depletion.

\newpage
\section{Velocity profiles}
\label{sect velocity profile}
\begin{figure}[h!]
\includegraphics[bb=61 238 402 767,clip,width=8.7cm]{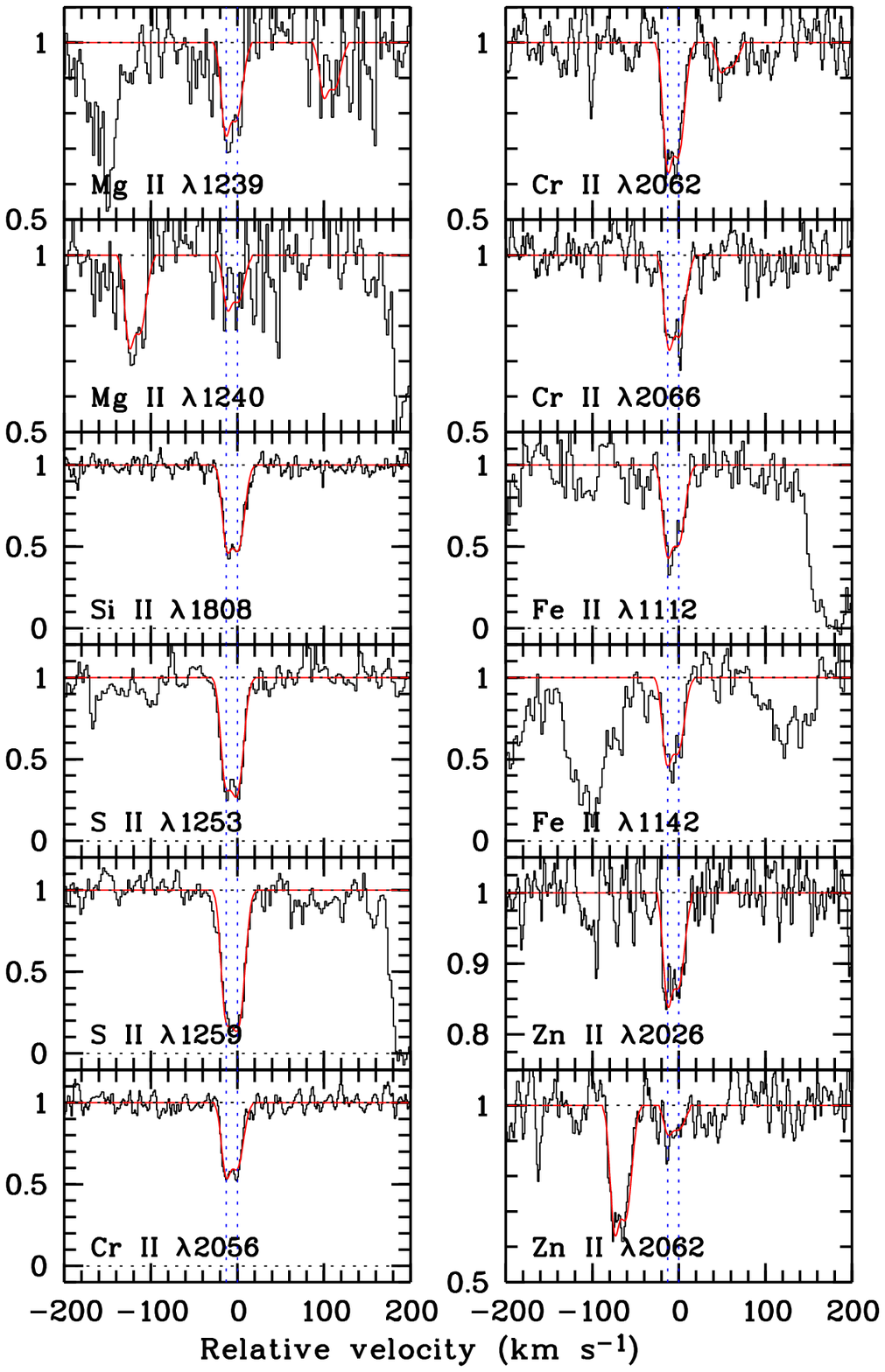}
\caption{Velocity profiles of selected low-ionization transition lines from the
DLA system at $z_{\rm abs}=2.025$ towards Q\,0010$-$002.}
\label{q0010_2.025}
\end{figure}

\begin{figure}
\includegraphics[bb=61 155 402 767,clip,width=8.7cm]{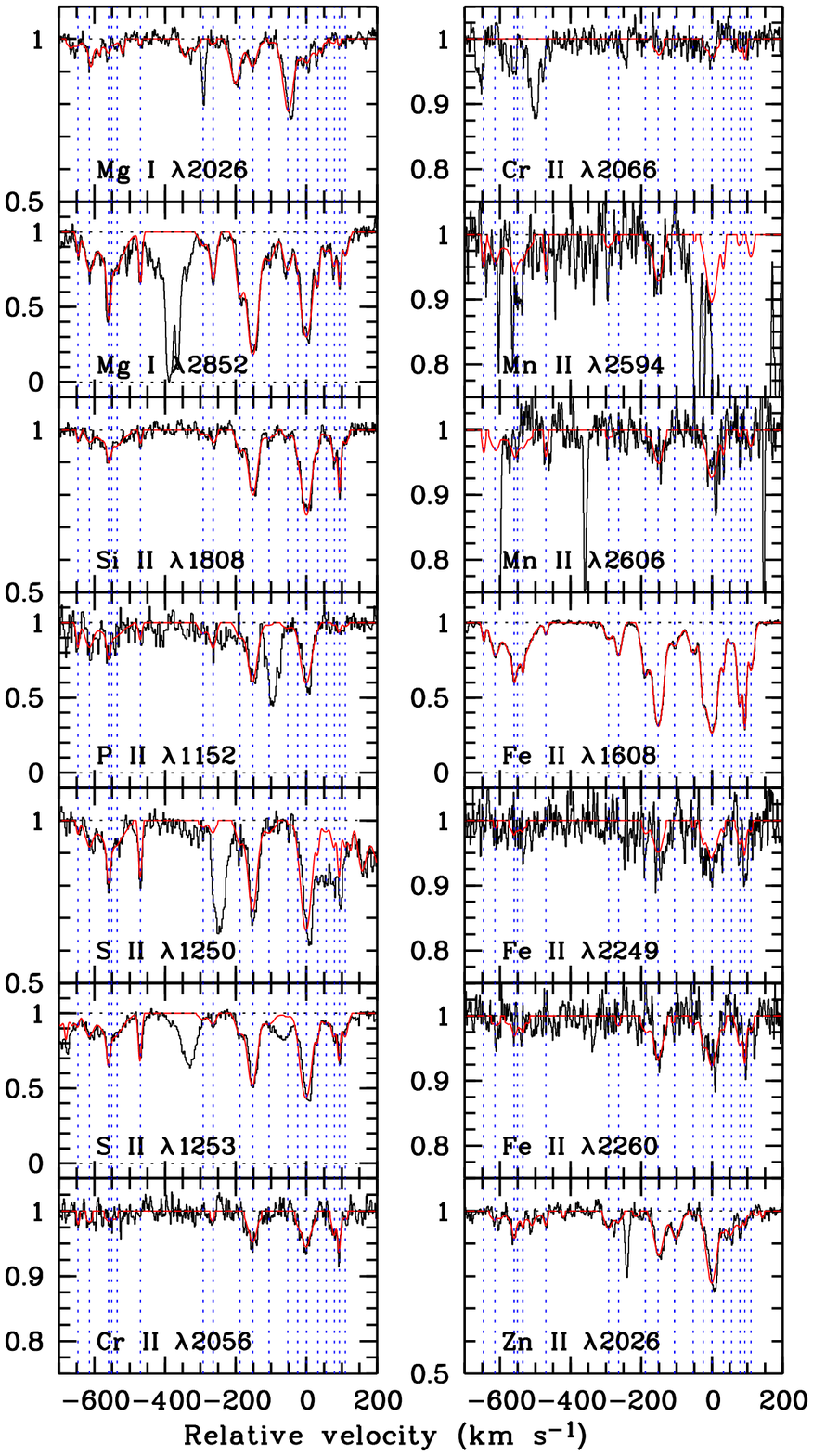}
\caption{Velocity profiles of selected low-ionization transition lines from the
DLA system at $z_{\rm abs}=1.973$ towards Q\,0013$-$004.}
\label{q0013_1.973}
\end{figure}

\clearpage

\begin{figure}
\includegraphics[bb=61 320 402 767,clip,width=8.7cm]{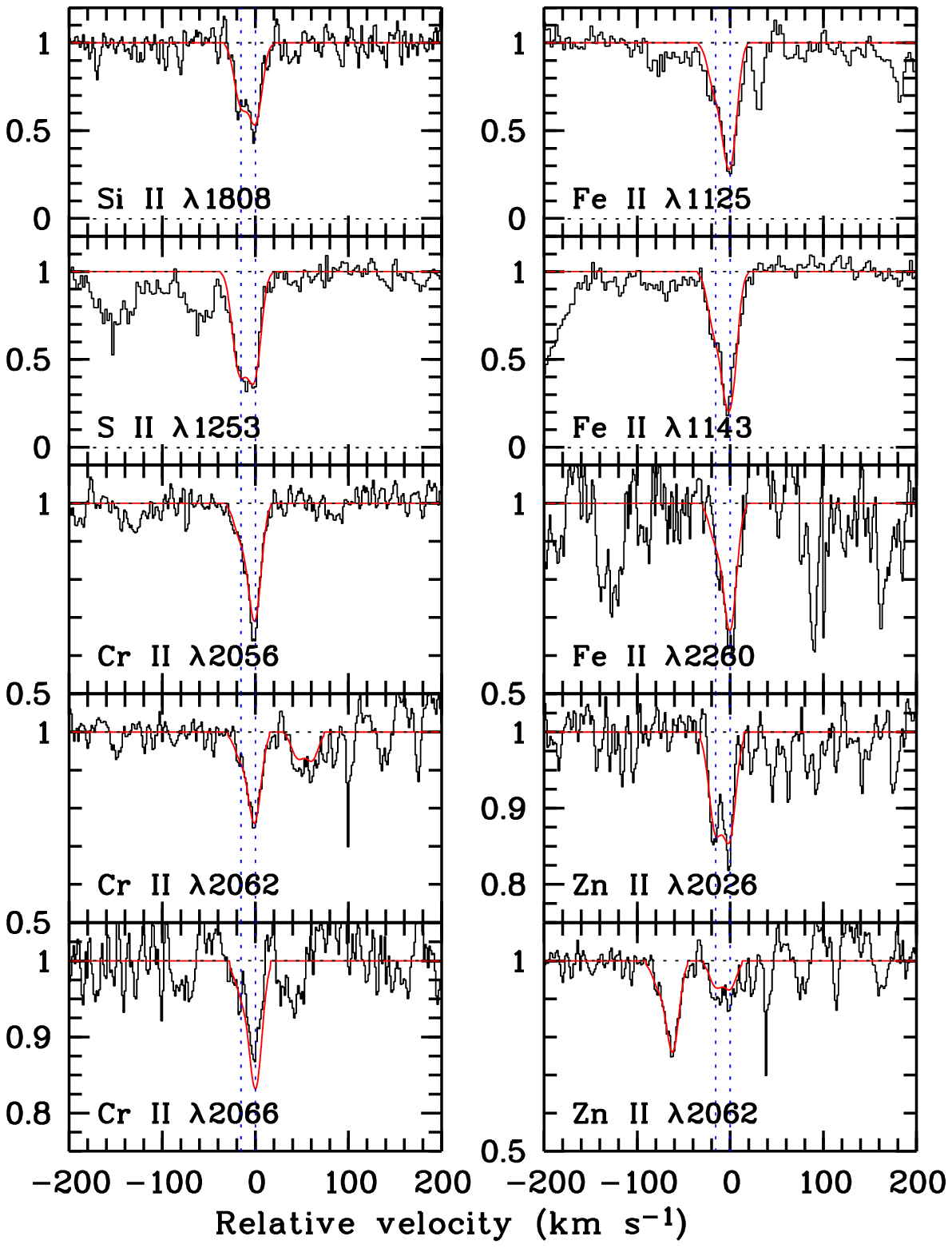}
\caption{Velocity profiles of selected low-ionization transition lines from the
DLA system at $z_{\rm abs}=2.671$ towards Q\,0058$-$292.}
\label{q0058_2.671}
\end{figure}

\begin{figure}
\includegraphics[bb=61 155 402 767,clip,width=8.7cm]{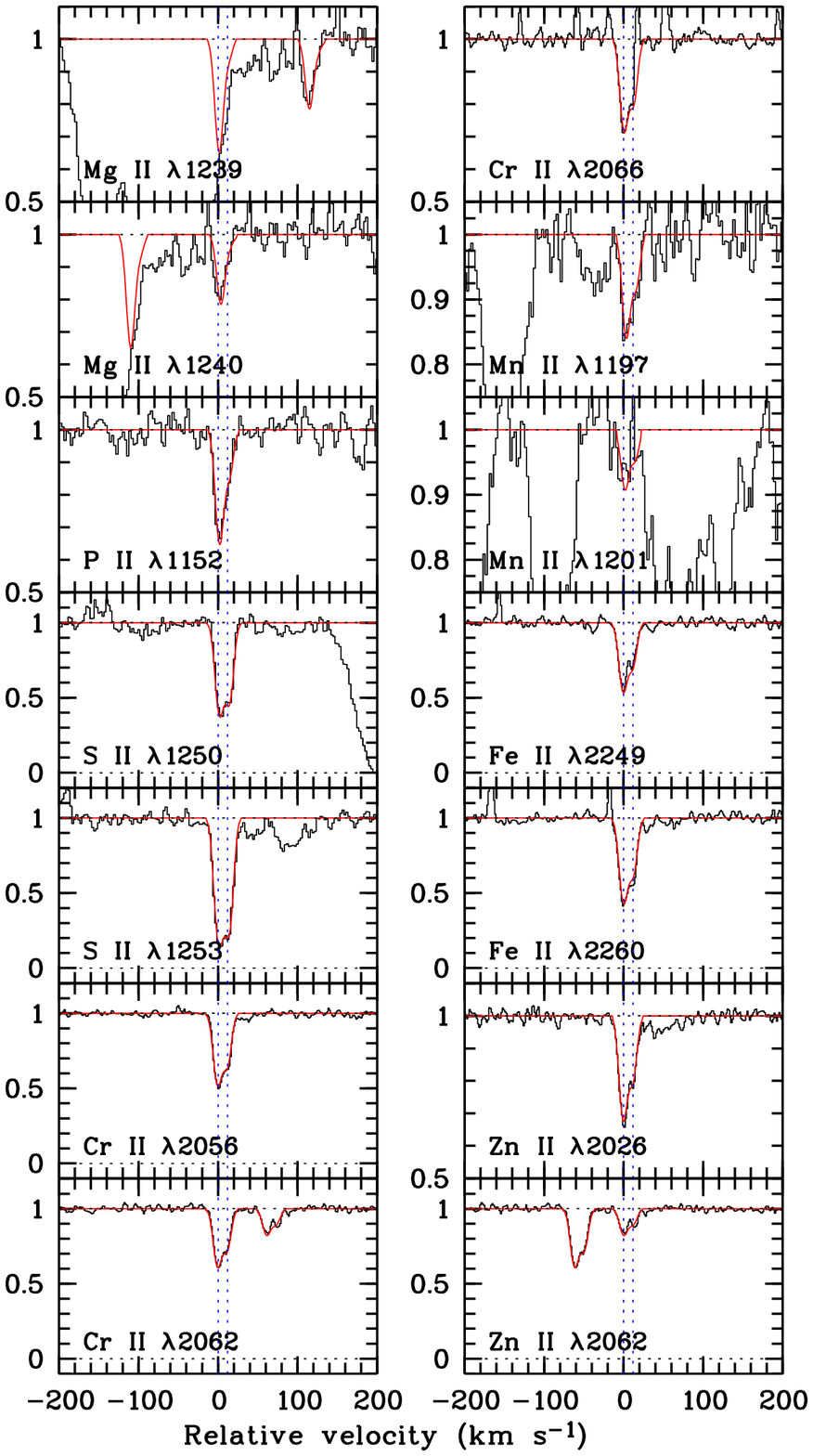}
\caption{Velocity profiles of selected low-ionization transition lines from the
DLA system at $z_{\rm abs}=2.309$ towards Q\,0100$+$130.}
\label{q0100_2.309}
\end{figure}

\clearpage

\begin{figure}
\includegraphics[bb=61 320 402 767,clip,width=8.7cm]{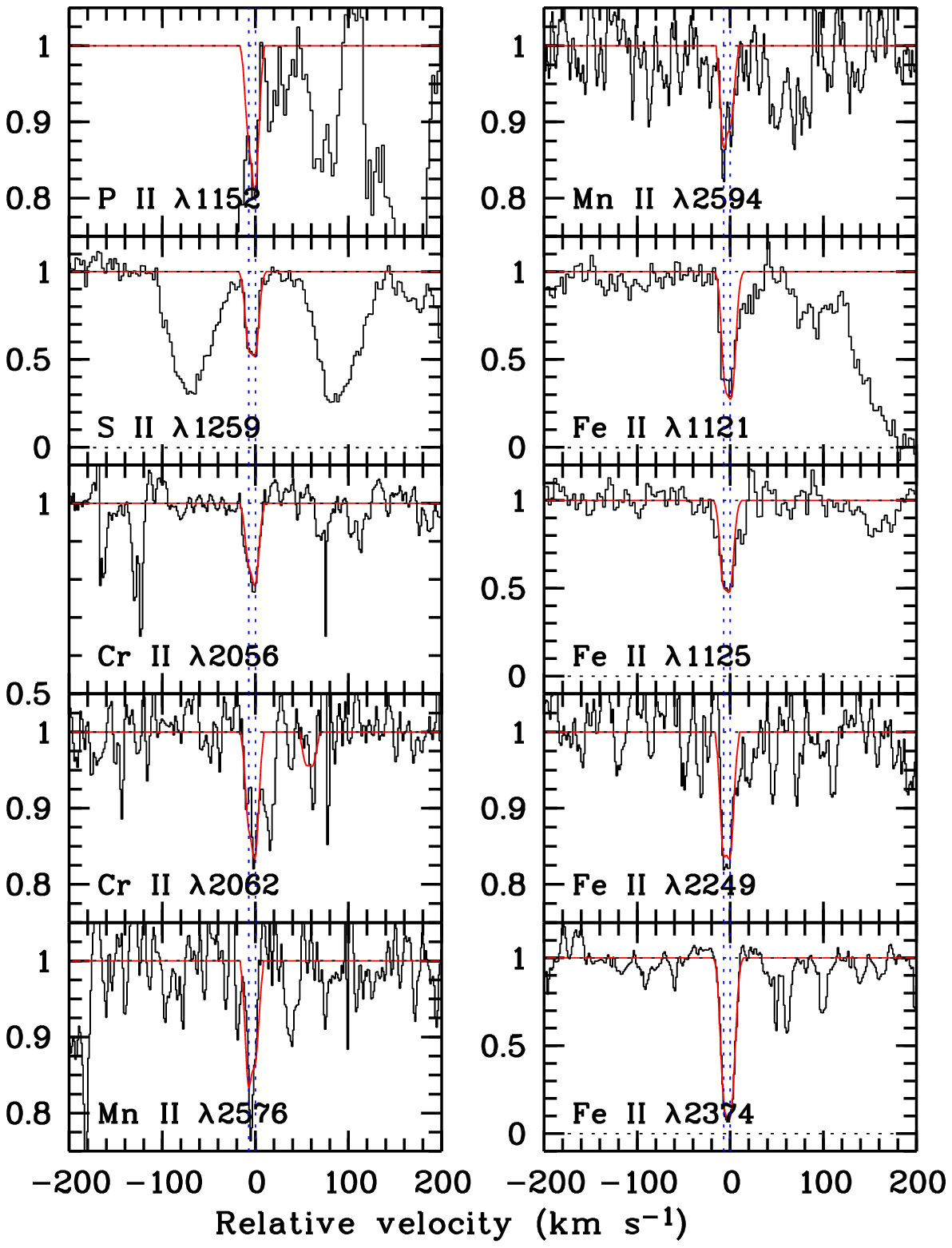}
\caption{Velocity profiles of selected low-ionization transition lines from the
DLA system at $z_{\rm abs}=2.370$ towards Q\,0102$-$190.}
\label{q0102_2.370}
\end{figure}

\begin{figure}
\includegraphics[bb=61 402 402 767,clip,width=8.7cm]{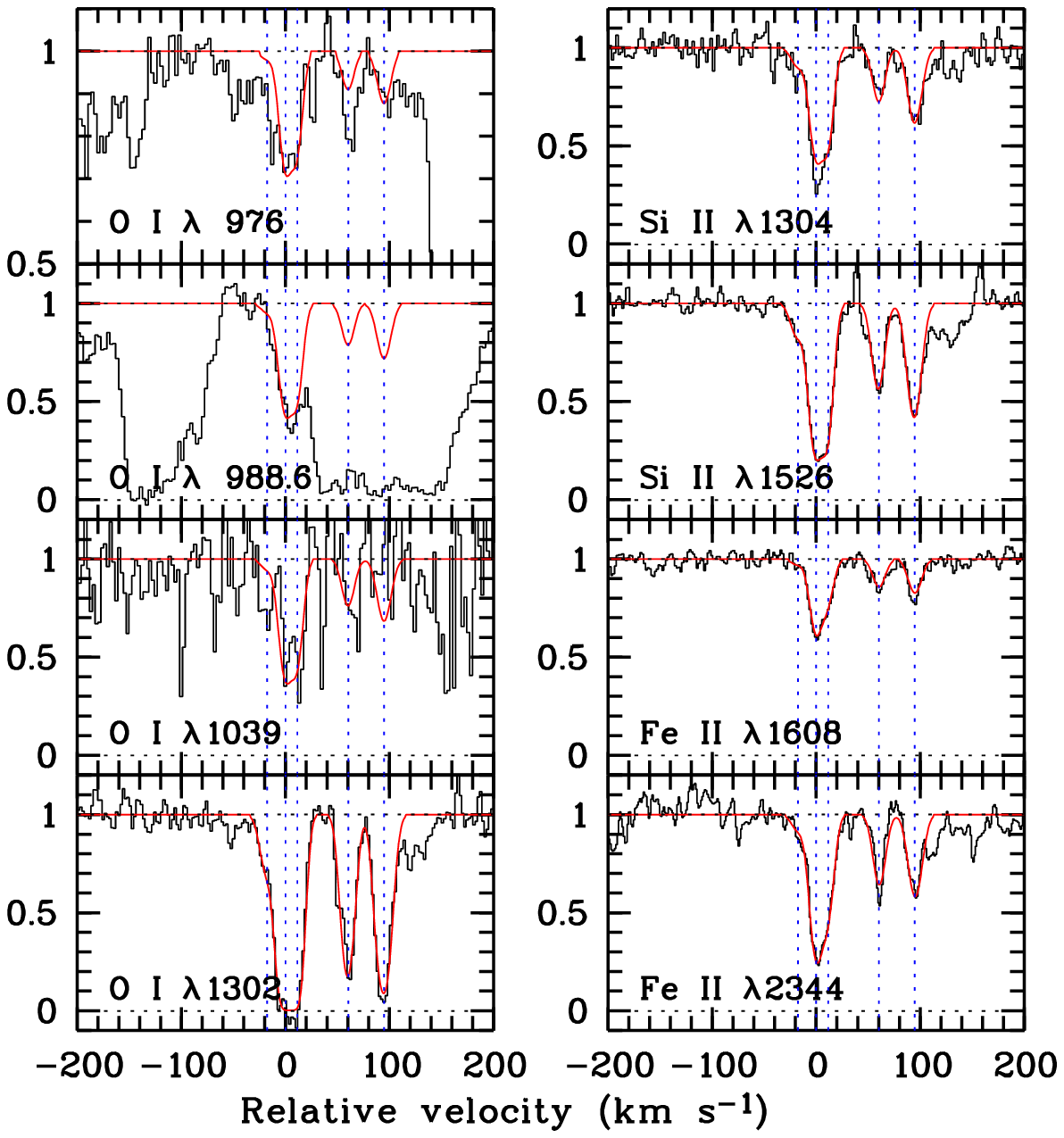}
\caption{Velocity profiles of selected low-ionization transition lines from the
sub-DLA system at $z_{\rm abs}=2.926$ towards Q\,0102$-$190.}
\label{q0102_2.926}
\end{figure}

\begin{figure}
\includegraphics[bb=61 567 402 767,clip,width=8.7cm]{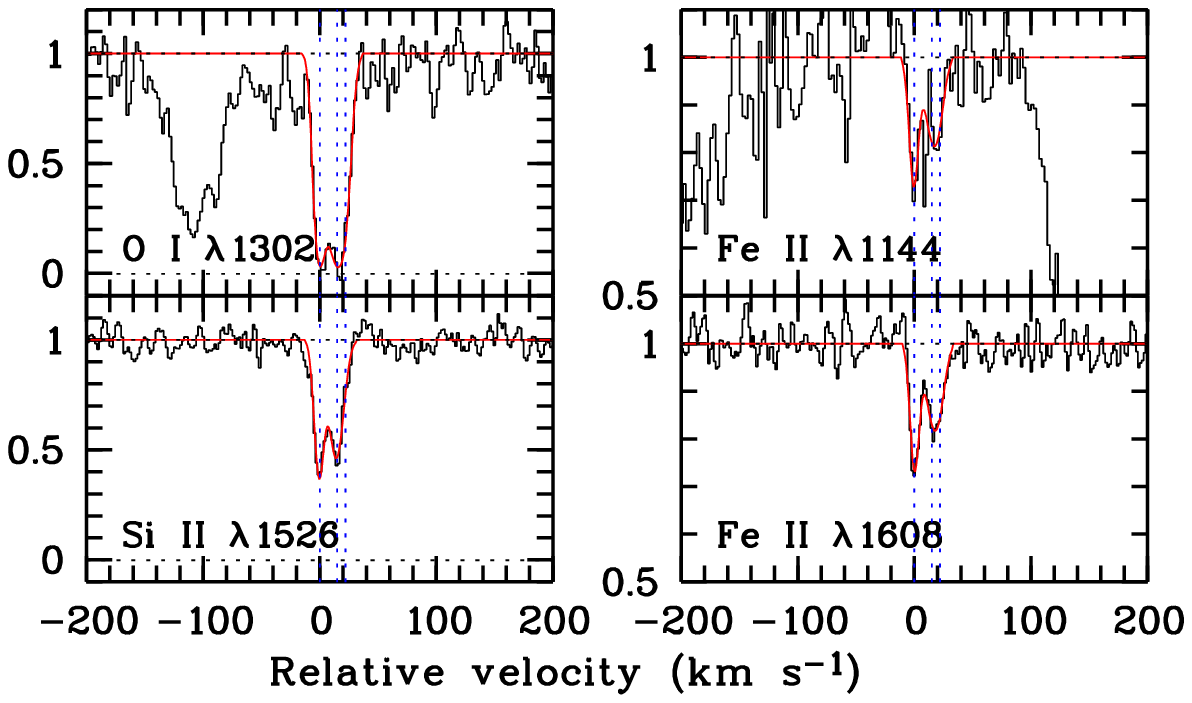}
\caption{Velocity profiles of selected low-ionization transition lines from the
DLA system at $z_{\rm abs}=2.418$ towards Q\,0112$-$306.}
\label{q0112_2.418}
\end{figure}

\begin{figure}
\includegraphics[bb=61 650 402 767,clip,width=8.7cm]{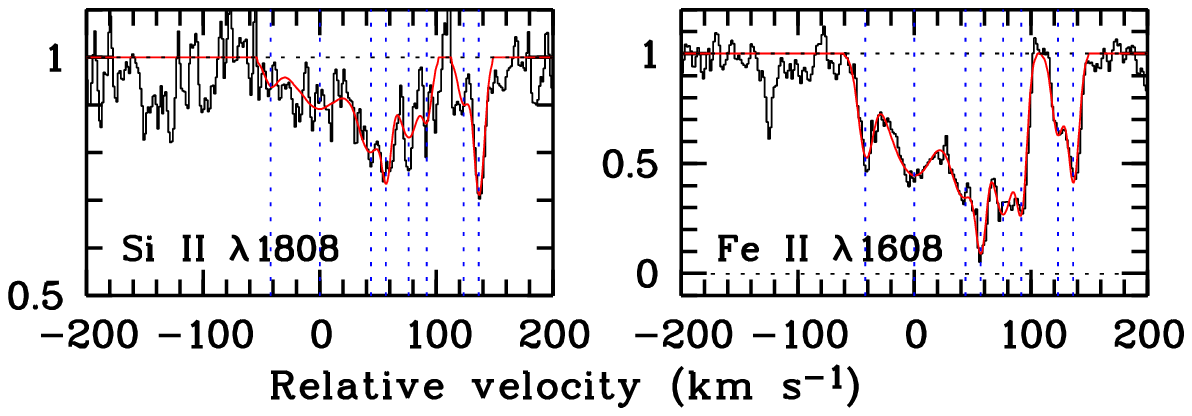}
\caption{Velocity profiles of selected low-ionization transition lines from the
DLA system at $z_{\rm abs}=2.702$ towards Q\,0112$-$306.}
\label{q0112_2.702}
\end{figure}

\begin{figure}
\includegraphics[bb=61 485 402 767,clip,width=8.7cm]{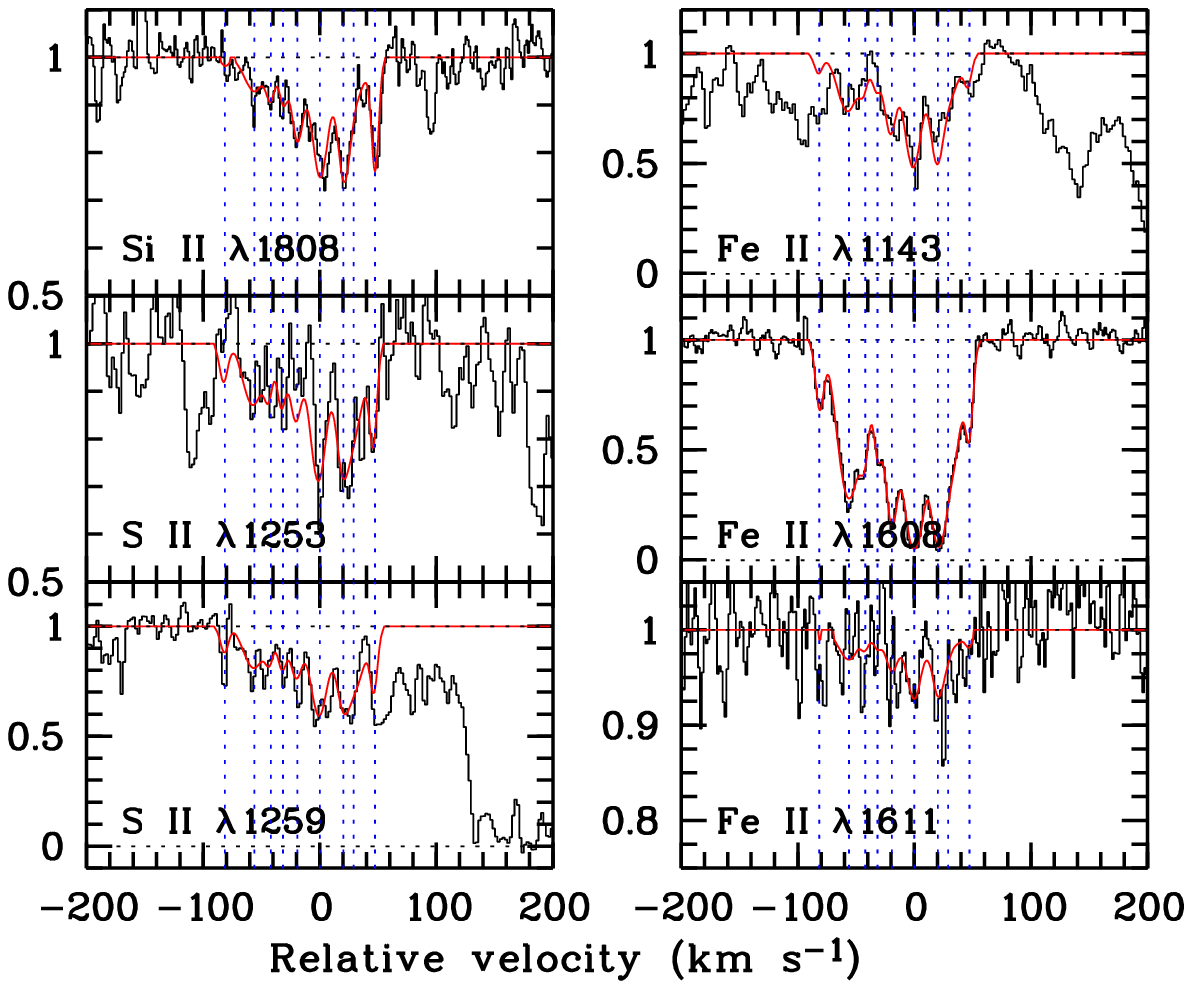}
\caption{Velocity profiles of selected low-ionization transition lines from the
DLA system at $z_{\rm abs}=2.423$ towards Q\,0112$+$030.}
\label{q0112g_2.423}
\end{figure}

\clearpage

\begin{figure}
\includegraphics[bb=61 320 402 767,clip,width=8.7cm]{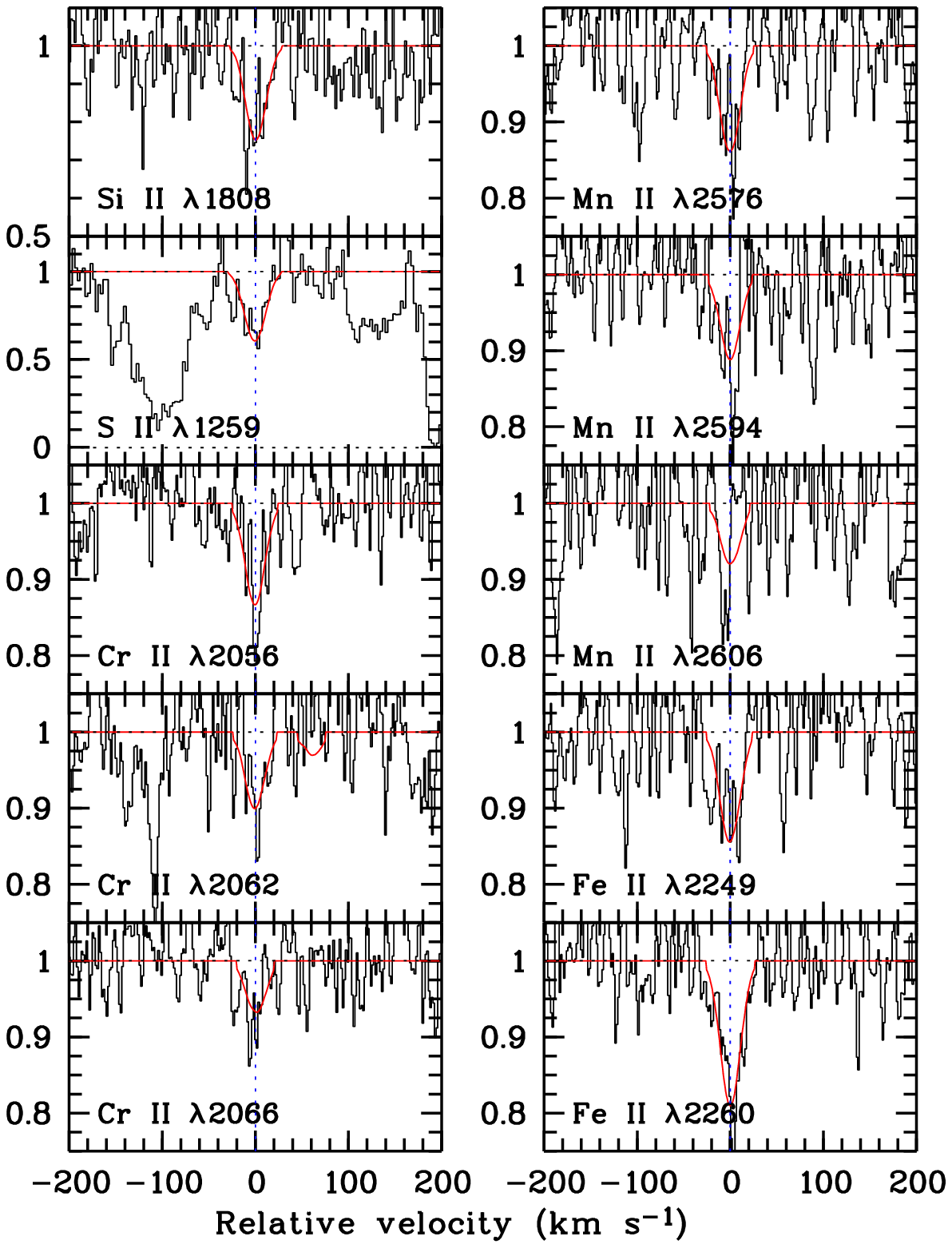}
\caption{Velocity profiles of selected low-ionization transition lines from the
DLA system at $z_{\rm abs}=2.107$ towards Q\,0135$-$273.}
\label{q0135_2.107}
\end{figure}

\begin{figure}
\includegraphics[bb=61 402 402 767,clip,width=8.7cm]{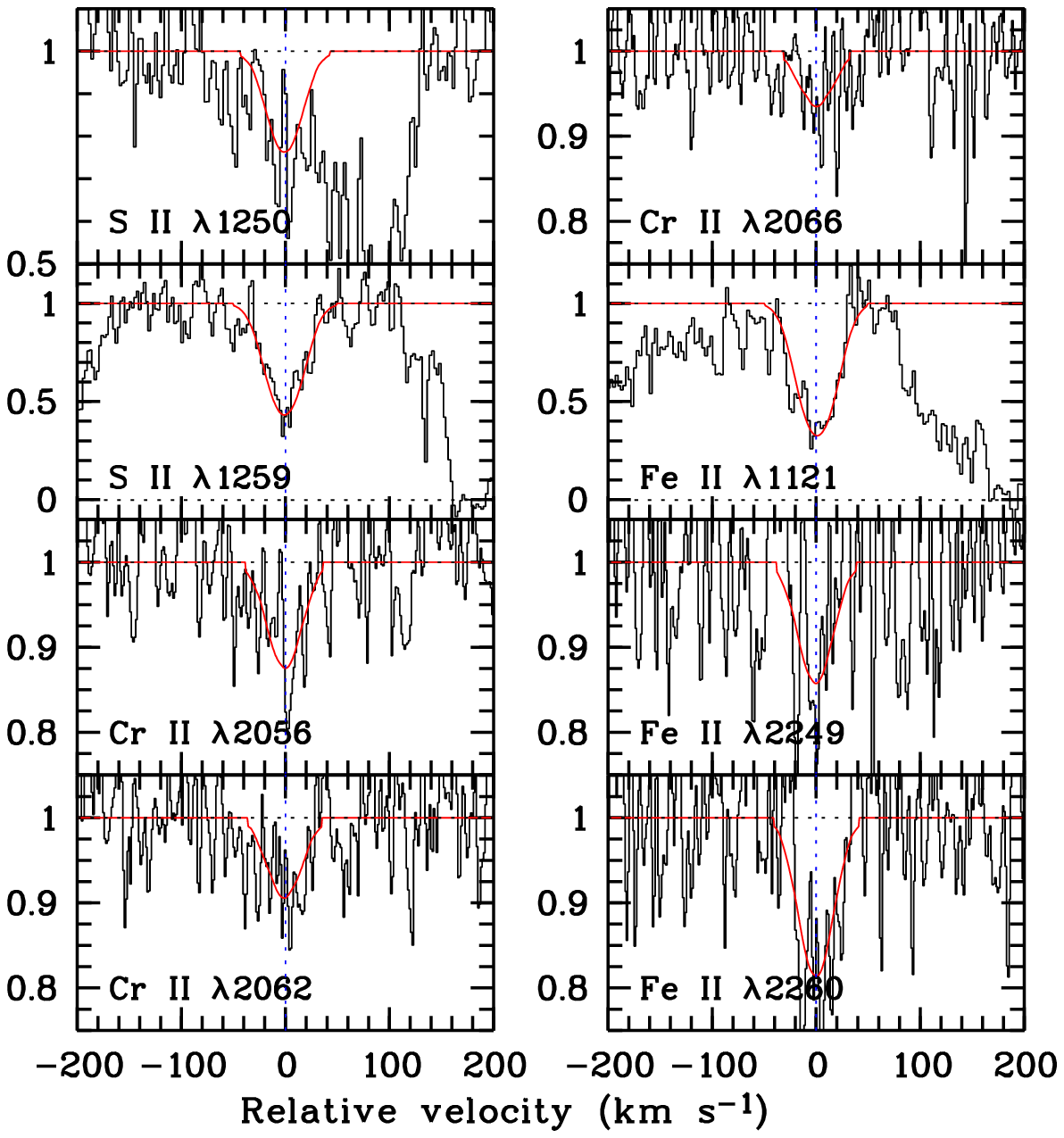}
\caption{Velocity profiles of selected low-ionization transition lines from the
DLA system at $z_{\rm abs}=2.800$ towards Q\,0135$-$273.}
\label{q0135_2.800}
\end{figure}

\begin{figure}
\includegraphics[bb=61 320 402 767,clip,width=8.7cm]{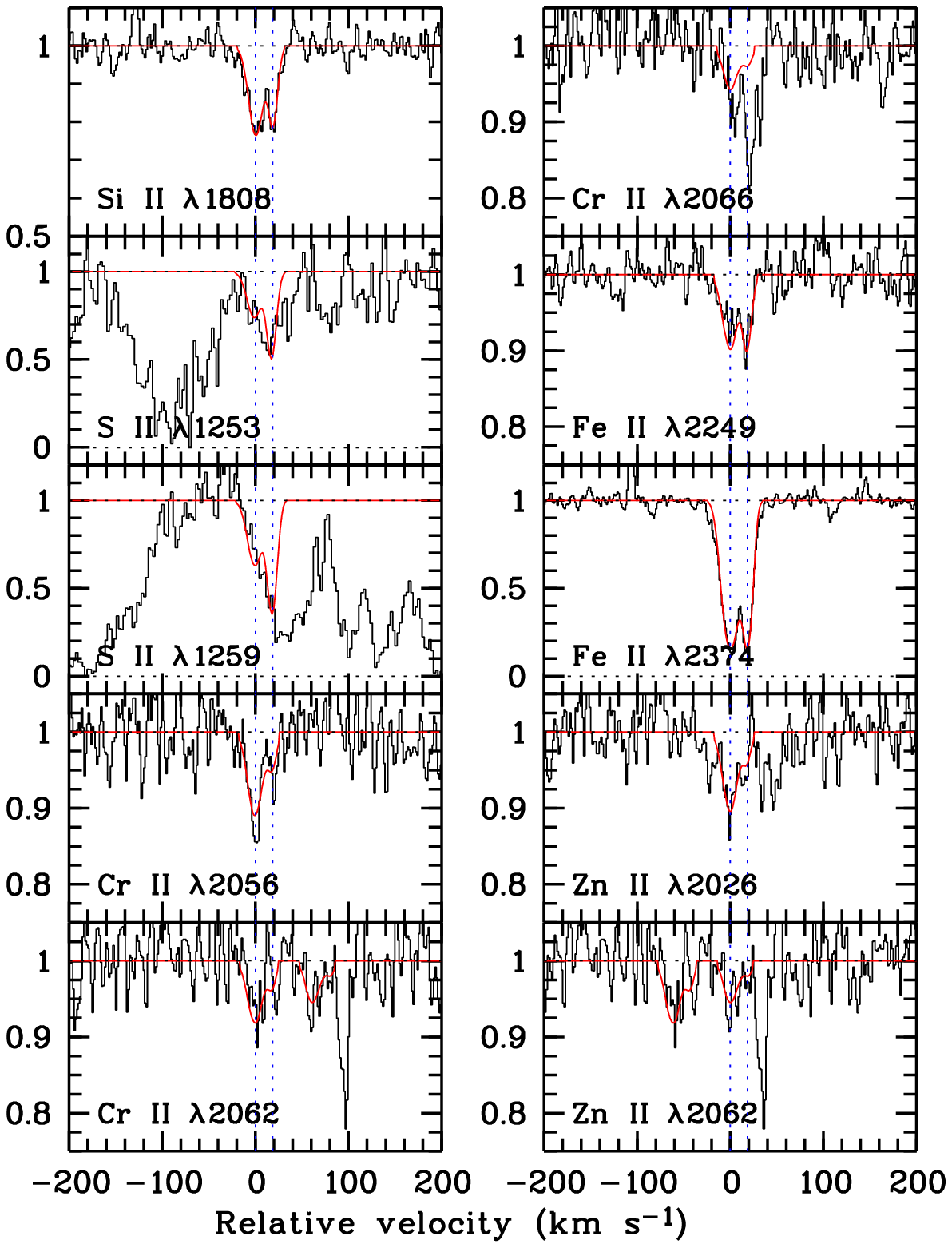}
\caption{Velocity profiles of selected low-ionization transition lines from the
DLA system at $z_{\rm abs}=1.769$ towards Q\,0216$+$080.}
\label{q0216_1.769}
\end{figure}

\begin{figure}
\includegraphics[bb=61 402 402 767,clip,width=8.7cm]{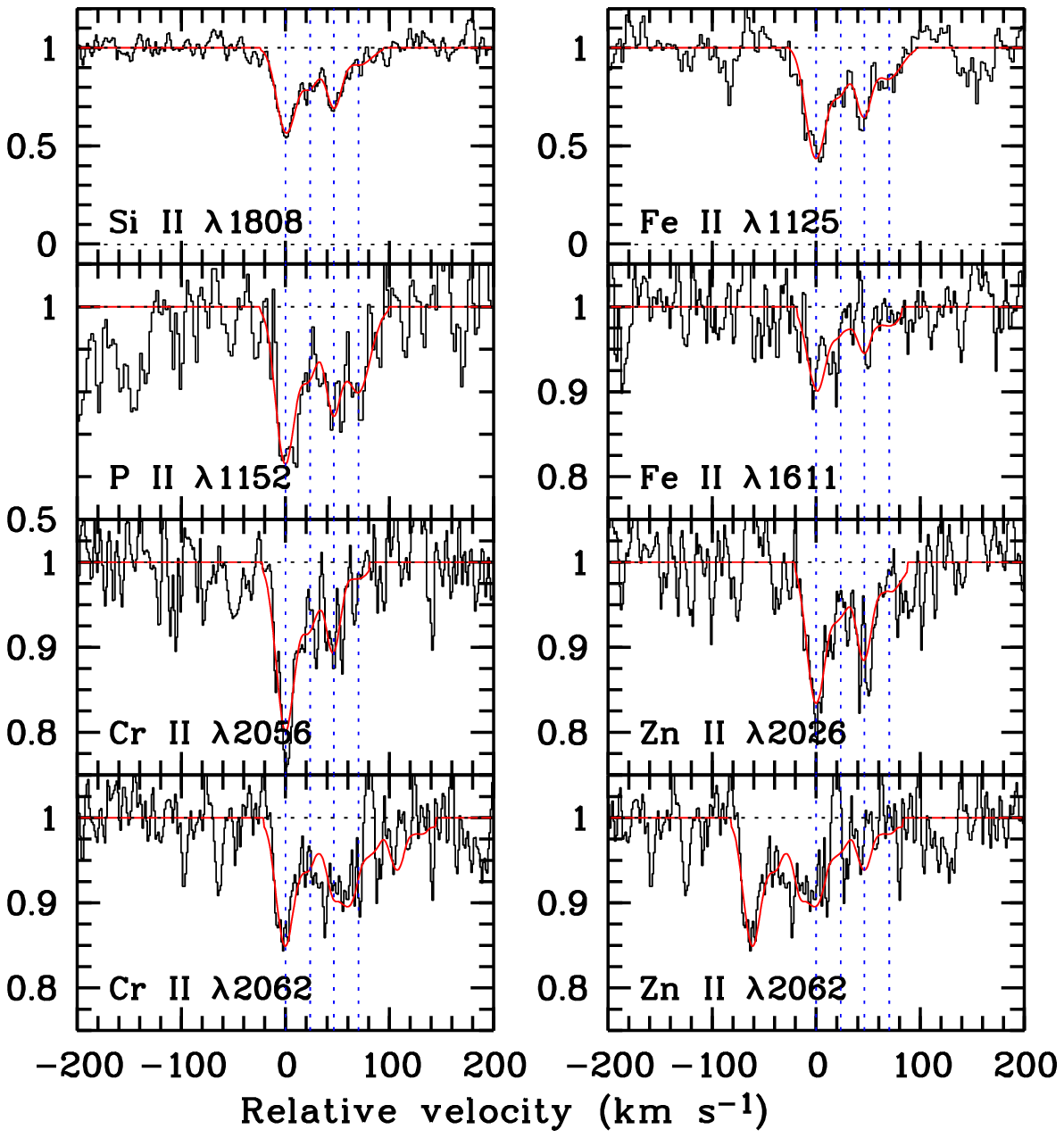}
\caption{Velocity profiles of selected low-ionization transition lines from the
DLA system at $z_{\rm abs}=2.293$ towards Q\,0216$+$080.}
\label{q0216_2.293}
\end{figure}

\clearpage

\begin{figure}
\includegraphics[bb=61 402 402 767,clip,width=8.7cm]{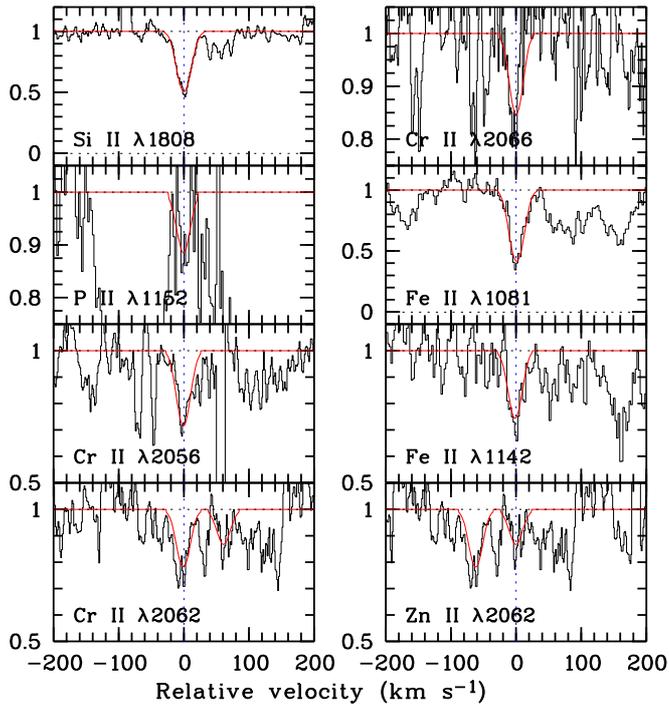}
\caption{Velocity profiles of selected low-ionization transition lines from the
DLA system at $z_{\rm abs}=3.062$ towards Q\,0336$-$017.}
\label{q0336_3.062}
\end{figure}

\begin{figure}
\includegraphics[bb=61 402 402 767,clip,width=8.7cm]{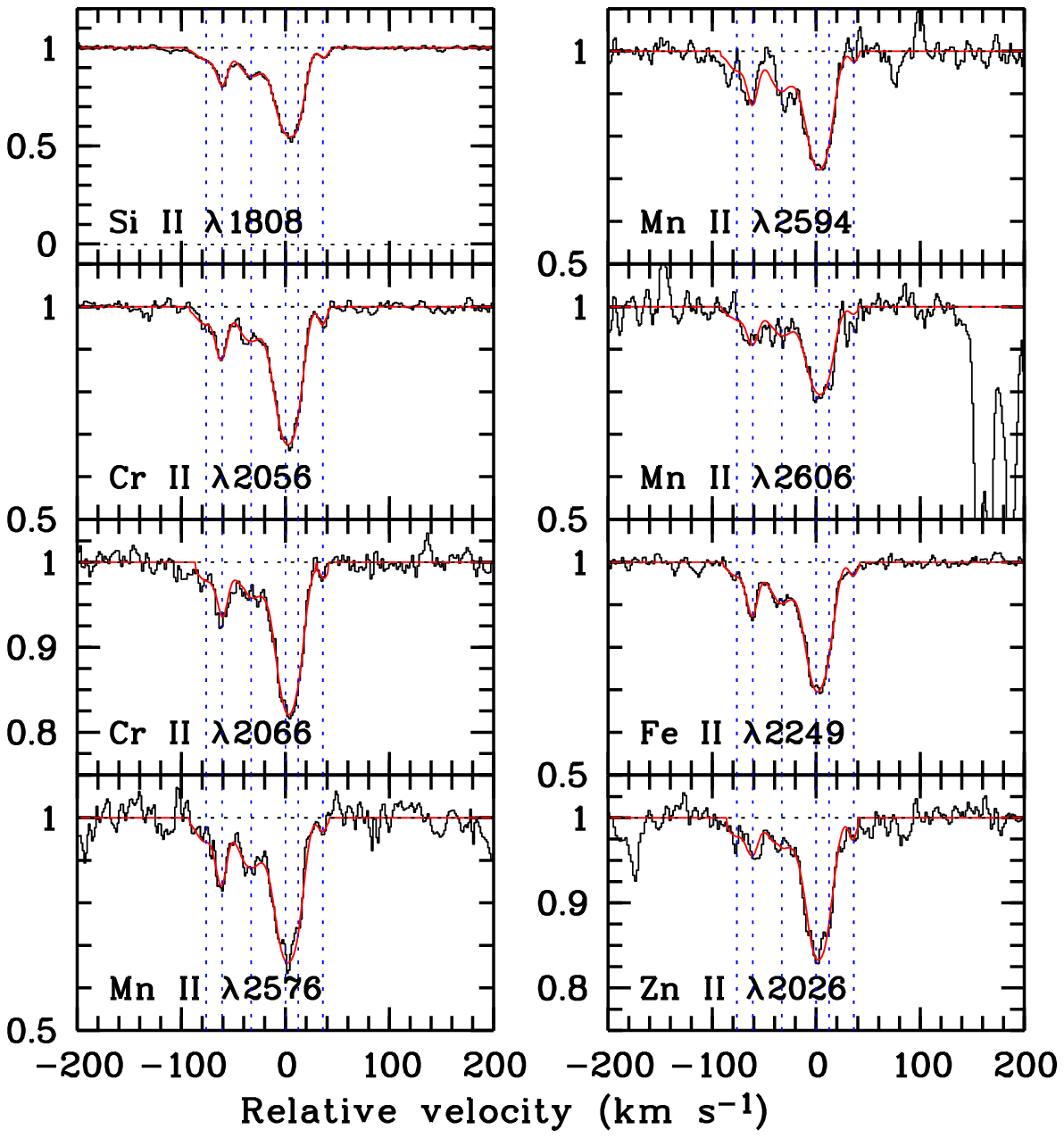}
\caption{Velocity profiles of selected low-ionization transition lines from the
DLA system at $z_{\rm abs}=1.913$ towards Q\,0405$-$443.}
\label{q0405_1.913}
\end{figure}

\begin{figure}
\includegraphics[bb=61 73 402 767,clip,width=8.7cm]{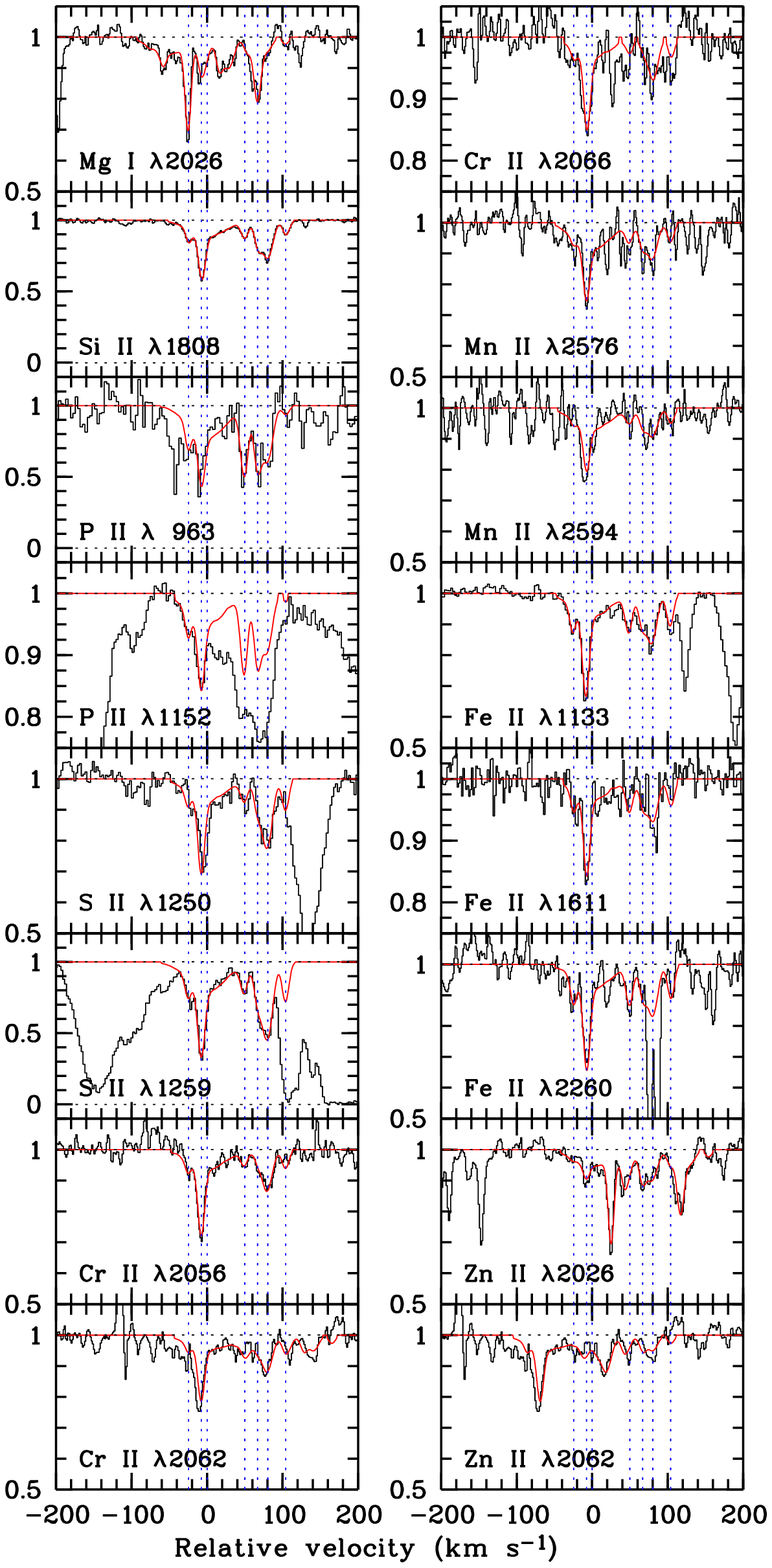}
\caption{Velocity profiles of selected low-ionization transition lines from the
DLA system at $z_{\rm abs}=2.550$ towards Q\,0405$-$443.}
\label{q0405_2.550}
\end{figure}

\clearpage

\begin{figure}
\includegraphics[bb=61 155 402 767,clip,width=8.7cm]{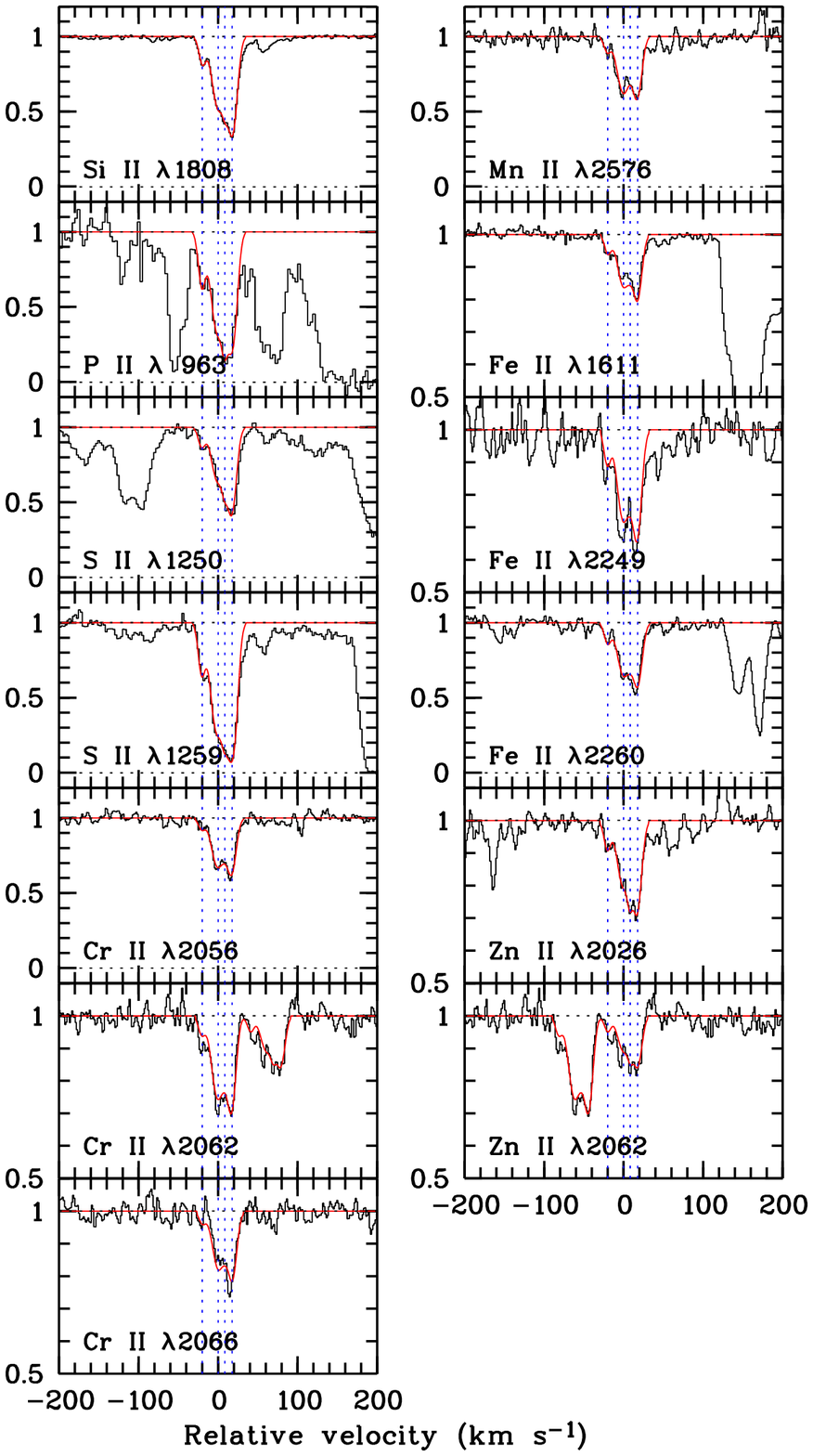}
\caption{Velocity profiles of selected low-ionization transition lines from the
DLA system at $z_{\rm abs}=2.595$ towards Q\,0405$-$443.}
\label{q0405_2.595}
\end{figure}

\begin{figure}
\includegraphics[bb=61 650 402 767,clip,width=8.7cm]{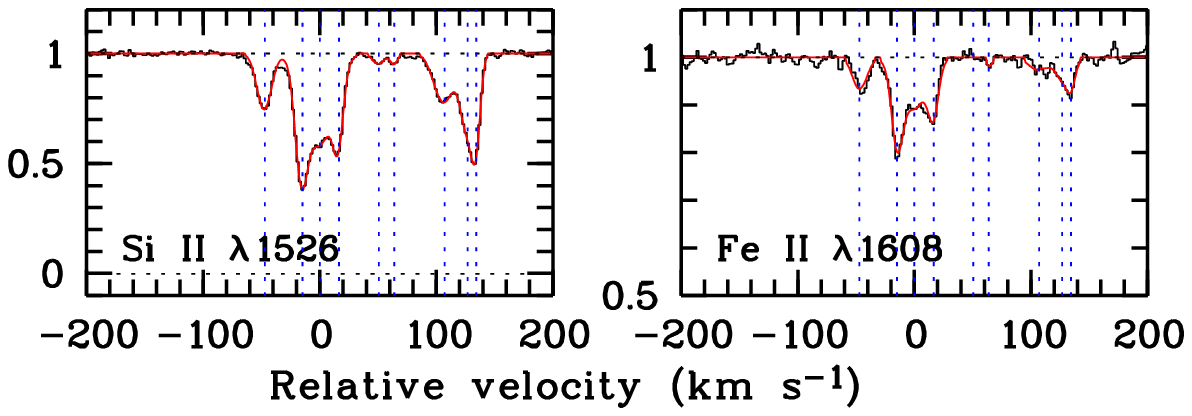}
\caption{Velocity profiles of selected low-ionization transition lines from the
DLA system at $z_{\rm abs}=2.622$ towards Q\,0405$-$443.}
\label{q0405_2.622}
\end{figure}

\begin{figure}
\includegraphics[bb=61 320 402 767,clip,width=8.7cm]{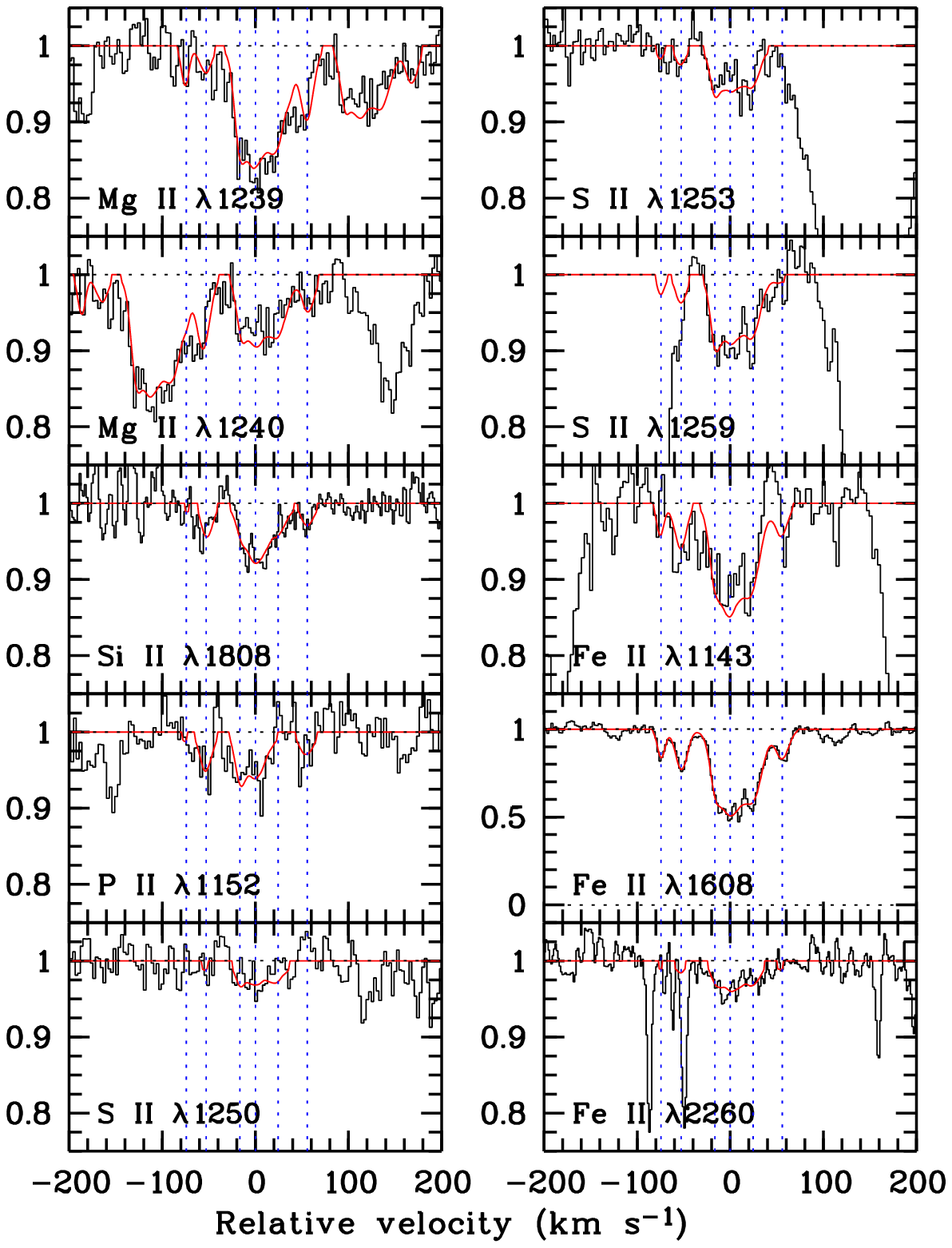}
\caption{Velocity profiles of selected low-ionization transition lines from the
DLA system at $z_{\rm abs}=2.067$ towards Q\,0450$-$131.}
\label{q0450_2.067}
\end{figure}

\clearpage

\begin{figure}
\includegraphics[bb=61 155 402 767,clip,width=8.7cm]{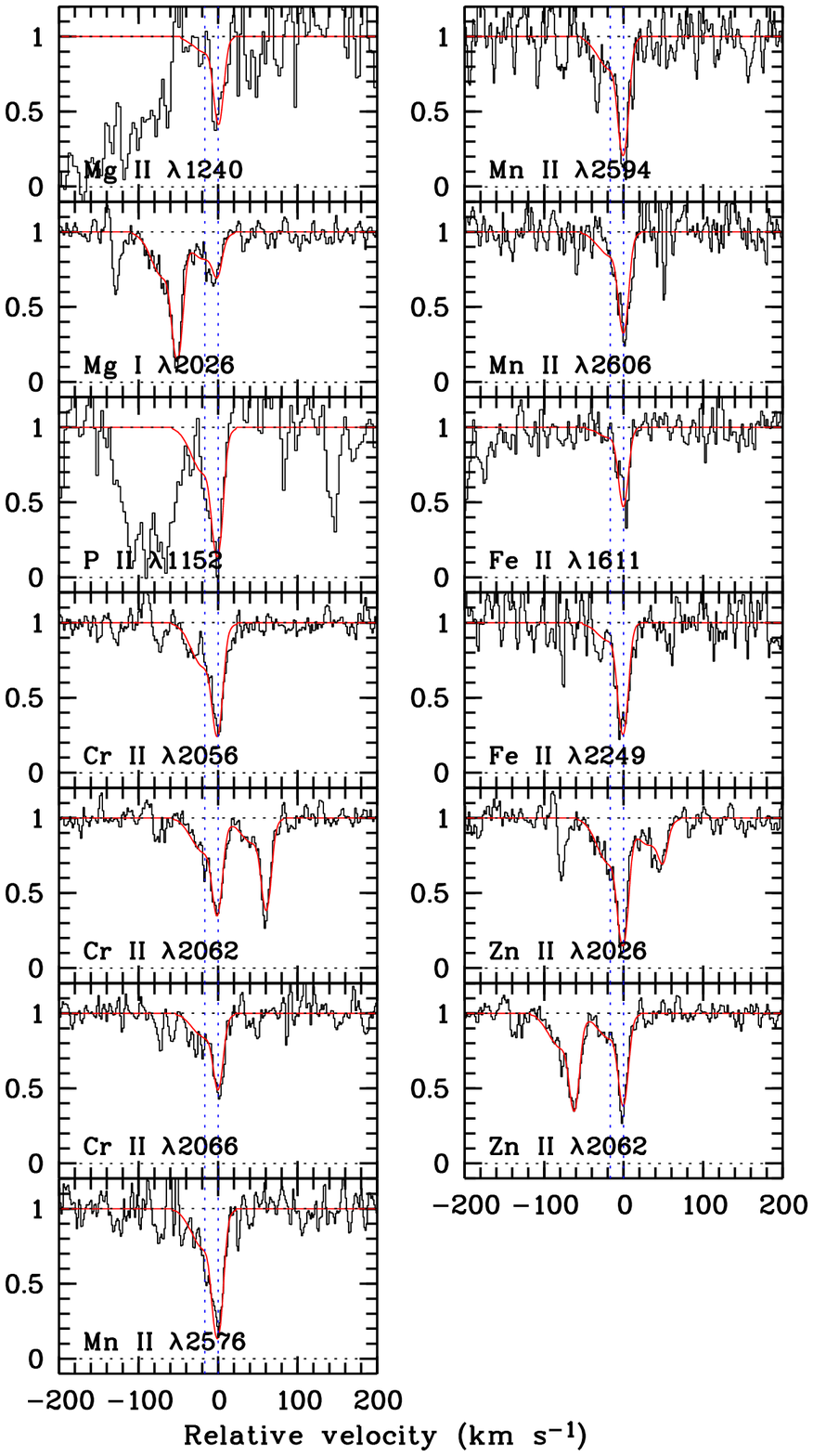}
\caption{Velocity profiles of selected low-ionization transition lines from the
DLA system at $z_{\rm abs}=2.040$ towards Q\,0458$-$020.}
\label{q0458_2.040}
\end{figure}

\begin{figure}
\includegraphics[bb=61 320 402 767,clip,width=8.7cm]{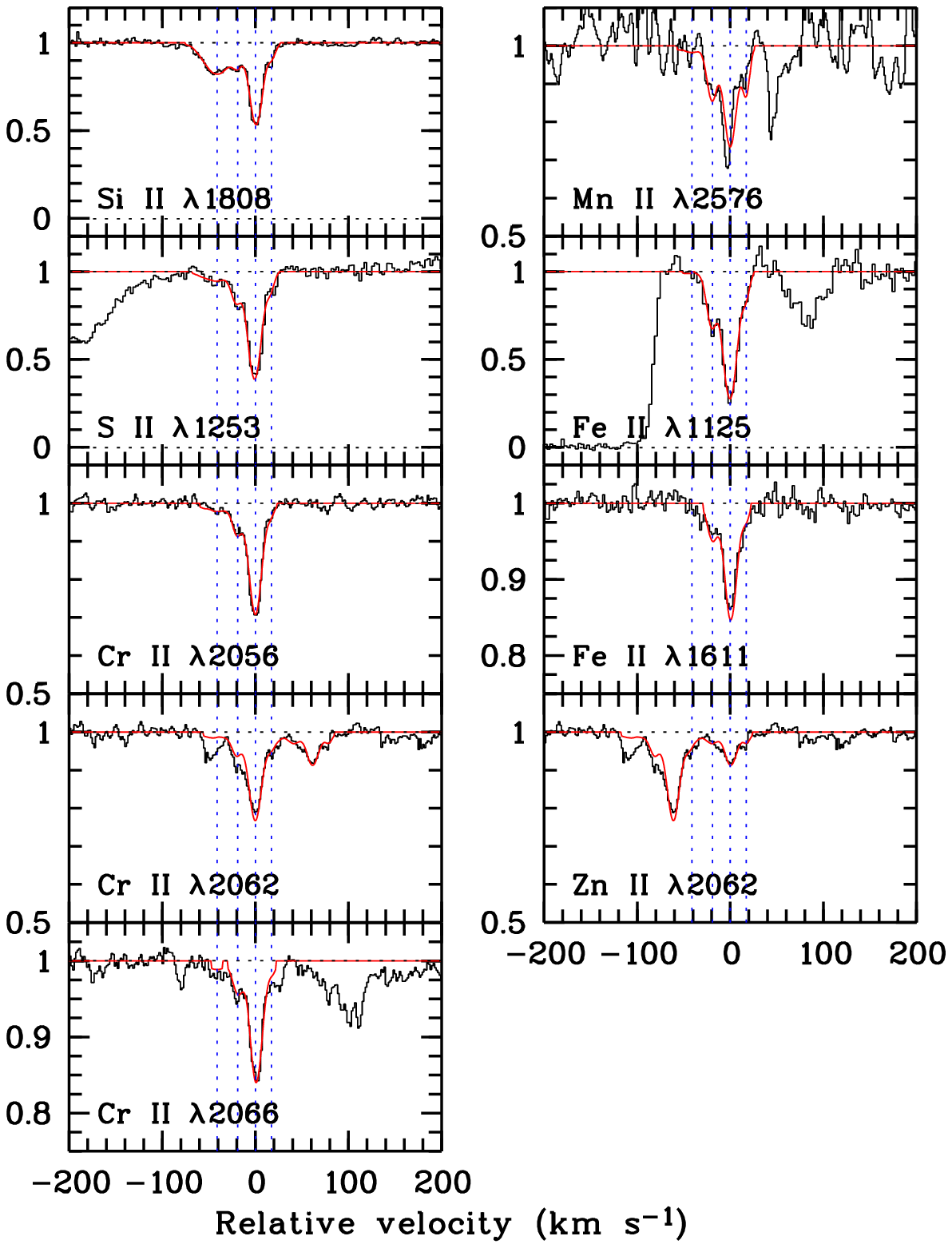}
\caption{Velocity profiles of selected low-ionization transition lines from the
DLA system at $z_{\rm abs}=2.141$ towards Q\,0528$-$250.}
\label{q0528_2.141}
\end{figure}

\clearpage

\begin{figure}
\includegraphics[bb=61 155 402 767,clip,width=8.7cm]{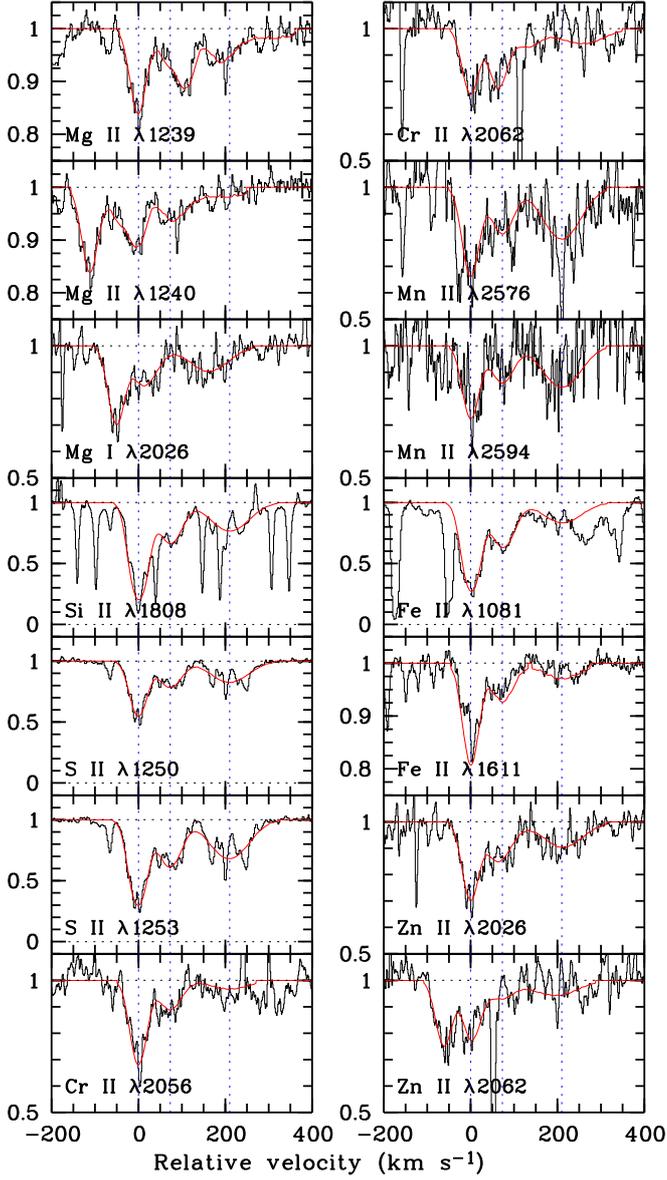}
\caption{
Velocity profiles of selected low-ionization transition lines from the DLA system at $z_{\rm abs}=2.811$ towards Q\,0528$-$250. There is a multi-component structure around 200 km~s$^{-1}$ in this system. However, these components are cleanly observed only in the \suii{} lines. For important transitions, such as for \znii{} and the other metals, the multi-component decomposition is not statistically significant because of the lower S/N, especially over weak transitions. We therefore adopted a single component to more robustly estimate the total column density of the majority of the metals. Our column density measurements are in good agreement (mostly within 0.05~dex, and within 0.1~dex for \siii{}) with the values in the literature \citep{Lu96,Centurion03}; see Sect. \ref{sect comp lit}.
}

\label{q0528_2.811}
\end{figure}

\begin{figure}
\includegraphics[bb=61 320 402 767,clip,width=8.7cm]{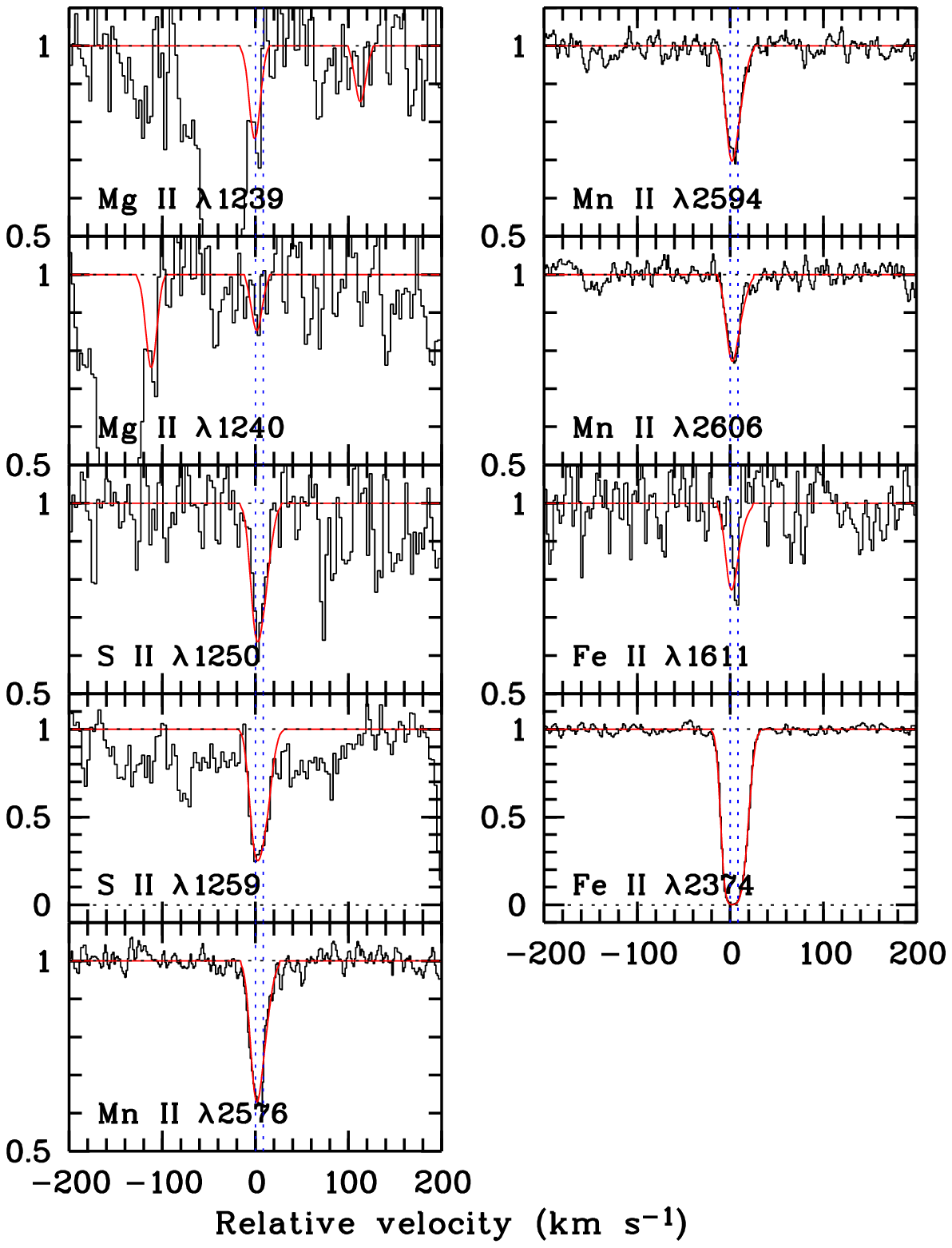}
\caption{Velocity profiles of selected low-ionization transition lines from the
DLA system at $z_{\rm abs}=1.864$ towards Q\,0841$+$129.}
\label{q0841_1.864}
\end{figure}

\clearpage

\begin{figure}
\includegraphics[bb=61 73 402 767,clip,width=8.7cm]{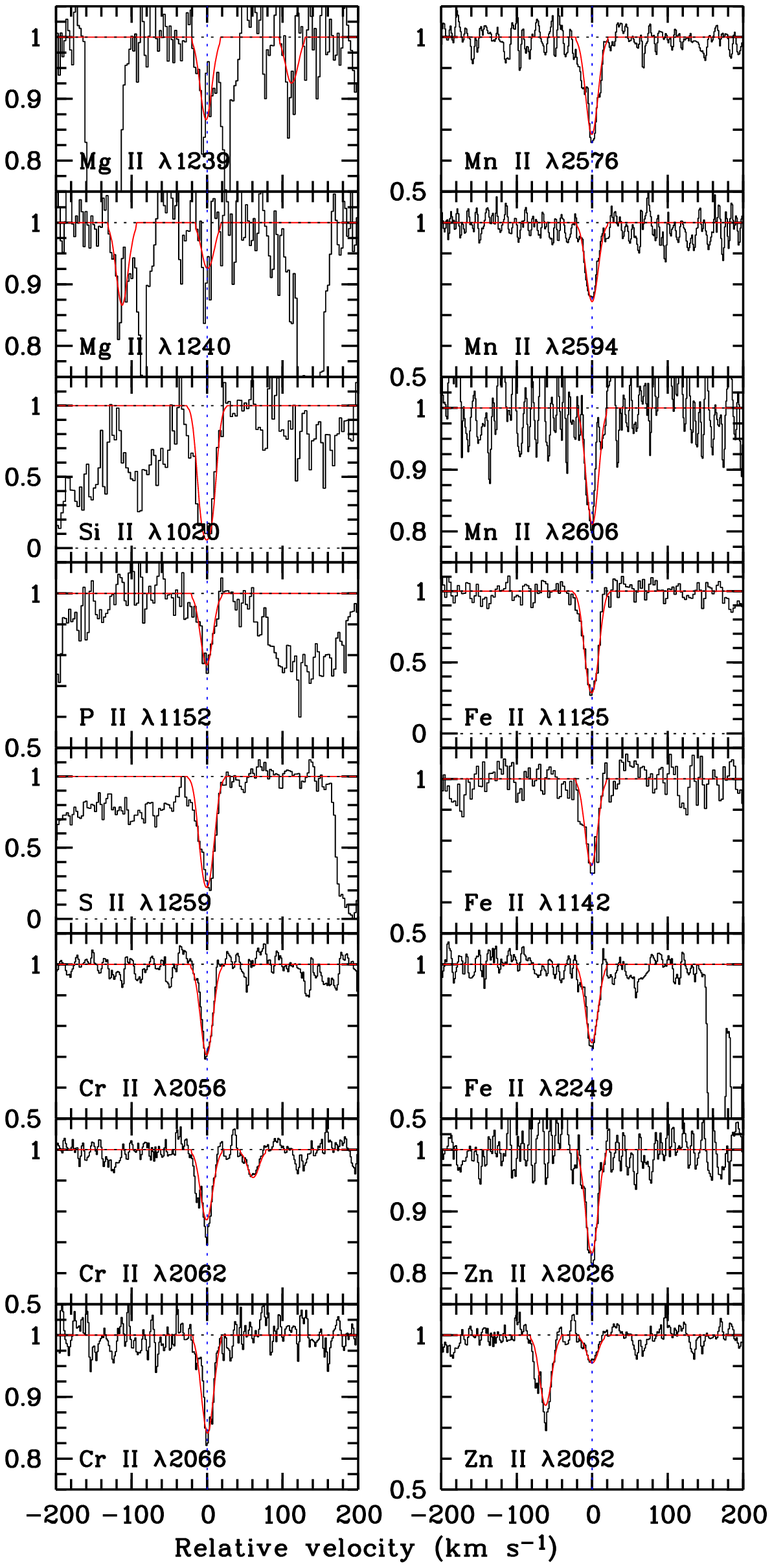}
\caption{Velocity profiles of selected low-ionization transition lines from the
DLA system at $z_{\rm abs}=2.375$ towards Q\,0841$+$129.}
\label{q0841_2.375}
\end{figure}

\begin{figure*}
\hbox{
\includegraphics[bb=61 320 402 767,clip,width=8.7cm]{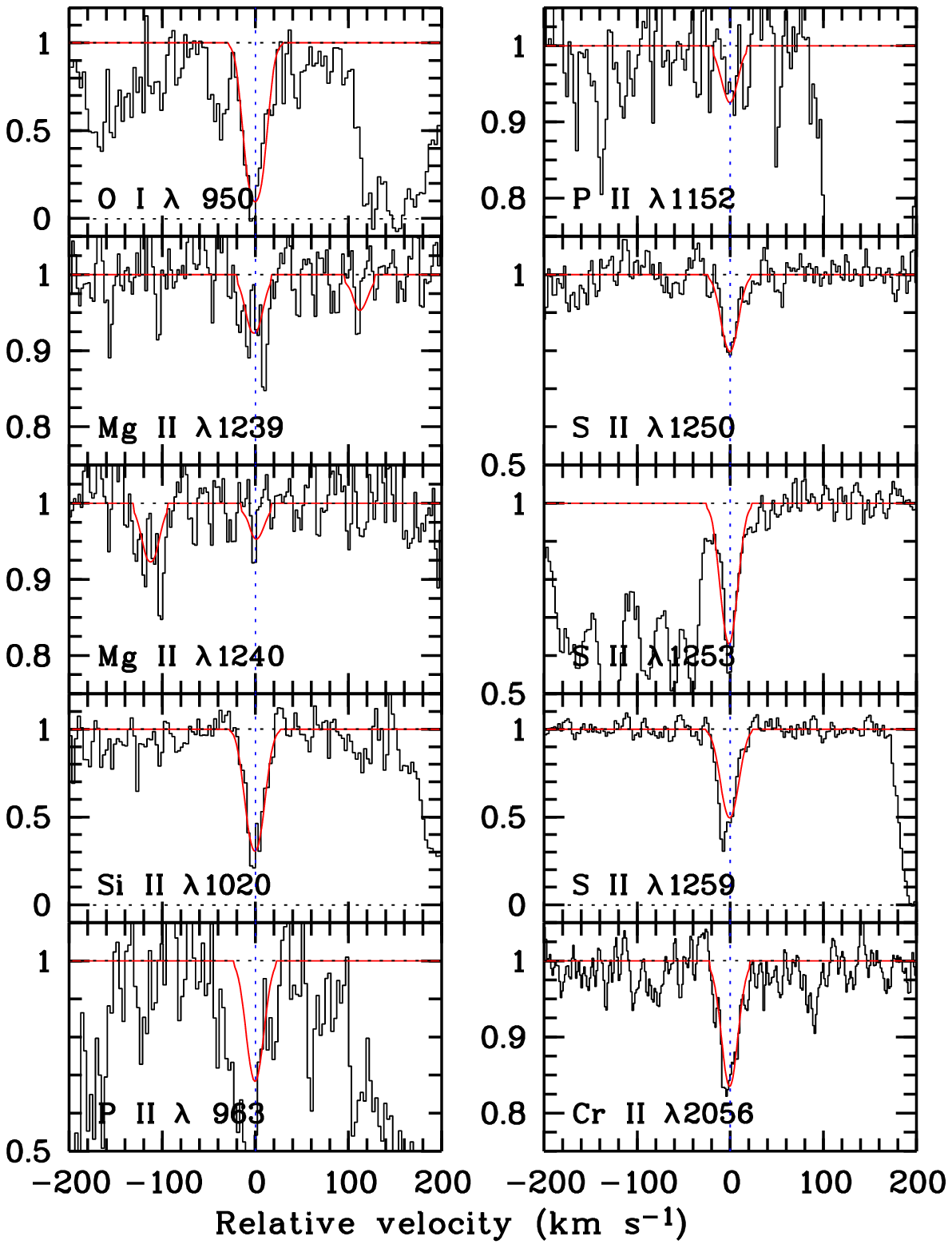}
\includegraphics[bb=61 320 402 767,clip,width=8.7cm]{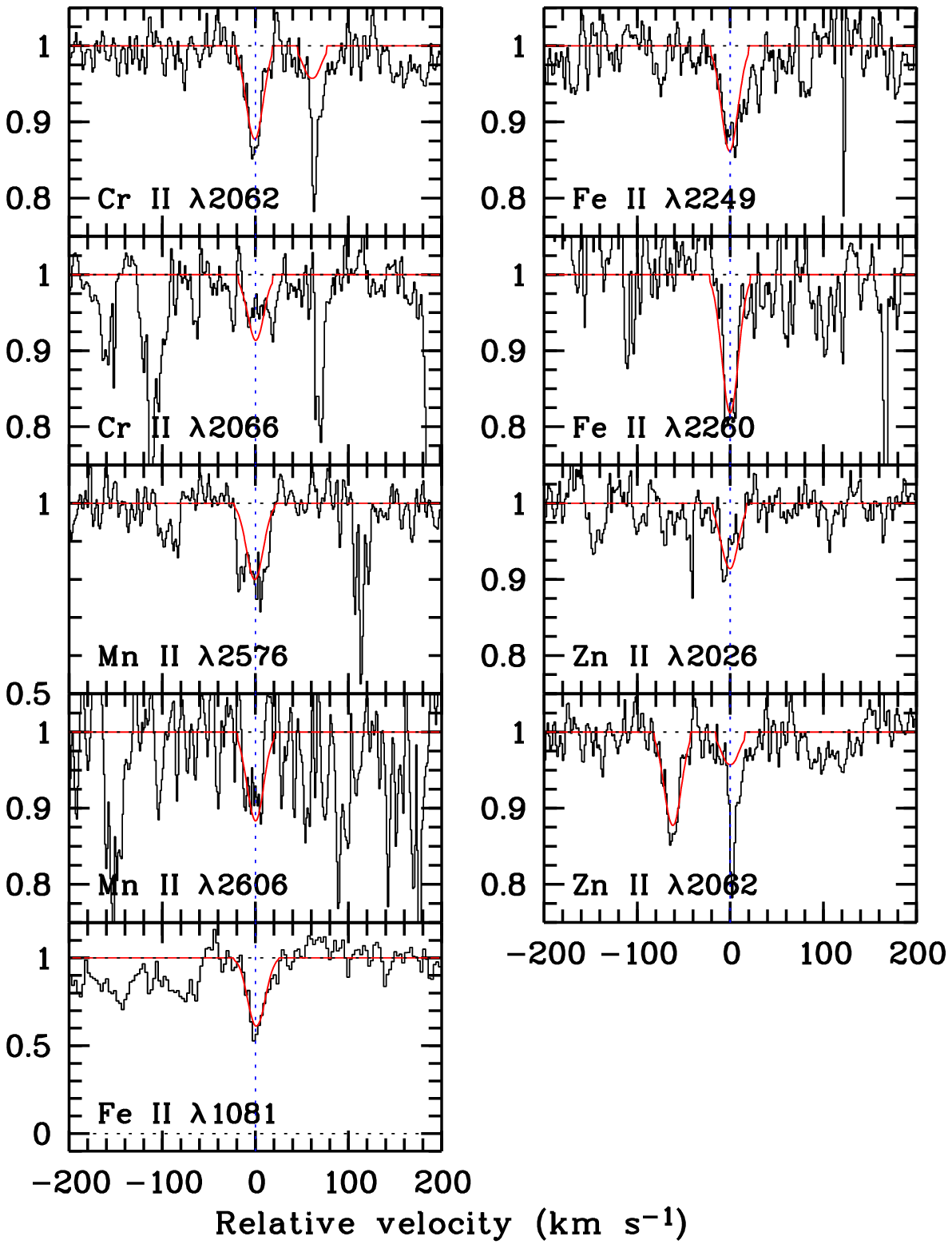}}
\caption{Velocity profiles of selected low-ionization transition lines from the
DLA system at $z_{\rm abs}=2.476$ towards Q\,0841$+$129.}
\label{q0841_2.476}
\end{figure*}

\begin{figure}
\includegraphics[bb=61 485 402 767,clip,width=8.7cm]{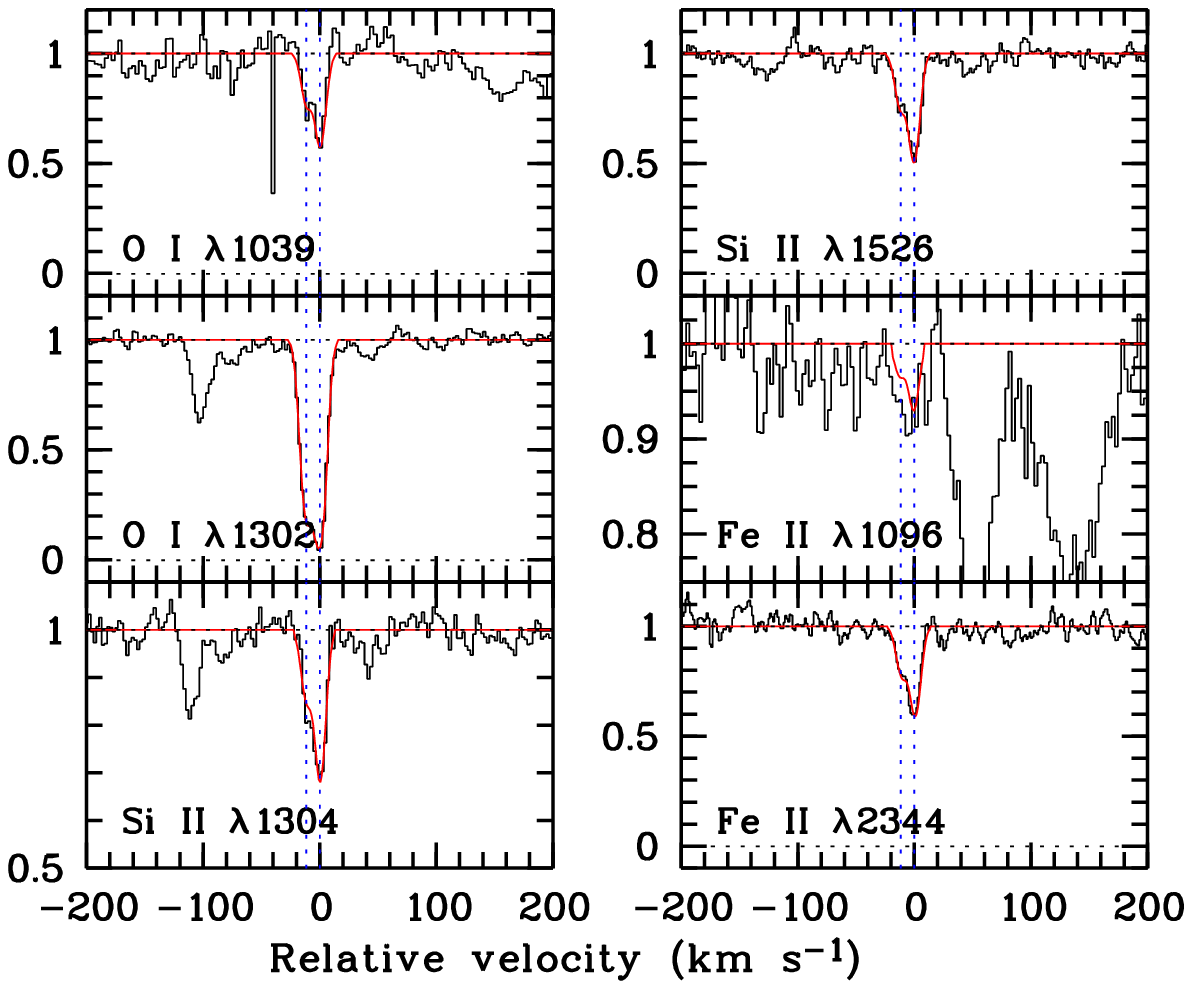}
\caption{Velocity profiles of selected low-ionization transition lines from the
DLA system at $z_{\rm abs}=2.618$ towards Q\,0913$+$072.}
\label{q0913_2.618}
\end{figure}

\begin{figure}
\includegraphics[bb=61 238 402 767,clip,width=8.7cm]{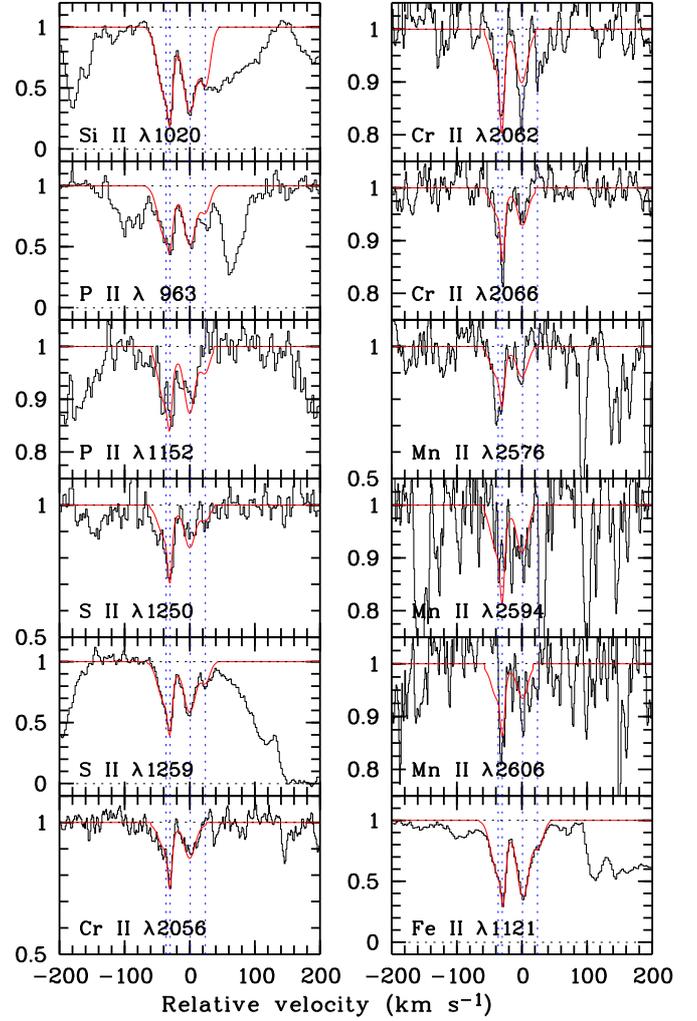}
\caption{Velocity profiles of selected low-ionization transition lines from the
DLA system at $z_{\rm abs}=2.778$ towards Q\,1036$-$229.}
\label{q1036_2.778}
\end{figure}

\clearpage

\begin{figure}
\includegraphics[bb=61 73 402 767,clip,width=8.7cm]{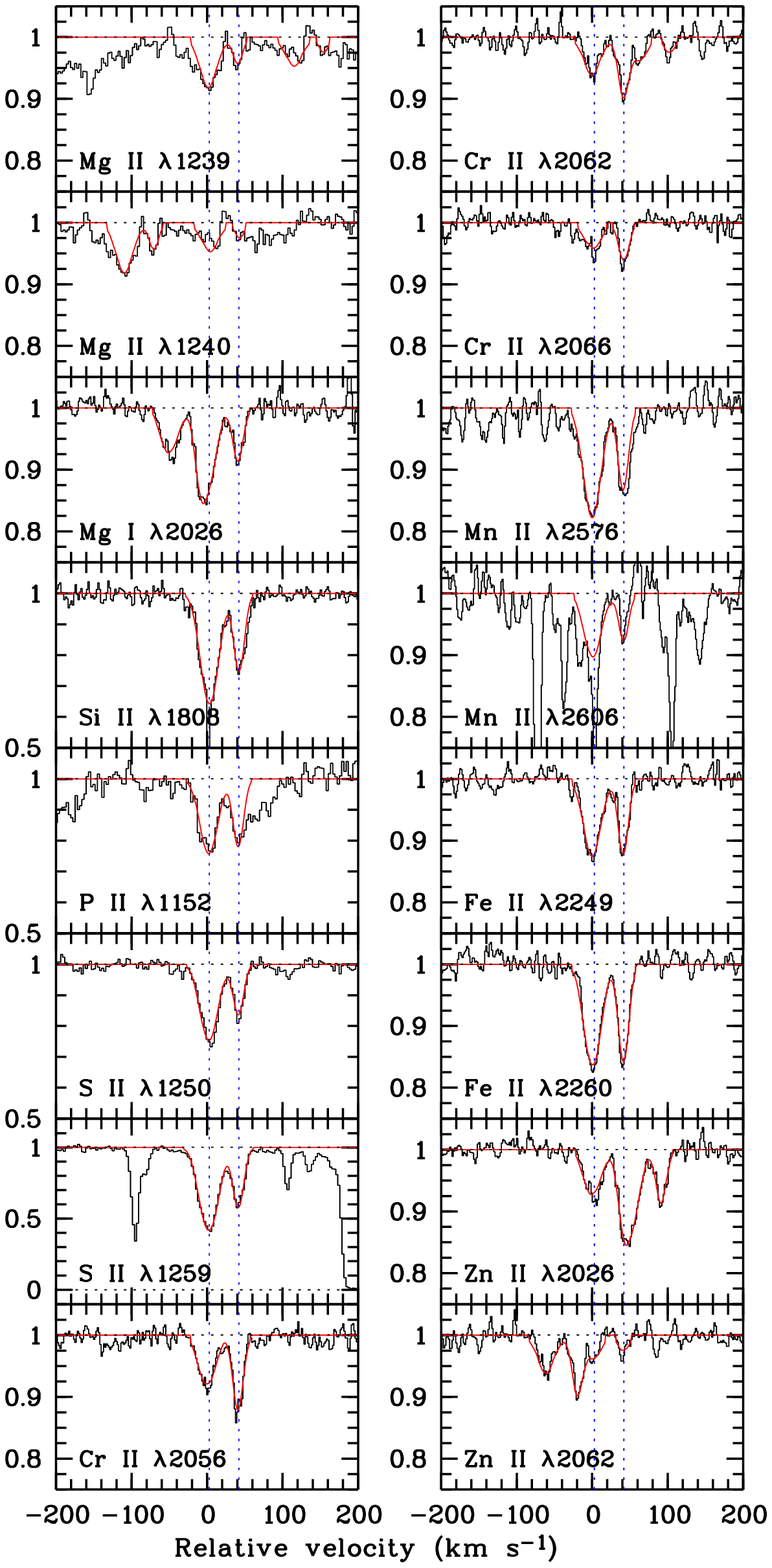}
\caption{Velocity profiles of selected low-ionization transition lines from the
sub-DLA system at $z_{\rm abs}=2.139$ towards Q\,1037$-$270.}
\label{q1037_2.139}
\end{figure}

\begin{figure}
\includegraphics[bb=61 650 402 767,clip,width=8.7cm]{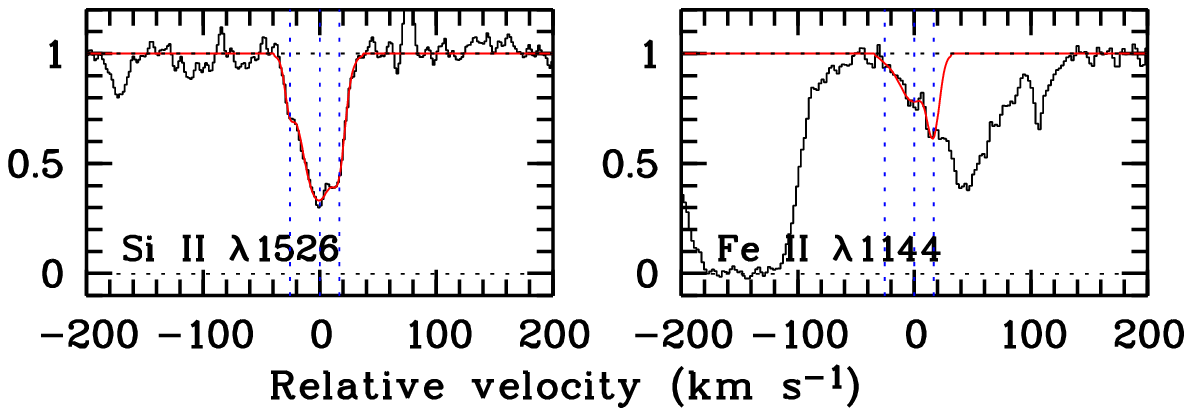}
\caption{Velocity profiles of selected low-ionization transition lines from the
sub-DLA system at $z_{\rm abs}=3.482$ towards Q\,1108$-$077.}
\label{q1108_3.482}
\end{figure}

\begin{figure}
\includegraphics[bb=61 485 402 767,clip,width=8.7cm]{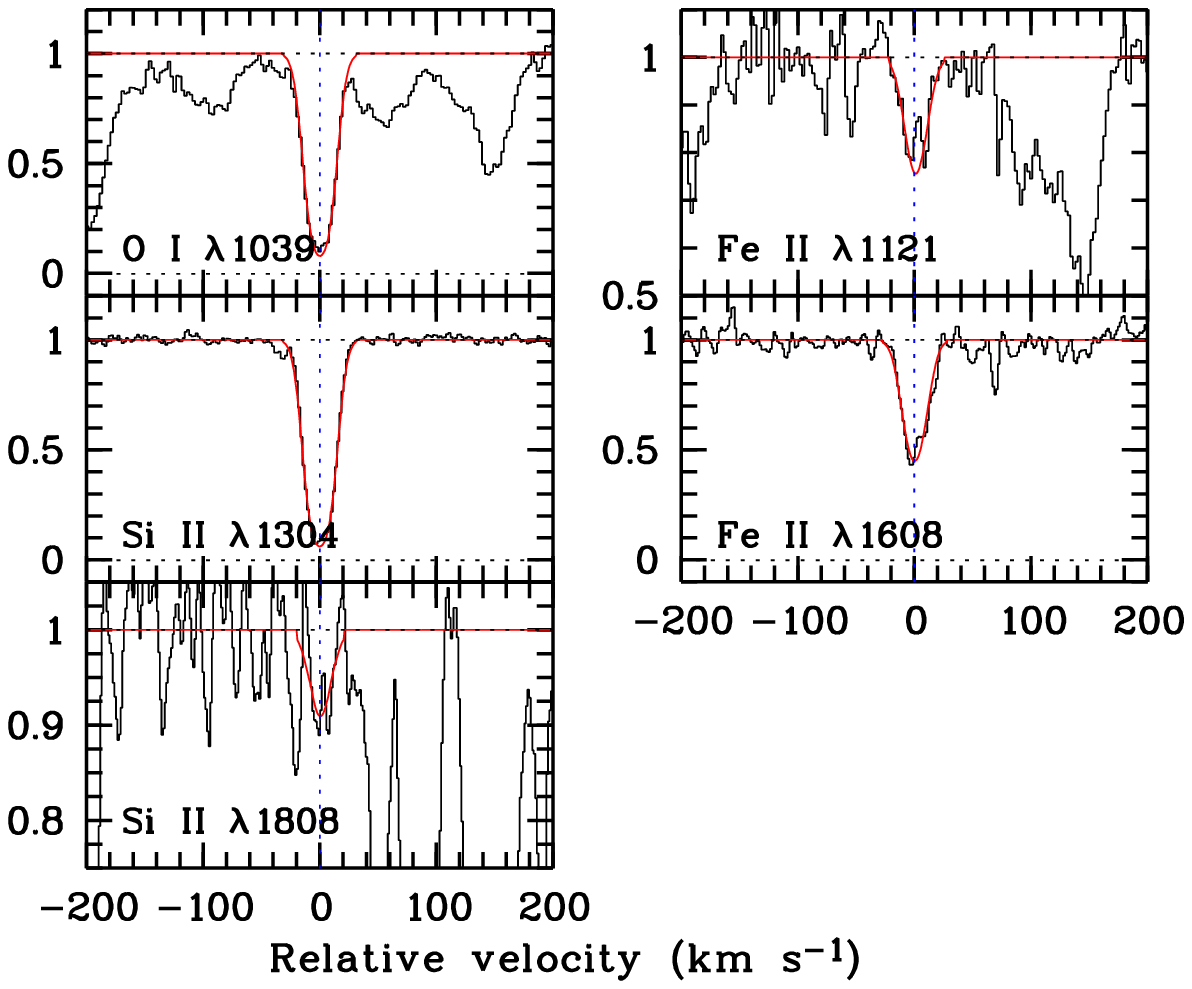}
\caption{Velocity profiles of selected low-ionization transition lines from the
DLA system at $z_{\rm abs}=3.608$ towards Q\,1108$-$077.}
\label{q1108_3.608}
\end{figure}

\begin{figure}
\includegraphics[bb=61 238 402 767,clip,width=8.7cm]{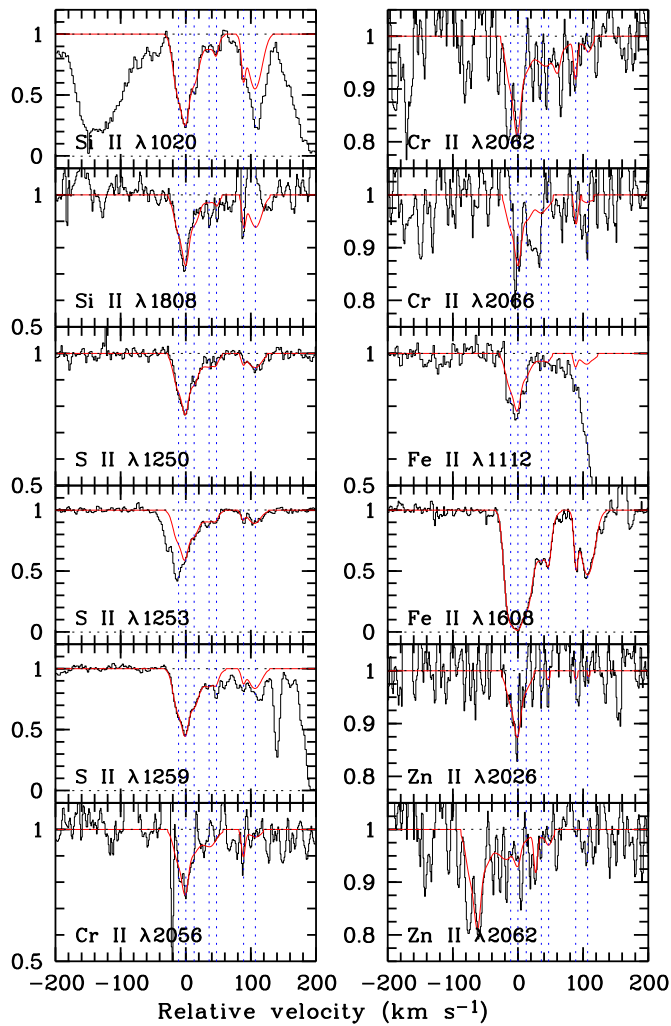}
\caption{Velocity profiles of selected low-ionization transition lines from the
DLA system at $z_{\rm abs}=3.266$ towards Q\,1111$-$152.}
\label{q1111_3.266}
\end{figure}

\clearpage

\begin{figure}
\includegraphics[bb=61 320 402 767,clip,width=8.7cm]{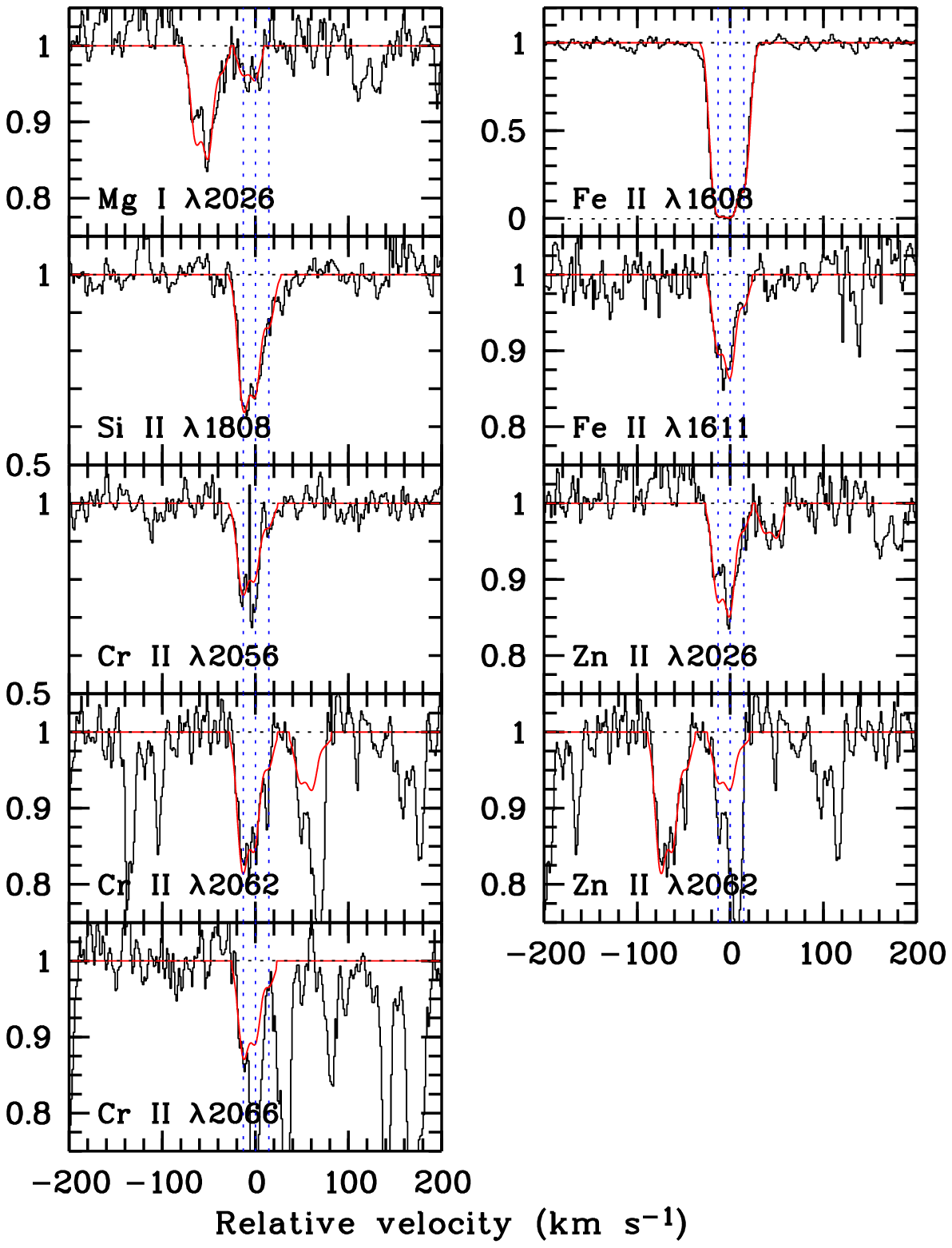}
\caption{Velocity profiles of selected low-ionization transition lines from the
DLA system at $z_{\rm abs}=3.350$ towards Q\,1117$-$134.}
\label{q1117_3.350}
\end{figure}

\begin{figure}
\includegraphics[bb=61 155 402 767,clip,width=8.7cm]{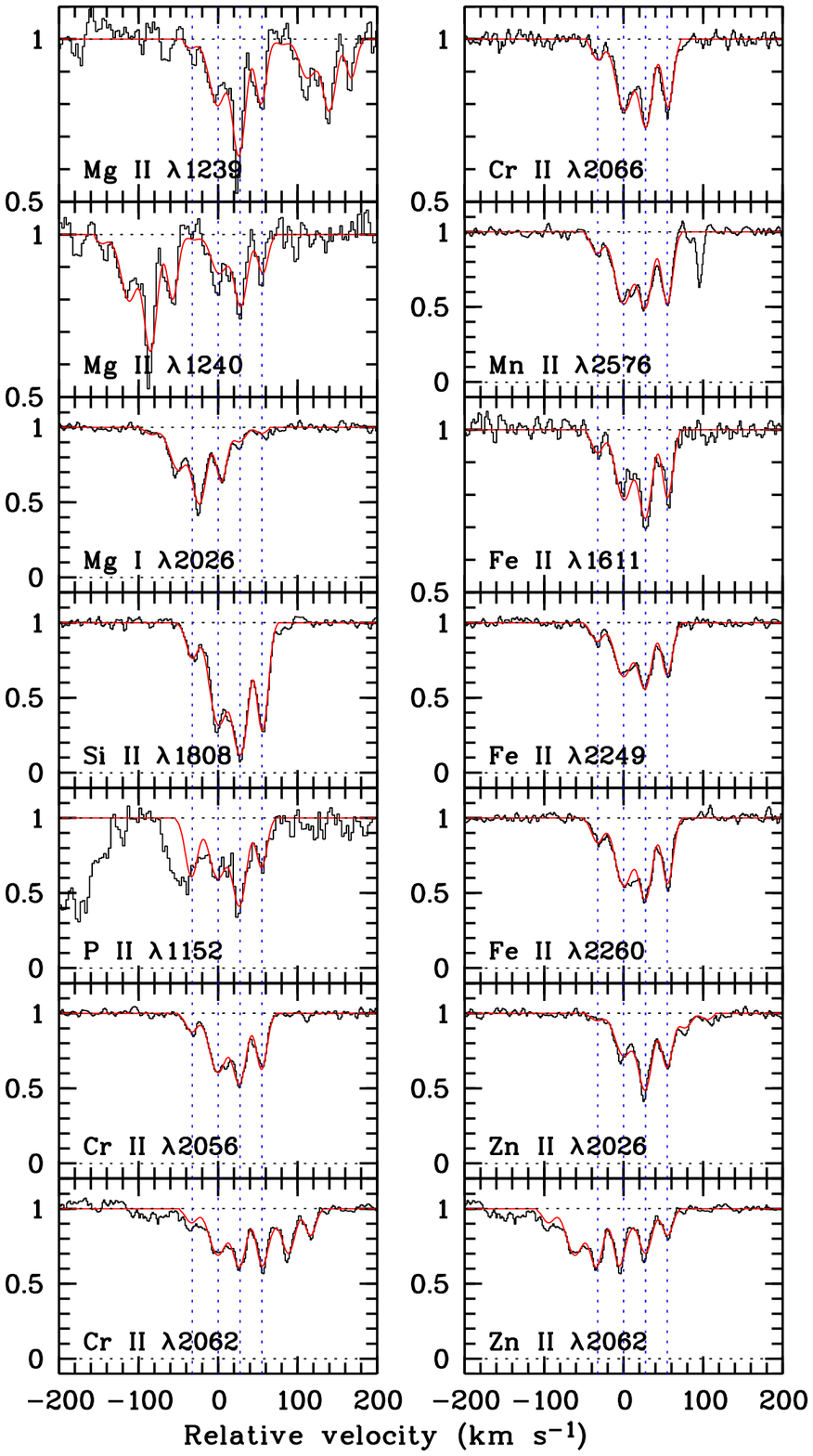}
\caption{Velocity profiles of selected low-ionization transition lines from the
DLA system at $z_{\rm abs}=1.943$ towards Q\,1157$+$014.}
\label{q1157_1.943}
\end{figure}

\clearpage

\begin{figure}
\includegraphics[bb=61 320 402 767,clip,width=8.7cm]{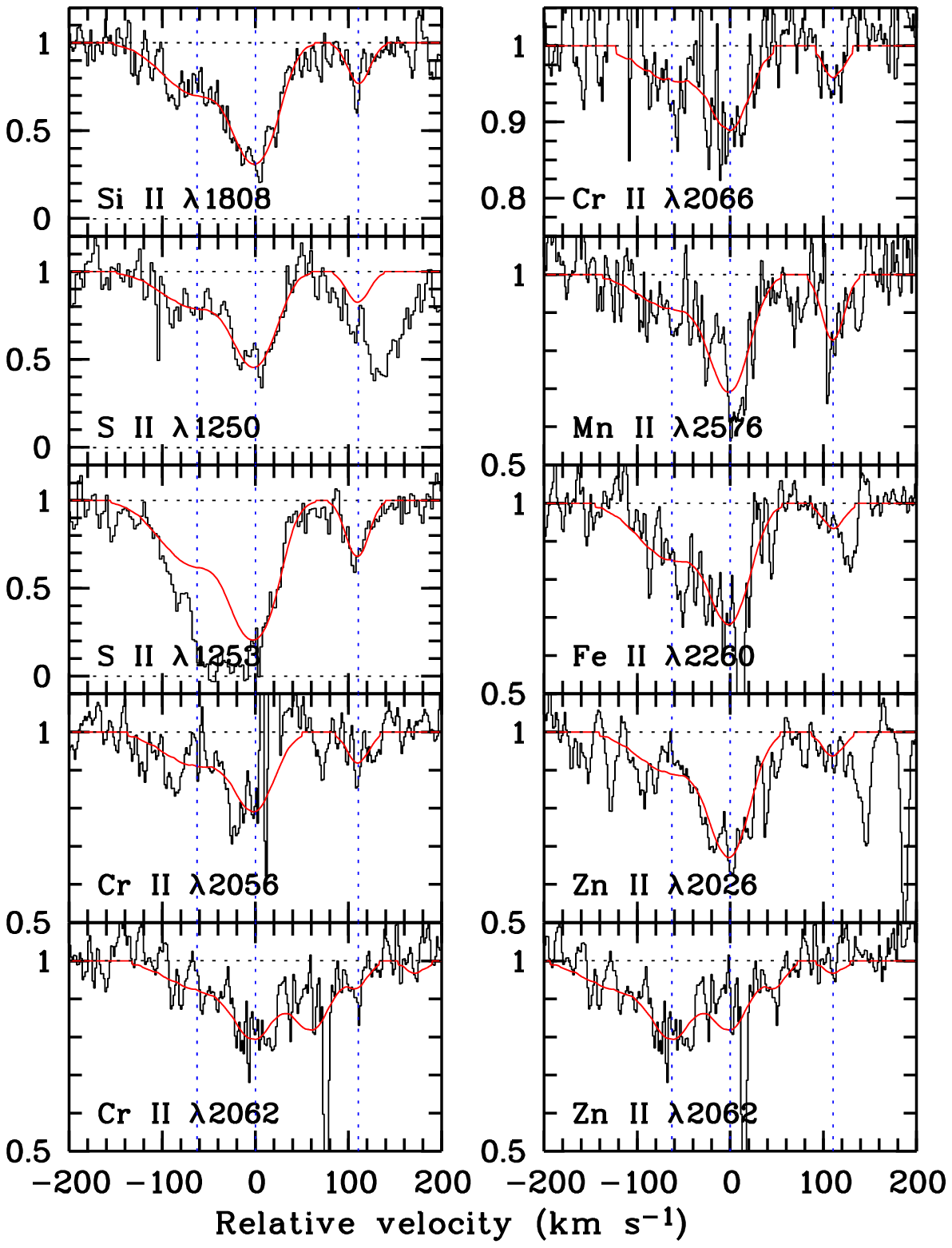}
\caption{Velocity profiles of selected low-ionization transition lines from the
DLA system at $z_{\rm abs}=2.584$ towards Q\,1209$+$093.}
\label{q1209_2.584}
\end{figure}

\begin{figure}
\includegraphics[bb=61 320 402 767,clip,width=8.7cm]{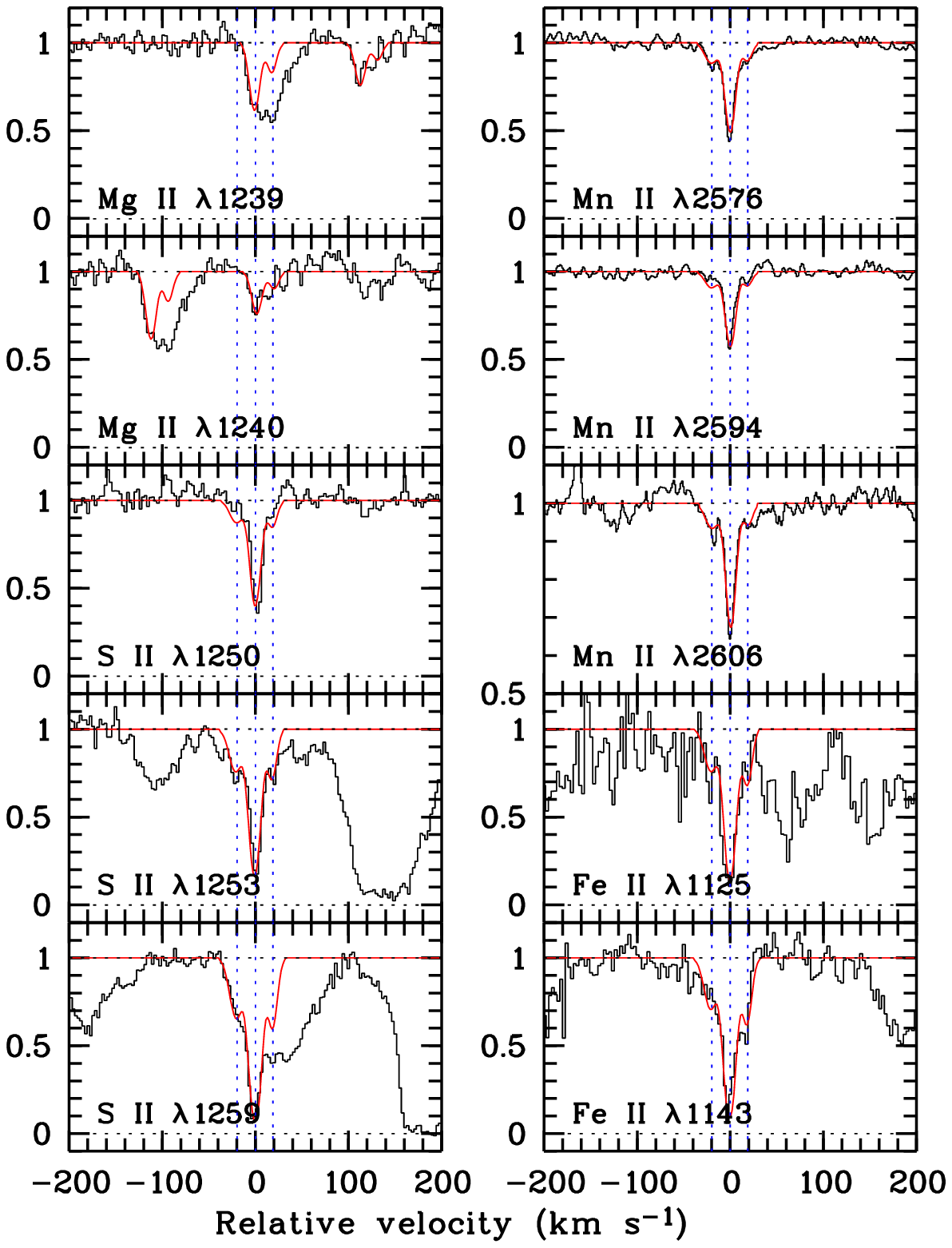}
\caption{Velocity profiles of selected low-ionization transition lines from the
DLA system at $z_{\rm abs}=1.892$ towards Q\,1210$+$175.}
\label{q1210_1.892}
\end{figure}

\begin{figure*}
\hbox{
\includegraphics[bb=61 402 402 767,clip,width=8.7cm]{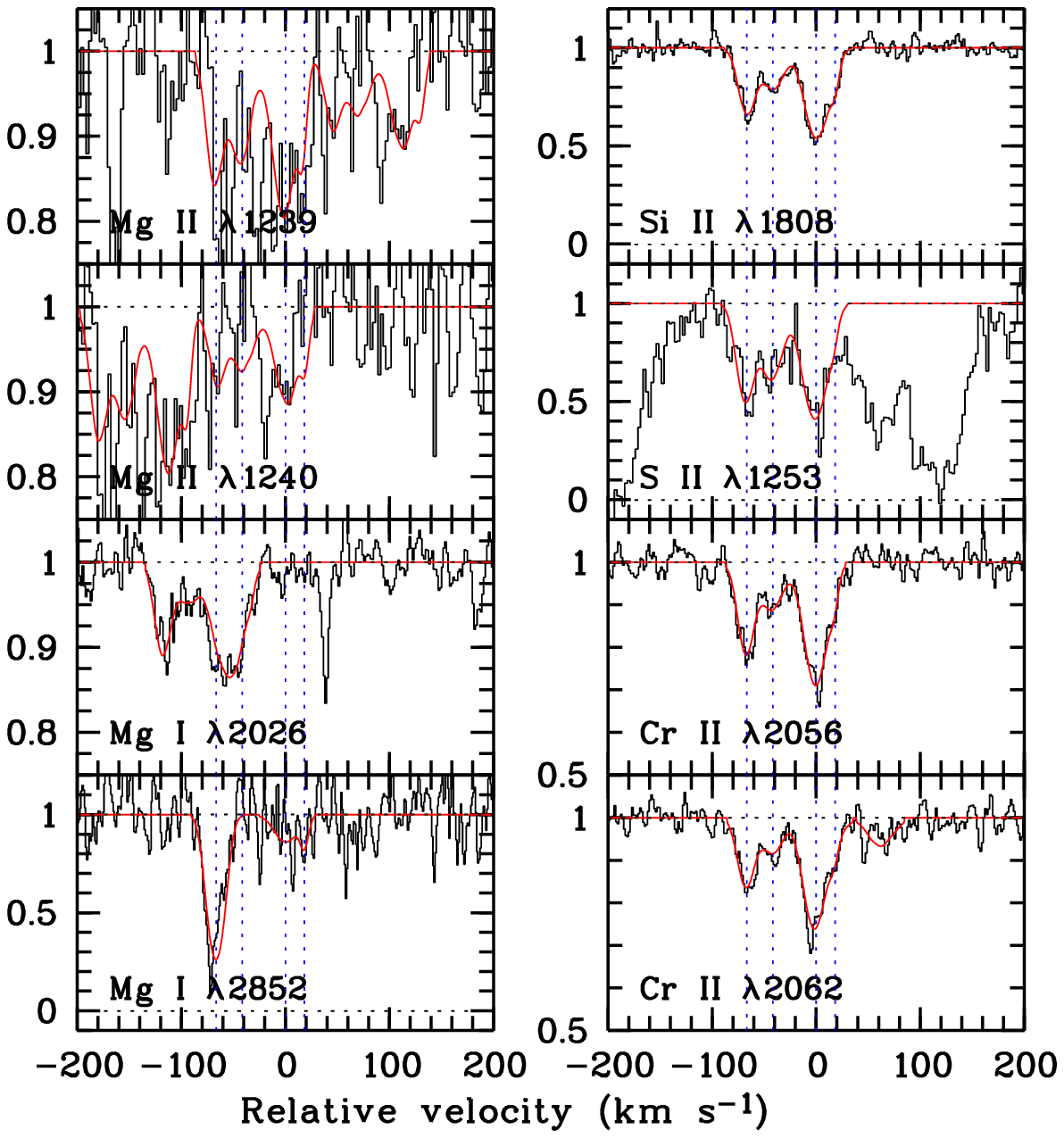}
\includegraphics[bb=61 402 402 767,clip,width=8.7cm]{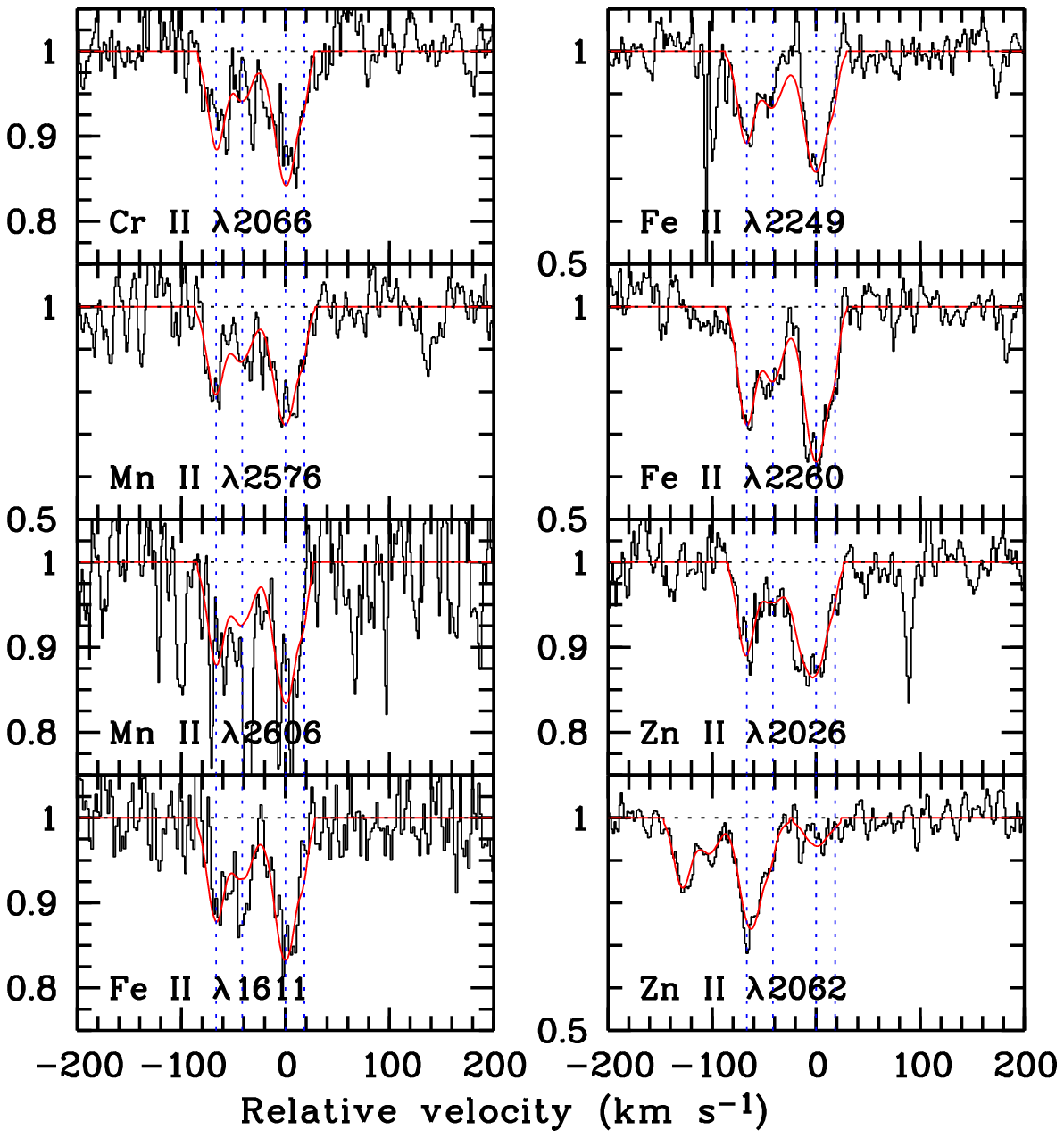}}
\caption{Velocity profiles of selected low-ionization transition lines from the
DLA system at $z_{\rm abs}=2.466$ towards Q\,1223$+$178.}
\label{q1223_2.466}
\end{figure*}

\begin{figure}
\includegraphics[bb=61 402 402 767,clip,width=8.7cm]{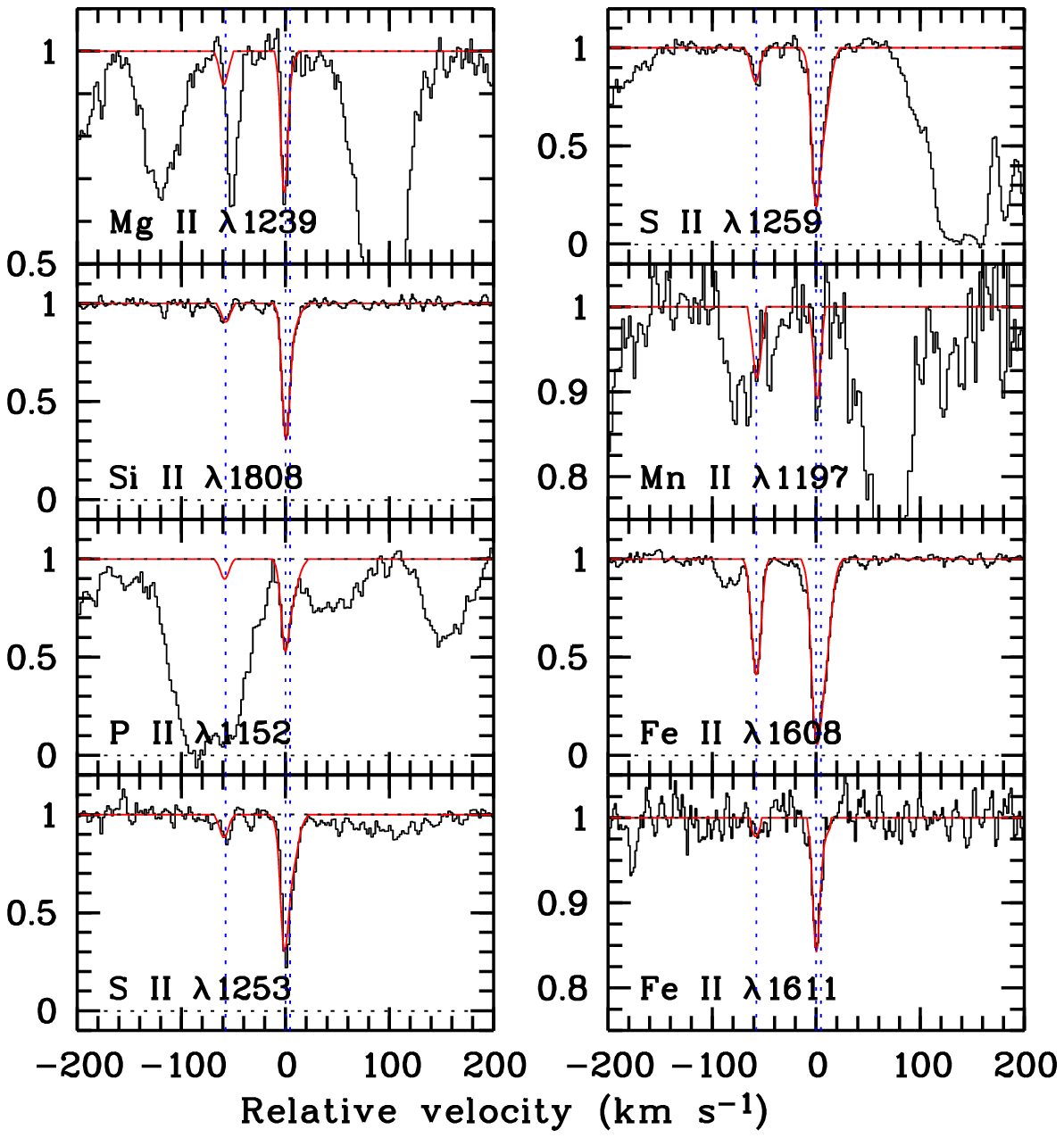}
\caption{Velocity profiles of selected low-ionization transition lines from the
DLA system at $z_{\rm abs}=2.338$ towards Q\,1232$+$082.}
\label{q1232_2.338}
\end{figure}

\clearpage

\begin{figure}
\includegraphics[bb=61 73 402 767,clip,width=8.7cm]{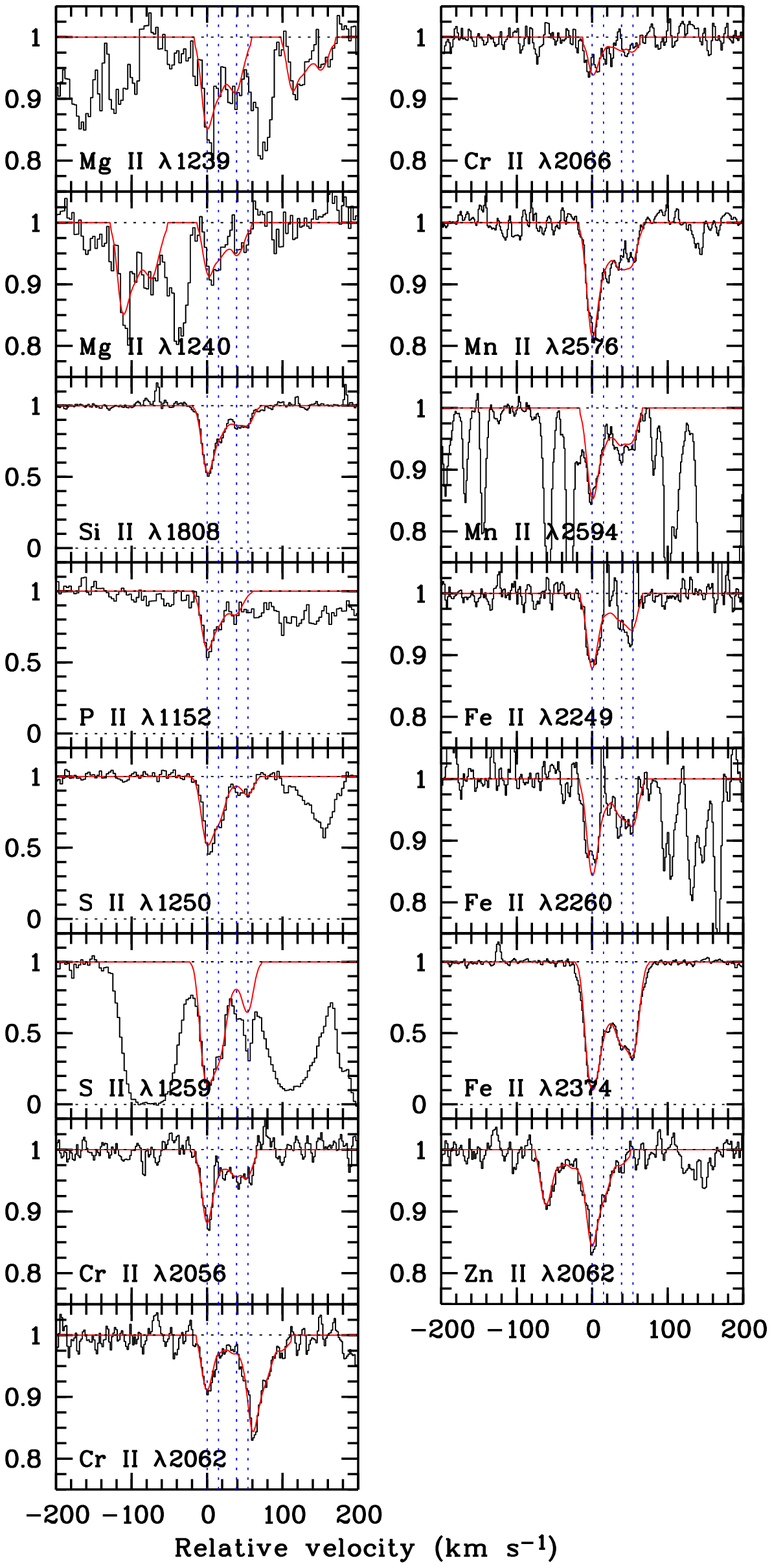}
\caption{Velocity profiles of selected low-ionization transition lines from the
DLA system at $z_{\rm abs}=1.776$ towards Q\,1331$+$170.}
\label{q1331_1.776}
\end{figure}

\begin{figure}
\includegraphics[bb=61 402 402 767,clip,width=8.7cm]{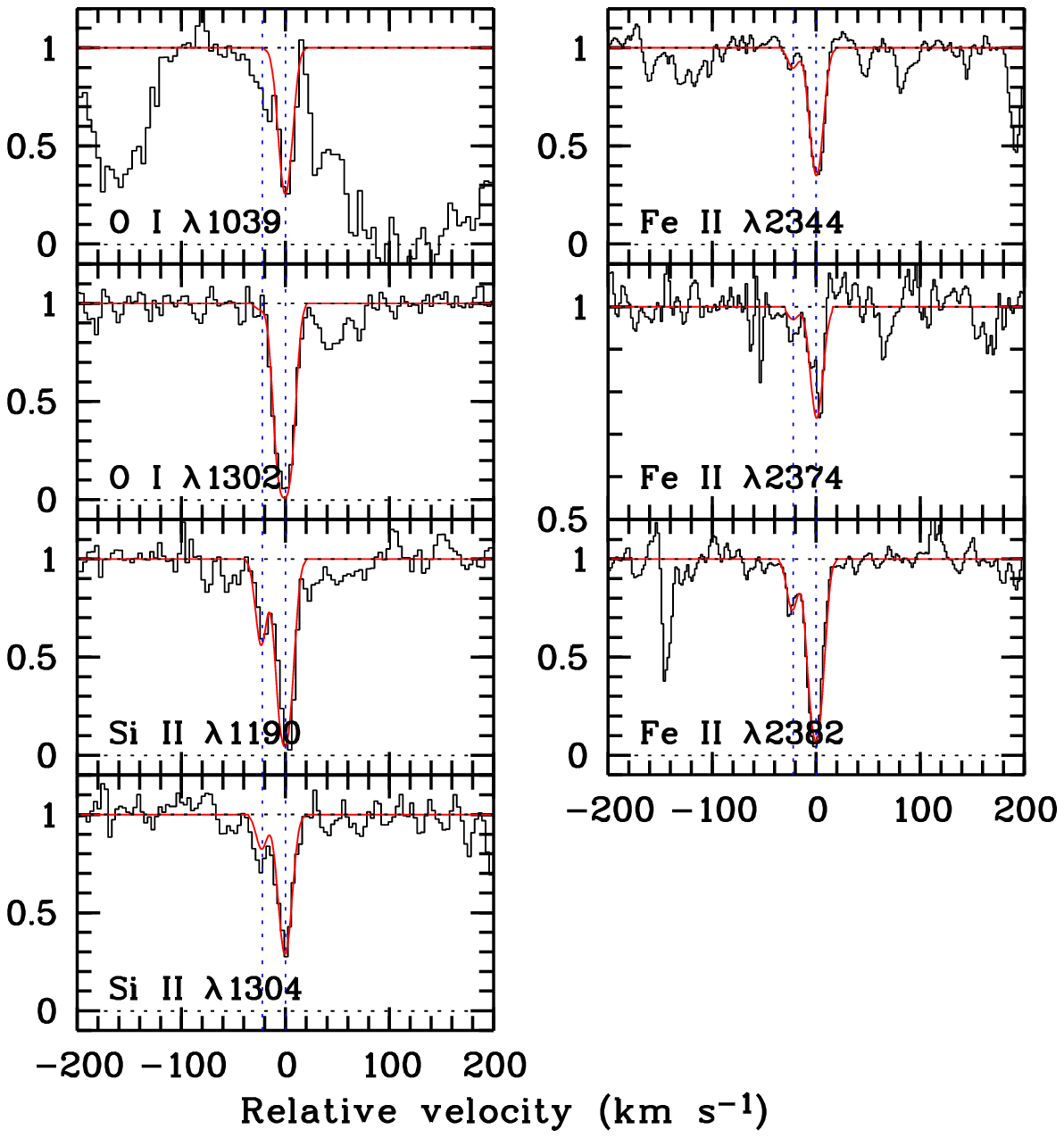}
\caption{Velocity profiles of selected low-ionization transition lines from the
sub-DLA system at $z_{\rm abs}=2.508$ towards Q\,1337$+$113.}
\label{q1337_2.508}
\end{figure}

\clearpage

\begin{figure}
\includegraphics[bb=61 238 402 767,clip,width=8.7cm]{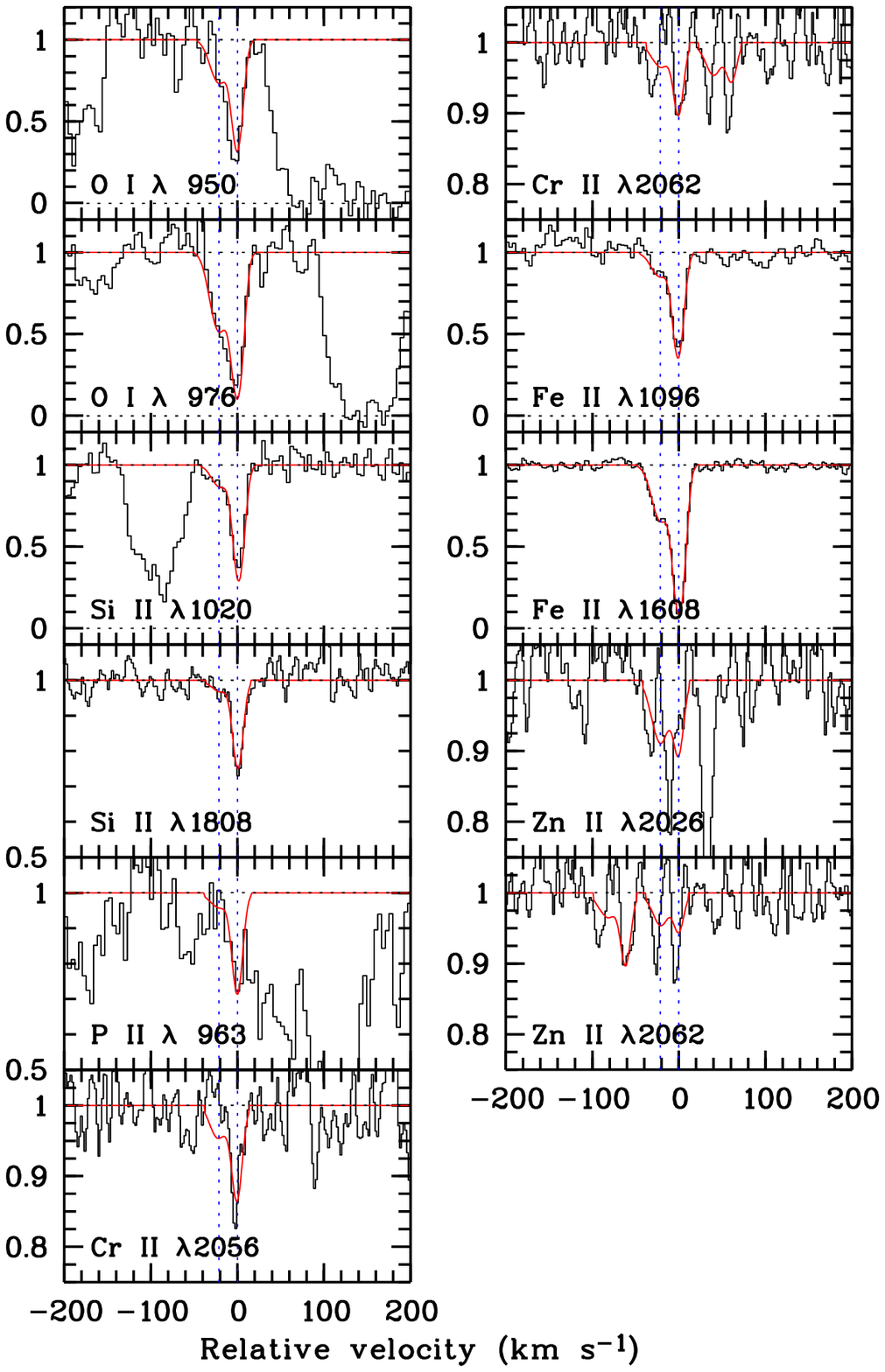}
\caption{Velocity profiles of selected low-ionization transition lines from the
DLA system at $z_{\rm abs}=2.796$ towards Q\,1337$+$113.}
\label{q1337_2.796}
\end{figure}

\begin{figure}
\includegraphics[bb=61 320 402 767,clip,width=8.7cm]{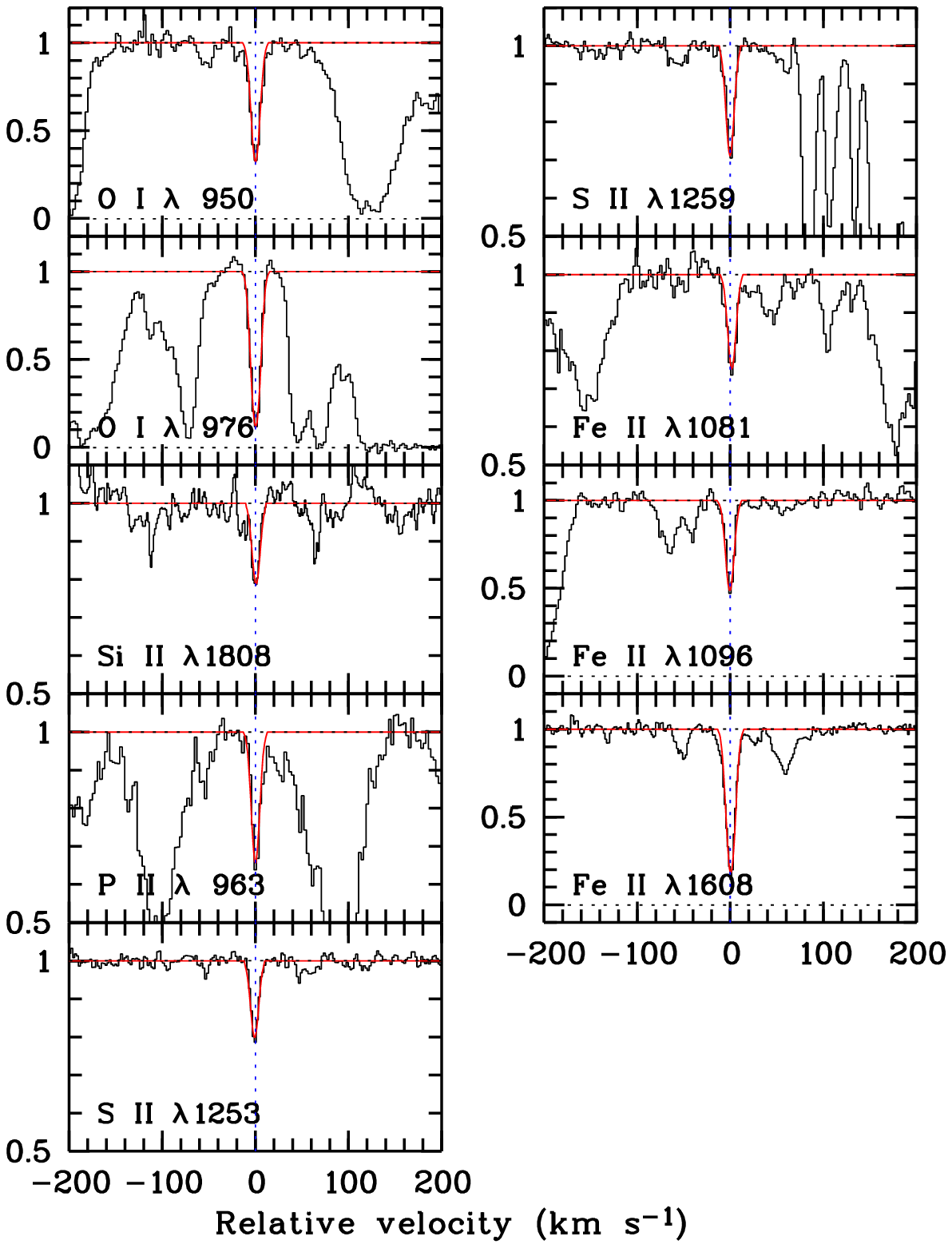}
\caption{Velocity profiles of selected low-ionization transition lines from the
sub-DLA system at $z_{\rm abs}=3.118$ towards Q\,1340$-$136.}
\label{q1340_3.118}
\end{figure}

\clearpage

\begin{figure}
\includegraphics[bb=61 238 402 767,clip,width=8.7cm]{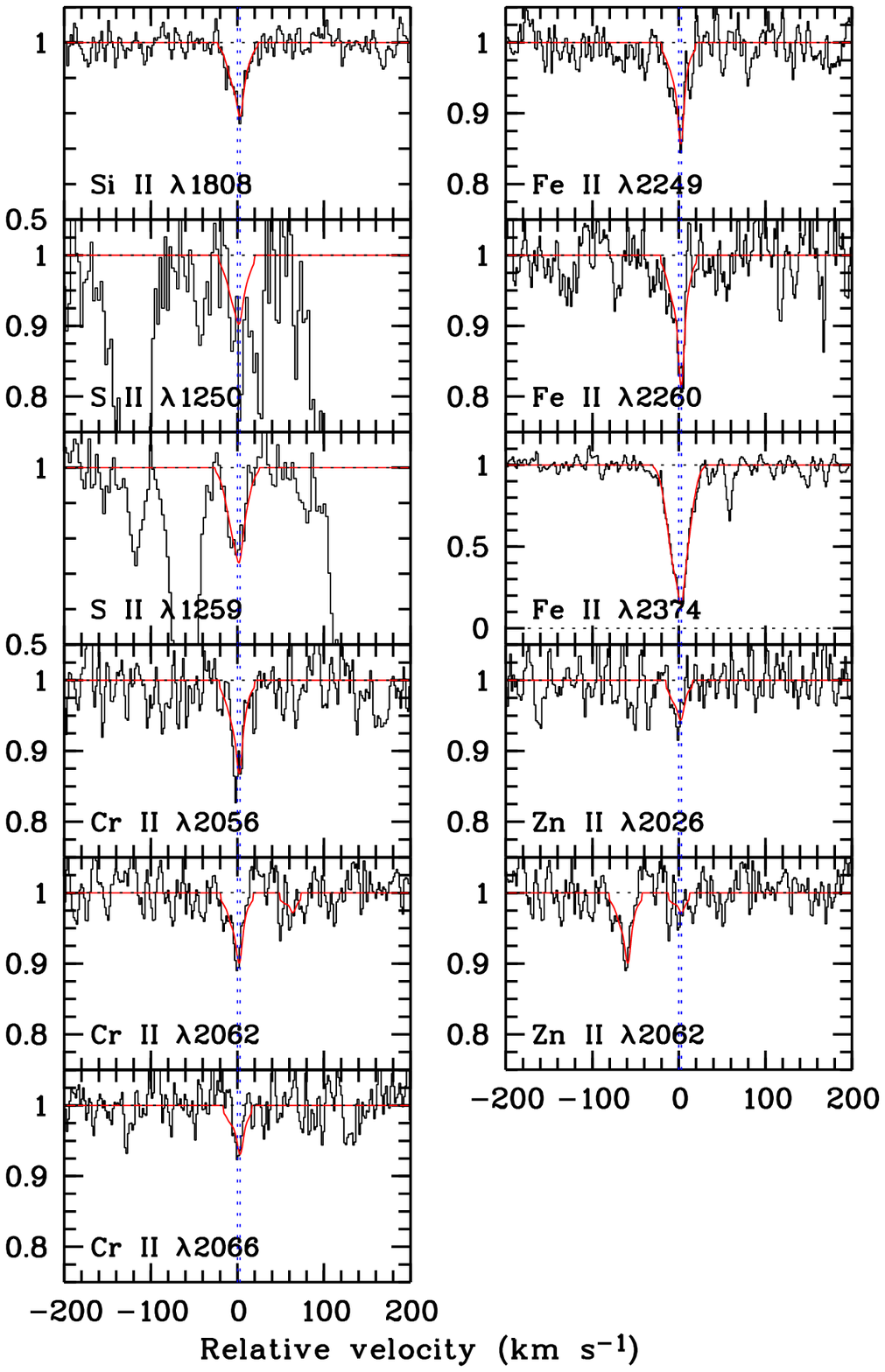}
\caption{Velocity profiles of selected low-ionization transition lines from the
DLA system at $z_{\rm abs}=2.019$ towards Q\,1409$+$095.}
\label{q1409_2.019}
\end{figure}

\begin{figure}
\includegraphics[bb=61 485 402 767,clip,width=8.7cm]{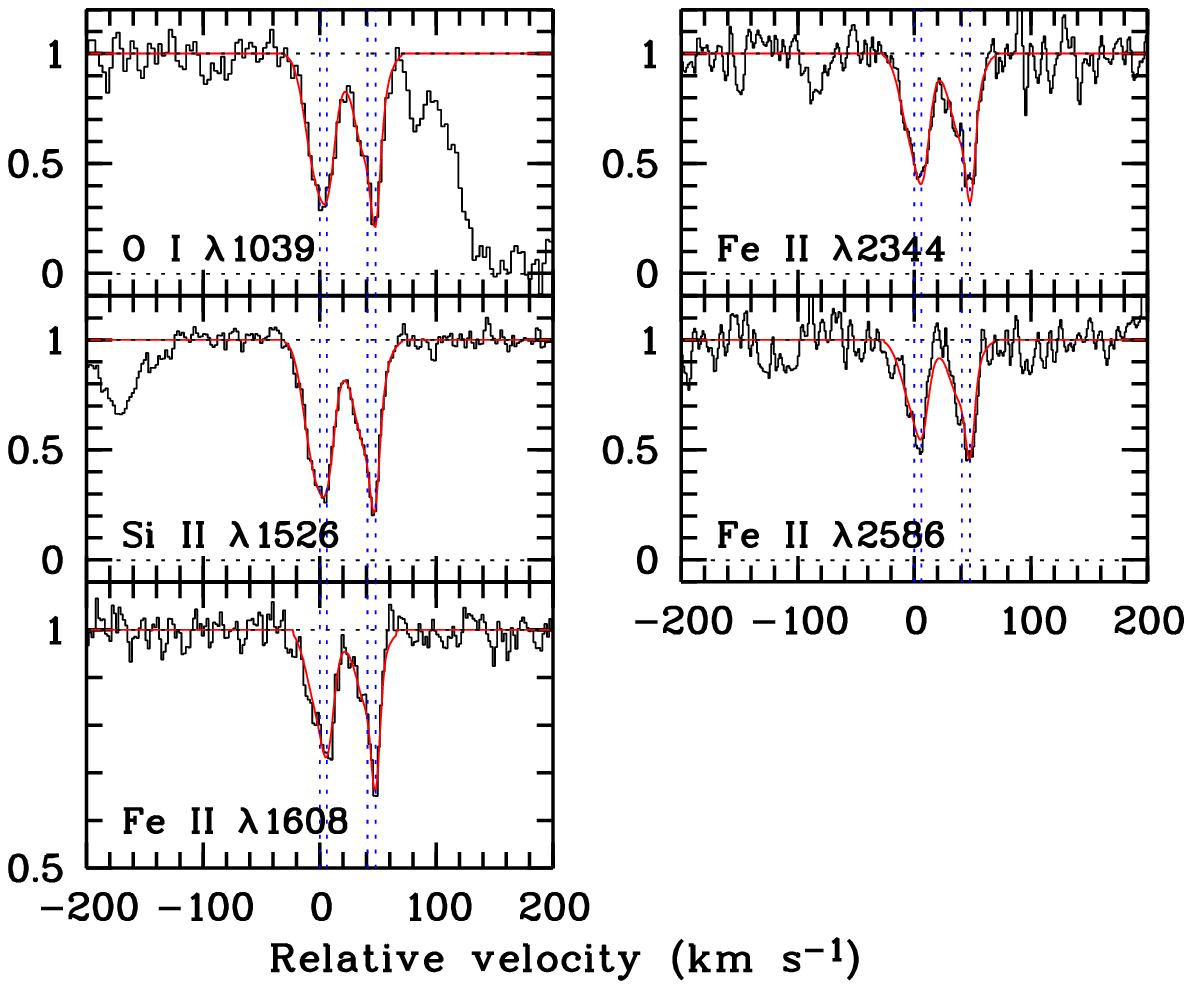}
\caption{Velocity profiles of selected low-ionization transition lines from the
DLA system at $z_{\rm abs}=2.456$ towards Q\,1409$+$095.}
\label{q1409_2.456}
\end{figure}

\begin{figure}
\includegraphics[bb=61 402 402 767,clip,width=8.7cm]{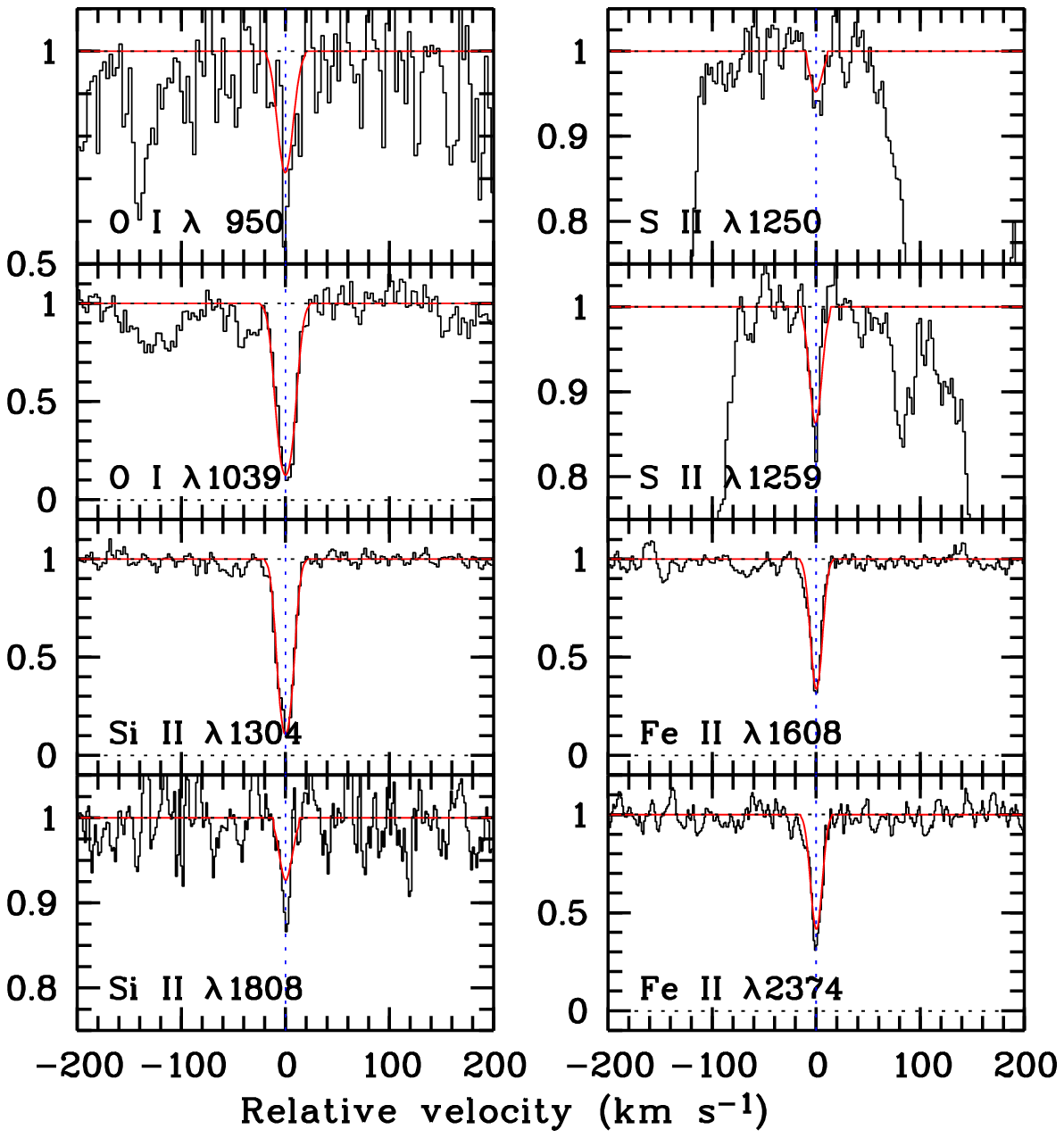}
\caption{Velocity profiles of selected low-ionization transition lines from the
sub-DLA system at $z_{\rm abs}=2.668$ towards Q\,1409$+$095.}
\label{q1409_2.668}
\end{figure}

\begin{figure}
\includegraphics[bb=61 320 402 767,clip,width=8.7cm]{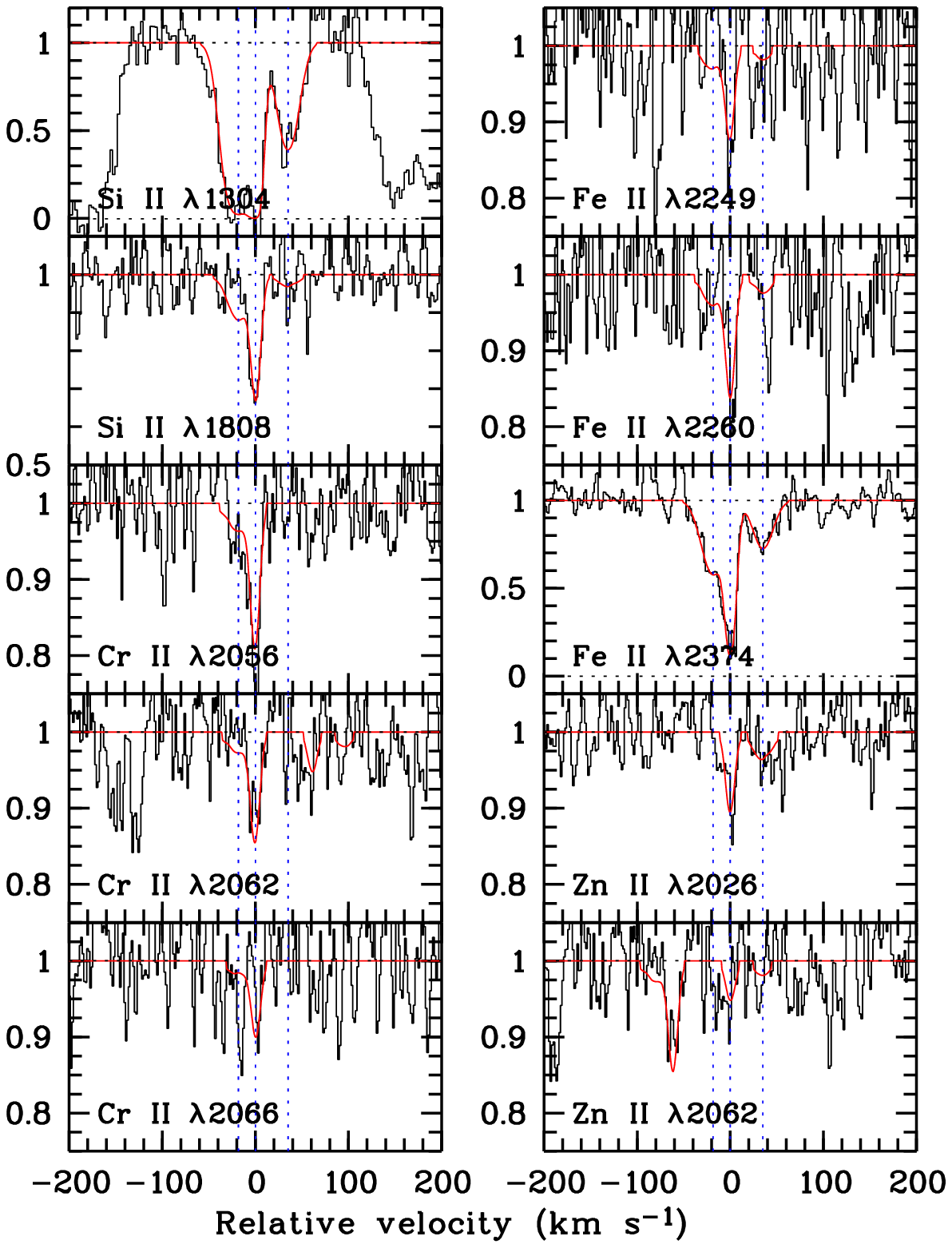}
\caption{Velocity profiles of selected low-ionization transition lines from the
DLA system at $z_{\rm abs}=2.255$ towards Q\,1451$+$123.}
\label{q1451_2.255}
\end{figure}

\clearpage

\begin{figure}
\includegraphics[bb=61 567 402 767,clip,width=8.7cm]{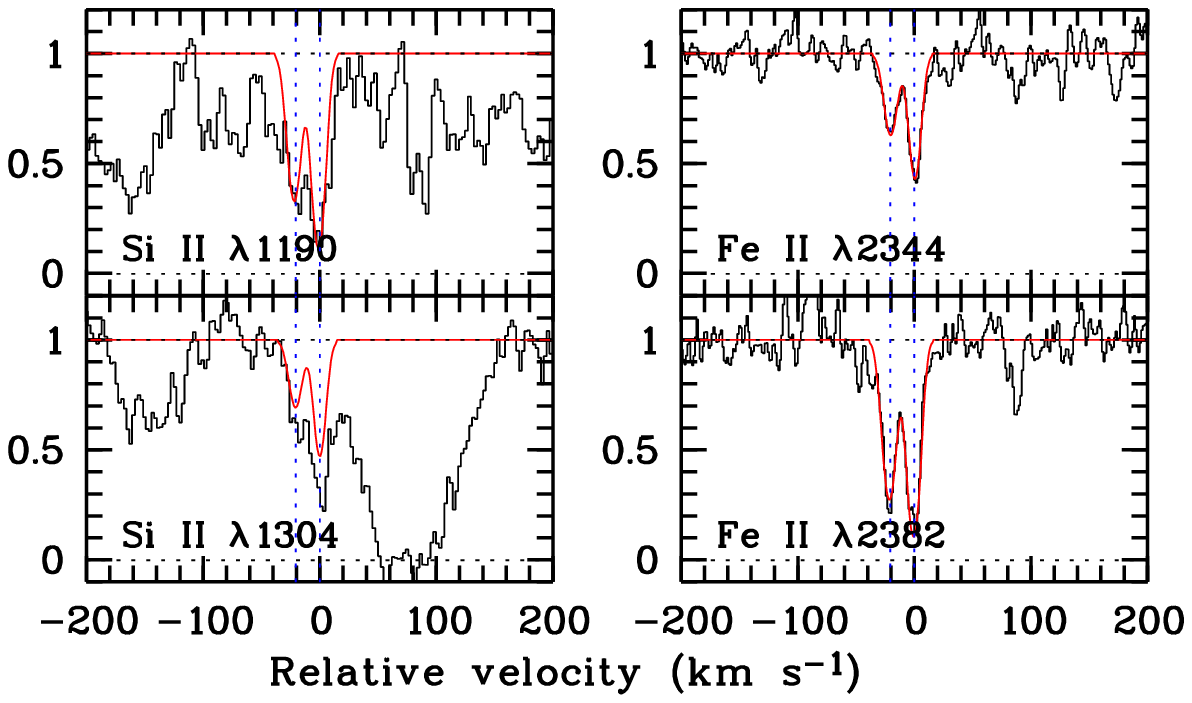}
\caption{Velocity profiles of selected low-ionization transition lines from the
DLA system at $z_{\rm abs}=2.469$ towards Q\,1451$+$123.}
\label{q1451_2.469}
\end{figure}

\begin{figure}
\includegraphics[bb=61 650 402 767,clip,width=8.7cm]{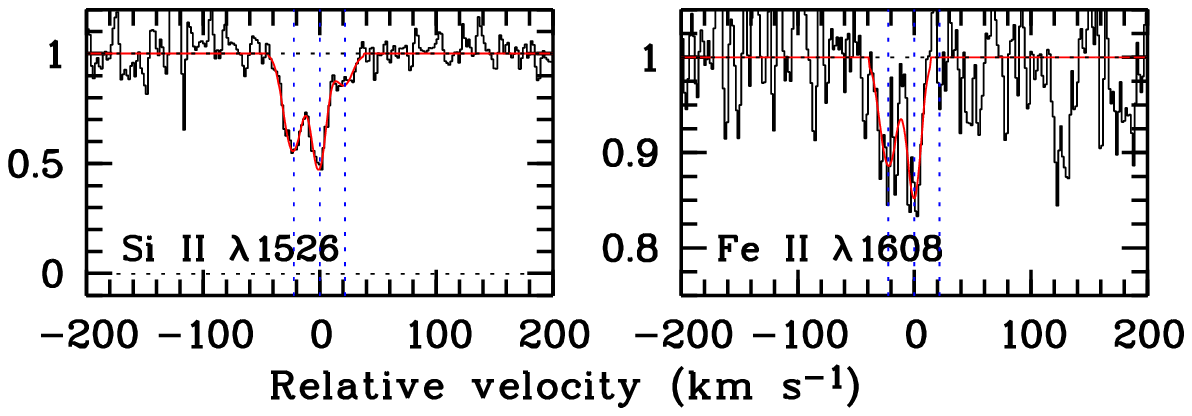}
\caption{Velocity profiles of selected low-ionization transition lines from the
sub-DLA system at $z_{\rm abs}=3.171$ towards Q\,1451$+$123.}
\label{q1451_3.171}
\end{figure}

\begin{figure}
\includegraphics[bb=61 402 402 767,clip,width=8.7cm]{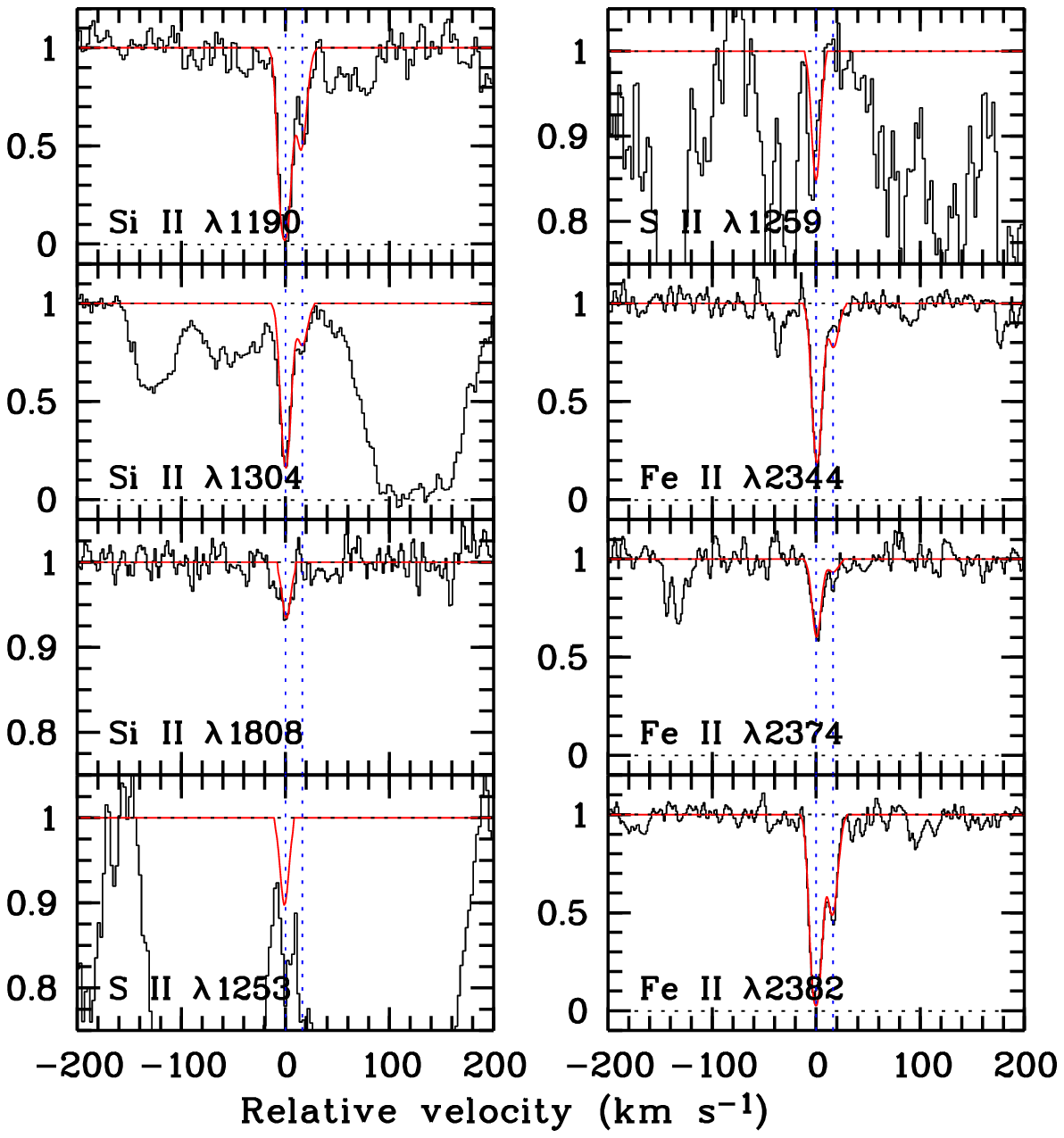}
\caption{Velocity profiles of selected low-ionization transition lines from the
sub-DLA system at $z_{\rm abs}=2.507$ towards Q\,2059$-$360.}
\label{q2059_2.507}
\end{figure}

\begin{figure}
\includegraphics[bb=61 320 402 767,clip,width=8.7cm]{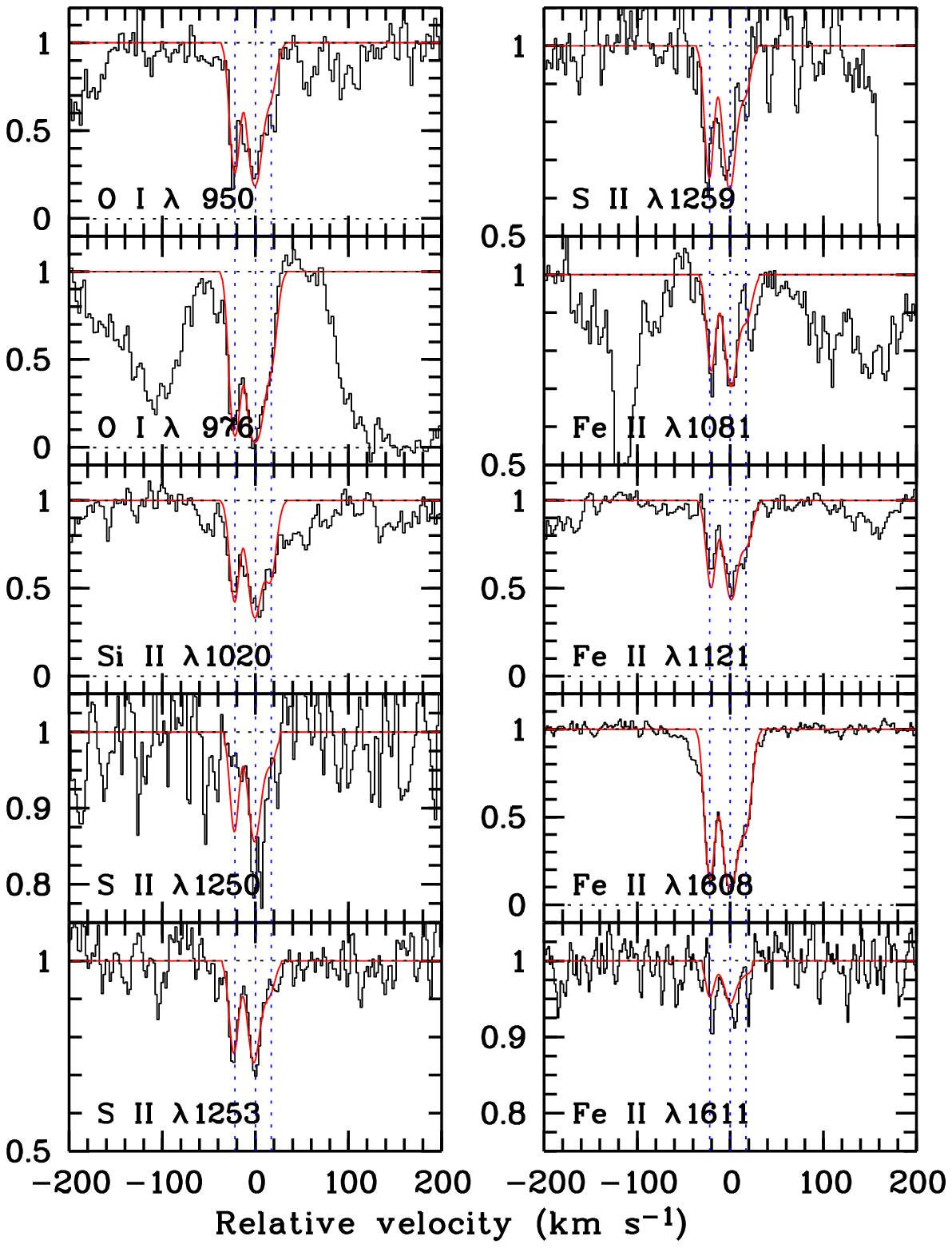}
\caption{Velocity profiles of selected low-ionization transition lines from the
DLA system at $z_{\rm abs}=3.083$ towards Q\,2059$-$360.}
\label{q2059_3.083}
\end{figure}

\clearpage

\begin{figure}
\includegraphics[bb=61 238 402 767,clip,width=8.7cm]{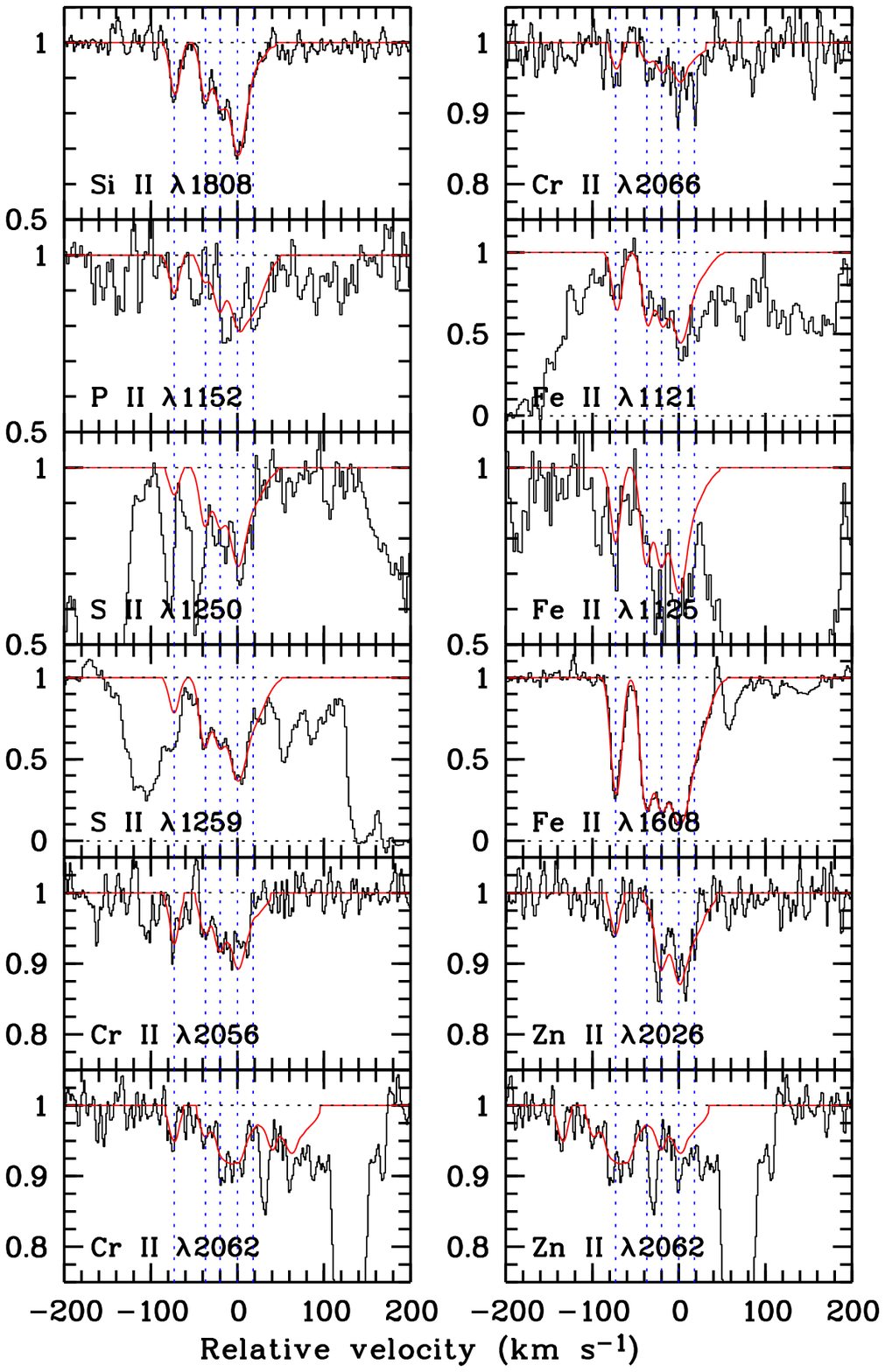}
\caption{Velocity profiles of selected low-ionization transition lines from the
sub-DLA system at $z_{\rm abs}=1.996$ towards Q\,2116$-$358.}
\label{q2116_1.996}
\end{figure}

\begin{figure}
\includegraphics[bb=61 155 402 767,clip,width=8.7cm]{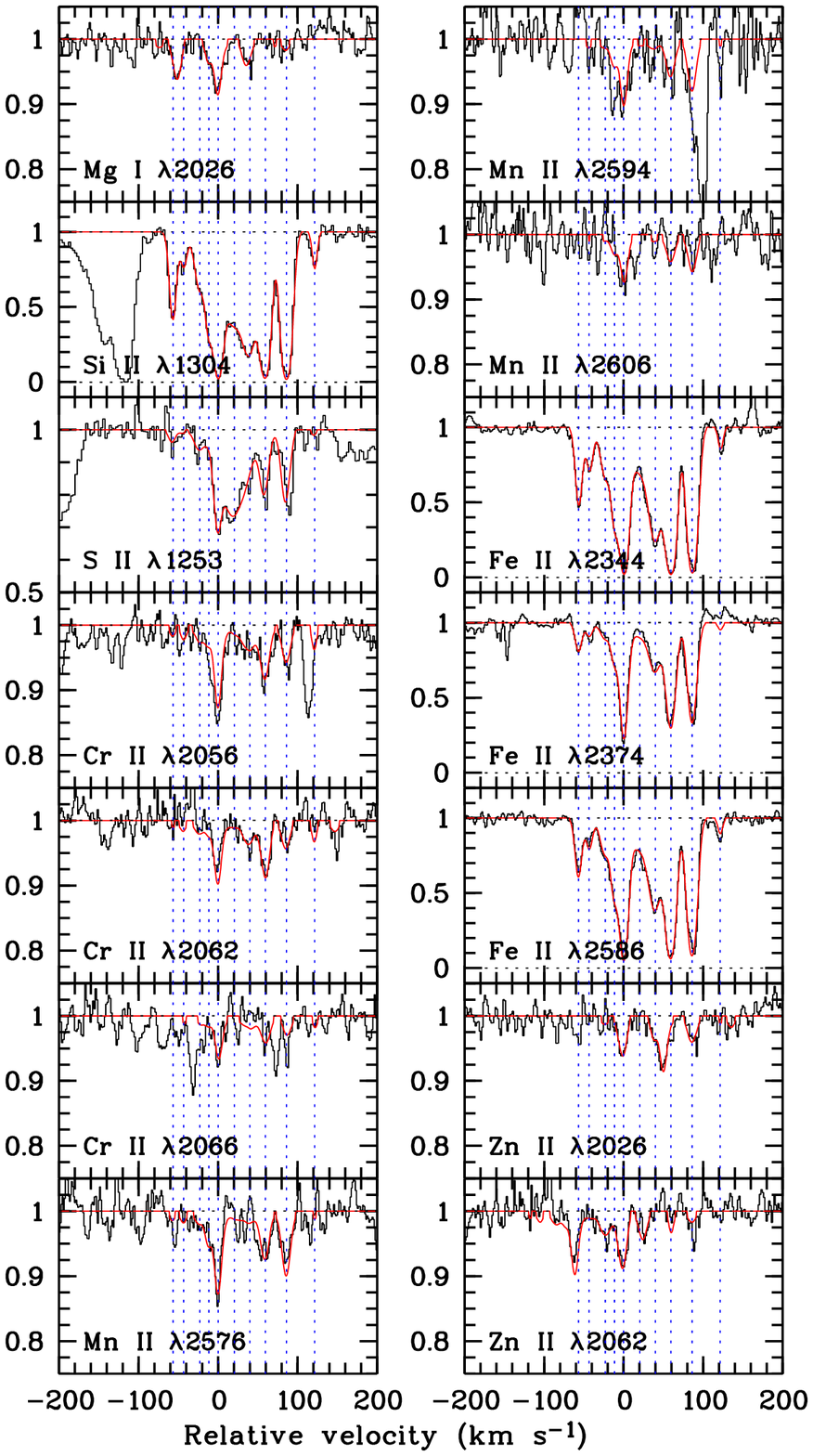}
\caption{Velocity profiles of selected low-ionization transition lines from the
DLA system at $z_{\rm abs}=2.383$ towards Q\,2138$-$444.}
\label{q2138_2.383}
\end{figure}

\clearpage

\begin{figure*}
\hbox{
\includegraphics[bb=61 320 402 767,clip,width=8.7cm]{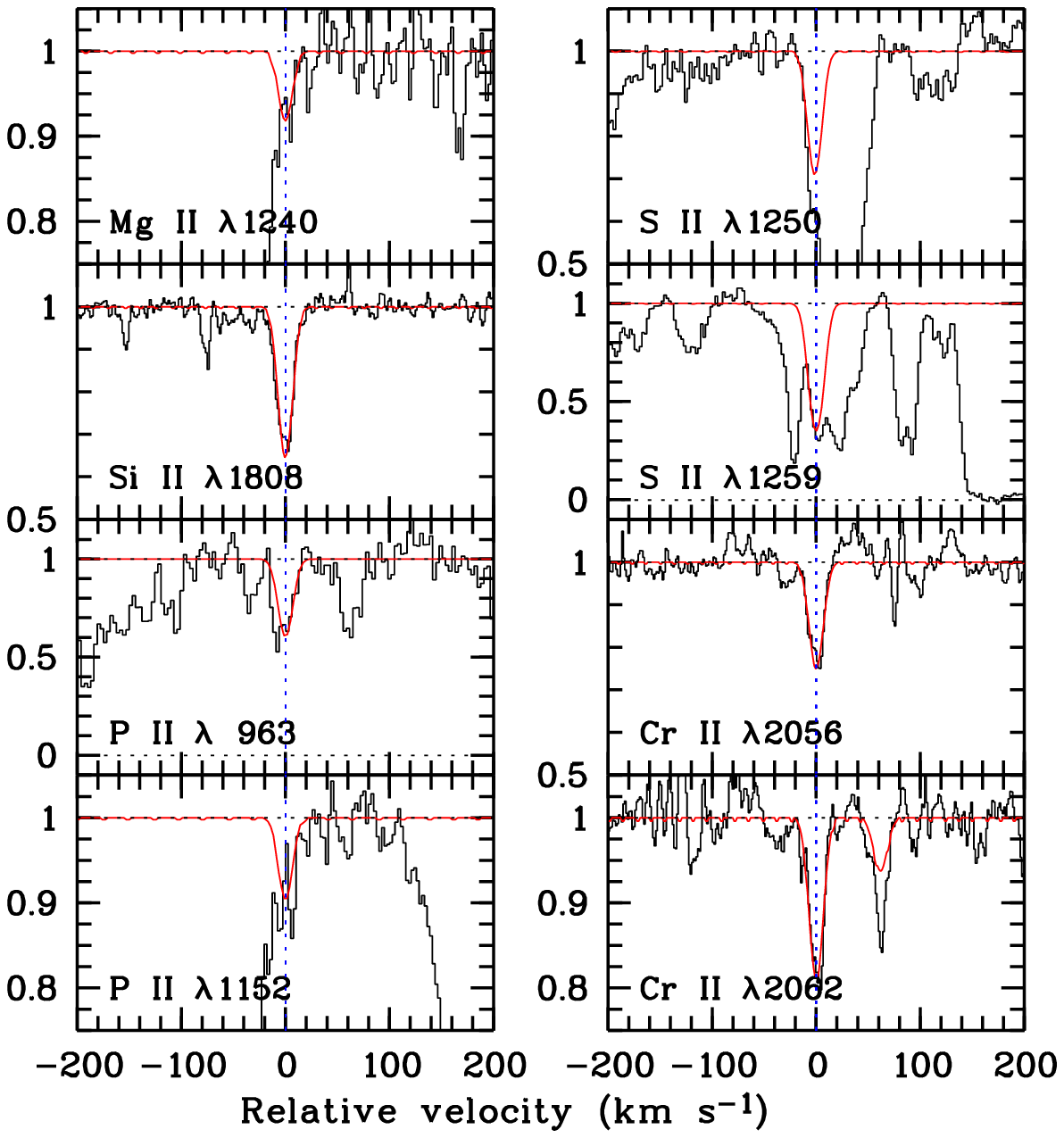}
\includegraphics[bb=61 320 402 767,clip,width=8.7cm]{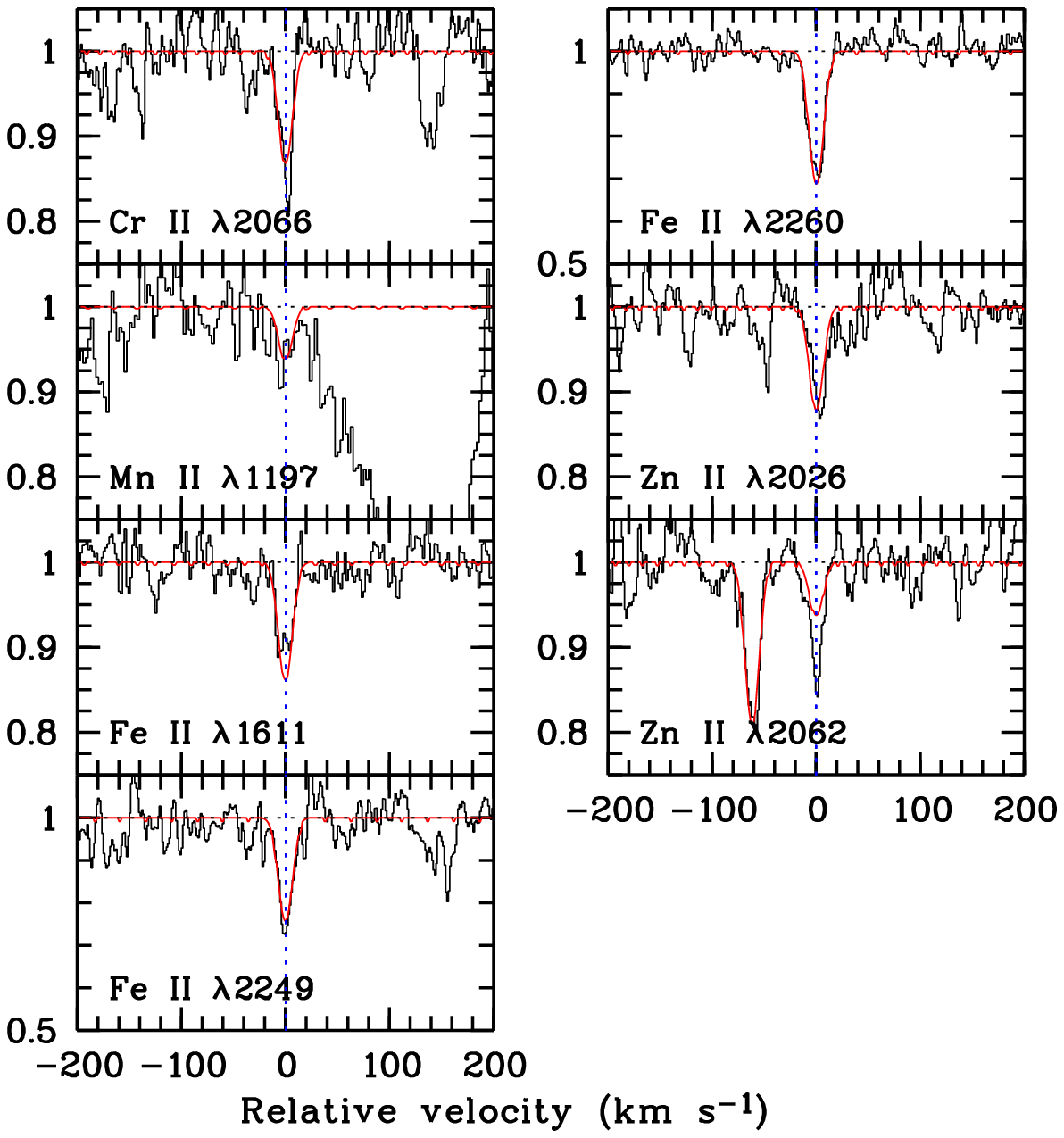}}
\caption{Velocity profiles of selected low-ionization transition lines from the
DLA system at $z_{\rm abs}=2.852$ towards Q\,2138$-$444.}
\label{q2138_2.852}
\end{figure*}

\begin{figure}
\includegraphics[bb=61 485 402 767,clip,width=8.7cm]{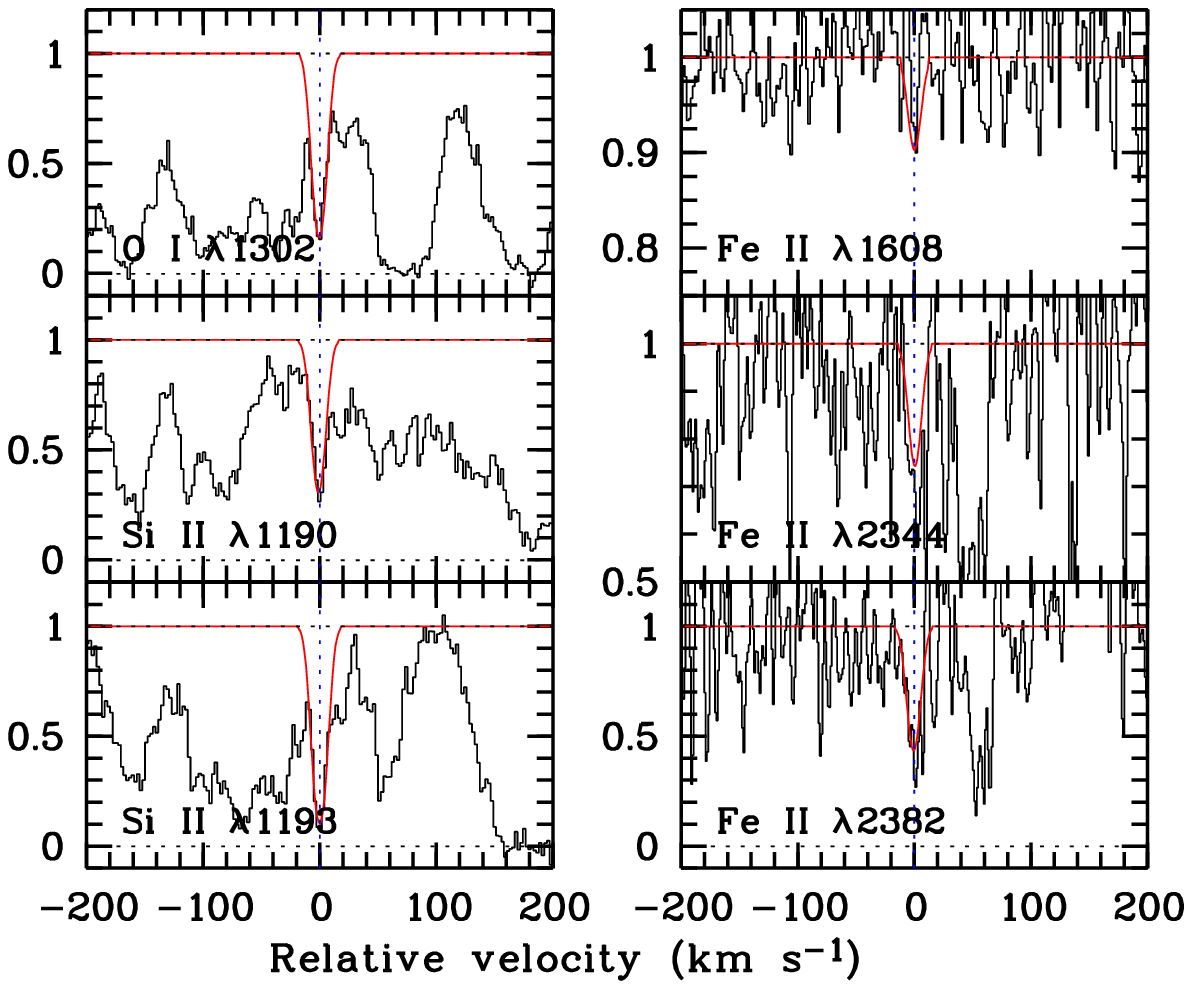}
\caption{Velocity profiles of selected low-ionization transition lines from the
sub-DLA system at $z_{\rm abs}=3.142$ towards Q\,2152$+$137.}
\label{q2152_3.142}
\end{figure}

\begin{figure}
\includegraphics[bb=61 567 402 767,clip,width=8.7cm]{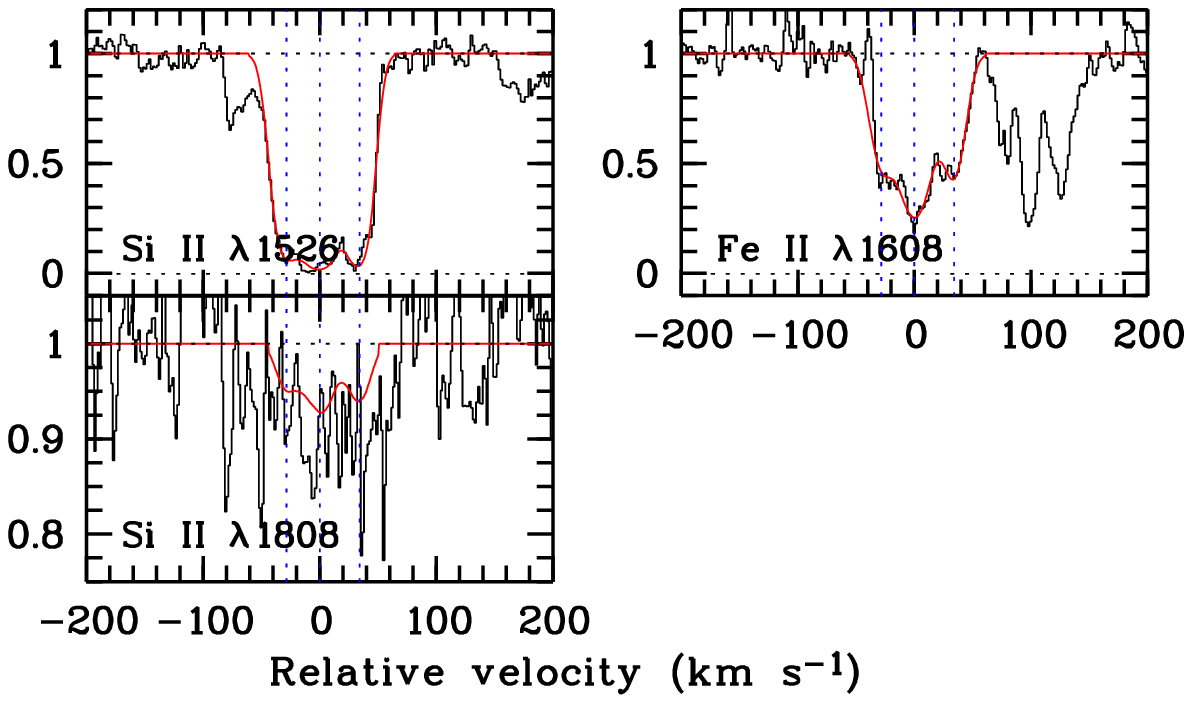}
\caption{Velocity profiles of selected low-ionization transition lines from the
DLA system at $z_{\rm abs}=3.316$ towards Q\,2152$+$137.}
\label{q2152_3.316}
\end{figure}

\begin{figure}
\includegraphics[bb=61 650 402 767,clip,width=8.7cm]{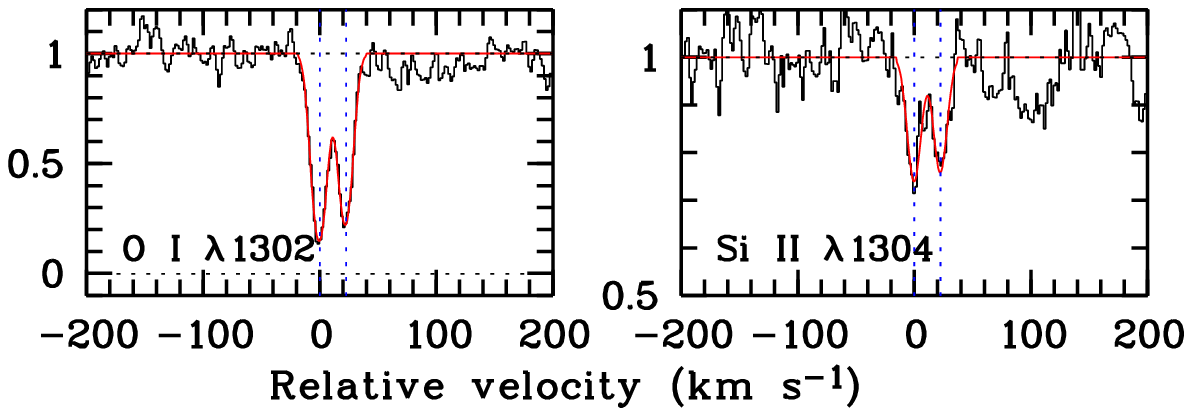}
\caption{Velocity profiles of selected low-ionization transition lines from the
sub-DLA system at $z_{\rm abs}=4.212$ towards Q\,2152$+$137.}
\label{q2152_4.212}
\end{figure}

\begin{figure}
\includegraphics[bb=61 73 402 767,clip,width=8.7cm]{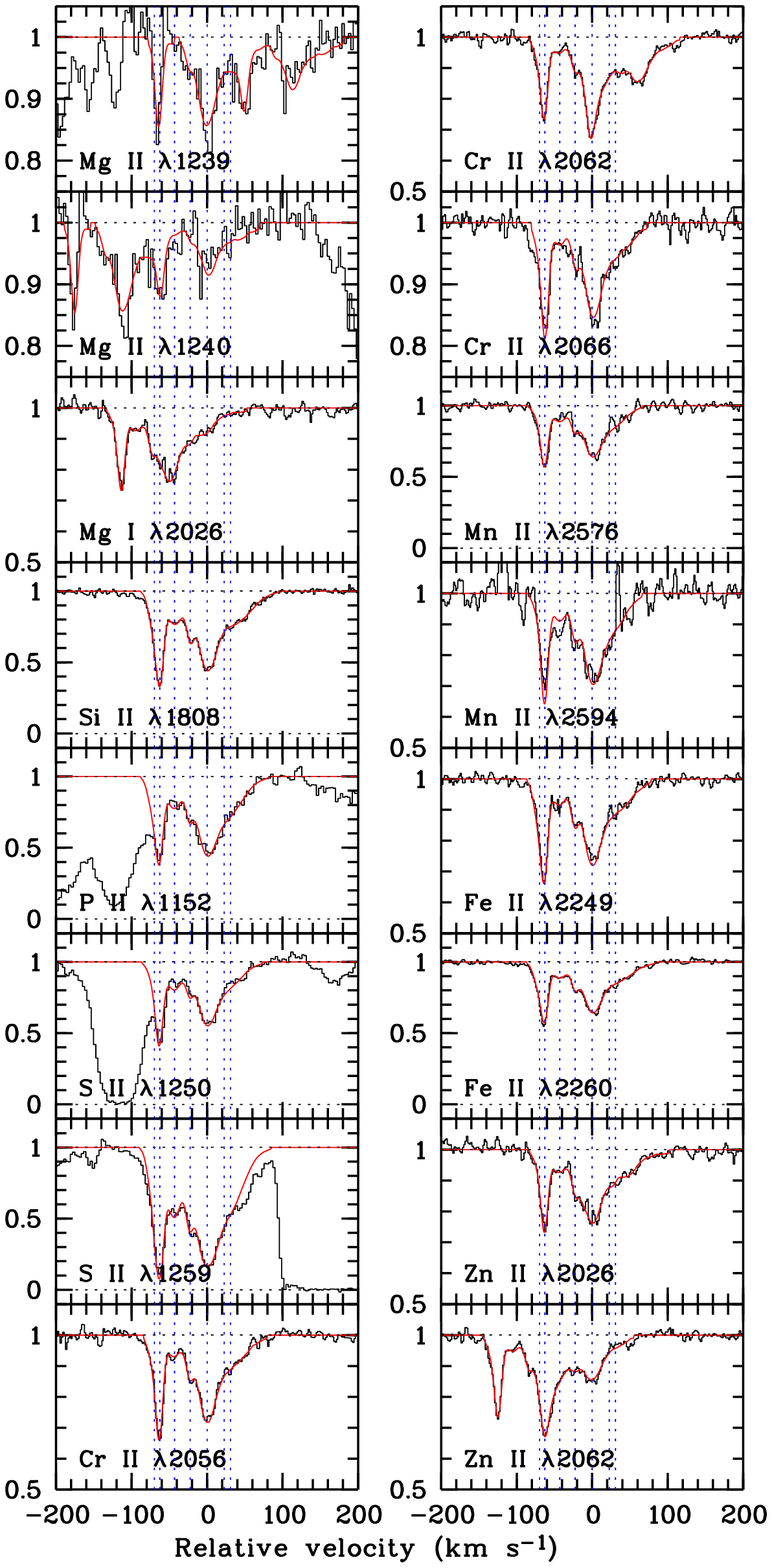}
\caption{Velocity profiles of selected low-ionization transition lines from the
DLA system at $z_{\rm abs}=1.921$ towards Q\,2206$-$199.}
\label{q2206_1.921}
\end{figure}

\clearpage

\begin{figure}
\includegraphics[bb=61 485 402 767,clip,width=8.7cm]{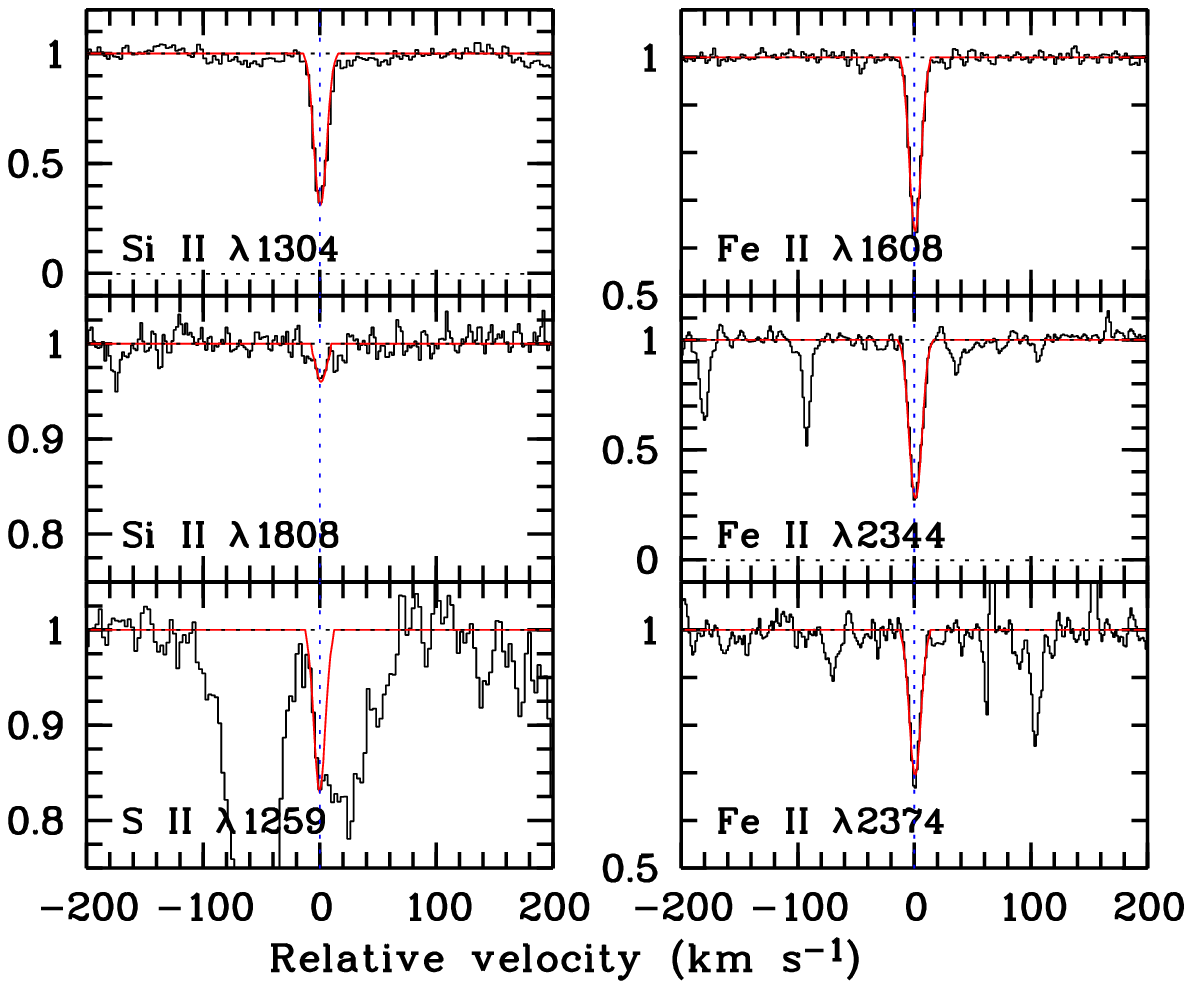}
\caption{Velocity profiles of selected low-ionization transition lines from the
DLA system at $z_{\rm abs}=2.076$ towards Q\,2206$-$199.}
\label{q2206_2.076}
\end{figure}

\begin{figure}
\includegraphics[bb=61 402 402 767,clip,width=8.7cm]{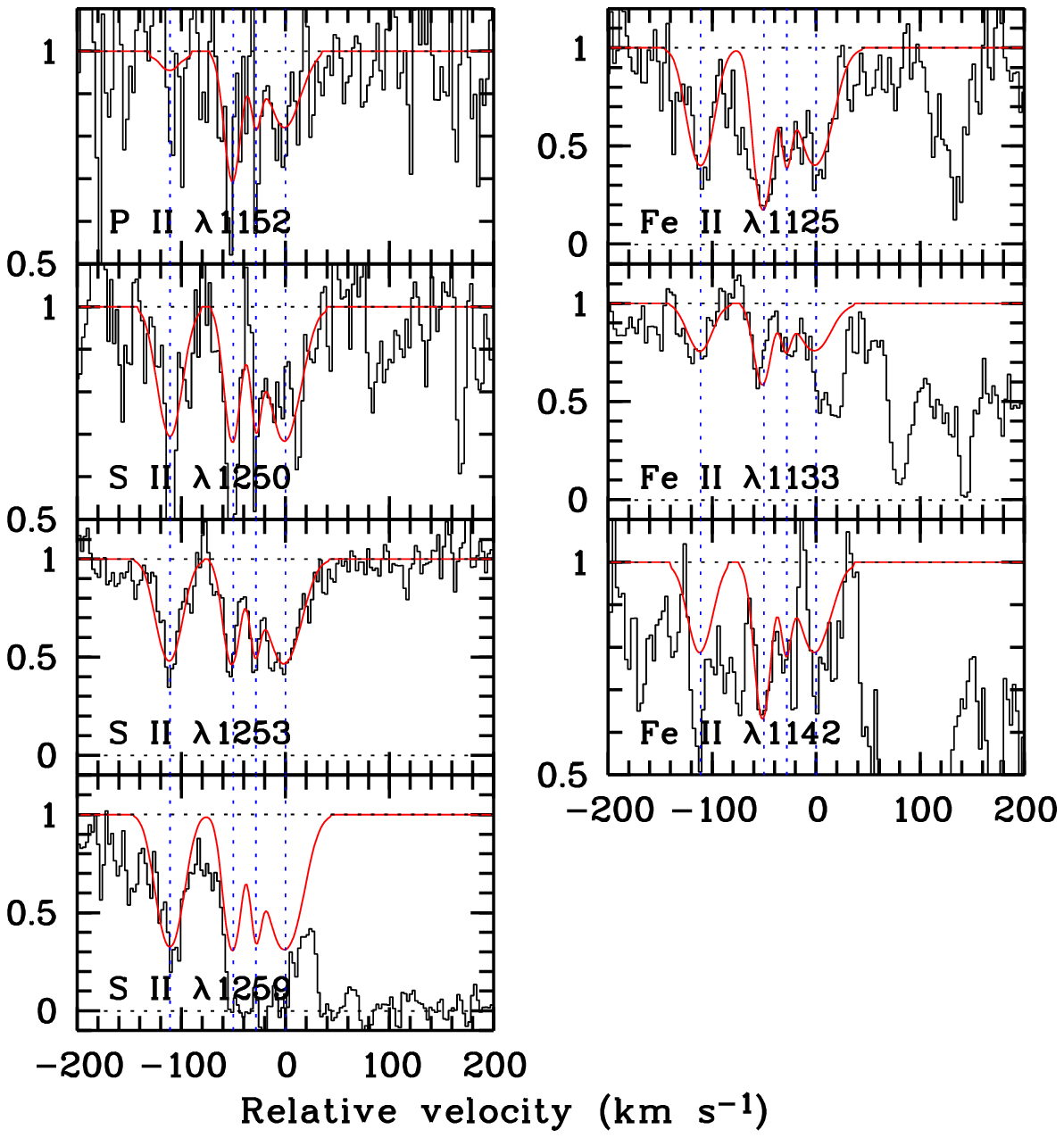}
\caption{Velocity profiles of selected low-ionization transition lines from the
DLA system at $z_{\rm abs}=1.864$ towards Q\,2230$+$025.}
\label{q2230_1.864}
\end{figure}

\begin{figure}
\includegraphics[bb=61 485 402 767,clip,width=8.7cm]{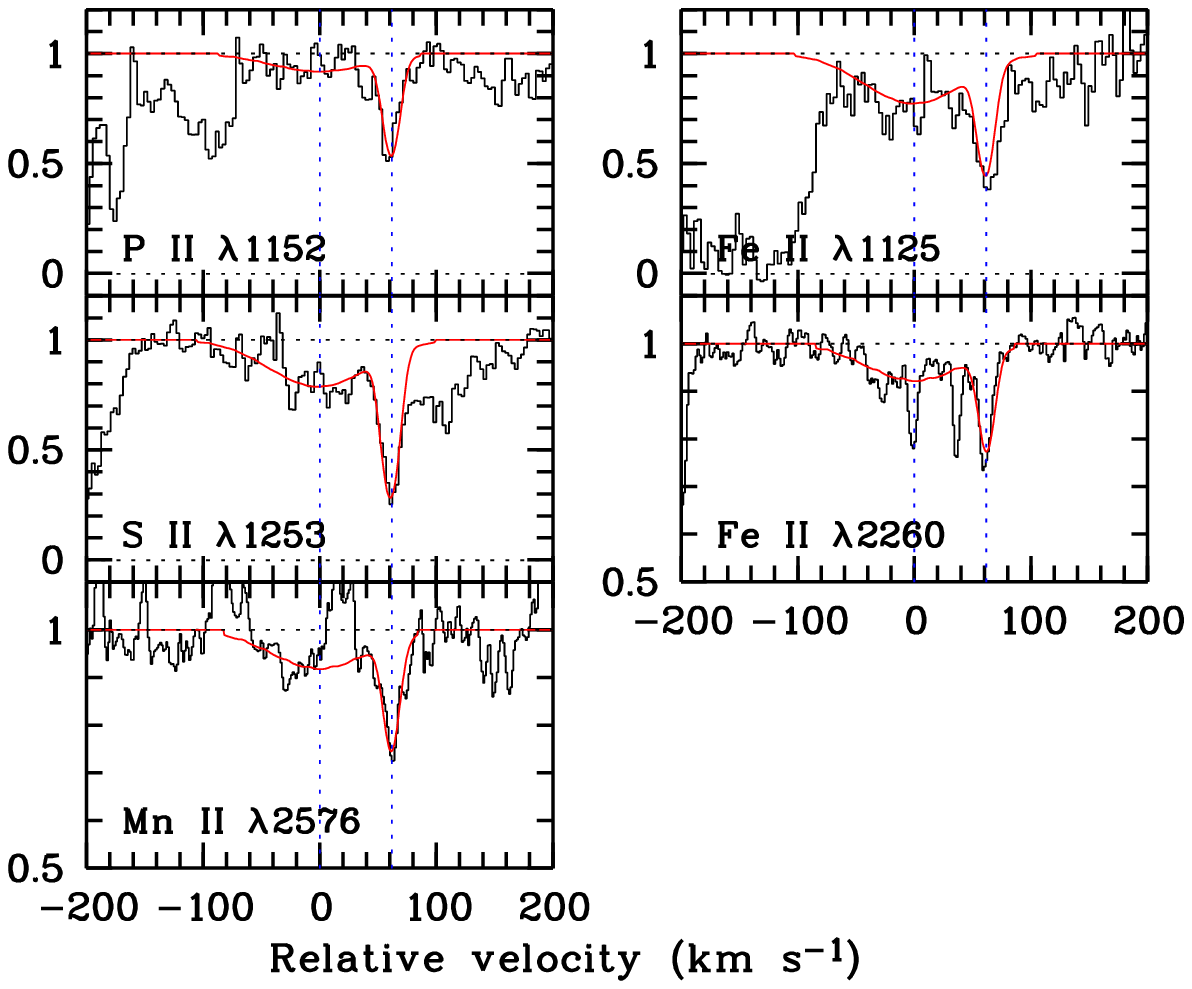}
\caption{Velocity profiles of selected low-ionization transition lines from the
DLA system at $z_{\rm abs}=2.066$ towards Q\,2231$-$002.}
\label{q2231_2.066}
\end{figure}

\begin{figure}
\includegraphics[bb=61 238 402 767,clip,width=8.7cm]{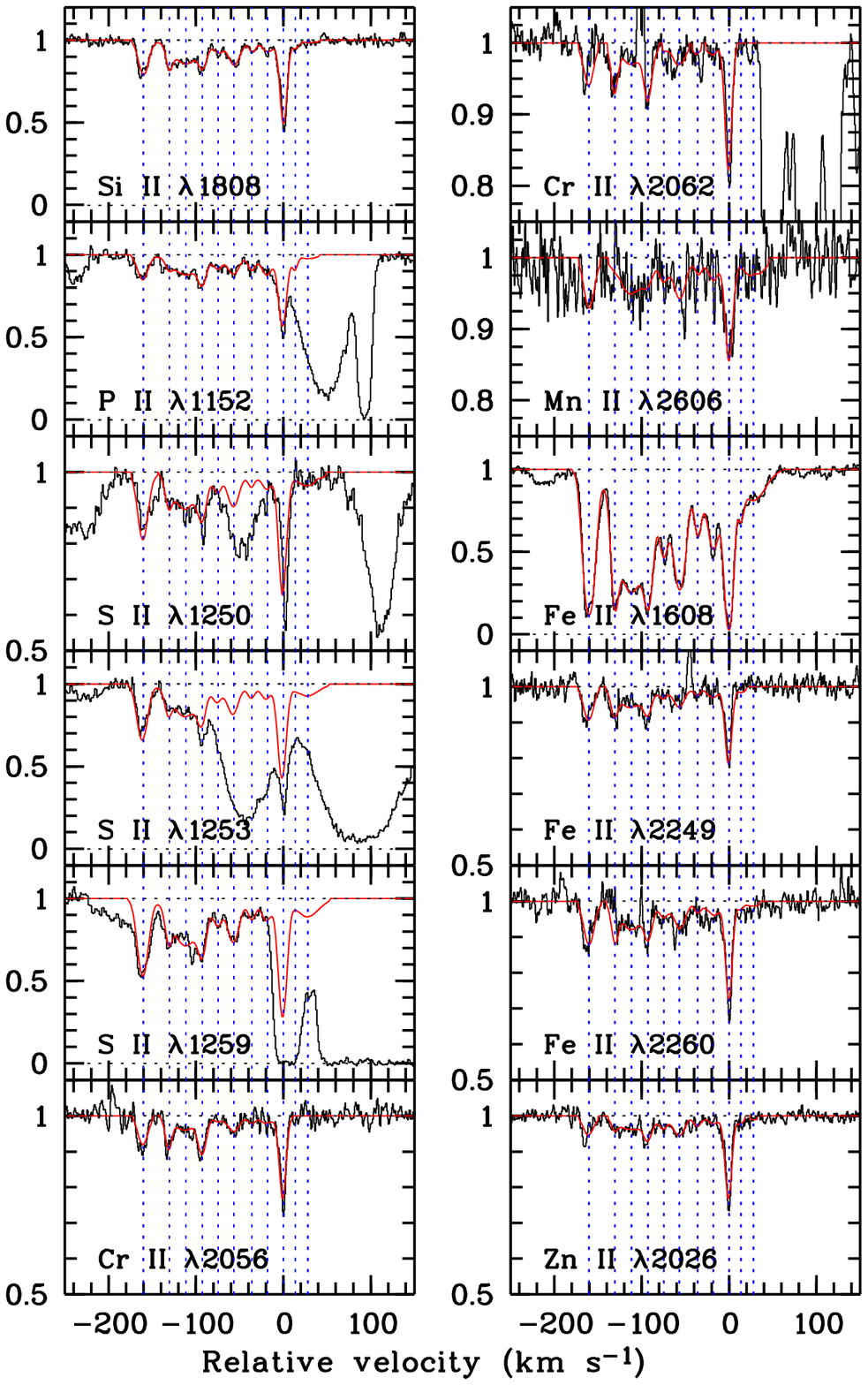}
\caption{Velocity profiles of selected low-ionization transition lines from the
DLA system at $z_{\rm abs}=2.331$ towards Q\,2243$-$605.}
\label{q2243_2.331}
\end{figure}

\clearpage

\begin{figure}
\includegraphics[bb=61 155 402 767,clip,width=8.7cm]{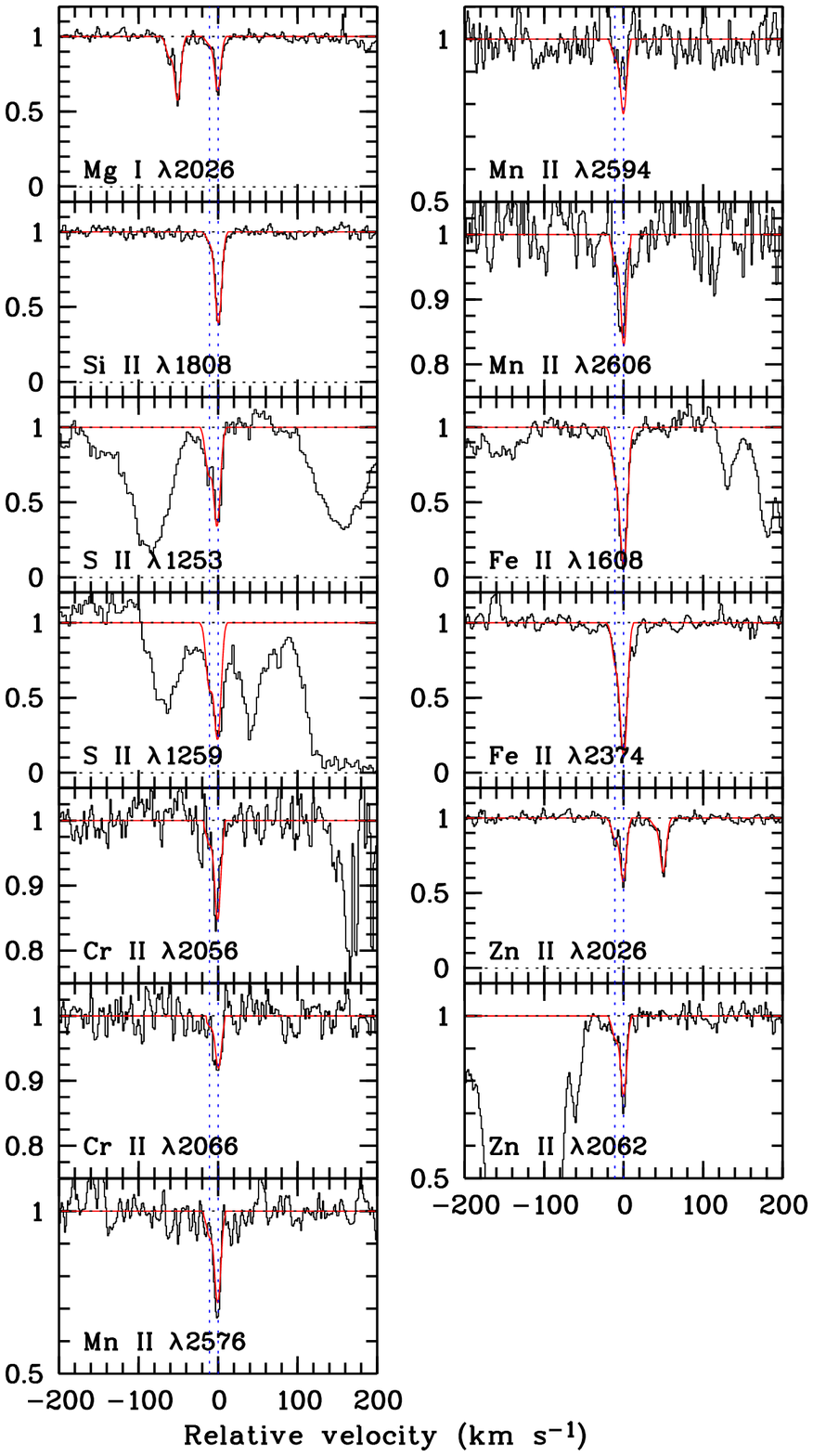}
\caption{Velocity profiles of selected low-ionization transition lines from the
sub-DLA system at $z_{\rm abs}=2.287$ towards Q\,2332$-$094.}
\label{q2332_2.287}
\end{figure}

\begin{figure}
\includegraphics[bb=61 567 402 767,clip,width=8.7cm]{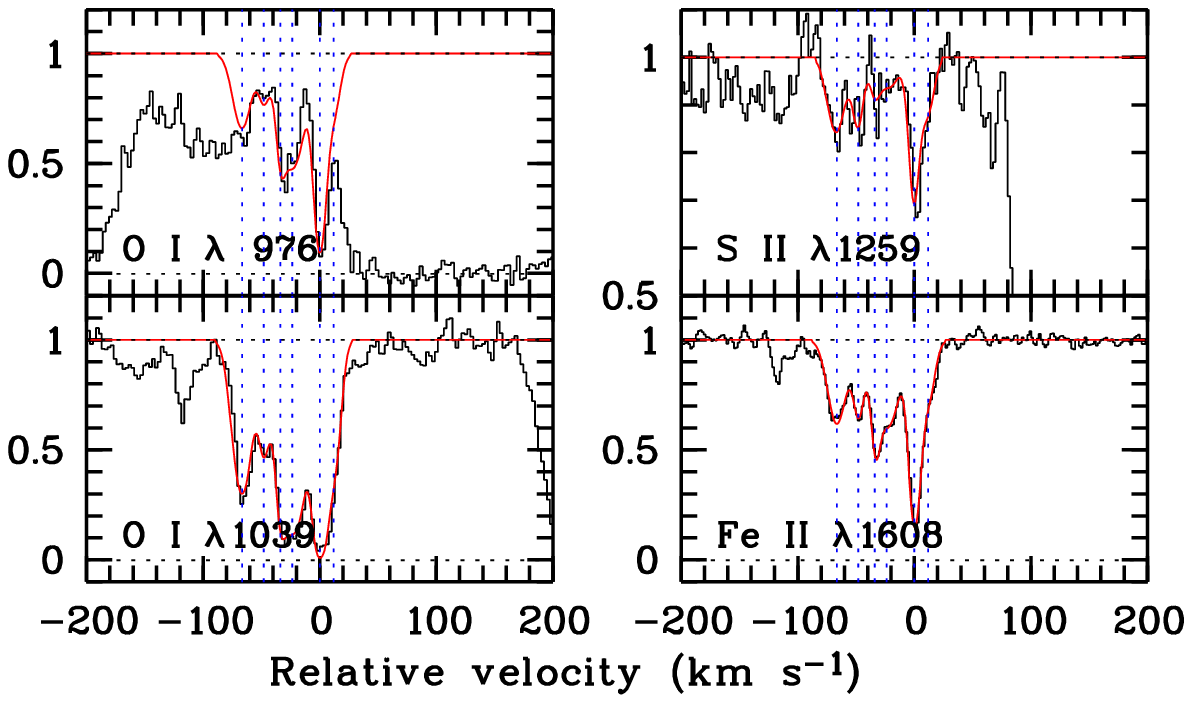}
\caption{Velocity profiles of selected low-ionization transition lines from the
DLA system at $z_{\rm abs}=3.057$ towards Q\,2332$-$094.}
\label{q2332_3.057}
\end{figure}

\begin{figure}
\includegraphics[bb=61 73 402 767,clip,width=8.7cm]{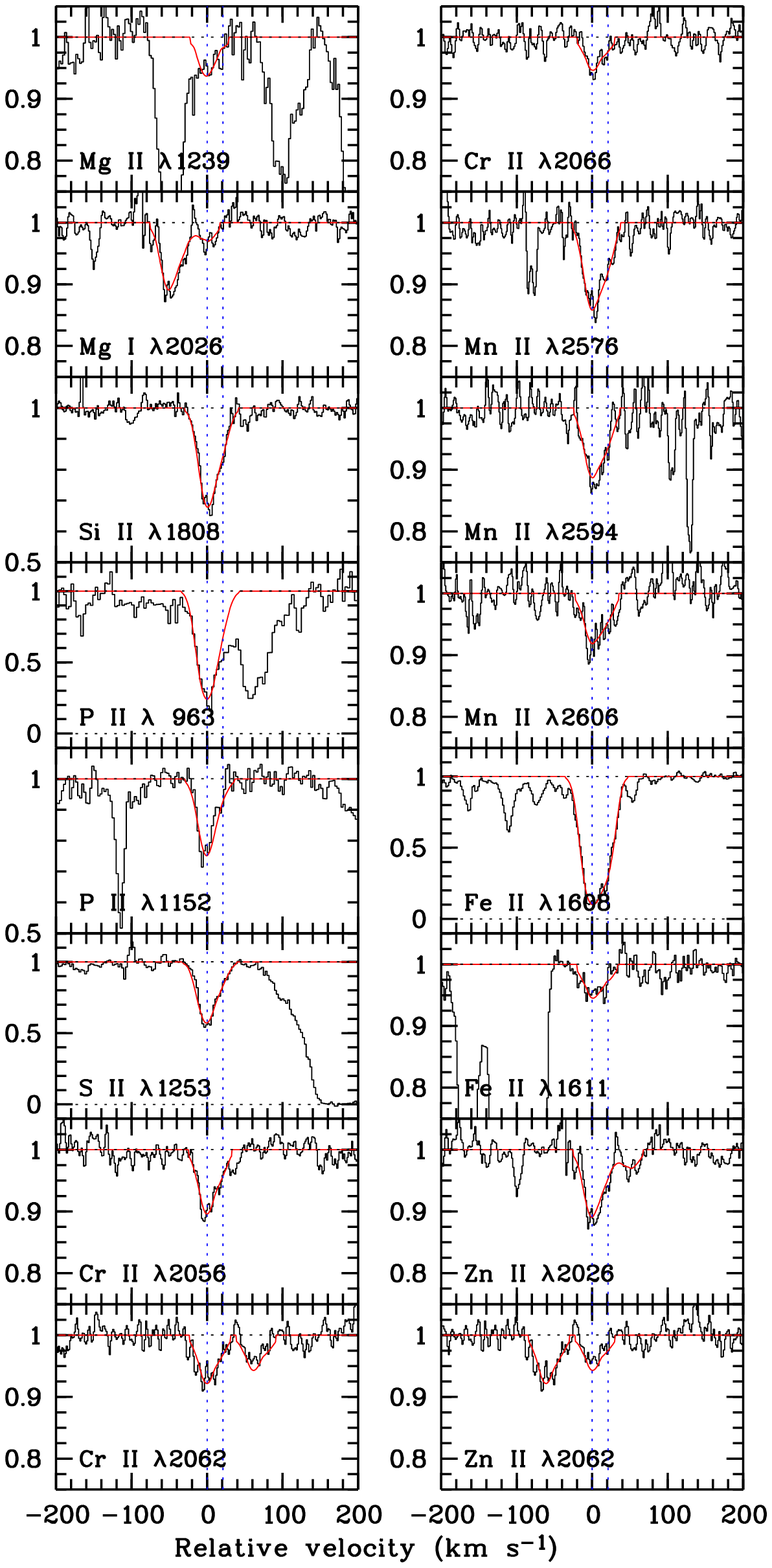}
\caption{Velocity profiles of selected low-ionization transition lines from the
DLA system at $z_{\rm abs}=2.431$ towards Q\,2343$+$125.}
\label{q2343_2.431}
\end{figure}

\clearpage

\begin{figure}
\includegraphics[bb=61 320 402 767,clip,width=8.7cm]{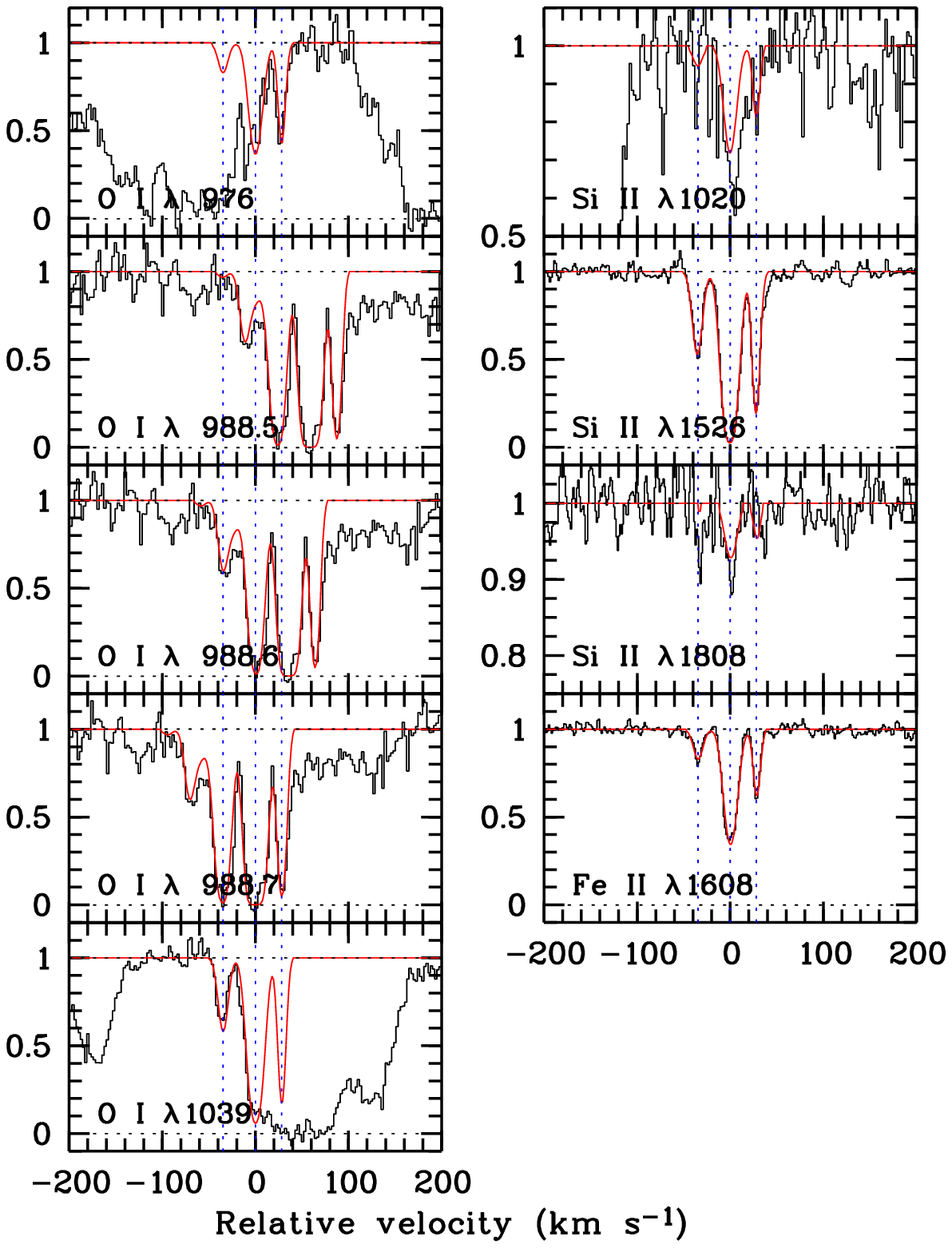}
\caption{Velocity profiles of selected low-ionization transition lines from the
DLA system at $z_{\rm abs}=2.538$ towards Q\,2344$+$125.}
\label{q2344_2.538}
\end{figure}

\begin{figure}
\includegraphics[bb=61 485 402 767,clip,width=8.7cm]{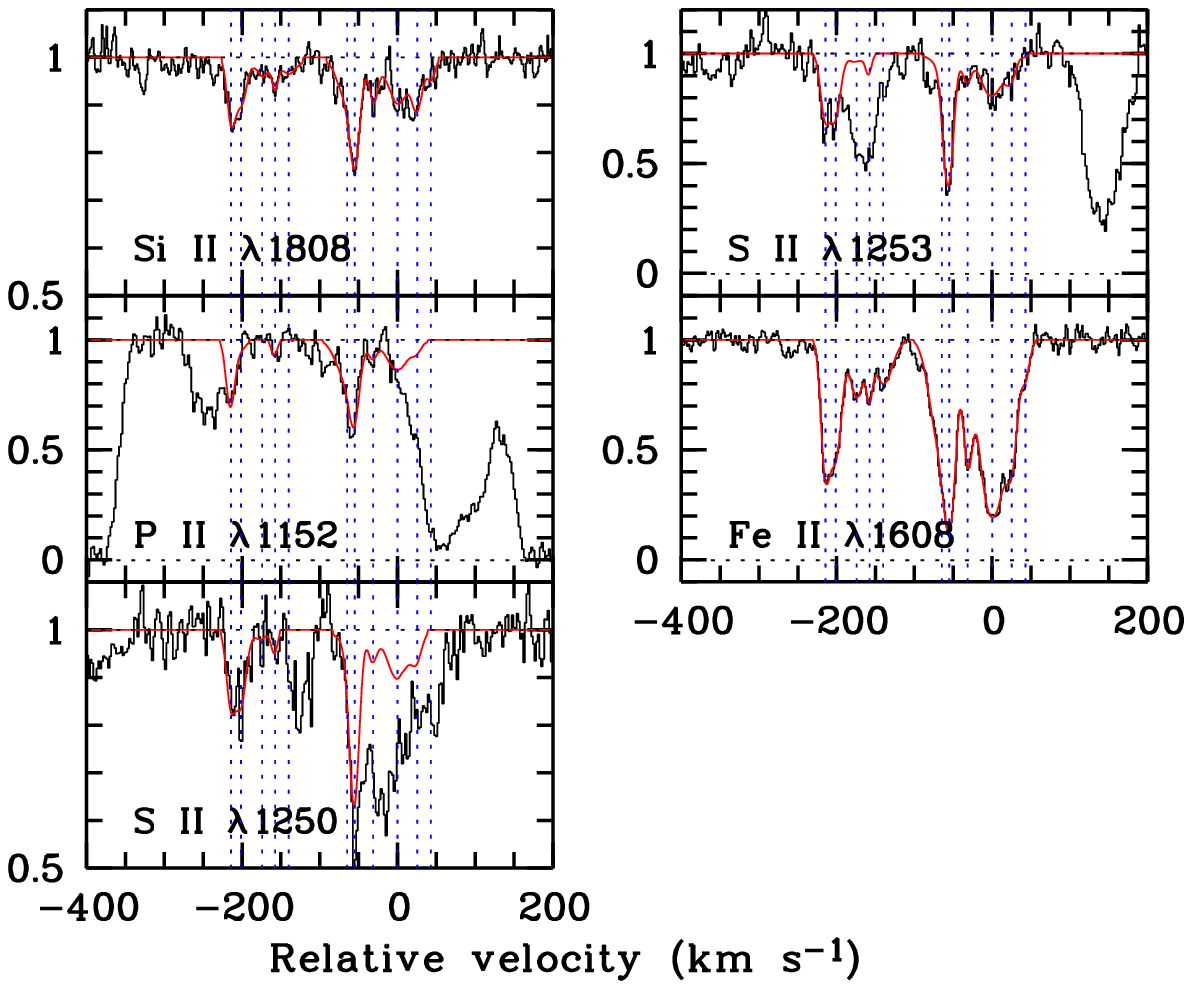}
\caption{Velocity profiles of selected low-ionization transition lines from the
DLA system at $z_{\rm abs}=2.427$ towards Q\,2348$-$011.}
\label{q2348_2.427}
\end{figure}

\begin{figure}
\includegraphics[bb=61 567 402 767,clip,width=8.7cm]{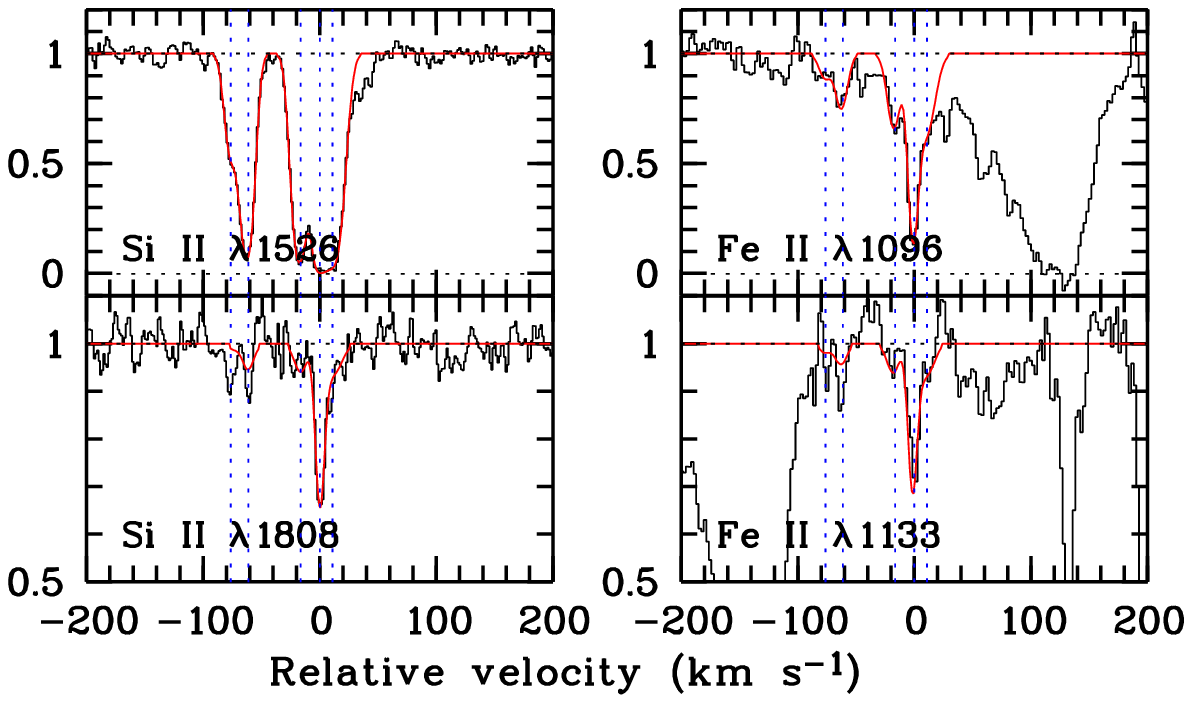}
\caption{Velocity profiles of selected low-ionization transition lines from the
DLA system at $z_{\rm abs}=2.615$ towards Q\,2348$-$011.}
\label{q2348_2.615}
\end{figure}

\begin{figure}
\includegraphics[bb=61 485 402 767,clip,width=8.7cm]{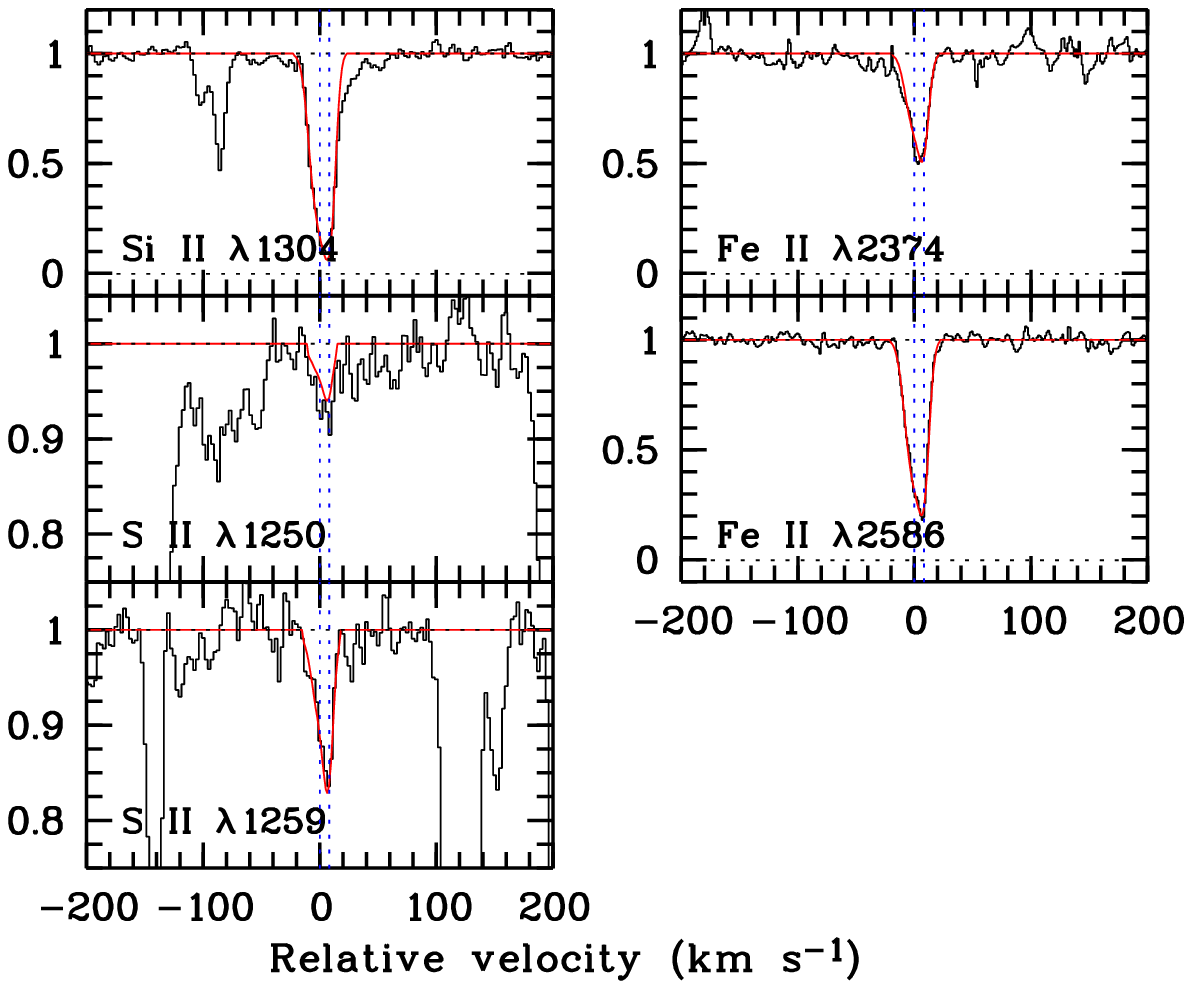}
\caption{Velocity profiles of selected low-ionization transition lines from the
DLA system at $z_{\rm abs}=2.279$ towards Q\,2348$-$147.}
\label{q2348-14_2.279}
\end{figure}

\clearpage

\begin{figure}
\includegraphics[bb=61 238 402 767,clip,width=8.7cm]{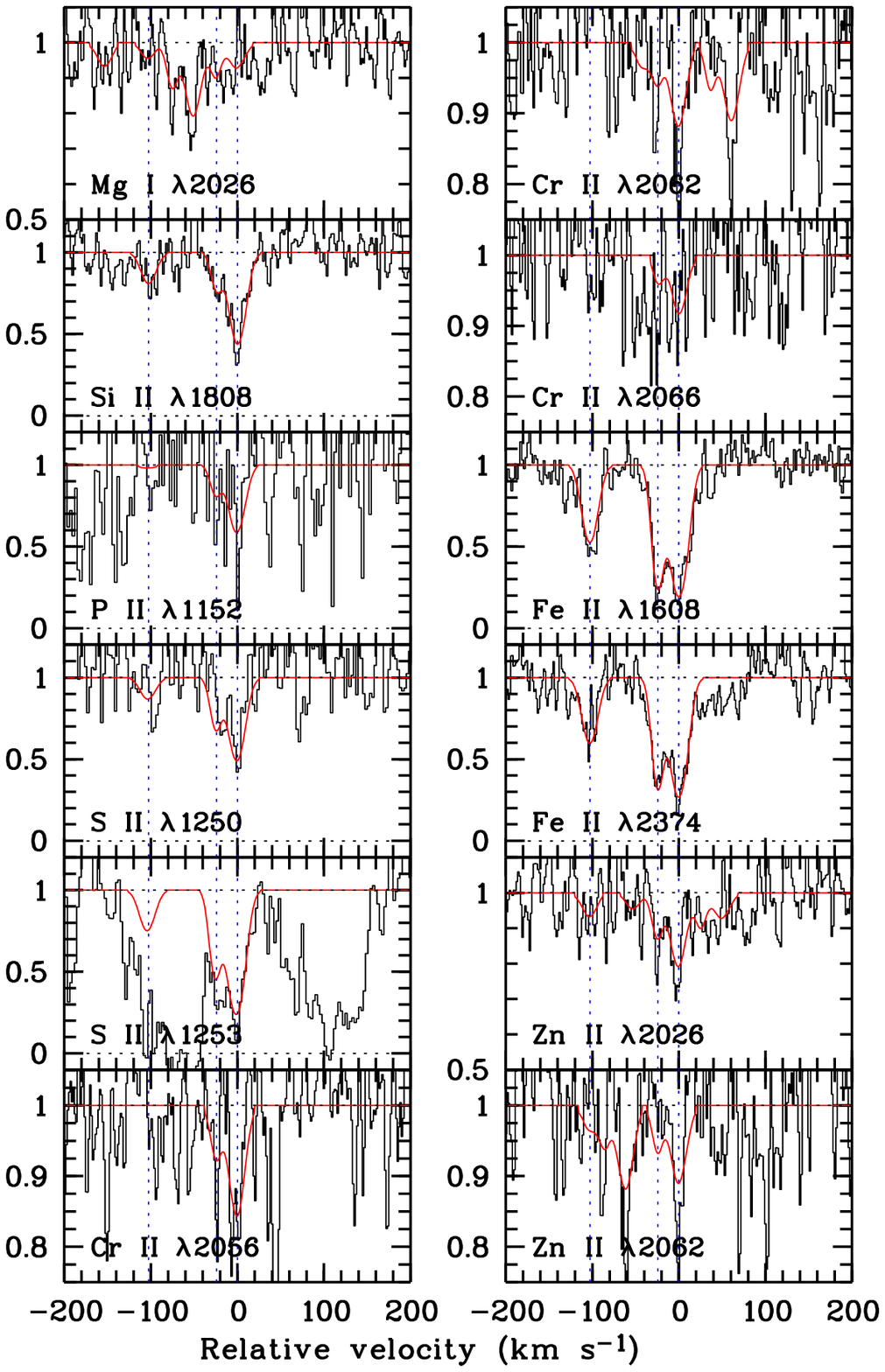}
\caption{Velocity profiles of selected low-ionization transition lines from the
DLA system at $z_{\rm abs}=2.095$ towards Q\,2359$-$022.}
\label{q2359_2.095}
\end{figure}

\begin{figure}
\includegraphics[bb=61 567 402 767,clip,width=8.7cm]{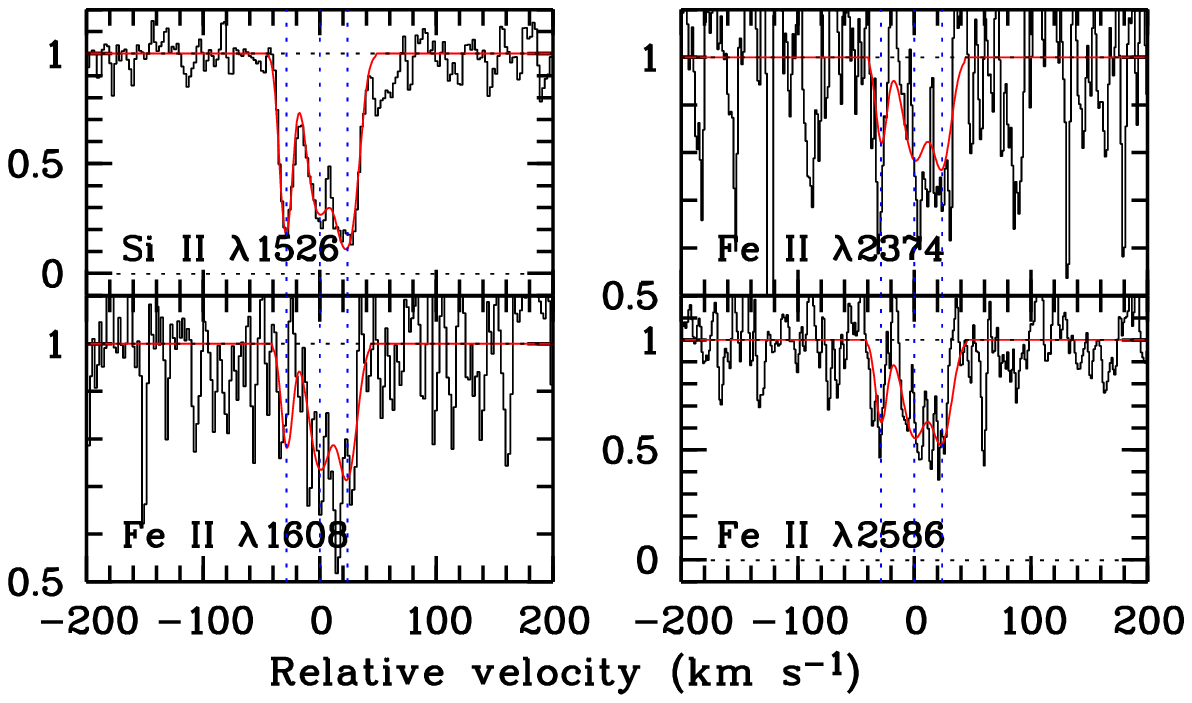}
\caption{Velocity profiles of selected low-ionization transition lines from the
DLA system at $z_{\rm abs}=2.154$ towards Q\,2359$-$022.}
\label{q2359_2.154}
\end{figure}

\section{Column densities and abundances of the DLA sample}

{\tiny
\begin{longtable}{@{}l@{\hspace{1.8mm}}r@{\hspace{1.8mm}}l@{\hspace{1.8mm}}c@{\hspace{1.8mm}}c@{\hspace{1.8mm}}c@{\hspace{1.8mm}}c@{\hspace{1.8mm}}c@{\hspace{1.8mm}}c@{\hspace{1.8mm}}c@{\hspace{1.8mm}}c@{\hspace{1.8mm}}c@{\hspace{1.8mm}}c@{}}
\caption{UVES DLA sample: Ionic column densities for
individual components $^1$}\\
\hline
\hline
Quasar & & $z_{\rm abs}$
   & $\log N($O\,{\sc i}$)$
   & $\log N($Mg\,{\sc ii}$)$
   & $\log N($Si\,{\sc ii}$)$
   & $\log N($P\,{\sc ii}$)$
   & $\log N($S\,{\sc ii}$)$
   & $\log N($Cr\,{\sc ii}$)$
   & $\log N($Mn\,{\sc ii}$)$
   & $\log N($Fe\,{\sc ii}$)$
   & $\log N($Zn\,{\sc ii}$)$
   & $b_{\rm turb}$\\
& n &&&&&&&&&&& [km s$^{-1}$]\\
\hline
\endfirsthead
\caption{continued}\\
\hline
\hline
Quasar & & $z_{\rm abs}$
   & $\log N($O\,{\sc i}$)$
   & $\log N($Mg\,{\sc ii}$)$
   & $\log N($Si\,{\sc ii}$)$
   & $\log N($P\,{\sc ii}$)$
   & $\log N($S\,{\sc ii}$)$
   & $\log N($Cr\,{\sc ii}$)$
   & $\log N($Mn\,{\sc ii}$)$
   & $\log N($Fe\,{\sc ii}$)$
   & $\log N($Zn\,{\sc ii}$)$
   & $b_{\rm turb}$\\
& n &&&&&&&&&&& [km s$^{-1}$]\\
\hline
\endhead
\hline
\endfoot
Q\,0000$-$263 & 1  & 3.39013
   & $16.42\pm 0.10$
   & ....
   & $15.06\pm 0.02$
   & $12.63\pm 0.04$
   & $14.74\pm 0.03$
   & $13.10\pm 0.03$
   & ....
   & $14.87\pm 0.03$
   & $11.91\pm 0.05$
   & $10.0\pm 0.3$\\
\hline
Q\,0010$-$002 & 1  & 2.02471
   & bld
   & $15.19\pm 0.08$
   & $14.90\pm 0.05$
   & $<14.20$ $^{\rm a}$
   & $14.54\pm 0.05$
   & $13.07\pm 0.04$
   & $<12.70$ $^{\rm a}$
   & $14.84\pm 0.06$
   & $11.73\pm 0.06$
   & $5.5\pm 0.4$\\
              & 2  & 2.02484
   & bld
   & $15.29\pm 0.07$
   & $15.09\pm 0.03$
   & $<14.20$ $^{\rm a}$
   & $14.81\pm 0.03$
   & $13.18\pm 0.03$
   & $<12.70$ $^{\rm a}$
   & $14.92\pm 0.05$
   & $11.84\pm 0.05$
   & $8.4\pm 0.5$\\
\hline
Q\,0013$-$004 & 1  & 1.96648
   & $<17.50$ $^{\rm a}$
   & bld
   & $<13.61$ $^{\rm a}$
   & $12.43\pm 0.05$
   & $13.45\pm 0.03$
   & $<11.89$ $^{\rm a}$
   & $<11.76$ $^{\rm a}$
   & $12.75\pm 0.01$
   & $<11.10$ $^{\rm a}$
   & $5.2\pm 0.4$\\
              & 2  & 1.96680
   & $<17.50$ $^{\rm a}$
   & bld
   & $13.97\pm 0.05$
   & $12.81\pm 0.03$
   & $14.12\pm 0.02$
   & $<11.89$ $^{\rm a}$
   & $11.78\pm 0.18$
   & $13.40\pm 0.01$
   & $11.34\pm 0.06$
   & $15.5\pm 0.4$\\
              & 3  & 1.96734
   & $<17.50$ $^{\rm a}$
   & bld
   & $13.82\pm 0.06$
   & $12.46\pm 0.06$
   & $13.98\pm 0.02$
   & $<11.89$ $^{\rm a}$
   & $<11.76$ $^{\rm a}$
   & $13.00\pm 0.01$
   & $11.31\pm 0.05$
   & $5.9\pm 0.4$\\
              & 4  & 1.96758
   & $<17.50$ $^{\rm a}$
   & bld
   & $<13.61$ $^{\rm a}$
   & $<12.30$ $^{\rm a}$
   & $<13.37$ $^{\rm a}$
   & $<11.89$ $^{\rm a}$
   & $<11.76$ $^{\rm a}$
   & $12.61\pm 0.01$
   & $<11.10$ $^{\rm a}$
   & $2.1\pm 0.7$\\
              & 5  & 1.96743
   & $<17.50$ $^{\rm a}$
   & bld
   & $14.60\pm 0.02$
   & $13.02\pm 0.04$
   & $14.60\pm 0.02$
   & $12.18\pm 0.09$
   & $12.20\pm 0.10$
   & $13.97\pm 0.01$
   & $11.70\pm 0.06$
   & $39.9\pm 0.6$\\
\end{longtable}
\flushleft
$^1$ Ionic column density measurements are from this work, except for
the systems at $z_{\rm abs}=3.390$
towards Q\,0000$-$263 \citep{Molaro01}, $z_{\rm abs}=3.025$ and 2.087
towards, respectively, Q\,0347$-$383
and Q\,1444$+$014 \citep{Ledoux03}, $z_{\rm abs}=1.962$
towards Q\,0551$-$366 \citep{Ledoux02b}, and $z_{\rm abs}=4.224$
towards Q\,1441$+$276 \citep{Ledoux06}.\\
Associated errors are $1\sigma$ standard deviations.\newline

$^{\rm a}$ $5\sigma$ optically thin limit for non-detection.\\
$^{\rm b}$ In this particular case, a mixture of turbulent and thermal
broadening [$b_{\rm T}($O\,{\sc i}$)=8.5\pm 0.3$ km s$^{-1}$]
was used.\\
``sat/bld'' means that most or all the components in the observed
transition lines from this ion are either saturated or blended, or both.\\
``....'' means that no transition line from this ion was observed.

}

{\tiny
\begin{longtable}{@{}l@{\hspace{1.8mm}}c@{\hspace{1.8mm}}c@{\hspace{1.8mm}}c@{\hspace{1.8mm}}c@{\hspace{1.8mm}}c@{\hspace{1.8mm}}c@{\hspace{1.8mm}}c@{\hspace{1.8mm}}c@{\hspace{1.8mm}}c@{\hspace{1.8mm}}c@{\hspace{1.8mm}}c@{}} 
\caption{UVES DLA sample: Integrated metal abundances $^1$}\\

\hline
\hline
\rule[-0.2cm]{0mm}{0.8cm}
Quasar & $z_{\rm abs}$ & $\log N($H\,{\sc i}$)$ & [O/H]  & [Mg/H] & [Si/H] & [P/H]  & [S/H] & [Cr/H] & [Mn/H] & [Fe/H] & [Zn/H]\\
\hline\hline
  Q0000-263 & 3.390 & $21.40\pm0.08$
 & $-1.67\pm0.13$
       &  ... 
 & $-1.85\pm0.08$
 & $-2.19\pm0.09$
 & $-1.80\pm0.09$
 & $-1.94\pm0.09$
       &  ... 
 & $-2.00\pm0.09$
 & $-2.12\pm0.09$
\\
  Q0010-002 & 2.025 & $20.95\pm0.10$
       &  ... 
 & $-0.97\pm0.11$
 & $-1.15\pm0.10$
     & $< 0.13$
 & $-1.09\pm0.10$
 & $-1.16\pm0.10$
     & $<-1.43$
 & $-1.24\pm0.11$
 & $-1.49\pm0.11$
\\
  Q0013-004 & 1.973 & $20.83\pm0.05$
     & $< 1.26$
       &  ... 
 & $-0.80\pm0.05$
 & $-0.30\pm0.05$
 & $-0.48\pm0.05$
 & $-1.28\pm0.06$
 & $-1.32\pm0.07$
 & $-1.40\pm0.05$
 & $-0.65\pm0.05$
\\
  Q0058-292 & 2.671 & $21.10\pm0.10$
       &  ... 
     & $<-1.36$
 & $-1.38\pm0.11$
     & $<-1.22$
 & $-1.27\pm0.10$
 & $-1.61\pm0.10$
     & $<-1.58$
 & $-1.82\pm0.10$
 & $-1.59\pm0.10$
\\
  Q0100+130 & 2.309 & $21.35\pm0.08$
       &  ... 
 & $-1.40\pm0.10$
       &  ... 
 & $-1.78\pm0.08$
 & $-1.36\pm0.08$
 & $-1.61\pm0.08$
 & $-2.16\pm0.09$
 & $-1.78\pm0.08$
 & $-1.64\pm0.08$
\\
 Q0102-190a & 2.370 & $21.00\pm0.08$
     & $< 0.21$
     & $<-1.30$
       &  ... 
     & $<-1.84$
 & $-1.79\pm0.08$
 & $-1.84\pm0.08$
 & $-2.44\pm0.09$
 & $-2.04\pm0.08$
     & $<-1.88$
\\
 Q0102-190b & 2.926 & $20.00\pm0.10$
 & $-1.55\pm0.10$
     & $< 0.22$
 & $-1.45\pm0.10$
     & $<-0.83$
     & $<-1.18$
     & $<-0.19$
     & $<-0.27$
 & $-1.67\pm0.10$
     & $< 0.02$
\\
 Q0112-306a & 2.418 & $20.50\pm0.08$
 & $-2.20\pm0.11$
     & $<-0.25$
 & $-2.37\pm0.08$
     & $<-0.83$
     & $<-0.98$
       &  ... 
     & $<-0.98$
 & $-2.55\pm0.09$
       &  ... 
\\
 Q0112-306b & 2.702 & $20.30\pm0.10$
     & $< 1.11$
     & $< 0.54$
 & $-0.44\pm0.11$
     & $<-0.17$
       &  ... 
       &  ... 
     & $<-0.08$
 & $-0.97\pm0.10$
       &  ... 
\\
  Q0112+030 & 2.423 & $20.90\pm0.10$
       &  ... 
     & $<-0.24$
 & $-1.15\pm0.10$
       &  ... 
 & $-1.21\pm0.11$
       &  ... 
     & $<-0.85$
 & $-1.53\pm0.10$
       &  ... 
\\
 Q0135-273a & 2.107 & $20.30\pm0.15$
       &  ... 
     & $<-0.55$
 & $-0.94\pm0.17$
       &  ... 
 & $-1.01\pm0.16$
 & $-1.12\pm0.15$
 & $-1.57\pm0.15$
 & $-1.19\pm0.15$
     & $<-1.21$
\\
 Q0135-273b & 2.800 & $21.00\pm0.10$
       &  ... 
     & $<-1.39$
       &  ... 
     & $<-0.18$
 & $-1.30\pm0.10$
 & $-1.65\pm0.10$
     & $<-1.73$
 & $-1.70\pm0.10$
     & $<-1.83$
\\
 Q0216+080a & 1.769 & $20.30\pm0.10$
       &  ... 
     & $<-0.20$
 & $-0.91\pm0.10$
       &  ... 
 & $-0.96\pm0.16$
 & $-1.24\pm0.12$
     & $<-0.30$
 & $-1.31\pm0.10$
 & $-1.06\pm0.12$
\\
 Q0216+080b & 2.293 & $20.50\pm0.10$
       &  ... 
       &  ... 
 & $-0.60\pm0.10$
 & $-0.46\pm0.12$
       &  ... 
 & $-0.94\pm0.11$
       &  ... 
 & $-1.10\pm0.10$
 & $-0.76\pm0.11$
\\
  Q0336-017 & 3.062 & $21.10\pm0.10$
       &  ... 
       &  ... 
 & $-1.36\pm0.10$
 & $-1.90\pm0.12$
       &  ... 
 & $-1.55\pm0.10$
     & $<-2.11$
 & $-1.78\pm0.10$
     & $<-1.40$
\\
  Q0347-383 & 3.025 & $20.73\pm0.05$
 & $-0.97\pm0.12$
       &  ... 
 & $-1.46\pm0.06$
 & $-1.43\pm0.11$
 & $-1.09\pm0.06$
 & $-1.53\pm0.07$
       &  ... 
 & $-1.86\pm0.05$
 & $-1.23\pm0.07$
\\
 Q0405-443a & 1.913 & $20.80\pm0.10$
       &  ... 
       &  ... 
 & $-0.88\pm0.10$
     & $< 0.41$
       &  ... 
 & $-0.97\pm0.10$
 & $-1.39\pm0.10$
 & $-1.11\pm0.10$
 & $-1.09\pm0.10$
\\
 Q0405-443b & 2.550 & $21.15\pm0.15$
     & $<-0.18$
       &  ... 
 & $-1.30\pm0.15$
 & $-1.46\pm0.15$
 & $-1.31\pm0.15$
 & $-1.51\pm0.15$
 & $-2.11\pm0.19$
 & $-1.57\pm0.15$
 & $-1.42\pm0.16$
\\
 Q0405-443c & 2.595 & $21.05\pm0.10$
       &  ... 
       &  ... 
 & $-1.07\pm0.10$
 & $-1.28\pm0.10$
 & $-1.03\pm0.10$
 & $-1.27\pm0.10$
 & $-1.69\pm0.10$
 & $-1.44\pm0.10$
 & $-1.18\pm0.10$
\\
 Q0405-443d & 2.622 & $20.45\pm0.10$
       &  ... 
     & $<-0.53$
 & $-1.99\pm0.10$
     & $<-0.97$
     & $<-1.13$
     & $<-1.01$
     & $<-1.06$
 & $-2.34\pm0.10$
     & $<-0.77$
\\
  Q0450-131 & 2.067 & $20.50\pm0.07$
       &  ... 
     & $<-0.29$
 & $-1.32\pm0.08$
 & $-1.16\pm0.11$
 & $-1.52\pm0.08$
     & $<-1.31$
     & $<-1.60$
 & $-1.73\pm0.07$
     & $<-1.20$
\\
  Q0458-020 & 2.040 & $21.70\pm0.10$
       &  ... 
 & $-1.19\pm0.12$
       &  ... 
 & $-1.40\pm0.11$
       &  ... 
 & $-1.61\pm0.10$
 & $-2.00\pm0.11$
 & $-1.84\pm0.11$
 & $-1.28\pm0.10$
\\
 Q0528-250a & 2.141 & $20.98\pm0.05$
     & $< 0.31$
     & $<-1.20$
 & $-1.21\pm0.05$
     & $<-0.30$
 & $-1.31\pm0.05$
 & $-1.49\pm0.05$
 & $-1.90\pm0.07$
 & $-1.67\pm0.05$
 & $-1.42\pm0.06$
\\
 Q0528-250b & 2.811 & $21.35\pm0.07$
     & $<-0.67$
 & $-1.04\pm0.07$
 & $-0.76\pm0.07$
       &  ... 
 & $-0.90\pm0.07$
 & $-1.28\pm0.07$
 & $-1.49\pm0.07$
 & $-1.35\pm0.07$
 & $-0.97\pm0.07$
\\
  Q0551-366 & 1.962 & $20.70\pm0.08$
       &  ... 
     & $<-0.47$
 & $-0.59\pm0.09$
 & $-0.33\pm0.18$
 & $-0.41\pm0.09$
 & $-1.07\pm0.09$
 & $-1.07\pm0.08$
 & $-1.12\pm0.08$
 & $-0.41\pm0.08$
\\
 Q0841+129a & 1.864 & $21.00\pm0.10$
       &  ... 
     & $<-1.13$
       &  ... 
     & $<-0.08$
 & $-1.40\pm0.11$
       &  ... 
 & $-1.94\pm0.11$
 & $-1.59\pm0.11$
       &  ... 
\\
 Q0841+129b & 2.375 & $21.05\pm0.10$
     & $<-0.01$
 & $-1.48\pm0.11$
 & $-1.46\pm0.10$
 & $-1.63\pm0.10$
 & $-1.41\pm0.10$
 & $-1.61\pm0.10$
 & $-2.05\pm0.10$
 & $-1.80\pm0.10$
 & $-1.65\pm0.10$
\\
 Q0841+129c & 2.476 & $20.80\pm0.10$
 & $-1.37\pm0.10$
 & $-1.41\pm0.11$
 & $-1.55\pm0.10$
 & $-1.87\pm0.11$
 & $-1.45\pm0.10$
 & $-1.60\pm0.10$
 & $-1.98\pm0.10$
 & $-1.80\pm0.10$
 & $-1.66\pm0.10$
\\
  Q0913+072 & 2.618 & $20.35\pm0.10$
 & $-2.46\pm0.10$
     & $<-0.66$
 & $-2.54\pm0.10$
     & $<-1.17$
     & $<-1.56$
     & $<-1.45$
     & $<-1.56$
 & $-2.73\pm0.10$
     & $<-1.22$
\\
  Q1036-229 & 2.778 & $20.93\pm0.05$
     & $<-0.24$
     & $<-0.89$
 & $-1.21\pm0.05$
 & $-1.36\pm0.05$
 & $-1.26\pm0.05$
 & $-1.41\pm0.06$
 & $-1.91\pm0.08$
 & $-1.64\pm0.05$
     & $<-1.16$
\\
 Q1108-077b & 3.608 & $20.37\pm0.07$
 & $-1.69\pm0.07$
       &  ... 
 & $-1.54\pm0.07$
     & $<-1.54$
       &  ... 
     & $<-1.16$
     & $<-1.65$
 & $-1.97\pm0.07$
       &  ... 
\\
  Q1111-152 & 3.266 & $21.30\pm0.05$
     & $<-1.09$
       &  ... 
 & $-1.71\pm0.05$
     & $<-1.72$
 & $-1.64\pm0.05$
 & $-1.65\pm0.07$
     & $<-1.53$
 & $-1.96\pm0.05$
 & $-1.71\pm0.11$
\\
  Q1117-134 & 3.350 & $20.95\pm0.10$
       &  ... 
     & $<-0.88$
 & $-1.36\pm0.10$
     & $<-0.24$
       &  ... 
 & $-1.49\pm0.11$
     & $<-1.23$
 & $-1.61\pm0.10$
 & $-1.47\pm0.11$
\\
  Q1157+014 & 1.944 & $21.80\pm0.10$
       &  ... 
 & $-1.41\pm0.10$
 & $-1.32\pm0.10$
 & $-1.48\pm0.11$
       &  ... 
 & $-1.64\pm0.10$
 & $-2.03\pm0.10$
 & $-1.81\pm0.10$
 & $-1.50\pm0.10$
\\
  Q1209+093 & 2.584 & $21.40\pm0.10$
       &  ... 
       &  ... 
 & $-0.94\pm0.10$
       &  ... 
 & $-0.95\pm0.10$
 & $-1.46\pm0.10$
 & $-1.78\pm0.10$
 & $-1.51\pm0.10$
 & $-1.07\pm0.10$
\\
  Q1210+175 & 1.892 & $20.70\pm0.08$
       &  ... 
 & $-0.66\pm0.09$
       &  ... 
       &  ... 
 & $-0.82\pm0.08$
       &  ... 
 & $-1.46\pm0.08$
 & $-1.28\pm0.08$
       &  ... 
\\
  Q1223+178 & 2.466 & $21.40\pm0.10$
     & $<-0.09$
 & $-1.14\pm0.11$
 & $-1.41\pm0.10$
     & $<-1.16$
 & $-1.35\pm0.10$
 & $-1.55\pm0.10$
 & $-2.03\pm0.10$
 & $-1.67\pm0.10$
 & $-1.69\pm0.10$
\\
  Q1232+082 & 2.338 & $20.90\pm0.08$
       &  ... 
 & $-1.14\pm0.09$
 & $-1.22\pm0.08$
 & $-1.09\pm0.18$
 & $-1.32\pm0.08$
       &  ... 
 & $-1.94\pm0.13$
 & $-1.86\pm0.08$
       &  ... 
\\
  Q1331+170 & 1.776 & $21.15\pm0.07$
     & $<-0.34$
 & $-1.19\pm0.08$
 & $-1.35\pm0.07$
 & $-1.18\pm0.07$
 & $-1.17\pm0.07$
 & $-1.91\pm0.07$
 & $-2.15\pm0.07$
 & $-2.00\pm0.07$
 & $-1.34\pm0.08$
\\
 Q1337+113a & 2.508 & $20.12\pm0.05$
 & $-1.87\pm0.09$
     & $<-0.46$
 & $-1.76\pm0.06$
     & $<-0.83$
       &  ... 
     & $<-0.94$
     & $<-1.10$
 & $-2.20\pm0.05$
     & $<-0.80$
\\
 Q1337+113b & 2.796 & $21.00\pm0.08$
 & $-1.89\pm0.11$
     & $<-1.41$
 & $-1.81\pm0.09$
     & $<-2.00$
       &  ... 
 & $-1.89\pm0.14$
     & $<-1.50$
 & $-2.15\pm0.08$
     & $<-1.58$
\\
  Q1340-136 & 3.118 & $20.05\pm0.08$
 & $-1.21\pm0.08$
       &  ... 
 & $-1.15\pm0.08$
 & $-1.37\pm0.08$
 & $-1.31\pm0.08$
     & $<-1.28$
     & $<-1.48$
 & $-1.59\pm0.08$
     & $<-1.04$
\\
 Q1409+095a & 2.019 & $20.65\pm0.10$
     & $< 0.61$
       &  ... 
 & $-1.52\pm0.11$
     & $< 0.03$
 & $-1.64\pm0.11$
 & $-1.65\pm0.12$
     & $<-1.94$
 & $-1.78\pm0.10$
 & $-1.68\pm0.16$
\\
 Q1409+095b & 2.456 & $20.53\pm0.08$
 & $-1.88\pm0.08$
     & $<-0.29$
 & $-2.01\pm0.08$
     & $< 0.55$
     & $<-1.34$
     & $<-1.31$
     & $<-1.41$
 & $-2.27\pm0.08$
     & $<-0.98$
\\
  Q1441+276 & 4.224 & $20.95\pm0.10$
     & $< 0.11$
 & $-0.54\pm0.10$
       &  ... 
 & $-0.37\pm0.12$
 & $-0.53\pm0.10$
       &  ... 
       &  ... 
 & $-1.09\pm0.10$
       &  ... 
\\
  Q1444+014 & 2.087 & $20.25\pm0.07$
       &  ... 
     & $<-0.51$
 & $-1.37\pm0.08$
 & $-0.63\pm0.09$
 & $-0.72\pm0.08$
 & $-1.51\pm0.13$
     & $<-0.69$
 & $-1.73\pm0.08$
 & $-0.86\pm0.09$
\\
 Q1451+123a & 2.255 & $20.35\pm0.10$
       &  ... 
     & $<-0.14$
 & $-0.90\pm0.13$
     & $< 0.75$
     & $<-0.77$
 & $-1.12\pm0.14$
     & $<-1.24$
 & $-1.40\pm0.11$
 & $-1.13\pm0.15$
\\
 Q1451+123b & 2.469 & $20.40\pm0.10$
       &  ... 
     & $<-0.45$
 & $-2.22\pm0.13$
       &  ... 
     & $<-1.64$
     & $<-1.19$
     & $<-1.30$
 & $-2.49\pm0.10$
     & $<-1.03$
\\
 Q1451+123c & 3.171 & $20.20\pm0.20$
     & $<-0.63$
       &  ... 
 & $-2.05\pm0.21$
     & $<-0.44$
       &  ... 
     & $<-0.61$
       &  ... 
 & $-2.33\pm0.22$
     & $<-0.35$
\\
 Q2059-360a & 2.507 & $20.29\pm0.07$
       &  ... 
     & $<-0.73$
 & $-1.89\pm0.08$
     & $<-0.81$
 & $-1.75\pm0.20$
     & $<-1.21$
     & $<-1.49$
 & $-2.27\pm0.08$
     & $<-1.20$
\\
 Q2059-360b & 3.083 & $20.98\pm0.08$
 & $-1.57\pm0.09$
     & $<-1.21$
 & $-1.63\pm0.09$
     & $<-1.64$
 & $-1.67\pm0.09$
     & $<-1.68$
       &  ... 
 & $-1.97\pm0.08$
     & $<-1.55$
\\
  Q2116-358 & 1.996 & $20.10\pm0.07$
     & $< 0.76$
     & $<-0.13$
 & $-0.37\pm0.09$
 & $-0.27\pm0.12$
 & $-0.25\pm0.10$
 & $-0.69\pm0.09$
     & $<-0.48$
 & $-0.85\pm0.08$
 & $-0.40\pm0.11$
\\
 Q2138-444a & 2.383 & $20.60\pm0.05$
       &  ... 
       &  ... 
 & $-1.15\pm0.05$
       &  ... 
 & $-0.99\pm0.11$
 & $-1.23\pm0.07$
 & $-1.64\pm0.07$
 & $-1.54\pm0.05$
 & $-1.21\pm0.10$
\\
 Q2138-444b & 2.852 & $20.98\pm0.05$
     & $<-0.75$
     & $<-1.45$
 & $-1.63\pm0.05$
 & $-2.04\pm0.07$
 & $-1.57\pm0.05$
 & $-1.69\pm0.05$
 & $-2.26\pm0.06$
 & $-1.82\pm0.05$
 & $-1.80\pm0.05$
\\
 Q2152+137b & 3.316 & $20.50\pm0.15$
       &  ... 
     & $<-0.39$
 & $-1.32\pm0.15$
     & $< 0.66$
       &  ... 
     & $<-0.91$
       &  ... 
 & $-1.48\pm0.15$
     & $<-0.77$
\\
 Q2206-199a & 1.921 & $20.67\pm0.05$
       &  ... 
 & $-0.53\pm0.06$
 & $-0.39\pm0.05$
 & $-0.20\pm0.06$
 & $-0.38\pm0.05$
 & $-0.70\pm0.05$
 & $-1.06\pm0.05$
 & $-0.80\pm0.05$
 & $-0.60\pm0.05$
\\
 Q2206-199b & 2.076 & $20.44\pm0.05$
       &  ... 
     & $<-1.51$
 & $-2.27\pm0.05$
     & $<-0.71$
     & $<-1.94$
     & $<-2.27$
     & $<-2.10$
 & $-2.58\pm0.05$
     & $<-2.11$
\\
  Q2230+025 & 1.864 & $20.90\pm0.10$
       &  ... 
     & $<-0.48$
       &  ... 
 & $-0.95\pm0.11$
 & $-0.71\pm0.10$
       &  ... 
     & $<-1.04$
 & $-1.03\pm0.10$
       &  ... 
\\
  Q2231-002 & 2.066 & $20.55\pm0.07$
     & $< 0.46$
       &  ... 
       &  ... 
 & $-0.57\pm0.08$
 & $-0.55\pm0.07$
       &  ... 
 & $-1.30\pm0.07$
 & $-1.10\pm0.07$
       &  ... 
\\
  Q2243-605 & 2.331 & $20.65\pm0.05$
     & $< 0.50$
       &  ... 
 & $-0.76\pm0.05$
 & $-0.59\pm0.05$
 & $-0.74\pm0.05$
 & $-1.08\pm0.05$
 & $-1.34\pm0.06$
 & $-1.17\pm0.05$
 & $-0.91\pm0.05$
\\
 Q2332-094a & 2.287 & $20.07\pm0.07$
       &  ... 
     & $<-0.23$
 & $-0.56\pm0.07$
       &  ... 
 & $-0.54\pm0.07$
 & $-1.18\pm0.09$
 & $-1.37\pm0.08$
 & $-1.31\pm0.08$
 & $-0.39\pm0.07$
\\
 Q2332-094b & 3.057 & $20.50\pm0.07$
 & $-1.31\pm0.07$
       &  ... 
       &  ... 
     & $<-0.86$
 & $-1.23\pm0.08$
     & $<-0.97$
       &  ... 
 & $-1.60\pm0.07$
     & $<-0.85$
\\
 Q2343+125a & 2.431 & $20.40\pm0.07$
       &  ... 
 & $-0.94\pm0.11$
 & $-0.80\pm0.07$
 & $-0.76\pm0.07$
 & $-0.81\pm0.07$
 & $-1.24\pm0.07$
 & $-1.54\pm0.07$
 & $-1.36\pm0.07$
 & $-0.98\pm0.07$
\\
 Q2344+125b & 2.538 & $20.50\pm0.10$
 & $-1.67\pm0.10$
       &  ... 
 & $-1.76\pm0.10$
     & $< 0.46$
       &  ... 
       &  ... 
       &  ... 
 & $-1.97\pm0.10$
       &  ... 
\\
 Q2348-011a & 2.427 & $20.50\pm0.10$
       &  ... 
     & $<-0.16$
 & $-0.75\pm0.10$
 & $-0.37\pm0.10$
 & $-0.54\pm0.10$
       &  ... 
       &  ... 
 & $-1.15\pm0.10$
       &  ... 
\\
 Q2348-011b & 2.615 & $21.30\pm0.08$
     & $<-0.23$
     & $<-0.85$
 & $-1.97\pm0.08$
     & $<-0.10$
       &  ... 
       &  ... 
     & $<-1.63$
 & $-2.12\pm0.08$
       &  ... 
\\
  Q2348-147 & 2.279 & $20.63\pm0.05$
       &  ... 
     & $<-1.29$
 & $-1.91\pm0.08$
     & $<-0.49$
 & $-1.99\pm0.08$
       &  ... 
     & $<-2.01$
 & $-2.33\pm0.08$
       &  ... 
\\
 Q2359-022a & 2.095 & $20.65\pm0.10$
       &  ... 
       &  ... 
 & $-0.72\pm0.10$
 & $-0.63\pm0.15$
 & $-0.61\pm0.11$
 & $-1.24\pm0.13$
     & $<-1.10$
 & $-1.65\pm0.10$
 & $-0.90\pm0.13$
\\
 Q2359-022b & 2.154 & $20.30\pm0.10$
     & $< 1.61$
       &  ... 
     & $<-1.57$
     & $<-0.13$
     & $<-0.85$
     & $<-0.73$
     & $<-0.95$
 & $-2.00\pm0.11$
     & $<-0.57$
\\

\hline
\hline

\label{tab abundances}
\end{longtable}
\vspace{-3mm}
\flushleft
\footnotesize{$^1$ Abundances [X/H$]\equiv\log [N($X$)/N($H$)]-\log [N($X$)/N($H$)]_\odot$
calculated using the Solar abundances given in Table~\ref{tab solar}, as
the sum of the column densities measured in individual components of the
profiles (Table~\ref{tab columns vel}). Letters appended to a QSO name refer to different absorption systems along the same line of sight. Associated errors are $1\sigma$ standard deviations.}
}

\end{document}